\documentclass[
 amsmath,amssymb,
prfluids,
longbibliography
]{revtex4-1}
\usepackage{color}
\usepackage{rotate}
\usepackage{graphicx}
\usepackage{dcolumn}
\usepackage{bm}
\usepackage{tikz}
\usepackage{pifont}

\newcommand\Nu{\text{Nu}}
\newcommand\Ra{\text{Ra}}
\newcommand\Ta{\text{Ta}}
\newcommand\Rey{\text{Re}}
\newcommand\Pran{\text{Pr}}

\newcommand\Ek{\text{Ek}}
\newcommand\Ro{\text{Ro}}
\newcommand\uu{{\bf u}}

\newcommand{\oo}{\color{magenta}}
\newcommand{\bb}{\color{black}}

\newcommand{\back}{\color{black}}

\definecolor{blue5}{RGB}{0, 13, 52}
\definecolor{blue4}{RGB}{0, 102, 155}
\definecolor{blue3}{RGB}{0, 137, 204} 
\definecolor{blue2}{RGB}{7, 170, 255}
\definecolor{blue1}{RGB}{130, 211, 255}
\definecolor{pink1}{RGB}{255, 117, 186}
\definecolor{pink2}{RGB}{255, 0, 124}
\definecolor{pink3}{RGB}{196, 0, 96} 
\definecolor{pink4}{RGB}{153, 0, 76}
\definecolor{pink5}{RGB}{52, 0, 13}
\definecolor{pink5}{RGB}{52, 0, 13}

\definecolor{gfblue4}{RGB}{0, 102, 155}
\definecolor{gfred4}{RGB}{153, 0, 76} 

\begin{document}

\title{Wall modes and the transition to bulk convection in rotating Rayleigh--B\'enard convection}

\author{Xuan Zhang$^{1}$}
\author{Philipp Reiter$^{1}$}
\author{Olga~Shishkina$^{1}$}
\author{Robert~E.~Ecke$^{1,2,3}$}

\affiliation{$^{1}$Max Planck Institute for Dynamics and Self-Organization, 37077 G\"ottingen, Germany}
\affiliation{$^{2}$Center for Nonlinear Studies, Los Alamos National Laboratory, Los Alamos, New Mexico 87545, USA}
\affiliation{$^{3}$Department of Physics, University of Washington, Seattle, WA 98195, USA}

\begin{abstract}

We investigate states of rapidly rotating Rayleigh--B\'enard convection in a cylindrical cell over a range of Rayleigh number $3\times10^5\leq \Ra \leq 5\times10^{9}$ and Ekman number $10^{-6} \leq \Ek \leq 10^{-4}$ for Prandtl number $\Pran = 0.8$ and aspect ratios $1/5 \leq \Gamma \leq 5$ using direct numerical simulations.  We characterize, for perfectly insulating sidewall boundary conditions, the first transition to convection via wall mode instability and the nonlinear growth and instability of the resulting wall mode states including a secondary transition to time dependence. We show how the radial structure of the vertical velocity $u_z$ and the temperature $T$ is captured well by the linear eigenfunctions of the wall mode instability where the radial width of $u_z$ is $\delta_{u_z} \sim \Ek^{1/3} r/H$ whereas $\delta_T \sim e^{-k r}$ ($k$ is the wavenumber of an laterally infinite wall mode state).  The disparity in spatial scales for $\Ek = 10^{-6}$ means that the heat transport is dominated by the radial structure of $u_z$ since $T$ varies slowly over the radial scale $\delta_{u_z}$.  We further describe how the transition to a state of bulk convection is influenced by the presence of the wall mode states.  We use temporal and spatial scales as measures of the local state of convection and the Nusselt number $\Nu$ as representative of global transport. 
Our results elucidate the evolution of the wall state of rotating convection and confirm that wall modes  are strongly linked with the boundary zonal flow (BZF) being the robust remnant of nonlinear wall mode states.  We also show how the heat transport ($\Nu$) contributions of wall modes and bulk modes are related and discuss approaches to disentangling their relative contributions. 

\end{abstract}

\maketitle

\date{\today}

\section{Introduction} \label{sec-Introduction}

Thermal convection driven by buoyancy and subject to rotation \cite{Kunnen2021,Ecke2023} is a phenomenon of great relevance in many physical disciplines, especially in geo- and astrophysics and also in engineering applications. In the Rayleigh--B\'enard (RBC) geometry a fluid with thermal expansion coefficient $\alpha$, kinematic viscosity $\nu$, and thermal diffusivity $\kappa$ is bounded above and below by isothermal boundaries separated by a height $H$ and heated from below so as to produce a temperature difference $\Delta$. 
In a gravitational field with acceleration $g$, the strength of thermal forcing is determined by 
the Rayleigh number $\Ra \equiv \alpha g \Delta H^3/(\kappa \nu)$, 
and the fluid type by the Prandtl number $\Pran = \nu/\kappa$.  
In the presence of rotation about an axis parallel to gravity, the influence of the Coriolis force is a balance of angular rotation rate $\Omega_d$ and viscous diffusion time $H^2/\nu$. 
There are many ways to represent this balance namely non-dimensional $\Omega \equiv \Omega_d H^2/\nu$ and 
\oo
the Taylor number $\Ta \equiv (2 \Omega)^{2}$.  
Another measure popular in geophysics that we use here is the Ekman number 
$\Ek \equiv 
\nu/\left ( 2 \Omega_D H^2 \right )$ 
\bb
that nicely captures the limit of rapid rotation as a small parameter, i.e., $\Ek \rightarrow 0$. 
A similar parameter that measures the balance of buoyancy and rotation is the convective Rossby number $\Ro \equiv \sqrt{\Ra/\left ( \Pran \Ta \right )} = \left (\Ra/\Pran \right )^{1/2} \Ek = \sqrt{g \alpha \Delta /H}/\left (2 \Omega_d \right )$ where the latter is the ratio of buoyancy frequency and rotation frequency. Note that $\Ro$ is independent of dissipation parameters.  Unless otherwise specified, we use length scale $H$, time scale $\sqrt{H/(g \alpha \Delta)}$, and temperature scale $\Delta$ for non-dimensional variables. 

In physical realizations of convection, there are solid horizontal boundaries, usually of cylindrical geometry with aspect ratio $\Gamma \equiv D/H$ ($D$ is the cylinder diameter), that have a finite thermal conductivity and diffusivity. 
In direct numerical simulations (DNS), these sidewall boundaries can be perfectly insulating or, less frequently, perfectly conducting.
Experimental boundaries are somewhere in between these limits and tend towards the insulating case so as to minimize heat transport through the sidewalls. We consider here the case of perfectly insulating sidewalls; results for perfectly conducting boundaries will be presented elsewhere.

The global response of the system is measured by the molecular-diffusion-normalized heat transport, Nusselt number $\Nu \equiv (\langle u_zT \rangle_{z}-\kappa\partial_z\langle {T}\rangle_{z})/(\kappa\Delta/H)$.
Here, $T$ denotes the temperature, $\bf{u}$ is the velocity field with component $u_z$ in the vertical direction, and $\langle \cdot\rangle_{z}$ denotes the average in time and over a horizontal cross-section at height $z$ from the bottom. 

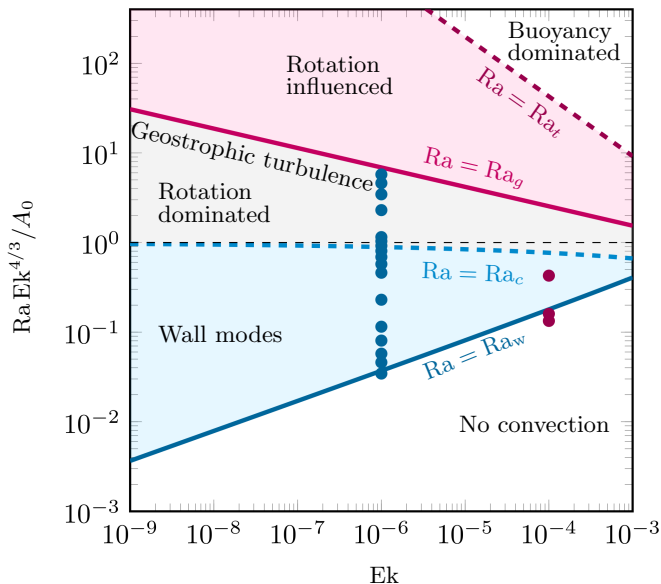
\begin{figure}[th]
\unitlength1truecm
\begin{picture}(18,8)
\put(5,0.5){\includegraphics[height=8cm]{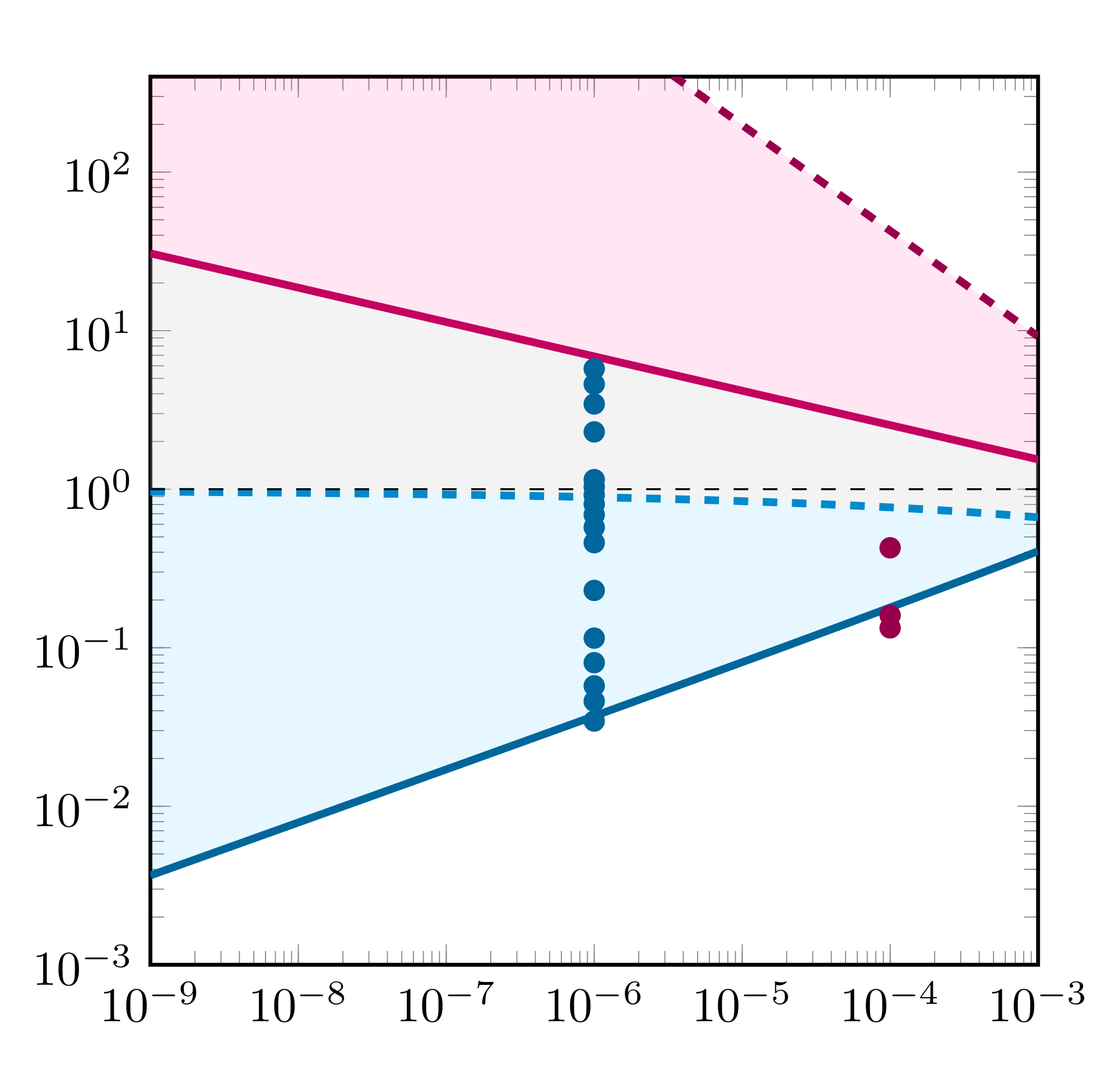}}
\put(4.5,3.5){\rotatebox{90}{ $\Ra\,\Ek^{4/3}/A_0$}}
\put(10.7,7){\rotatebox{-35}{\color{pink4}{$\Ra=\Ra_t$}}}
\put(10,5.9){\rotatebox{-13}{\color{pink3}{$\Ra=\Ra_g$}}}
\put(10,4.35){\rotatebox{-2}{\color{blue3}{$\Ra=\Ra_c$}}}
\put(10,3){\rotatebox{22}{\color{blue4}{$\Ra=\Ra_\text{w}$}}}
\put(6.1,6.25){\rotatebox{-13}{\color{black}{Geostrophic turbulence}}}
\put(9.2,0.3){ $\Ek$}
\put(11.15,7.55){Buoyancy}
\put(11.15,7.25){dominated}
\put(8.2,7.1){Rotation}
\put(8.2,6.8){influenced}
\put(6.5,5.4){Rotation}
\put(6.5,5.1){dominated}
\put(6.5,3.5){Wall modes}
\put(10.5,2.3){No convection}
\end{picture}
\vskip-5mm
\caption{Schematic phase diagram of states of rotating convection (perfectly insulating sidewall boundary conditions): $\Ra \Ek^{4/3}/A_0$ 
\oo 
(dashed black line) where $A_0=8.7$ is the asymptotic $\Ek \rightarrow 0$ value versus $\Ek$. 
The lines for $\Ra_\text{c}$ (dashed blue line) \cite{Chandrasekhar1961,Niiler1965,Homsy1971} and $\Ra_\text{w}$ (solid blue line) \cite{Herrmann1993} 
\bb 
are from linear stability analysis with idealized boundary conditions on laterally infinite domains.  
$\Ra_g$ marks the approximate boundary of the geostrophic, rotation-dominated region and $\Ra_\text {t}$ denotes the transition to buoyancy-dominated flow  \cite{Wedi2021,Kunnen2021,Ecke2023}.  
The precise details of this diagram depend on parameters including $\Pran$ and $\Gamma$ and are not included here. 
The data points show the parameter values reported in this paper.
}
\label{PhaseDiag}
\end{figure}

The multitude of states of rotating Rayleigh--B\'enard convection (RRBC) in a fluid layer bounded above and below by isothermal boundaries have been elucidated over many decades going back to early linear stability calculations \cite{Chandrasekhar1953, Chandrasekhar1961} and preliminary experiments \cite{Nakagawa1955} (see also \cite{Ecke2023}). 
They found that the onset of bulk convection was strongly suppressed by the rotation-induced Coriolis force which acts to inhibit the vertical motions associated with thermal convection via mechanisms associated with the Taylor--Proudman constraint on slow, inviscid flows \cite{Greenspan1968}. 
In particular, the critical Rayleigh number $\Ra_\text{c}$ was found to vary approximately as $\Ek^{-4/3}$ for rapid rotation. 
Whereas the form of convection without rotation is in the form of convection rolls with characteristic length $\lambda \approx 2H$ (for a roll pair), rotation induces vortical motions near onset \cite{Veronis1959} with a corresponding characteristic length $\lambda \sim \Ek^{1/3}$ \cite{Chandrasekhar1953}.  
Experimental heat transport measurements \cite{Rossby1969, Lucas1983, Pfotenhauer1984, Pfotenhauer1987} confirmed the suppression of the convective onset by rotation, noted that some manner of convection ensued prior to strong bulk flow of linear theory, and revealed that rotation could, over some ranges in $\Ra$ and $\Pran$, enhance heat transport over non-rotating turbulent RBC.  The nature of the ``subcritical instability'' identified in the heat transport measurements was partially explained by linear stability calculations of stationary wall-mode states \cite{Buell1983, Pfotenhauer1987} but simultaneous heat transport and shadowgraph visualization experiments \cite{Zhong1991, Zhong1993} together with theory \cite{Ecke1992, Goldstein1993} showed that these wall-mode states precess in the rotating frame and have 
\oo
precession frequency $\omega$ and integer (resulting from the cylindrically periodic azimuthal symmetry) mode number $m$.
\bb  
Further experiments \cite{Ning1993, Liu1997, Liu1999} and linear stability calculations \cite{Goldstein1993, Herrmann1993, Kuo1993} showed beautiful correspondence between theory and experiment --- here we denote $\Ra_\text{w} \sim \Ek^{-1}$ as the critical value for the transition from the diffusive conduction state to one of wall-mode convection.  
A phase diagram showing the boundaries of different convective states in RRBC is presented in Fig.\ \ref{PhaseDiag} where we plot $\Ra/\Ra_\text{c}(\Ek=0)$ versus $\Ek$.

The enhancement of heat transport ($\Nu$) in RRBC compared to its value in non-rotating RBC ($\Nu_0$) has been observed in many experimental studies 
\oo 
\cite{Rossby1969, Zhong1993, Liu1997, Zhang2009, King2009, Weiss2016, Cheng2015, Cheng2020} and DNS \cite{Julien1996, Kunnen2006, Stevens2010b, Horn2014, Kunnen2016}
\bb
and has been attributed to Ekman pumping from the thermal boundary layers. 
This enhancement occurs over a range of $\Ra$, $\Pran$ and $\Ek$, with states at low $\Pran$ or small $\Ek$ not showing an enhancement but rather a decrease in $\Nu$ for all $\Ra$ \cite{Lucas1983, Pfotenhauer1987, Niemela2010, Ecke2014, Horn2015, Wedi2021, Lu2021}.  
We denote the value $\Ra_{\text{t}}$ at fixed $\Ek$ as the transition from buoyancy-dominated flow to rotation-influenced behavior.

Another well-defined region of rapidly rotating convection is the geostrophic regime where the flow maintains geostrophic balance and is rotation-dominated.  
The importance of this regime to geophysical and astrophysical situations was articulated through the derivation \cite{Sprague2006} and simulation \cite{Julien2012a, Julien2012b, Stellmach2014} of a reduced set of equations valid in the limit $\Ek \rightarrow 0$ and $\Ra \rightarrow \infty$ such that the scaled variable $\widetilde{\Ra} \equiv \Ra \Ek^{4/3}$ remains finite. The evolution of states from the onset of bulk convection at  
$\widetilde {\Ra}_\text{c} \approx 8.7$
depends on $\Pran$ and includes cellular patterns, convective Taylor vortices, plume states, and geostrophic turbulence.  
The range $1 \leq \Ra/\Ra_\text{c} \lesssim 20$ was explored using these reduced equations and a prediction was made that in the geostrophic turbulence regime 
$\Nu \sim \widetilde\Ra^{3/2}$ where it is controlled by the interior 
\oo
quasi-geostrophic (QG)
\bb
flow rather than being boundary layer controlled, a result that was demonstrated in the simulations only for $\Pran = 1$ over a short range in $\Ra \Ek^{4/3}$.
The experimental and DNS test of these predicted states \cite{King2009, Ecke2014, Cheng2015, Cheng2020} and their $\Nu$ scalings is challenging \cite{Cheng2018} owing to the need for very small $\Ek < 10^{-6}$, large $\Ra$, and the experimental resolution to measure rather small deviations from $\Nu \sim 1$, i.e., over a range  $1 \leq \Nu \lesssim 20$. 

The flow structure of rotating convection for sufficiently small $\Ek \lesssim 0.02$ starts with wall modes that arise as a supercritical bifurcation from the no-flow base state \cite{Zhong1991, Ecke1992, Zhong1993, Kuo1993, Goldstein1993, Herrmann1993}.  These initial flow states persist to higher $\Ra$ and serve as the nonlinear base state for the onset of fully three-dimensional bulk convection.  
In the idealized case usually considered in DNS, the sidewall is perfectly adiabatic. In experiments, however, this is not the case. 
The details of the onset and evolution of wall modes depend on the ratio of thermal conductivity of the sidewalls to that of the non-convecting fluid.  
Often the sidewalls are acrylic or polycarbonate with thermal conductivity of order 0.2 W/(m$\cdot$K) compared to, for example, water with 0.6 W/(m$\cdot$K).  
These conditions yield small (order 10\%) modification of the asymptotic critical parameters for insulating sidewall boundary conditions \cite{Kuo1993} with higher Rayleigh number for the onset of the wall modes $\Ra_\text{w}$ (74.4 versus 63.6), lower procession frequency at the onset $\omega_{\kappa_\text c} = \omega_\text c H^2/\kappa = $ (57.7 versus 66.1), and lower wave number $k_\text{w}$ (5.55 versus 6.07). Here we consider the idealized case of perfectly insulating sidewalls. 

\subsection{Direct Numerical Simulations}\label{subsec-Direct}

We consider rotating Rayleigh--B\'enard convection (RRBC) in a cylindrical container with radius $R$ and height $H$ ($\Gamma\equiv 2R/H$).
The governing equations for RRBC in the Oberbeck--Boussinessq approximation include the continuity equation ${\bf \nabla}\cdot\uu=0$ and the momentum equation and heat equations:
\begin{eqnarray}
 \partial_t \uu + (\uu\cdot{\bf \nabla})\uu &=&- \nabla p+ \nu{\bf \nabla}^2 \uu  - 2\Omega_d{\bf e}_z \times \uu+\alpha(T-T_0)g{\bf e}_z,\label{eq:2}\\
 \partial_t T + (\uu\cdot{\bf \nabla}) T &=& \kappa{\bf \nabla}^2 T.\label{eq:3}
\end{eqnarray}
Here, $p$ denotes the reduced kinematic pressure, ${\bf e}_z$ is the unit vector pointing upward,  and $T_0\equiv(T_++T_-)/2$.
We do not include any centrifugal acceleration because, in laboratory realizations of RRBC, it is negligible compared to the gravitational acceleration, i.e., the  Froude number is very small, $Fr\equiv\Omega_d^2 R/g\ll1$.
We apply the standard boundary conditions for RBC which are no-slip for the velocity ($\uu=0$) at all walls, isothermal at the bottom plate ($T=T_+$ at $z=0$) and top plate ($T=T_-$ at $z=H$, where $T_-<T_+$) and  adiabatic ($\partial T/\partial{\bf n}=0$) at the sidewall for the temperature.
Here ${\bf n}$ denotes a unit vector orthogonal to the boundary surface.

Taking as the reference quantities $\tau=H/\sqrt{\alpha g \Delta H}$ for time, ${\Delta}$ for temperature, $H$ for length, $\sqrt{\alpha g \Delta H}$ for velocity, and $\alpha g \Delta H$ for the reduced kinematic pressure, from equations (\ref{eq:2})--(\ref{eq:3}) we obtain the following dimensionless governing equations 
\begin{eqnarray}
 \partial_t \uu + (\uu\cdot{\bf \nabla})\uu &=&- \nabla p+ ({\Pran/\Ra})^{1/2}{\bf \nabla}^2 \uu  - \Ro^{-1}{\bf e}_z\times\uu +T{\bf e}_z,\label{eq:22}\\
 \partial_t T + (\uu\cdot{\bf \nabla}) T &=& ({\Pran\Ra})^{-1/2}{\bf \nabla}^2 T.\label{eq:33}
\end{eqnarray}
The resulting set of equations  (\ref{eq:22})--(\ref{eq:33}) 
\oo 
together with ${\bf \nabla}\cdot\uu=0$
\bb
is solved numerically, using the direct numerical solver {\sc goldfish} \cite{Shishkina2015, Kooij2018} in its latest version \citep{Reiter2021a, Reiter2022}, which is optimized for efficient computation with massive parallelization.
The computational code uses a fourth-order finite-volume discretization on staggered grids and a third-order Runge--Kutta time integration scheme. 
The direct numerical solver {\sc goldfish} utilizes a two-dimensional pencil decomposition employing the {\sc 2DECOMP} library \cite{Li2010} and a {\sc hdf5} file management. 

Unless otherwise specified, all quantities are made dimensionless by $\tau$,  $H$, and $\Delta$ (velocity by $H/\tau$). The exception is the radial distance which we explicitly scale with the radius $R$, i.e., $r/R = (2/\Gamma)r/H$.

\subsection{Linear Stability Theory}\label{subsec-Linear}

One can establish using linear stability calculations the onset of RRBC of a laterally-infinite fluid layer  \cite{Chandrasekhar1961, Niiler1965, Homsy1971} and of  precessing wall modes \cite{Goldstein1993, Herrmann1993, Kuo1993, Zhang2009} that arise owing to the existence of sidewalls. Here we summarize the asymptotic results (see Appendix for empirical fits for larger $\Ek$) for
$\Pran\gtrsim0.68$ and $\Ek \rightarrow 0$.
For the bulk instability the onset Rayleigh number $\Ra_\text{c}$, critical wave number $k_ \text{c}$ and characteristic length $\lambda_\text{c}$ are:
\begin{eqnarray}
\Ra_\text{c} &\approx &
8.70\ \Ek^{-4/3}(1-1.11 \Ek^{1/6} + 0.153 \Ek^{1/3}),
\\
k_ \text{c} &\approx & 1.31\ \Ek^{-1/3} (1-0.554 \Ek^{1/6} - 0.345 \Ek^{1/3}),\\
\lambda_\text{c} &\approx & 4.80\ \Ek^{1/3} (1+0.554 \Ek^{1/6} + 0.345 \Ek^{1/3}),
\end{eqnarray}
where $k_\text{c}$ and $\lambda_\text{c}$ are normalized by $H^{-1}$ and $H$, respectively, and $\lambda_\text{c} k_ \text{c}=2\pi$.

For the precessing wall modes with perfectly insulating boundaries, one has for a planar wall (no curvature) and to second order in $\Ek$ (see also Appendix) \cite{Herrmann1993, Kuo1993, Zhang2009}:
\begin{eqnarray}
\Ra_\text{w} &\approx&\pi^2\sqrt{6\sqrt{3}}\ \Ek^{-1} + 46.5\ \Ek^{-2/3}\approx 31.8\ \Ek^{-1} + 46.5\ \Ek^{-2/3},\\
k_\text{w} &\approx& \pi \sqrt{2+\sqrt{3}} - 35.0\ \Ek^{1/3} \approx 6.07 - 35.0\ \Ek^{1/3} ,
\\
m_\text{w} &=& (\Gamma/2) k_\text{w},
\\
\omega_{\kappa_\text {w}} &\equiv& \omega_\text {w}  H^2/\kappa = \omega_{ff_\text{w}} \Ra^{1/2} \Pran^{1/2}\approx 2 \pi^2 \sqrt{6+3\sqrt{3}} -732\ \Ek^{1/3}\approx66.0 - 732\ \Ek^{1/3}, \\
\omega_{\text d_\text{w} } &\equiv&{\omega_{\text w}}/{\Omega}= 2 \omega_{\kappa_{\text w}} \Ek\,\Pran^{-1}\approx \Ek \,\Pran^{-1}(4\pi^2\sqrt{6+3\sqrt{3}}-1465\Ek^{1/3})
\approx
132.1\ \Ek\,\Pran^{-1} - 1465\ \Ek^{4/3}\,\Pran^{-1},
\end{eqnarray} 
where the spatial scale is $H$ and the precession frequency $\omega_\text{w}$ is made dimensionless with the time scale $H^2/\kappa$ for $\omega_{\kappa_\text {w}}$, $\sqrt{H/(g \alpha \Delta)}$ for $\omega_{ff_\text {w}}$, and $\Omega^{-1}$ for $\omega_{\kappa_\text {c}}$ and $\omega_{\text d_\text{w}}$, respectively. Recall that for realistic sidewall boundary conditions \cite{Kuo1993} these asymptotic values are modified by about 10\%; something that needs to be considered when comparing experimental and DNS results. For finite cylindrical geometries, especially for small $\Gamma$, the values also shift slightly with smaller $\Ra_\text{w}$ for 
$m=1, 2$ 
\cite{Goldstein1993,Herrmann1993} compared to their planar wall values.

The paper is organized to generally track the evolution of the RRBC state for $\Ek=10^{-6}$ and $\Pran=0.8$ over the range $3 \times 10^7 \leq \Ra \leq 5 \times 10^9$.  
\oo
In Section~\ref{sec-Wall}, we describe the properties of wall modes including (\ref{subsec-Effect}) effects of aspect ratio, (\ref{subsec-Wall}) their spatial and temporal structure in terms of linear state eigenfunctions, and (\ref{subsec-Evolution}) the evolution of the nonlinear state through a subcritical bifurcation. 
We also consider (\ref{subsec-Transition}) the transition to the bulk state and (\ref{subsec-Length}) the associated spatial and temporal properties of the coexisting wall mode and bulk mode states where we use the term boundary zonal flow (BZF) to label the wall-localized state in the presence of bulk flow.  
In subsection~\ref{subsec-Heat}, we focus on the heat transport ($\Nu$).  
We provide conclusions in Section~\ref{sec-Conclusion} and end with acknowledgements in Section~\ref{sec-Acknowledgement}.
The Appendix has details of the data presented here, some extra features of wall modes, and the empirical fits of critical parameters for larger $\Ek$.  
\bb  

\section{Wall modes}\label{sec-Wall}

The goal of this paper is to address the evolution of the states of rotating convection at small $\Ek$ from the onset of wall modes through the onset of bulk convection and up to the upper threshold of the geostrophic regime (Fig.\ \ref{PhaseDiag}). We begin with a description of pure wall modes for $\Pran = 0.8$,  $\Ek =10^{-6}$ and $10^{-4}$, and for a range of $1/5 \leq \Gamma \leq 5$ (details of these parameters are presented in Tables\ \ref{TABEk10m6} in the Appendix). 
Within this range we identify the critical onset $\Ra_\text{w}$, the mode number $m$ (and corresponding azimuthal wavenumber $k$), and different measures of temporal and spatial degrees of freedom. Owing to the previous lack of data for wall modes at small $\Ek$, we systematically consider aspects of the onset of wall modes with a particular emphasis on the effects of small $\Gamma < 1$.  
We then address the nonlinear evolution of the wall modes including a subcritical transition to time dependence and the mechanisms for that transition. We identify three distinct regions.  First, one has steady (in the precessing frame) wall mode states for $3 \times 10^7 \lesssim Ra \lesssim 3 \times 10^8$. 
For $4 \times 10^8 \lesssim Ra \leq 9 \times 10^8$, there is a subcritical bifurcation to time dependence followed by increasingly nonlinear states.  
The mechanism for this bifurcation is the lateral ejection of plumes originating near the azimuthal zero-crossings of temperature and $u_z$ and propagating into the interior, see also \cite{Favier2020}.  
Finally, one gets a transition at $\Ra \approx 10^9$ to bulk convection where the wall mode states are in a dynamic balance with the bulk states in the form of a BZF.  We characterize this later state in the range $10^9 \leq \Ra \leq 5 \times 10^9$.  Some aspects of these states emphasizing the connection between wall modes and BZF were reported earlier by us \cite{Ecke2022}. 

\subsection{Effect of aspect ratio $\Gamma$}\label{subsec-Effect}

\begin{figure}[th]
\unitlength1truecm
\begin{picture}(18,7)
\put(3,0.8){\includegraphics[height=6cm]{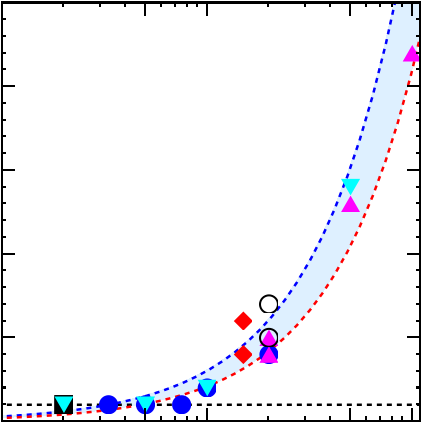}}
\put(10.5,3.25){\includegraphics[width=1cm]{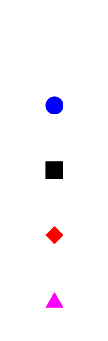}}
\put(10.5,1.8){\includegraphics[width=1cm]{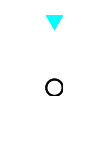}}
\put(11.5,5.3){$\Pran=0.8$ \cite{Zhang2020, Zhang2021}}
\put(11.5,4.7){$\Pran=5.2$ \cite{Wit2020}}
\put(11.5,4.1){$\Pran=1$  \cite{Favier2020}}
\put(11.5,3.5){$\Pran=6.4$ \cite{Ning1993, Zhong1993, Liu1999}}
\put(11.5,2.9){$\Pran=0.8$ \cite{Ecke2022}}
\put(11.5,2.3){$\Pran=6.4$ \cite{Horn2017}}
\put(2.85,0.4){0.1}
\put(4.85,0.4){0.5}
\put(5.85,0.4){1}
\put(6,0){$\Gamma$}
\put(7.9,0.4){5}
\put(8.65,0.4){10}
\put(2.55,6.7){25}
\put(2.55,5.5){20}
\put(2,3.6){$m$}
\put(2.55,4.3){15}
\put(2.55,3.1){10}
\put(2.7,1.925){5}
\put(2.7,0.75){0}
\put(7.3,1.2){$m=1$}
\end{picture}
\caption{
Azimuthal mode number $m$ vs. $\Gamma$. 
The shaded region between dashed lines corresponds to the dependence for $m(\Gamma)$ between the critical values for $\Ek=0$ (upper, asymptotic, 
\oo blue dashed curve)
\bb
and for $\Ek=10^{-5}$ (lower, 
\oo
red dashed curve).
\bb
Blue circles:  $\Pran=0.8$ \cite{Zhang2020, Zhang2021},
black squares:  $\Pran=5.2$ \cite{Wit2020},
red diamonds:  $\Pran=1$  \cite{Favier2020},
magenta upward triangles:  $\Pran=6.4$ \cite{Ning1993, Zhong1993, Liu1999},
cyan downward triangles:  $\Pran=0.8$ \cite{Ecke2022},
open circles: $\Pran=6.4$ \cite{Horn2017}.
}
\label{PICSM1}
\end{figure}
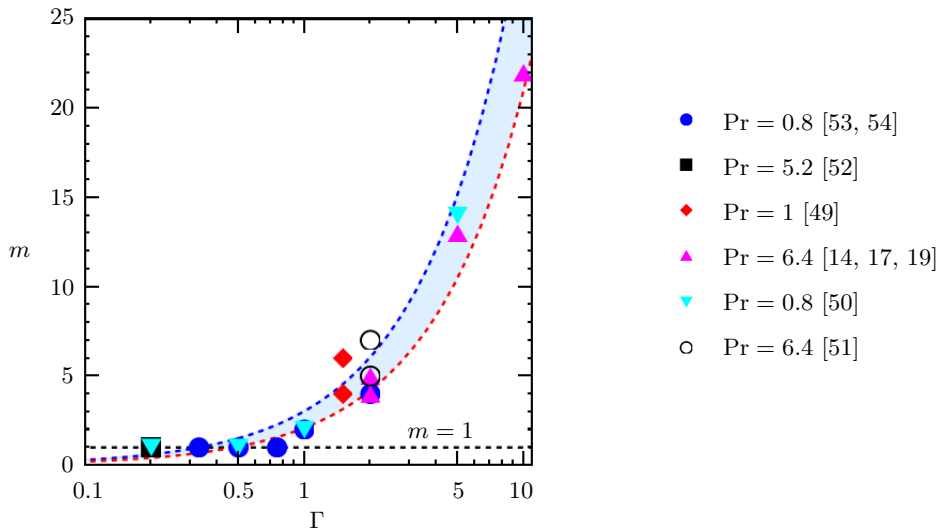

In Fig.~\ref{PICSM1}, we show $m$ for a number of data sets with different $\Gamma$ and $\Ek$ with corresponding values of $\Gamma$ of 2 \cite{Zhong1993}, 5 \cite{Ning1993}, 10 \cite{Liu1999}, 2 \cite{Horn2017}, 1/5 \cite{Wit2020}, 3/2 \cite{Favier2020}, and \{1/3, 1/2, 3/4, 1, 2\} \cite{Zhang2021}.  The first four examples are for $\Pran = 6.4$ and  primarily involve the properties of the wall-mode state. Most of the other values are for the BZF state where the wall localization coexists with a turbulent interior bulk mode.  
Because of the 
\oo
azimuthal periodicity,
\bb
$m$ is constrained to integer values so that for small $\Gamma < 1$, one has $m=1$.  
The BZF state has  systematically smaller values of $m \approx 2 \Gamma$ than the asymptotic ($\Ek \rightarrow 0$) condition $m_\text{w} \approx 3.03 \Gamma$ although the $\Ek$ correction in (1) is substantial in the range $10^{-6} \lesssim \Ek \lesssim 10^{-4}$, corresponding to 
$2.1 \lesssim m_\text{w} \lesssim 2.83$. 
(For $\Ek > 10^{-4}$ the expansion to second order in $\Ek$ is inaccurate, see Appendix).  In Fig.~\ref{PICSM1}, we show as a shaded region the predicted range of $m_\text{w}$ over the range $0 < \Ek < 10^{-5}$.  The trends and approximate range of $m$ are consistent with a strong connection between wall mode states and the BZF.  We next consider the onset of wall modes in more detail.

\begin{figure}[th]
\unitlength1truecm
\begin{picture}(18,7)
\put(2,0.8){\includegraphics[height=6cm]{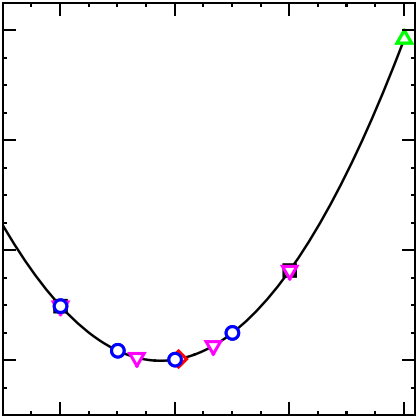}}
\put(10,0.8){\includegraphics[height=6cm]{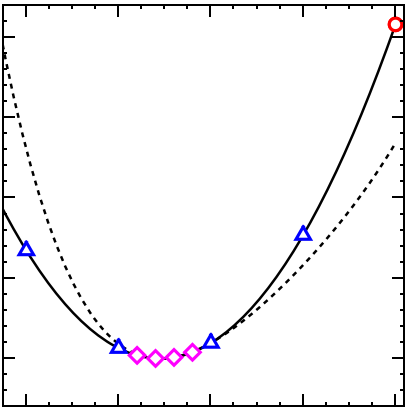}}
\put(0.75,3.3){\rotatebox{90}{$\Ra_\text{M}/\Ra_\text{w}$}}
\put(8.75,3.3){\rotatebox{90}{$\Ra_\text{M}/\Ra_\text{w}$}}
\put(1.4,6.3){1.6}
\put(1.4,4.7){1.4}
\put(1.4,3.1){1.2}
\put(1.4,1.55){1.0}
\put(7.65,0.4){10}
\put(5.2,0){$k$}
\put(4.2,6){$\Ek=10^{-6}$}
\put(6.9,6.3){\color{green}{$\left(\frac{1}{5},1\right)$}}
\put(2.7,5.5){$\Ra_\text{M}/\Ra_\text{w}=1+\xi_0^2(2m/\Gamma-k_\text{w})^2$}
\put(4.25,5){$\xi_0=0.18$}
\put(4.2,4.5){$k_\text{w}=5.75$}
\put(6.1,0.4){8}
\put(4.45,0.4){6}
\put(2.75,0.4){4}
\put(2.2,2){\color{blue}{$\left(2,4\right)$}}
\put(2.9,1.5){\color{blue}{$\left(2,5\right)$}}
\put(4,1.2){\color{blue}{(2,6)}}
\put(5.5,1.8){\color{blue}{(2,7)}}
\put(6.2,2.5){\color{blue}{(2,8)}}
\put(3,2.5){\color{magenta}{$\left(\frac{3}{2},3\right)$}}
\put(3.5,2.05){\color{magenta}{$\left(\frac{3}{2},4\right)$}}
\put(4.4,2.1){\color{magenta}{$\left(\frac{3}{2},5\right)$}}
\put(5.25,3){\color{magenta}{$\left(\frac{3}{2},6\right)$}}
\put(2.5,3){\color{black}{$\left(\frac{1}{2},1\right)$}}
\put(6.4,3){\color{black}{$\left(\frac{1}{2},2\right)$}}
\put(4.65,1.3){\color{red}{$\left(\frac{1}{3},1\right)$}}
\put(9.35,6.15){2.00}
\put(9.35,5){1.75}
\put(9.35,3.8){1.50}
\put(9.35,2.6){1.25}
\put(9.35,1.4){1.00}
\put(15.65,0.4){10}
\put(14.4,0.4){8}
\put(13.2,0){$k$}
\put(12.2,6){$\Ek=10^{-4}$}
\put(14.9,6.3){\color{red}{$\left(\frac{1}{5},1\right)$}}
\put(10.7,5.5){$\Ra_\text{M}/\Ra_\text{w}=1+\xi_0^2(2m/\Gamma-k_\text{w})^2$}
\put(12.25,5){$\xi_0=0.2$}
\put(12.2,4.5){$k_\text{w}=4.9$}
\put(13.02,0.4){6}
\put(11.65,0.4){4}
\put(10.3,0.4){2}
\put(11.5,1.1){\color{magenta}{(5,11)--(5,15)}}
\put(10.1,2.35){\color{blue}{(1,1)}}
\put(10.8,1.5){\color{blue}{(1,2)}}
\put(12.7,2.05){\color{blue}{(1,3)}}
\put(13.6,3.1){\color{blue}{(1,4)}}
\put(0.7,6.5){(a)}
\put(8.7,6.5){(b)}
\end{picture}
\caption{
$\Ra_\text{M}/\Ra_\text{w}$ versus $k$ 
\oo
where $\Ra_\text{M}$ is the value of $\Ra$ along the marginal stability curve given by
\bb
(solid black) $\Ra_\text{M}/\Ra_\text{w} = 1 + \xi_0^2\left (k-k_\text{w} \right )^2$ where, for periodic azimuthal conditions, one has discrete $k = 2 m/\Gamma$. 
(a) $\Ek = 10^{-6}$ with corresponding $k_\text{w} = 5.75$ and $\xi_0 = 0.18$ \cite{Kuo1993}, $\Pran \sim 1$ and data points ($\Gamma$, m) = (1/5,1), red; (1/3,1), green; (1/2, 1) and (1/2, 2), blue; (3/2,~3) to (3/2,~6), black \cite{Favier2020}; (2,~4) to (2,~8), magenta. 
(b) $\Ek = 10^{-4}$ with corresponding $k_\text{w} = 5.75$ and $\xi_0 = 0.18$ \cite{Kuo1993} and data points ($\Gamma$, m) = (1/5,~1), red; (1,~2) to (1,~4), blue; (5,~11) to (5,~15), magenta. 
Dashed black line \cite{Scheel2003} shows computed effects of higher order terms at $\Ek \approx 2 \times 10^{-3}$ and corresponding data points ($\Gamma$, m) = (1/5,1), red; (1, 1) and (1, 4), blue.
}
\label{RatRaMarg}
\end{figure}
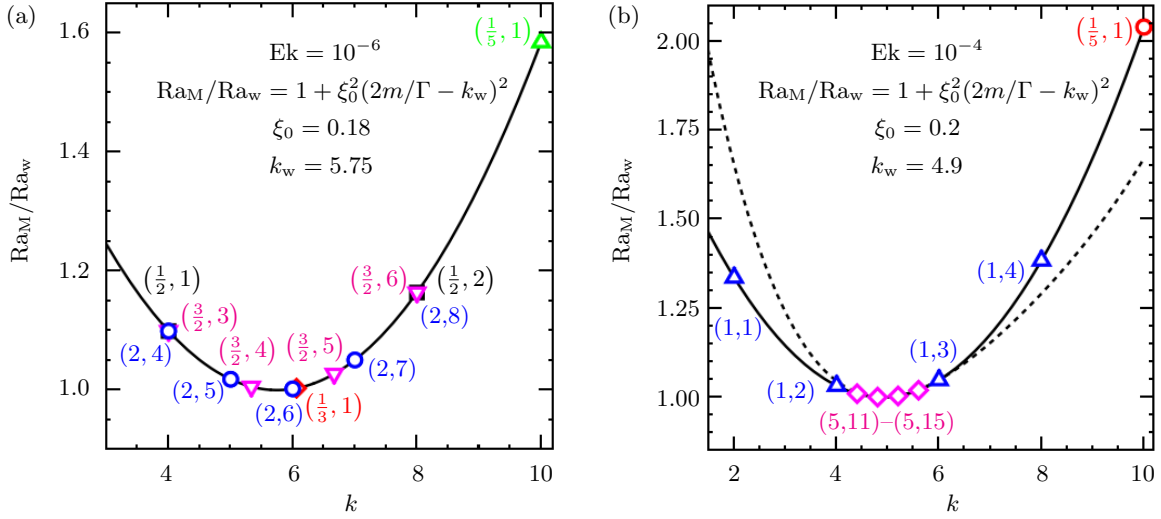

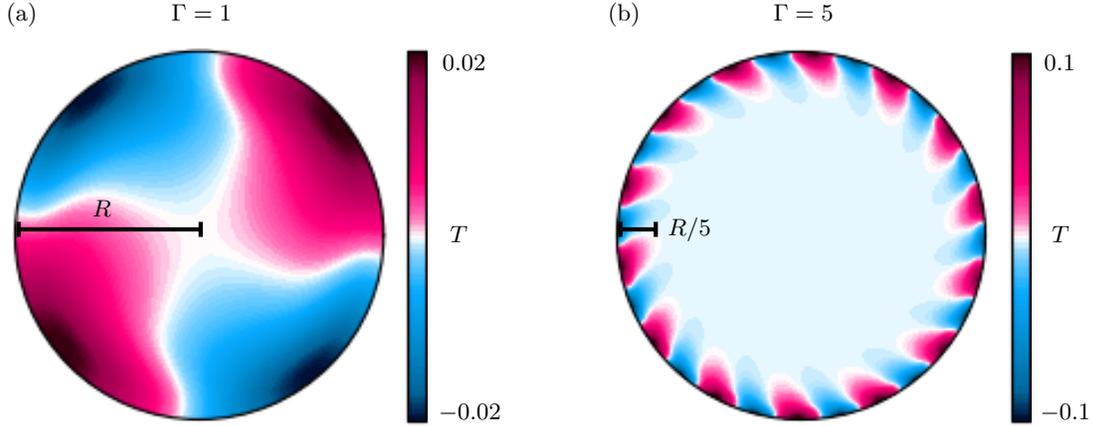
\begin{figure}[th]
\unitlength1truecm
\begin{picture}(18,6)
\put(1.5,2.5){
\begin{tikzpicture}[style=ultra thick]
\put(2.5,0.5){\includegraphics[height=5cm]{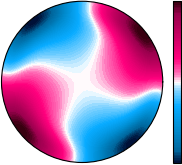}}
\put(10.5,0.5){\includegraphics[height=5cm]{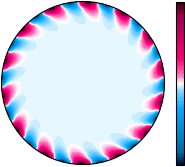}}
    \draw (2.55,3.1) -- (5,3.1) ;
    \draw (2.58,3.0) -- (2.58,3.2) ;
    \draw (5,3.0) -- (5, 3.2) ;
    \draw (10.55,3.1) -- (11.05,3.1) ;
    \draw (10.58,3.0) -- (10.58,3.2) ;
    \draw (11.05,3.0) -- (11.05,3.2) ;
    \end{tikzpicture}
}
\put(1.5,5.4){(a)}
\put(3.7,5.4){$\Gamma=1$}
\put(2.65,2.8){$R$}
\put(9.5,5.4){(b)}
\put(11.7,5.4){$\Gamma=5$}
\put(10.3,2.55){$R/5$}
\put(7.4,2.45){$T$}
\put(15.4,2.45){$T$}
\put(7.3,4.75){$0.02$}
\put(7.25,0.1){$-0.02$}
\put(23.4,2.45){$T$}
\put(15.3,4.75){$0.1$}
\put(15.25,0.1){$-0.1$}
\end{picture}
\caption{
Instantaneous temperature field at mid-height for $\Ek=10^{-4}$, $\Pran=0.8$. $(a)$ $\Gamma=1$, $\Ra=3\times10^5$, $\Nu \approx 1$, $m=2$, and $(b)$ $\Gamma=5$, $\Ra=3\times10^5$, $\Nu=1.03$, $m=14$.  Horizontal black bars show scale $R/\Gamma$. 
}
\label{PIC02}
\end{figure}

The critical $\Ra_\text{w}$ for $\Gamma = 1/5, 1, 5$ are determined from the DNS with $\Pran = 0.8$ at $\Ek = 10^{-4}$ and  $\Gamma = 1/2, 2$  at $\Ek = 10^{-6}$, see Table\ \ref{TAB1}. 
One can understand the dependence of $\Ra_\text{w}$ on $\Gamma$ by considering the discrete wavenumber $k_\text{m}$ of the corresponding mode number $m$ ($k_\text{m} = 2m/\Gamma$) which is selected owing to 
\oo
azimuthal periodicity
\bb
and comparing it to the critical wavenumber $k_\text{w}$ for a planar (flat) sidewall boundary.  The marginal stability curve, Fig.\ \ref{RatRaMarg}, 
\oo
is defined by the $k$ dependence of the onset $\Ra$ designated as the marginal stability Rayleigh Number $\Ra_\text{M} (k)$. For small differences $k-k_\text{w}$ the dependence of $\Ra_\text{M}$ is quadratic,
\bb
i.e., $\epsilon_\text{M}  \equiv \Ra_\text{M}(m)/\Ra_\text{w} - 1 = \xi_0^2 \left (k -k_\text{w} \right )^2$ \cite{Ning1993, Liu1999, Kuo1993}. 
The quadratic approximation is valid only for small $k - k_\text{w}$ and $\epsilon$ whereas for larger values there are terms higher order in 
$k - k_\text{w}$ 
(see, e.g., \cite{Scheel2003}). Nevertheless, this representation gives a reasonable semi-quantitative description of the variation of $\Ra_\text{M}$ for different parameter values.
\begin{table}[h]
\begin{center}
\begin{tabular}[t]{lccccc}
\toprule
 $\Ek$ 	& $\Gamma$ &  	    $\Ra_\text{w}$ 	&   $\Ro_\text{w}$	&  $m$	\\
\hline
$10^{-6}$	&	1/2	&	$2.8 \times 10^7$ 	& 	0.07 		&  1	\\
$10^{-6}$	&	2	&	$3. \times 10^7$ 	& 	0.07 		&  3	\\
$10^{-4}$	&	1/5	&	$8.3 \times 10^5$ 	&      	0.10		&  1	\\
		&	1 	&  	$3. \times 10^5$ 	&	0.03 		&  2	\\
		&	5	&  	$3.3 \times 10^5$ 	& 	0.06 		& 14	\\
\hline
 \end{tabular}
\caption{Data from DNS with $\Pran = 0.8$ showing $\Ek$, $\Gamma$, $\Ra_\text{w}$, $\Ro_\text{w}$, and $m$.}
\label{TAB1}
\end{center}
\end{table}
The difference  $k -k_\text{w}$ is particularly important for small $\Gamma$ as we will see below. The coefficient $\xi_0^2$ was evaluated for experimentally realized \cite{Zhong1993,Ning1993} sidewall thermal conductivity as a function of $\Ek$ \cite{Kuo1993} (see Appendix).  In Fig.\ \ref{RatRaMarg}(a), the curve is evaluated for $\Ek = 10^{-6}$ with $k_\text{w} = 5.75$ and $\xi_0 = 0.18$, and the labeled data points are the discrete values $k_\text{m}$ for $\Gamma$ = 1/5, 1/3, 1/2, 3/2 \cite{Favier2020},  and 2  which have the lowest values of $\Ra_{M}/\Ra_\text{w}$. One can see the large and unintuitive dependence on $\Gamma \lesssim 1/2$ where $m =1$: for $\Gamma = 1/5$, $\Ra_{M}/\Ra_\text{w} = 1.6$ whereas $\Gamma = 1/3$ yields $\Ra_{M}/\Ra_\text{w} = 1.003$ and $\Gamma=1/2$ has $\Ra_{M}/\Ra_\text{w} = 1.1$. 
The increase in $\Ra_\text{M}/\Ra_\text{w}$ is non-monotonic in $\Gamma$ and depends sensitively on $k_\text{w}$.  Another observation concerns the predicted lowest value for $\Gamma = 3/2$ which is $m =4$ whereas $m = 6$ was observed in DNS \cite{Favier2020} upon increasing $\Ra$ but $m=4$ when decreasing $\Ra$.  
This would be consistent with the $m=6$ mode resulting from starting the DNS for $\Ra \gtrsim \Ra_\text{w}$ from random initial conditions at $\Ra = 0$ so that the state would evolve to an $m \neq m_\text{w}$ where $m_\text{w}$ is the value corresponding to the smallest $k_\text{m} - k_\text{w}$. 
The effect of a sudden increase in $\Ra$ was used in experiments on wall modes \cite{Zhong1993, Ning1993, Liu1999} to access a wide range of available $m$.    
In Fig.\ \ref{RatRaMarg}(b), we consider $\Ek = 10^{-4}$ where $k_\text{w} = 4.9$ and  $\xi_0 = 0.2$ for $\Gamma = 1/5, 1, 5$.  
Here the shift in $\Ra_{M}/\Ra_\text{w}$ is even larger for $\Gamma = 1/5$ than for $\Ek = 10^{-6}$ but as $\Gamma$ increases the density of discrete $k_\text{m}$ increases as $\Delta k \equiv k_\text{m+1} - k_\text{m}  = 2/\Gamma$ so that they increasingly cluster near $\Ra_\text{M}/\Ra_\text{w} \approx 1$. For example, for $\Gamma = 5$, we observe modes $m=14, 15$ compared to the predicted values with the lowest $\Ra_{M}$, $m= 12,13$ (a small shift in $k_\text{w}$ would account for that difference). More importantly, we again appear to find that the selected $m$ can be different than the one closest to critical $m_\text{w}$ owing to starting from random initial conditions; $m=14$ near onset whereas $m=15$ above onset and similarly for $\Gamma = 1$, $m=2$ near onset and $m=3$ above onset as shown in Fig.\ \ref{PIC02}. When we slowly raise $\Ra$ from near onset, the selected $m$ near onset is preserved.

\begin{figure}[th]
\unitlength1truecm
\begin{picture}(18,6)
\put(3,0.5){\includegraphics[height=5cm]{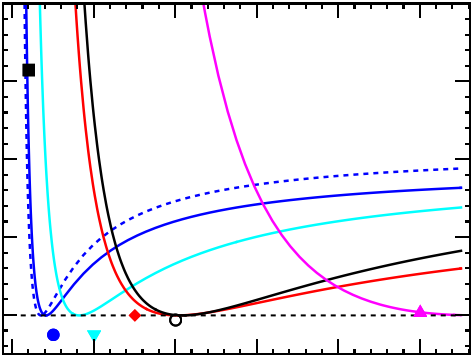}}
\put(10.5,1.2){\includegraphics[height=3.25cm]{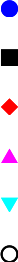}}
\put(1.75,2.5){\rotatebox{90}{$\Ra_\text{M}/\Ra_\text{w}$}}
\put(2.4,5.325){3.0}
\put(2.4,4.3){2.5}
\put(2.4,3.2){2.0}
\put(2.4,2.1){1.5}
\put(2.4,1.0){1.0}
\put(8.8,0.1){5}
\put(6.5,-0.3){$\Gamma$}
\put(7.65,0.1){4}
\put(6.5,0.1){3}
\put(5.35,0.1){2}
\put(4.25,0.1){1}
\put(3.1,0.1){0}
\put(11,4.2){$\Ek=10^{-6}$, $\Pran=0.8$}
\put(11,3.6){$\Ek=10^{-4}$, $\Pran=0.8$}
\put(11,3.0){$\Ek=10^{-6}$, $\Pran=1$ \cite{Favier2020}}
\put(11,2.4){$\Ek=10^{-4}$, $\Pran=0.8$}
\put(11,1.8){$\Ek=10^{-4}$, $\Pran=0.8$}
\put(11,1.2){$\Ek = 2 \times 10^{-3}$, $\Pran=6.7$ \cite{Goldstein1993}}
\put(1,5.5){(a)}
\put(6.3,4.5){\rotatebox{-75}{\color{magenta}$m=14$}}
\put(4.5,4.5){\rotatebox{-85}{\color{black}$m=6$}}
\put(3.8,4.5){\rotatebox{-85}{\color{red}$m=4$}}
\put(5.5,2.15){\rotatebox{10}{\color{blue}$m=1$}}
\put(5.6,1.6){\rotatebox{13}{\color{cyan}$m=2$}}
\end{picture}
\begin{picture}(18,6.5)
\put(3,0.8){\includegraphics[height=5cm]{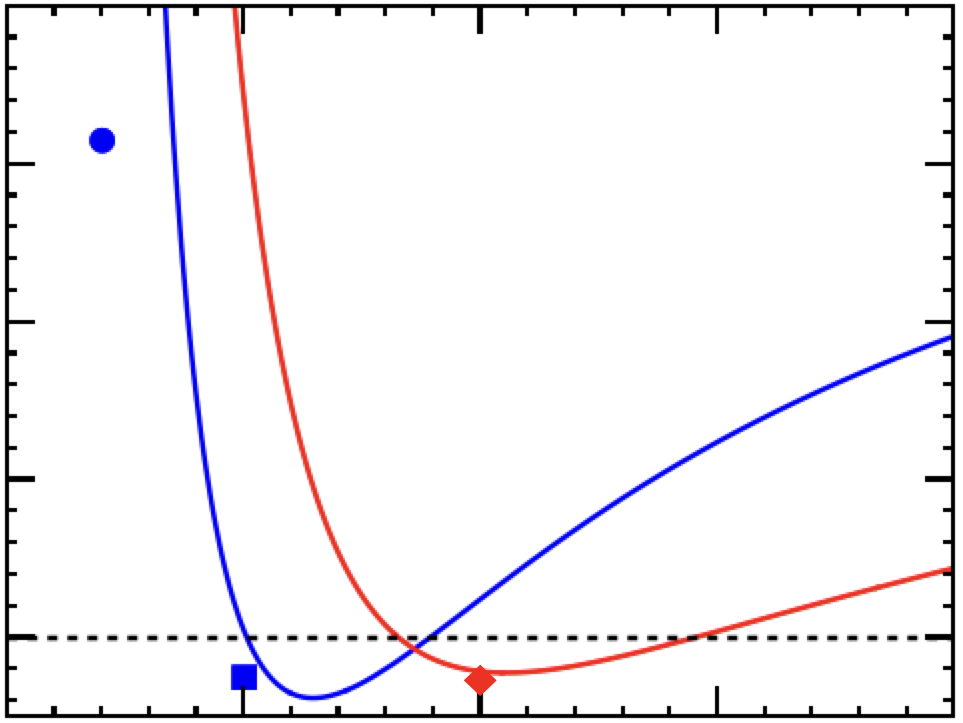}}
\put(10.5,2.65){\includegraphics[height=1.4cm]{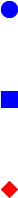}}
\put(1.75,2.5){\rotatebox{90}{$\Ra_\text{M}/\Ra_\text{w}$}}
\put(1,5.5){(b)}
\put(2.4,5.625){3.0}
\put(2.4,4.6){2.5}
\put(2.4,3.5){2.0}
\put(2.4,2.4){1.5}
\put(2.4,1.3){1.0}
\put(6.3,0){$\Gamma$}
\put(3,0.4){0}
\put(4.5,0.4){0.5}
\put(6.15,0.4){1.0}
\put(7.8,0.4){1.5}
\put(9.45,0.4){2.0}
\put(11,3.8){$\Ek=10^{-4}$, $\Pran=0.8$}
\put(11,3.2){$\Ek=10^{-6}$, $\Pran=0.8$}
\put(11,2.6){$\Ek=10^{-4}$, $\Pran=0.8$}
\put(6.7,2.3){\rotatebox{30}{\color{blue}$m=1$}}
\put(7.8,1.5){\rotatebox{20}{\color{red}$m=1$}}
\end{picture}
\caption{$\Ra_\text{M}/\Ra_\text{w}$ versus $\Gamma$. (a) The lines are calculations using values for $m$, $k_\text{w} (\Ek)$, and $\xi_0 (\Ek)$. Different $m =1, 2, 4, 6, 14$ are shown with corresponding colors (blue, red, orange, magenta, black); the dashed blue curve is evaluated for $\Ek = 10^{-6}$ whereas the solid blue is for $\Ek = 10^{-4}$. The data points are: solid circle: $\Ek = 10^{-4}$, $\Pran=0.8$; solid square: $\Ek = 10^{-6}$, $\Pran=0.8$; triangle: $\Ek = 10^{-6}$, $\Pran=1$ \cite{Favier2020}; diamond: $\Ek = 2 \times 10^{-3}$, $\Pran=6.7$ \cite{Goldstein1993}.  (b) Dotted lines show calculations \cite{Goldstein1993} for $m=1, 2$ taking into account the finite curvature of the cylindrical container for specific $\Gamma$.}
\label{RawGamma}
\end{figure}

To directly evaluate the observed and predicted onset values 
$\Ra_\text{M}$, we plot in 
Fig.\ \ref{RawGamma}(a),  $\Ra_\text{M}/\Ra_\text{w} = \epsilon_\text{M}+1$ for different observed values of $m$ at corresponding $\Gamma$.  The dramatic effect for $\Gamma = 1/5$ where $m=1$ noted above is well captured by the marginal stability model with an observed factor of 2.6 in the onset $\Ra_\text{M}$ compared to a predicted ratio of about 2.1.  For $\Gamma = 1/2, 1$, there is actually a decrease relative to $\Ra_\text{w}$ which is captured in calculations for finite $\Gamma$ \cite{Goldstein1993}, see Fig.\  \ref{RawGamma}(b).  As $\Gamma$ is further increased, the observed $m$ values yield $k \approx k_\text{w}$ so there is little or no shift in the onset $\Ra_\text{M}$. Given that $\epsilon_\text{M}$ is not precisely quadratic except for small $\epsilon$, $\xi_0$ is computed for imperfectly insulating sidewall 
boundary conditions, and  the computations for $\Ra_\text{w}$ and $k_\text{w}$ do not include rigid top/bottom 
boundary conditions,
the agreement is quite satisfactory.

\subsection{Wall mode spatial and temporal structure}\label{subsec-Wall}

We now turn to the spatial and temporal structure of the wall mode states. In previous work \cite{Zhang2020,Zhang2021,Ecke2022}, we considered the radial boundary length scales of wall modes and BZF states defined by $\delta_{u_z} =\max{ \langle u_z^2 \rangle_{\phi,t}^{1/2}}$ and the zero crossing $\delta_0$ of $\langle u_\phi \rangle_{\phi,t}$,
i.e., $\delta_0$ is the distance to the sidewall where $\langle u_\phi \rangle_{\phi,t}=0$, 
where $\langle\cdot\rangle_{\phi,t}$ means averaging in the azimuthal direction and in time. As we discuss below, $\delta_0$ may not be a good measure of radial extent owing to the midplane symmetry of $u_\phi$. 
Here we consider characteristics of the stationary and subcritical wall modes as represented in the fluid fields: temperature $T$, vertical velocity $u_z$, azimuthal velocity $u_\phi$, radial velocity $u_r$, and vertical vorticity $\omega_z$.   

\begin{figure}[th]
\unitlength1truecm
\begin{picture}(18,6)
\put(1.5,1.5){
\begin{tikzpicture}[style=ultra thick]
\put(1,0.5){\includegraphics[height=4.55cm]{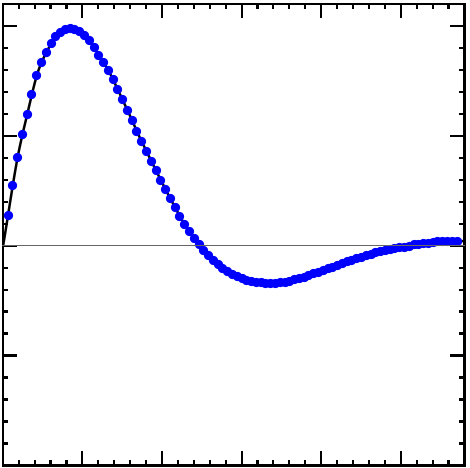}}
\put(7,0.5){\includegraphics[height=4.5cm]{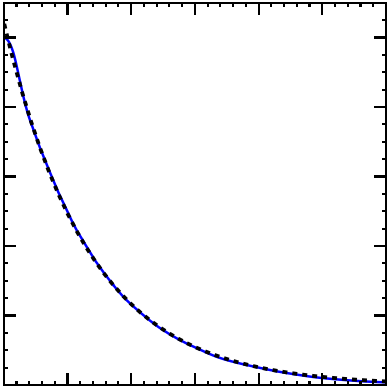}}
\put(8.7,2.2){\includegraphics[height=2.5cm]{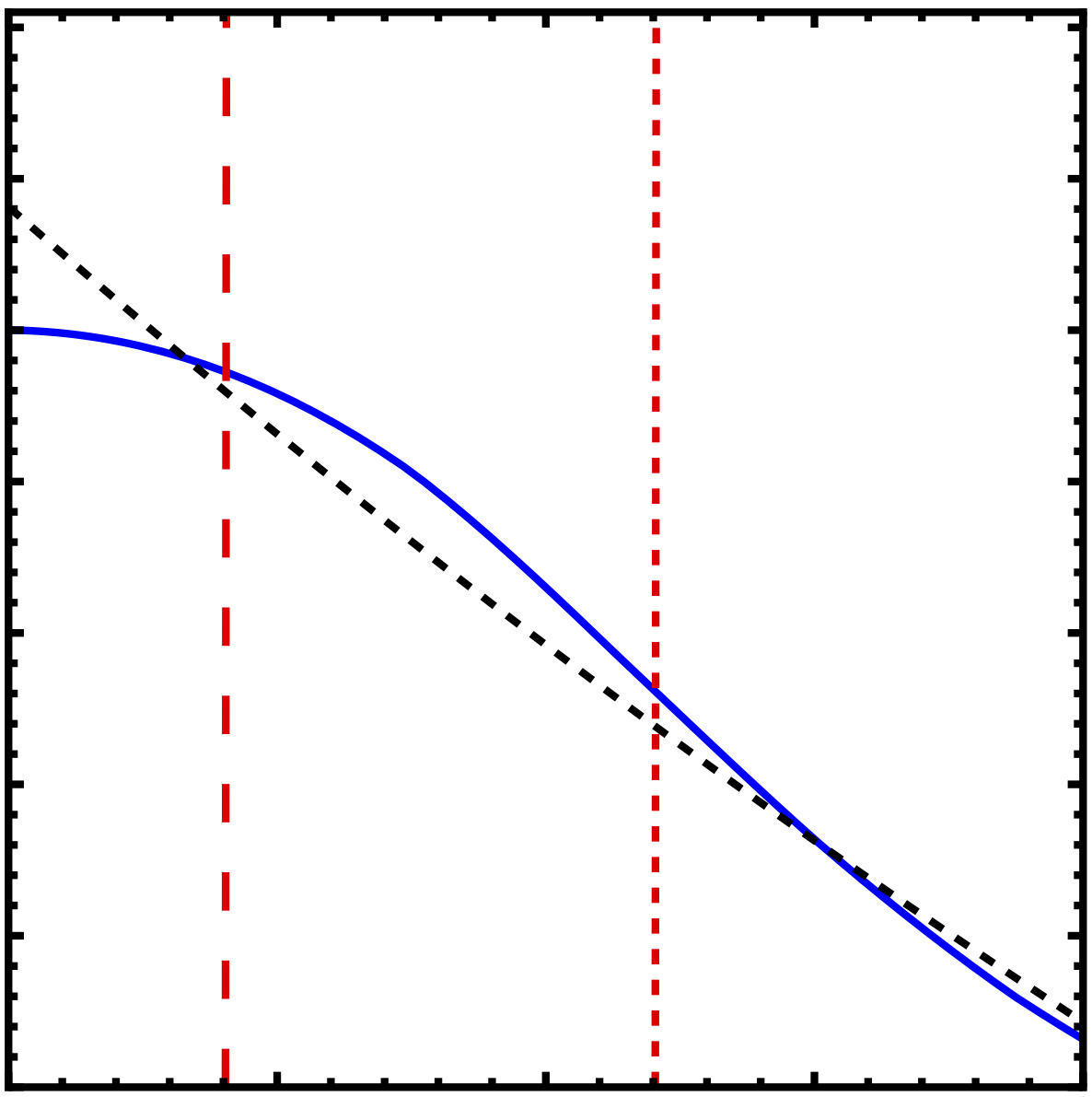}}
\put(13,0.5){\includegraphics[height=4.5cm]{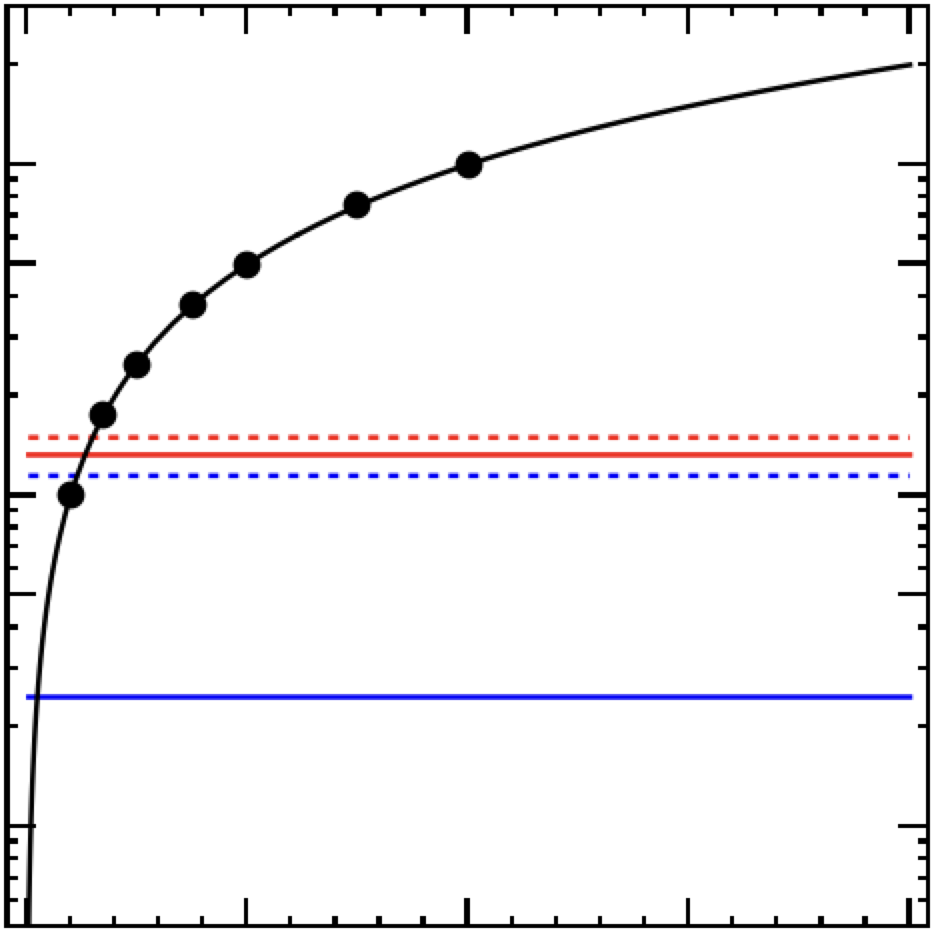}}
    \draw [-stealth](1.7,3.5) -- (1.7,4.5) ;
    \draw [stealth-](2.95,2.4) -- (2.95,1.4) ;
    \draw [stealth-](3.6,2) -- (3.6,1) ;
    \end{tikzpicture}
}
\put(0.2,5.6){(a)}
\put(6.2,5.6){(b)}
\put(12.2,5.6){(c)}
\put(0,3.1){$u_z$}
\put(0.7,5.2){1}
\put(0.5,4.15){$0.5$}
\put(0.7,3.05){0}
\put(0.2,2){$-0.5$}
\put(0.5,1){$-1$}
\put(0.9,0.7){0}
\put(1.65,0.7){1}
\put(2.425,0.7){2}
\put(3.2,0.7){3}
\put(4,0.7){4}
\put(4.75,0.7){5}
\put(2.4,0.3){$\Ek^{-1/3}x$}
\put(1.4,3.8){$\delta_{u_{z_p}}$}
\put(2.6,1.5){$\delta_{u_{z}}$}
\put(3.3,1.3){$\delta_{u_{z_\text{m}}}$}
\put(6.4,5){1.0}
\put(6.4,4.2){0.8}
\put(6.4,3.4){0.6}
\put(6.4,2.6){0.4}
\put(6.4,1.8){0.2}
\put(6.4,1){0.0}
\put(6.7,0.75){0.0}
\put(7.5,0.75){0.1}
\put(8.25,0.75){0.2}
\put(8.95,0.75){0.3}
\put(9.7,0.75){0.4}
\put(10.5,0.75){0.5}
\put(11.3,0.75){0.6}
\put(6,3.1){$T$}
\put(8.8,0.3){$x$}
\put(8.6,2.5){\scriptsize 0}
\put(9.0,2.5){\scriptsize 0.01}
\put(10.2,2.5){\scriptsize 0.03}
\put(8.2,5.1){\scriptsize 1.1}
\put(8.2,4.4){\scriptsize 1.0}
\put(8.2,3.71){\scriptsize 0.9}
\put(8.2,3.05){\scriptsize 0.8}
\put(14.05,0.75){1}
\put(15.15,0.75){2}
\put(16.2,0.75){3}
\put(17.25,0.75){4}
\put(15.15,0.3){$\Gamma$}
\put(12.7,4.75){1}
\put(12.45,4.25){0.5}
\put(12,3.1){$L$}
\put(12.45,3.15){0.1}
\put(12.3,2.6){0.05}
\put(12.3,1.5){0.01}
\put(13.2,4.2){\rotatebox{25}{$L_R=R=\Gamma/2$}}
\put(15,3.7){\color{red}{$\delta_T$}}
\put(15,2.4){\color{blue}{$\delta_{u_z}$}}
\put(14.5,1.4){$\Ek=10^{-6}$}
\end{picture}
\caption{
\oo
Linear eigenfunctions (discretized from data in Figures 4 and 5 of \cite{Herrmann1993} for $\Ek = 10^{-6}$ and $\Pran=7$): 
(a) Eigenfunction of $u_z$  versus $\Ek^{-1/3} x$, where $x$ reflects the distance from a flat sidewall. 
Characteristic lengths are indicated by the location of the peak $\delta_{u_{z_p}}$, the first zero crossing $\delta_{u_{z}}$, and the minimum $\delta_{u_{z_\text{m}}}$. 
(b) Eigenfunction of $T$ versus $x$.
The dashed line is an exponential fit to the data $1.05 e^{-x/0.13}$ that fits extremely well over the whole range except very near the sidewall on the order of the $u_z$ radial width (see Inset where short red (long) dashed vertical line is $r_{\max}$ ($r_0$) of $u_z$). 
(c) The geometric length scale $L_R = R = \Gamma/2$ is plotted versus values of $\Gamma$ considered (solid black circles) and compared to the temperature length scale $\delta_T$ and the vertical velocity length scale $\delta_{u_{z}}$ evaluated at $\Ek=10^{-4}$ (dashed blue) and $\Ek=10^{-6}$ (solid blue). 
\bb
}
\label{uzTEigen}
\end{figure}
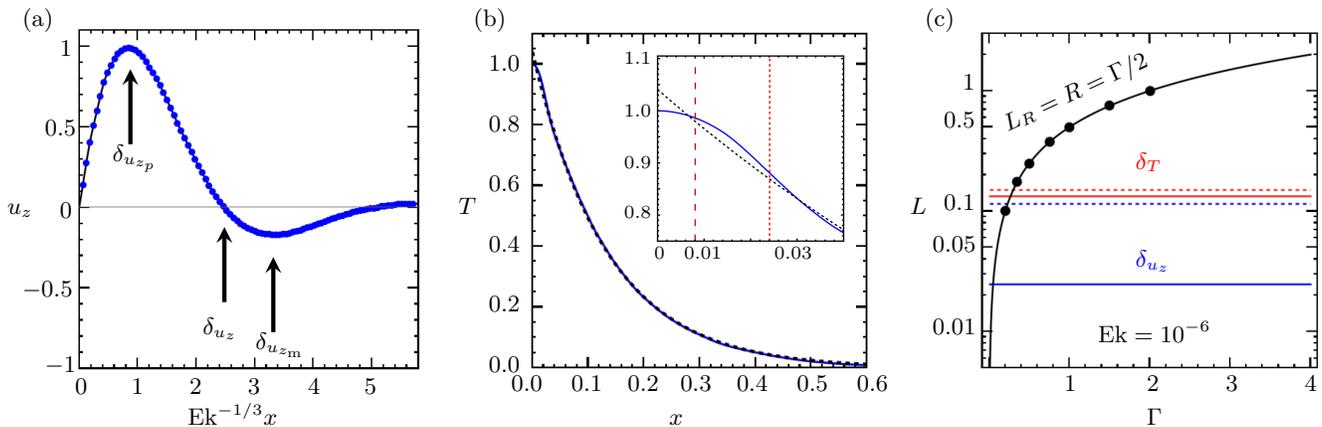

One feature of wall modes that is perhaps not appreciated is the decoupling of the radial length scales of the temperature field $T$ from the vertical velocity $u_z$ and vertical vorticity $\omega_z$ fields with decreasing $\Ek$ \cite{Herrmann1993}.  In Figs.\ \ref{uzTEigen}(a,b), we show, respectively, the radial eigenfunctions of $u_z(r)$ and $T(r)$ (interpolated from \cite{Herrmann1993} for $\Ek=10^{-6}$) where a Fourier--Bessel expansion is an excellent representation of $u(z)$ (anticipating the results in a cylindrical cell) and $T$ is well fit by an exponential function. 
The inset of Fig.\ \ref{uzTEigen}(b) shows that the necessary condition at the wall, $\partial T/\partial r (r=R) =0$, relaxes quickly with $r$ on length scales of order $\delta_{u_z}$.  The decoupling of length scales results from the $T$ length scale $\delta_\text{T} \sim k_\text{w}^{-1}$ 
($1/e$ decay length) 
being dependent on $\Ek$ only through its weak dependence $k_\text{w} (\Ek)$ whereas the $u_z$ length scale (first zero crossing) $ {\delta_{u_z} \sim \Ek^{1/3}}$ has an explicit $\Ek$ scaling. 
This separation of scales has important implications for RRBC in small aspect ratio containers with rapid rotation and for understanding the heat transport properties of wall modes and the BZF.  

In Fig.\ \ref{uzTEigen}(c), we consider the different length scales associated with the eigenfunctions of $u_z$ and $T$.  We take the geometric length scale $L_\text{R}$ of the container as $R$ so that $L_\text{R} = \Gamma/2$ and compare with the length scales $\delta_\text{T}$, and $\delta_{{u_z}_0}$ evaluated at $\Ek = 10^{-6}$ and $10^{-4}$. One obtains $\delta_\text{T} \approx 0.13$ (consistent with \cite{Herrmann1993} which predicts scaling of order  $k_\text{w}^{-1} = 0.17$), and $\delta_{u_z} = 2.5 \Ek^{1/3}$ which yields values of 0.12 for $\Ek=10^{-4}$ and 0.025 for $\Ek=10^{-6}$.  Thus, for $\Ek = 10^{-4}$, $\delta_T \approx \delta_{u_z}$ whereas for $\Ek = 10^{-6}$, $\delta_\text{T} \approx 5 \delta_{u_z}$. 
For $\Gamma \approx 1/3$, $\delta_\text{T}\approx L_R$.
Thus, the temperature field is highly constrained by the geometry for small $\Gamma \lesssim 1$. On the other hand, for $\Ek=10^{-6}$,  $\delta_{{u_z}_0} \ll \delta_\text{T}$ and only becomes of order $R$ for very small $\Gamma \sim 1/20$.  Thus, $u_z$ should not be affected by geometry and should have a form in cylindrical geometry similar to that of a planar boundary.  

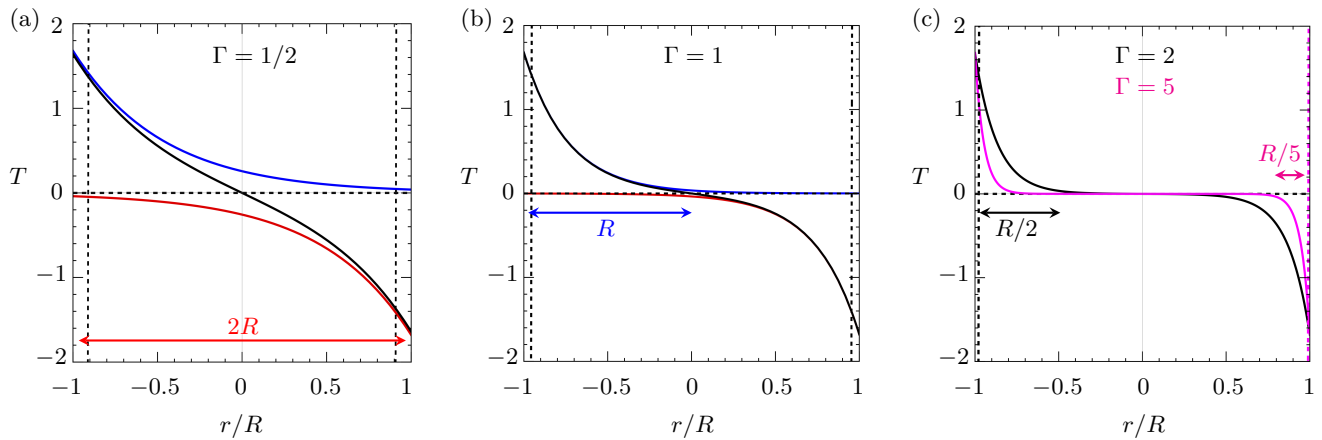
\begin{figure}[th]
\unitlength1truecm
\begin{picture}(18,6)
\put(1,1.5){
\begin{tikzpicture}[style=thick]
\put(1,0.5){\includegraphics[height=4.55cm]{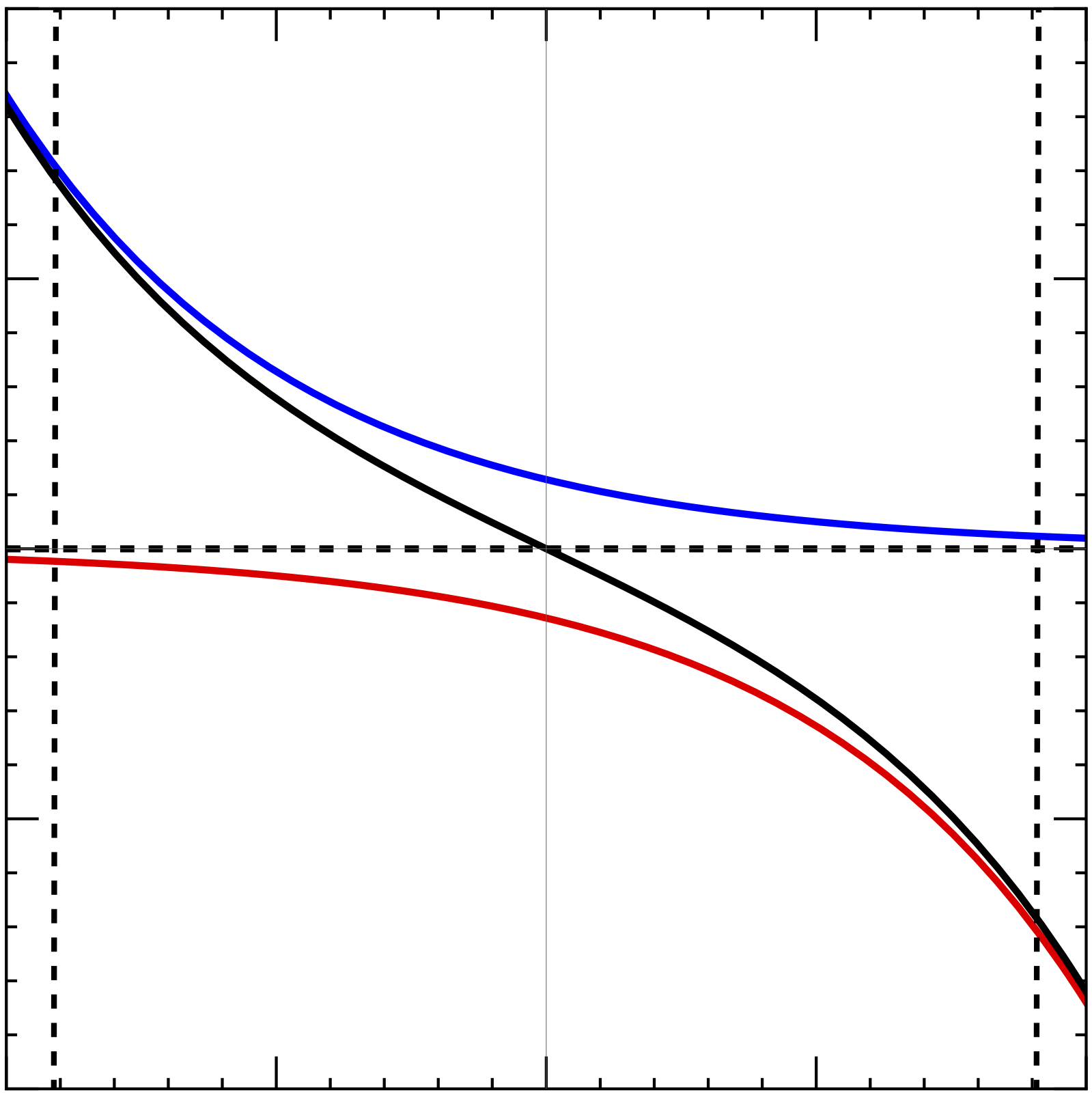}}
\put(7,0.5){\includegraphics[height=4.5cm]{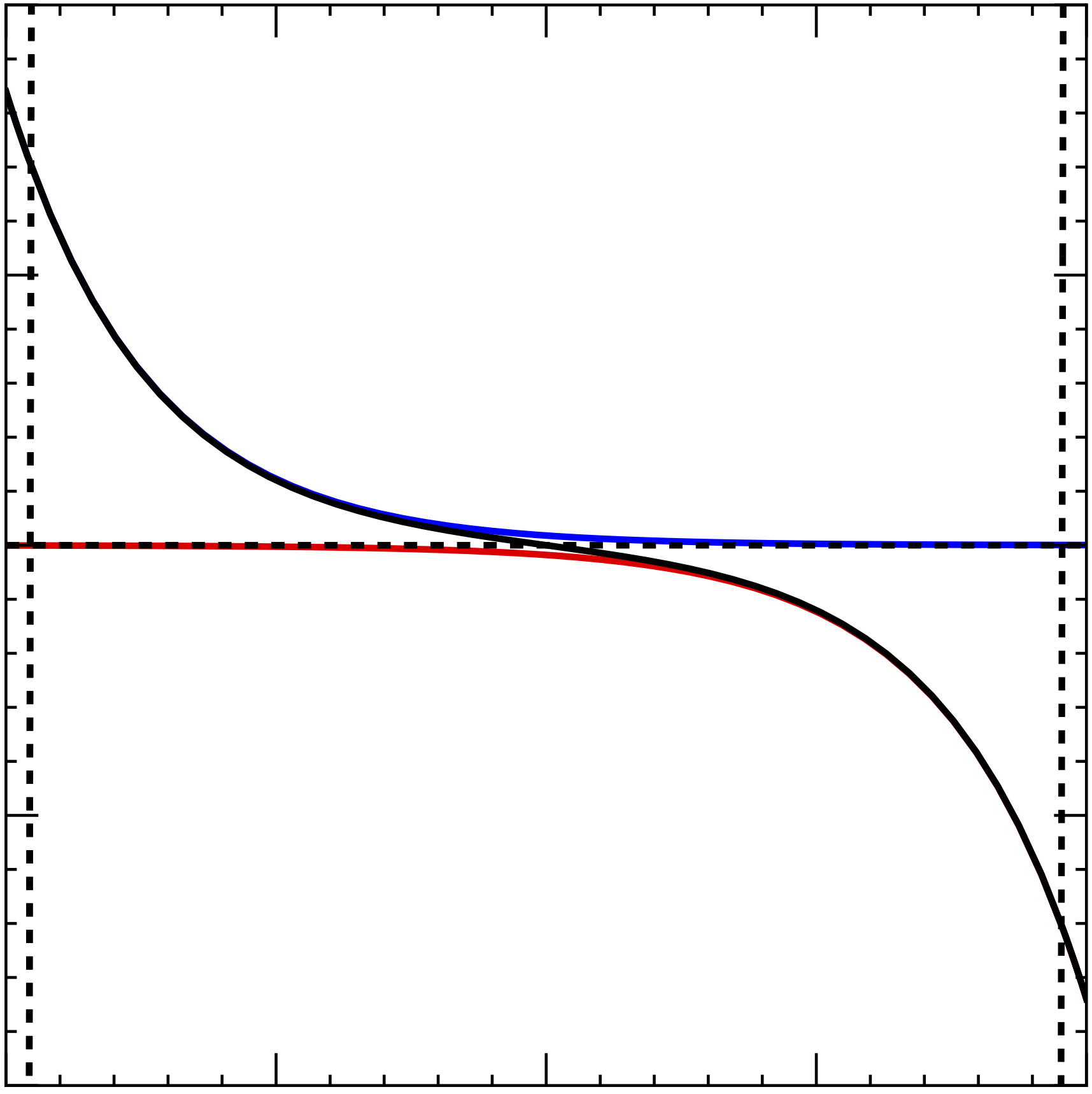}}
\put(13,0.5){\includegraphics[height=4.5cm]{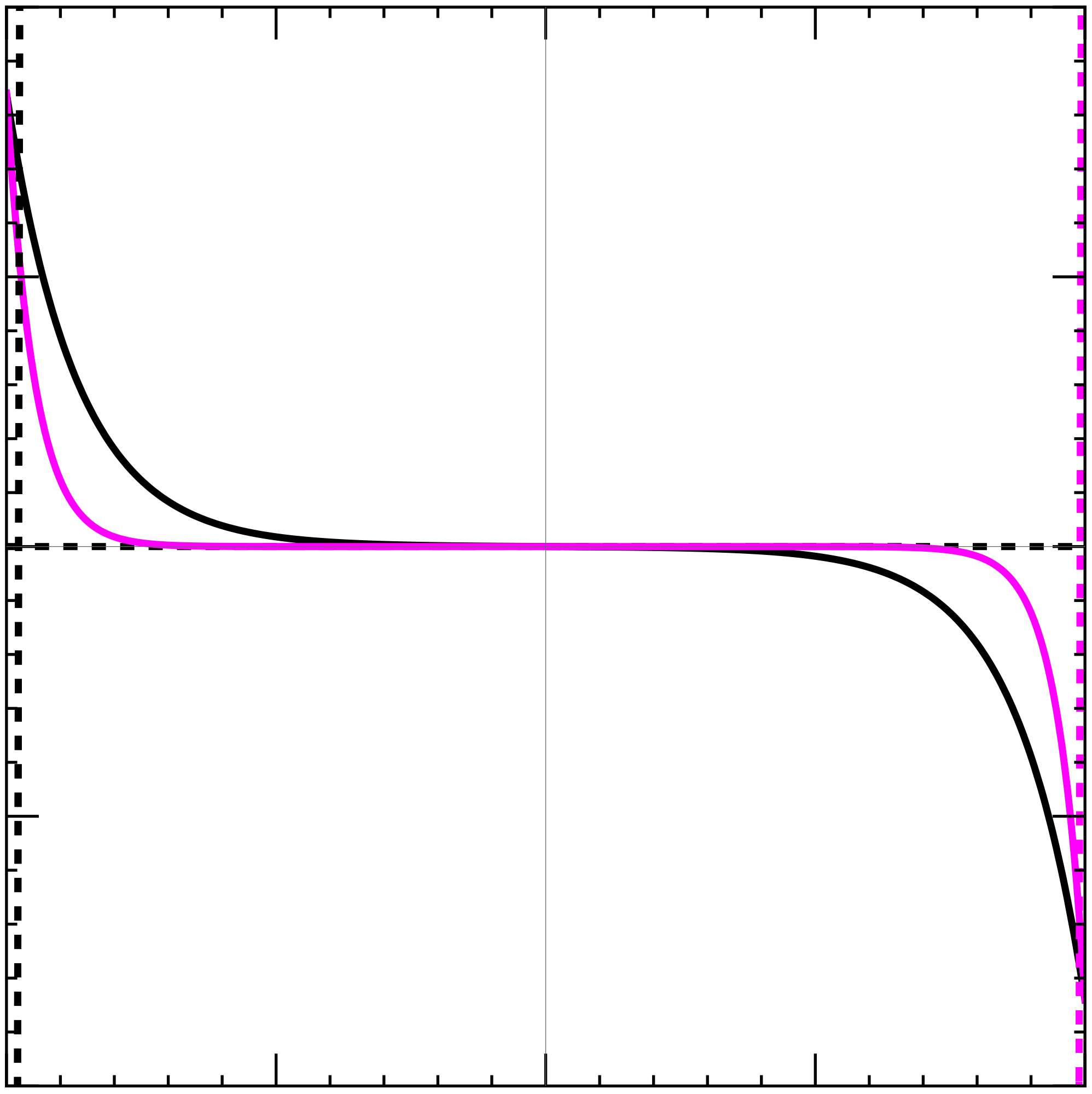}}
    \draw [stealth-stealth, color=red](1.1,0.8) -- (5.45,0.8) ;
    \draw [stealth-stealth, color=blue](7.075,2.5) -- (9.25,2.5) ;
    \draw [stealth-stealth](13.075,2.5) -- (14.15,2.5) ;
    \draw [stealth-stealth, color=magenta](17,3) -- (17.4,3) ;
    \end{tikzpicture}
}
\put(0.2,5.7){(a)}
\put(6.2,5.7){(b)}
\put(12.2,5.7){(c)}
\put(0.2,3.6){$T$}
\put(0.75,5.6){2}
\put(0.75,4.5){1}
\put(0.75,3.4){0}
\put(0.55,2.3){$-1$}
\put(0.55,1.2){$-2$}
\put(0.75,0.8){$-1$}
\put(1.8,0.8){$-0.5$}
\put(3.2,0.8){0}
\put(4.2,0.8){0.5}
\put(5.4,0.8){1}
\put(3,0.3){$r/R$}
\put(6.2,3.6){$T$}
\put(6.75,5.6){2}
\put(6.75,4.5){1}
\put(6.75,3.4){0}
\put(6.55,2.3){$-1$}
\put(6.55,1.2){$-2$}
\put(6.75,0.8){$-1$}
\put(7.8,0.8){$-0.5$}
\put(9.2,0.8){0}
\put(10.2,0.8){0.5}
\put(11.4,0.8){1}
\put(9,0.3){$r/R$}
\put(12.2,3.6){$T$}
\put(12.75,5.6){2}
\put(12.75,4.5){1}
\put(12.75,3.4){0}
\put(12.55,2.3){$-1$}
\put(12.55,1.2){$-2$}
\put(12.75,0.8){$-1$}
\put(13.8,0.8){$-0.5$}
\put(15.2,0.8){0}
\put(16.2,0.8){0.5}
\put(17.4,0.8){1}
\put(15,0.3){$r/R$}
\put(3.1,1.6){\color{red}{$2R$}}
\put(8,2.9){\color{blue}{$R$}}
\put(13.3,2.9){\color{black}{$R/2$}}
\put(16.8,3.9){\color{magenta}{$R/5$}}
\put(2.9,5.2){$\Gamma=1/2$}
\put(8.9,5.2){$\Gamma=1$}
\put(14.9,5.2){$\Gamma=2$}
\put(14.9,4.8){\color{magenta}{$\Gamma=5$}}
\end{picture}
\caption{Positive (red) and negative (blue) linear eigenfunctions of temperature $T$ as in Fig.~\ref{uzTEigen}b and their sum (black) for planar wall modes at $\Ek = 10^{-6}$ and $\Pran = 7$ \cite{Herrmann1993} plotted to 
\oo 
approximate a radial profile at mid-height in a confined cylindrical geometry with an odd $m$ so that $T(r)$ is radially asymmetric about $r=0$. 
\bb
(a) $\Gamma=1/2$, (b) $\Gamma=1$, (c) $\Gamma=2$ and $\Gamma=5$ (magenta).
}
\label{ThetaEigenGamma}
\end{figure}

We expand on the disparity in length scales by considering in Figures \ref{ThetaEigenGamma}(a-c), the wall-mode linear temperature eigenfunction of a planar wall mode state with perfectly insulating sidewall 
boundary conditions
plotted versus $r/R$ for different $\Gamma$: (a) 1/2, (b) 1, and (c) 2, 5. The positive and negative representations corresponding to opposite sides of the cylinder and the sum of the two as an approximation of the effects of finite geometry (see also \citep{Goldstein1993}) are illustrated.  For $\Gamma = 1/2$, there is considerable overlap of the two functions.  For $\Gamma =1$, one has just about achieved separation into individual contributions but one needs $\Gamma \gtrsim 2$ for a clean separation of the thermal field.  For comparison, we have plotted vertical dashed lines showing $\delta_{u_z}$.

\begin{figure}[th]
\unitlength1truecm
\begin{picture}(18,8.5)
\put(14,3.05){
\begin{tikzpicture}[style=thick]
\put(1,8.5){\includegraphics[height=6cm]{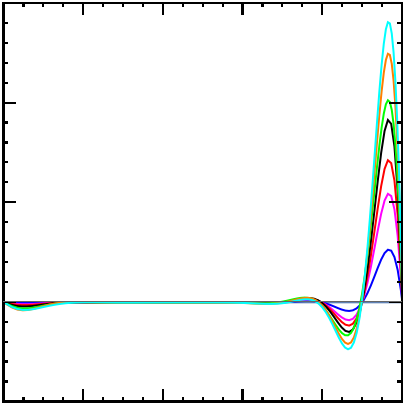}}
\put(2.8,11.2){\includegraphics[height=3cm]{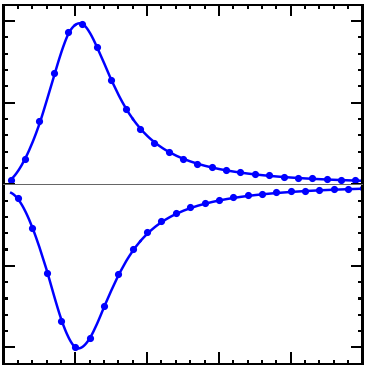}}
\put(9,8.5){\includegraphics[height=6cm]{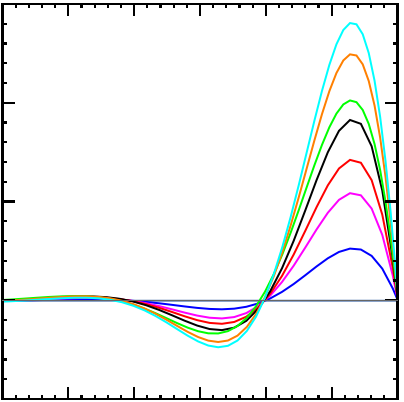}}
\put(10.5,11.7){\includegraphics[height=2.5cm]{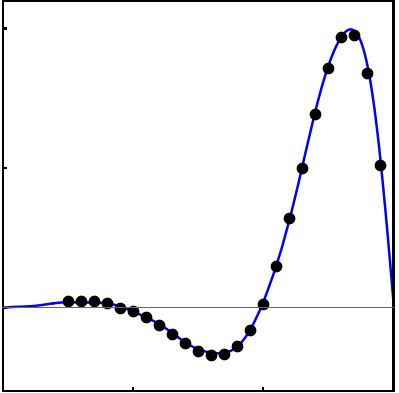}}
   \draw [stealth-stealth, color=black](13,9.6) -- (14.9,9.6) ;
\end{tikzpicture}
}
\put(1,-6.5){
\put(-0.2,14.8){(a)}
\put(4,13.7){\scriptsize Fourier--Bessel}
\put(4.3,13.4){\scriptsize coefficients}
\put(2.5,13.95){\scriptsize 1.0}
\put(2.5,13.3){\scriptsize $0.5$}
\put(2.5,12.65){\scriptsize $0.0$}
\put(2,12.65){$c_n$}
\put(2.3,11.975){\scriptsize $-0.5$}
\put(2.3,11.3){\scriptsize $-1.0$}
\put(2.8,10.9){\scriptsize 0.0}
\put(3.35,10.9){\scriptsize 0.2}
\put(3.95,10.9){\scriptsize 0.4}
\put(4.45,10.5){$n$}
\put(4.55,10.9){\scriptsize 0.6}
\put(5.15,10.9){\scriptsize 0.8}
\put(5.7,10.9){\scriptsize 1.0}
\put(0.5,14.3){0.15}
\put(0.5,12.9){0.10}
\put(0.5,11.4){0.05}
\put(-0.2,11.4){$u_z$}
\put(0.5,9.9){0.00}
\put(0.25,8.5){$-0.05$}
\put(0.95,8.1){0.0}
\put(2.15,8.1){0.2}
\put(3.35,8.1){0.4}
\put(3.9,7.6){$r/R$}
\put(4.5,8.1){0.6}
\put(5.7,8.1){0.8}
\put(6.85,8.1){1.0}

\put(7.8,14.8){(b)}
\put(8.5,14.3){0.15}
\put(8.5,12.9){0.10}
\put(8.5,11.4){0.05}
\put(7.8,11.4){$u_z$}
\put(8.5,9.9){0.00}
\put(8.25,8.5){$-0.05$}
\put(9.,8.1){0.7}
\put(10.93,8.1){0.8}
\put(12.9,8.1){0.9}
\put(12.1,7.6){$r/R$}
\put(14.9,8.1){1.0}
\put(10.6,11.35){\scriptsize 0.7}
\put(11.4,11.35){\scriptsize 0.8}
\put(11.7,10.95){$r/R$}
\put(12.2,11.35){\scriptsize 0.9}
\put(13,11.35){\scriptsize 1.0}
\put(10.2,12.15){\scriptsize 0.0}
\put(10.2,13.05){\scriptsize 0.5}
\put(9.7,13.05){$\widetilde u_z$}
\put(10.2,13.95){\scriptsize 1.0}
\put(14.1,9.2){$\delta_{u_z}$}
\put(13.8,10.1){\color{blue}{$3\times10^7$}}
\put(14.3,11.2){\color{magenta}{$4$}}
\put(14.3,11.8){\color{red}{$5$}}
\put(14.3,12.3){\color{black}{$7$}}
\put(14.2,13.1){\color{green}{$10$}}
\put(14.2,13.75){\color{orange}{$20$}}
\put(14.5,14.1){\color{cyan}{$30$}}
}
\end{picture}
\unitlength0.75truecm
\begin{picture}(18,7.5)
\put(2.5,2){
\begin{tikzpicture}[style=thick]
\put(1.4,0.6){\includegraphics[height=4.5cm]{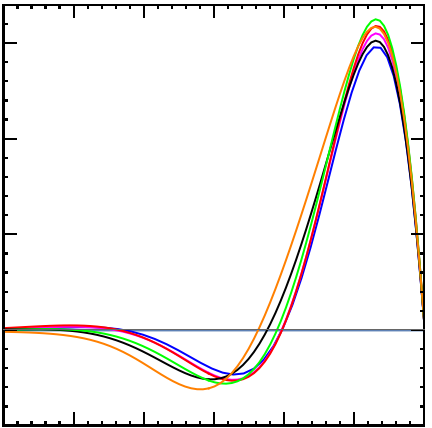}}
\put(1.9,3.3){\includegraphics[height=1.9cm]{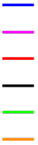}}
\put(9.45,0.6){\includegraphics[height=4.5cm]{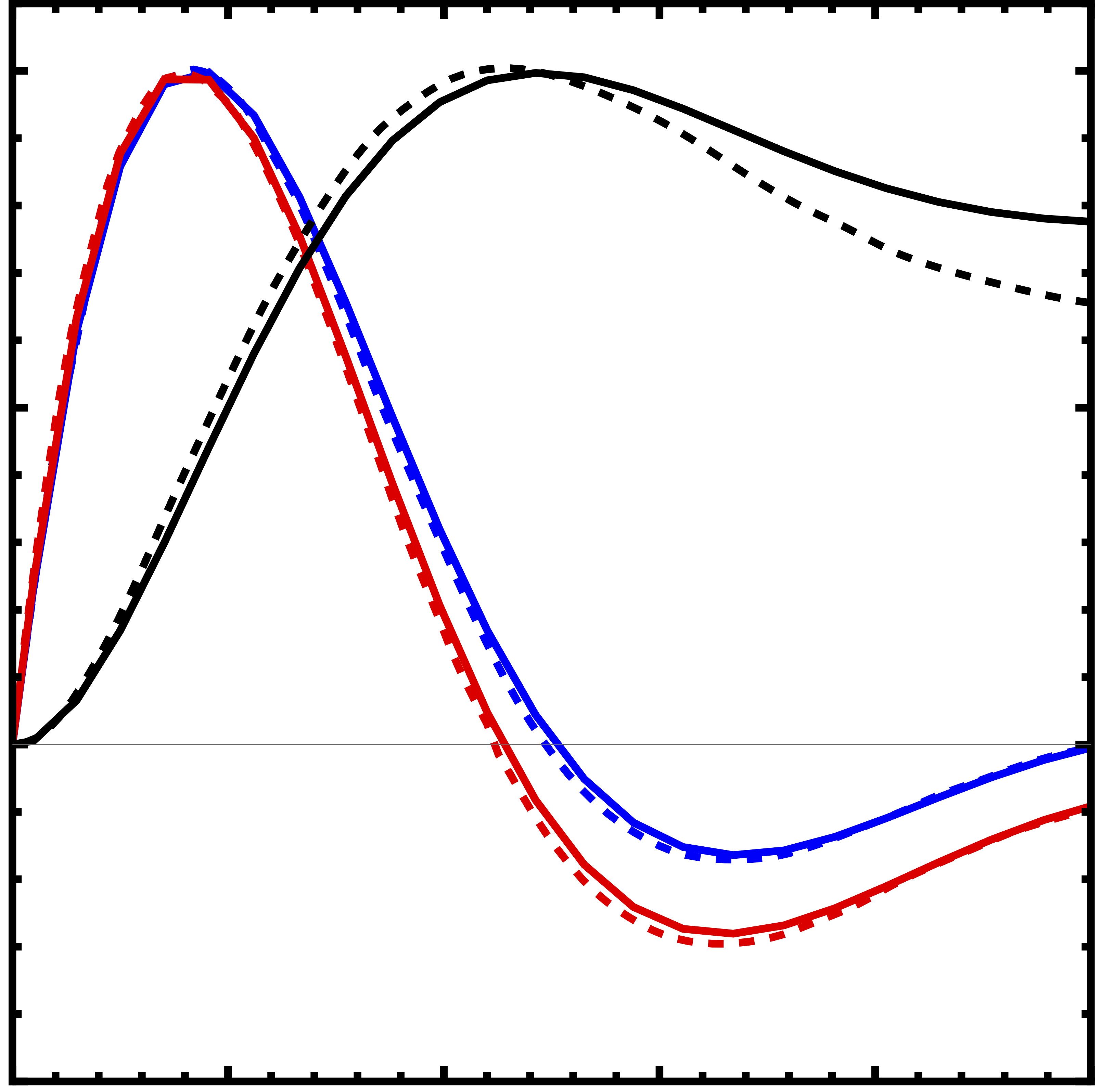}}
\put(17.5,0.6){\includegraphics[height=4.5cm]{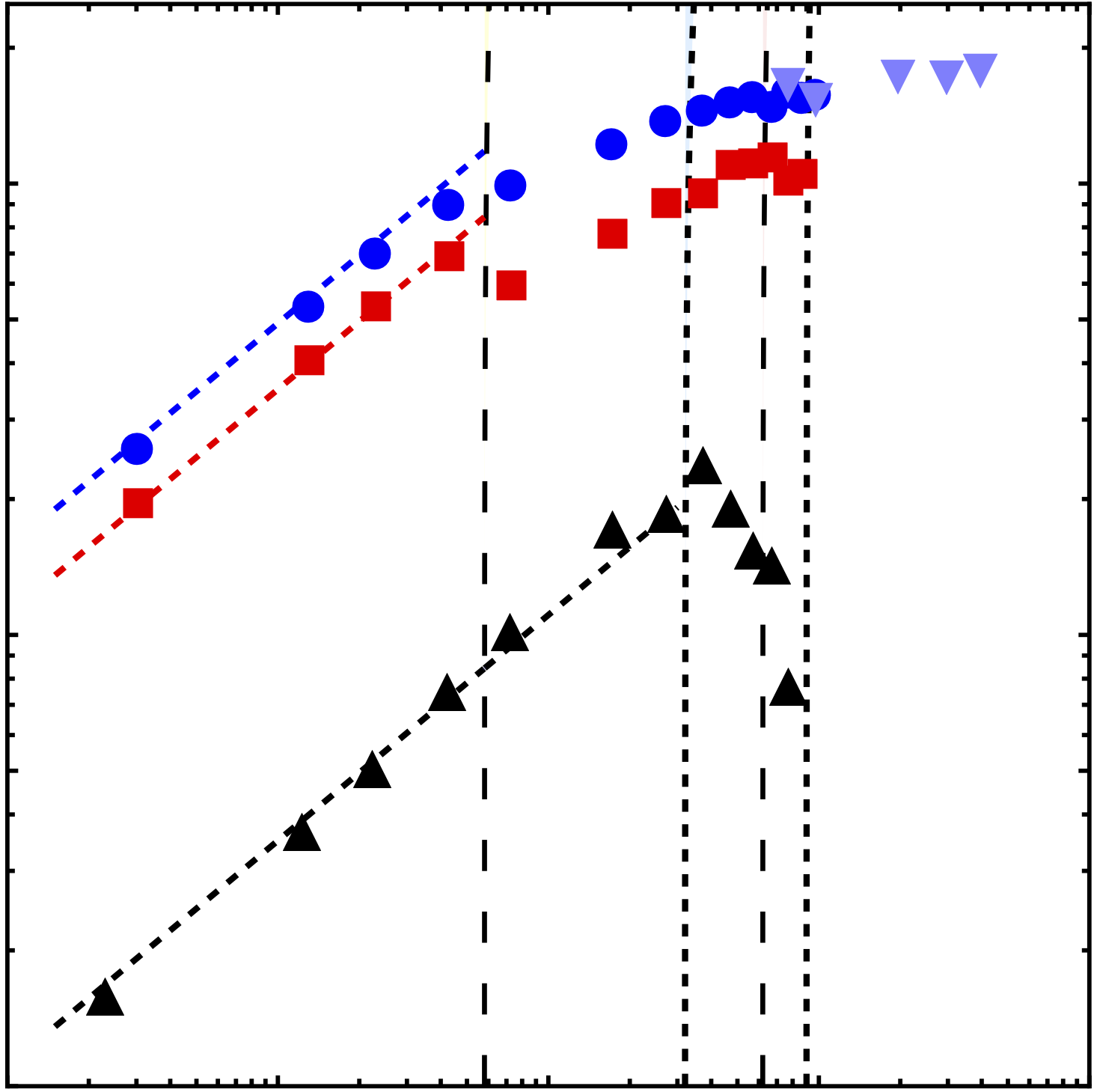}}
 \unitlength0.75truecm
   \draw [stealth-stealth, color=black](4,1.2) -- (5.45,1.2) ;
\end{tikzpicture}
}
\put(-2.5,0.5){
\put(-0.2,6.8){(c)}
\put(0.4,5.8){0.15}
\put(0.4,4.45){0.10}
\put(0.4,3.1){0.05}
\put(-0.2,3.7){$u_z$}
\put(0.4,1.8){0.00}
\put(0.15,0.5){$-0.05$}
\put(1.0,0){0.7}
\put(2.95,0){0.8}
\put(4.95,0){0.9}
\put(4,-0.5){$r/R$}
\put(6.95,0){1.0}
\put(6.1,1.1){$\delta_{u_z}$}
\put(2.7,5.5){$4\times10^8$}
\put(2.7,5.05){$5\times10^8$}
\put(2.7,4.6){$6\times10^8$}
\put(2.7,4.15){$7\times10^8$}
\put(2.7,3.7){$8\times10^8$}
\put(2.7,3.2){$9\times10^8$}

\put(7.8,6.8){(d)}
\put(8.55,6){1.0}
\put(8.0,2.8){\rotatebox{90}{$u_{z}$, $u_{\phi}$, $u_r$}}
\put(11,2.9){\color{red}{$u_{\phi}$}}
\put(12.3,2.9){\color{blue}{$u_{z}$}}
\put(14.5,5.6){\color{black}{$u_{r}$}}
\put(8.55,4.15){0.5}
\put(8.55,2.2){0.0}
\put(8.3,0.45){$-0.5$}
\put(9.2,0){0}
\put(10.4,0){1}
\put(11.6,0){2}
\put(12.8,0){3}
\put(14,0){4}
\put(11.8,-0.5){$\Ek^{-1/3}x$}
\put(15.15,0){5}

\put(8.05,0){
\put(7.8,6.8){(e)}
\put(8.0,1.7){\rotatebox{90}{$u_{z_{\max}}$, $u_{\phi_{\max}}$, $u_{r_{\max}}$}}
\put(10.1,1.7){\rotatebox{45}{\color{black}{$u_{r_{\max}}$}}}
\put(10.3,3.4){\rotatebox{45}{\color{red}{$u_{\phi_{\max}}$}}}
\put(10.1,4.7){\rotatebox{45}{\color{blue}{$u_{z_{\max}}$}}}
\put(8.4,5.4){$10^{-1}$}
\put(8.4,2.9){$10^{-2}$}
\put(8.4,0.4){$10^{-3}$}
\put(9.1,0){$10^{6}$}
\put(10.5,0){$10^{7}$}
\put(12,0){$10^{8}$}
\put(13.45,0){$10^{9}$}
\put(11.3,-0.5){$\Ra-\Ra_\text{w}$}
\put(14.9,0){$10^{10}$}
}}
\end{picture}
\caption{
Mid-plane $u_z$ (averaged in precessing frame at maximum in $\phi$)  versus $r/R$ for (a) $\Ra/10^7$ = 3, 4, 5, 7, 10, 20, 30 where dark (light) blue corresponds to the smallest (largest) $\Ra$.  Inset of (a) shows  Fourier--Bessel coefficients $c_n$, normalized by $c_{n_{\max}}(\Ra)$ and averaged over the set of $\Ra$ in (a). 
(b) Expanded version of (a) with $0.7 \leq r/R \leq 1.0$. Inset of (b) is $\widetilde{u}_z(r)$ reconstructed from the average of $c_n/c_{n_{\max}}$ in inset of (a).  
(c) $u_z$ versus $r/R$ of time-dependent wall-mode state for $4 \times 10^8 \leq \Ra \leq 9 \times 10^8$ (colors per legend) with saturating peak amplitude and wider spread in $\delta_{u_z}$. (d) $\Ra=3\times 10^7$  average radial structure (solid line) compared to linear eigenfunctions (dashed) 
\cite{Herrmann1993}  for $u_\phi$ (red)  $u_z$ (blue),  and $u_r$ (black). The first zero crossing of $u_z$ is at about the same $r$ as the maximum of $u_r$.  All profiles are scaled such that the maximum values are 1. 
(e) Variation of radial maximum values $u_{z_{\max}}$ (blue solid circles), $u_{z_{\text{rms},\max}}$ (light blue inverted triangles),  $u_{\phi_{\max}}$ (red solid squares), and  $u_{r_{\max}}$ (black solid triangles) vs $\Ra-\Ra_\text{w}$. Linear scaling $u \sim (\Ra-\Ra_\text{w})^{1/2}$ (blue, red, and black dashed lines, respectively). Vertical dashed lines from smaller to larger $\Ra$: end of quasi-linear regime (long-dashed) $\Ra \approx 9\times10^7$ - note decrease in $u_\phi$, subcritical instability to lateral jet ejection (dashed), transition to aperiodic (chaotic) jet ejection (long-dashed), and bulk onset (dashed).}
\label{uzFBessel}
\end{figure}
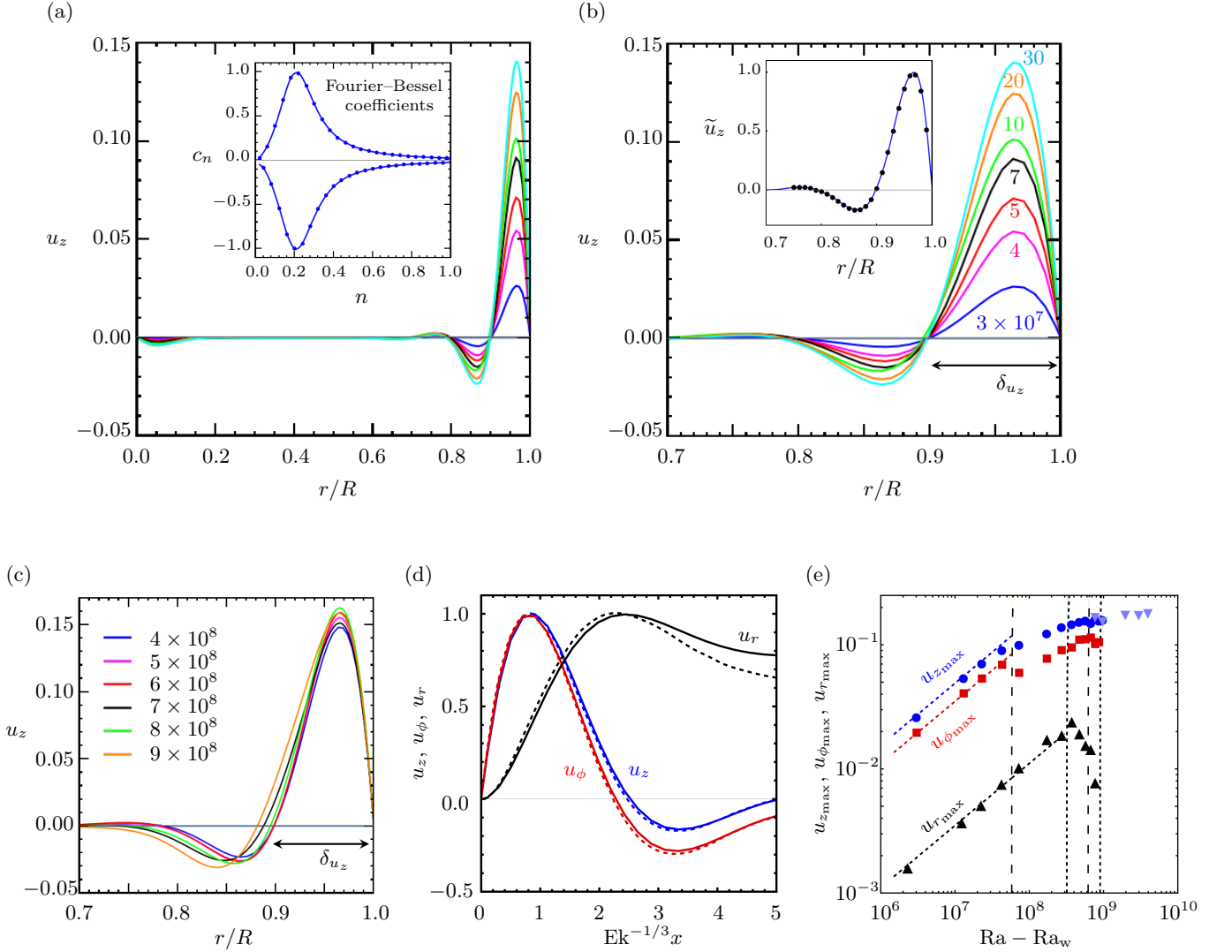

Whereas the analysis above was for a planar wall \cite{Herrmann1993}, our finite cylindrical convection cell suggests that the appropriate representation in that geometry are radial eigenfunctions using sums of Bessel functions of the first kind,
$J_\text{m}(kr)$ of order $m$ \cite{Goldstein1993}. It is therefore natural to consider $u_z (r)$ fields as Fourier--Bessel transforms with $m=1$:
\begin{eqnarray*}
c_n &=& 2/(J_{2}(j_{1n}))^2 \int_0^1 r u_z (r) J_1(j_{1n} r) dr,\\
u_z (r)  &\approx& \sum_1^N c_n J_1(j_{1n} r),
\end{eqnarray*}
where $u_z (r)$ is the radial component of the vertical velocity field and $j_{1n}$ are the $n$-th order zeros of $J_1(x)$.  Empirically, we find that $N=50$ yields excellent fits to $u_z (r)$ which is primarily consequential in the advection of heat and the determination of the global heat transport $\Nu$.  
In Fig.~\ref{uzFBessel}(a), we show the radial dependence of mean $u_z$ for $\Pr = 0.8$, $\Gamma = 1/2$, and $\Ek = 10^{-6}$ evaluated at its azimuthal maximum (in the precessing frame) and at the mid-plane ($z=1/2$) height for $3 \times 10^7 \leq \Ra \leq 3 \times 10^8$.  The inset is the normalized Fourier--Bessel decomposition coefficients $c_n$ which has a simple form and is quite independent (modulo an overall scale factor that increases with $\Ra$ for $\Ra < 3 \times 10^8$, i.e., for steady wall modes).  An expanded view in Fig.\ \ref{uzFBessel}(b) shows the details of the profiles near the sidewall and the definition of the radial extent of the main wall peak --- here we take the first zero crossing of $u_z(r)$ as the characteristic width $\delta_{u_z}$. 
The inset shows the radial profile obtained by averaging the normalized $c_n$ values over the $\Ra$ range.  
The data points are from the amplitude-normalized planar eigenfunction of $u_z$ shown in Fig.\ \ref{uzTEigen}(a) and match almost perfectly, indicating again that cylindrical curvature is not important here for $u_z$. For larger $\Ra \geq 4 \times 10^8$, the profiles change slightly as shown in Fig.\ \ref{uzFBessel}(b) with a larger spread in $\delta_{u_z}$. This change is associated with a secondary bifurcation of the wall modes as discussed below. 
It is quite surprising that the radial profile of $u_z$ maintains a close correspondence with the linear eigenfunction over such a large range of $\Ra$. In Fig.~\ref{uzFBessel}(d), we compare the linear eigenfunctions (dashed lines) for $u_z$, $u_\phi$, and $u_r$ with data for $\Ra = 3 \times 10^7$ (solid lines) where one sees very close correspondence between the eigenfunctions and the data for $u_z$ and $u_\phi$ with a slightly smaller first zero crossing for $u_\phi$. $u_r$ has a very different shape with a maximum value at a radius $r_{\max}$ (we define the characteristic length of $u_r$ as $\delta_{u_r} = 1-r_{\max}$) near the first zero crossing of $u_z$, and there is a distinct difference between the eigenfunction and the data indicating that $u_r$ is more affected by finite wall curvature than the other convective fields.  Figure~\ref{uzFBessel}(e) shows the peak amplitudes of $u_z$, $u_\phi$, and $u_r$ plotted as a function of $\Ra-\Ra_\text{w}$. 
Close to onset $u \sim (\Ra-\Ra_\text{w})^{1/2}$ as expected for a supercritical bifurcation. The monotonic increase in $u_z$ and $u_r$ appears to saturate near the onset of bulk convection at $\Ra \approx 10^9$ whereas $u_{r_{\max}}$ drops rapidly after the onset of the subcritical instability described below.  Finally, there is a small drop in $u_{\phi_{\max}}$ at $\Ra \approx 9 \times 10^7$ which signals the end of the weakly nonlinear wall mode regime.  This signature shows up in measures of $T$. 

\begin{figure}[th]
\unitlength1truecm
\begin{picture}(18,5)
\put(2,0){\includegraphics[height=4.2cm]{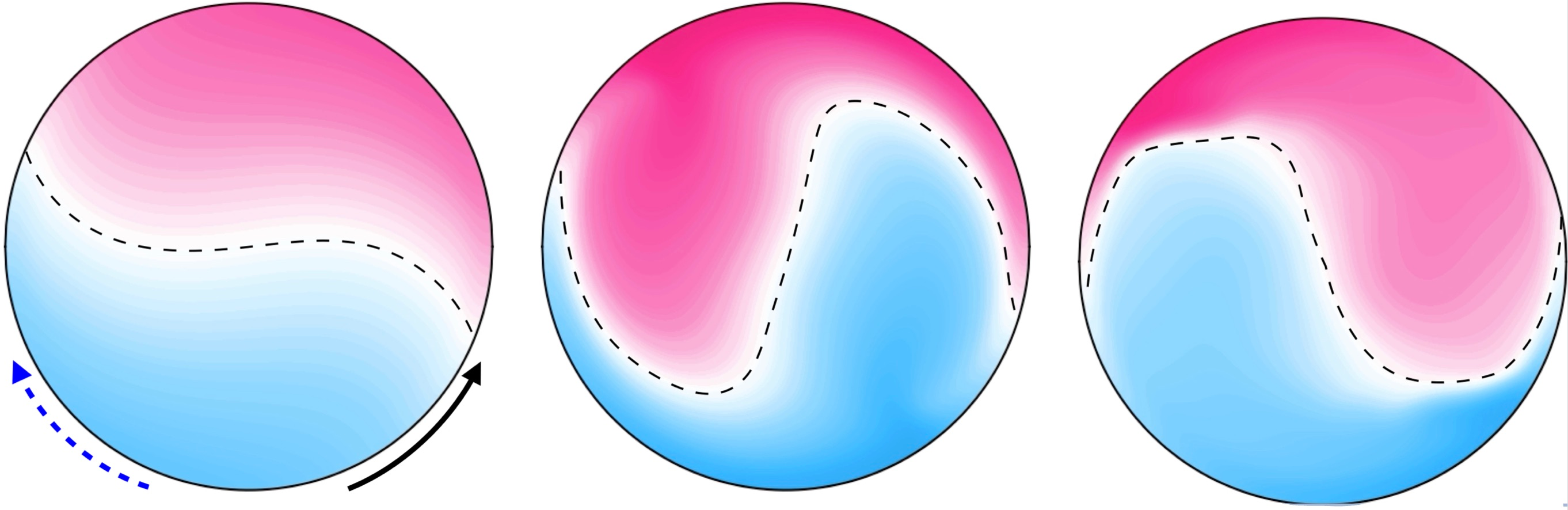}}
\put(2.2,0.3){\color{blue}$u_p$}
\put(5.6,0.3){\color{black}$\Omega$}
\put(1.8,3.5){(a)}
\put(6.3,3.5){(b)}
\put(10.8,3.5){(c)}
\end{picture}
\caption{
Mid-plane $T$ fields for $\Ra$: (a) $3 \times 10^7$, (b) $2 \times 10^8$, and (c) $5 \times 10^8$.  
Dashed lines show contours of the arithmetic mean of the top and bottom temperatures.  
Directions of rotation and precession are indicated in (a).
} 
\label{Contours}
\end{figure}
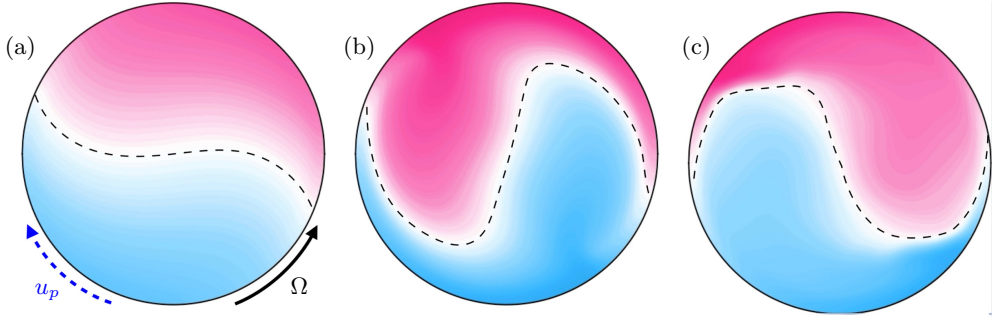

\begin{figure}[th]
\unitlength1truecm
\begin{picture}(18,6)
\put(4.8,1.2)
{
\begin{tikzpicture}[style=thick]
\put(1,0.5){\includegraphics[height=4.55cm]{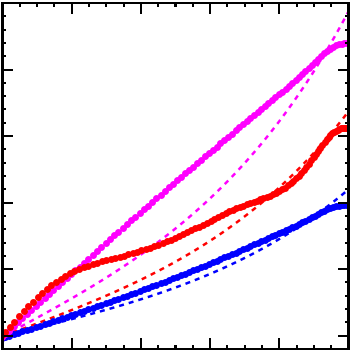}}
\put(7,0.5){\includegraphics[height=4.5cm]{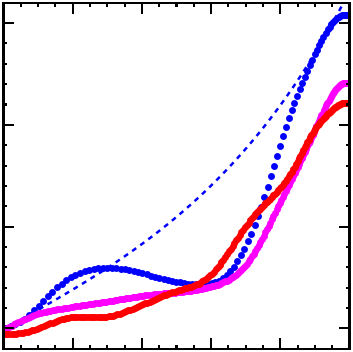}}
\put(13,0.5){\includegraphics[height=4.5cm]{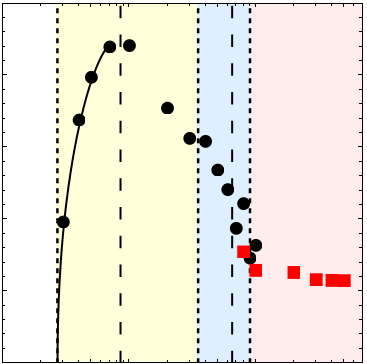}}
    \draw [dashed](5.0,0.5) -- (5.0,5) ;
    \draw [stealth-stealth](5,1) -- (5.45,1) ;
    \draw [dashed](11.0,0.5) -- (11.0,5) ;
    \draw [stealth-stealth](11,1) -- (11.45,1) ;
    \draw [|-|, color=black](4.925,0.75) -- (5.075,0.75) ;
    \draw [|-|, color=black](10.925,0.75) -- (11.075,0.75) ;
    \end{tikzpicture}
}
\put(-0.2,5.7){(a)}
\put(5.8,5.7){(b)}
\put(11.8,5.7){(c)}
\put(-0.2,3.4){$T$}
\put(5.9,3.4){$T$}
\put(11.8,3.1){\rotatebox{90}{$T_{\max}$}}
\put(0.35,5.5){0.25}
\put(0.35,4.75){0.20}
\put(0.35,3.9){0.15}
\put(0.35,3.05){0.10}
\put(0.35,2.2){0.05}
\put(0.35,1.35){0.00}
\put(1,0.8){0}
\put(1.75,0.8){0.2}
\put(2.65,0.8){0.4}
\put(3.55,0.8){0.6}
\put(4.45,0.8){0.8}
\put(5.4,0.8){1}
\put(3,0.3){$r/R$}
\put(3,3.5){\rotatebox{45}{\color{magenta}{$10^8$}}}
\put(4,3.2){\rotatebox{30}{\color{red}{$3\times10^8$}}}
\put(4.2,2.1){\rotatebox{27}{\color{blue}{$3\times10^7$}}}

\put(6.35,5.35){0.15}
\put(6.35,4.05){0.10}
\put(6.35,2.75){0.05}
\put(6.35,1.45){0.00}
\put(7,0.8){0}
\put(7.75,0.8){0.2}
\put(8.65,0.8){0.4}
\put(9.55,0.8){0.6}
\put(10.45,0.8){0.8}
\put(11.4,0.8){1}
\put(9,0.3){$r/R$}
\put(10.2,4.5){\rotatebox{60}{\color{blue}{$4\times10^8$}}}
\put(8.5,1.3){\rotatebox{20}{\color{red}{$8\times10^8$}}}
\put(10.2,2){\rotatebox{60}{\color{magenta}{$6\times10^8$}}}

\put(12.35,5.6){0.25}
\put(12.35,4.7){0.20}
\put(12.35,3.8){0.15}
\put(12.35,2.9){0.10}
\put(12.35,2){0.05}
\put(12.35,1.15){0.00}

\put(13,0.8){1}
\put(14.05,0.8){5}
\put(14.4,0.8){10}
\put(15.5,0.8){50}
\put(15.9,0.8){100}
\put(17,0.8){500}
\put(15,0.3){$\Ra/10^7$}
\put(5.05,1.9){$\delta_{u_{z}}$}
\put(11.05,1.9){$\delta_{u_{z}}$}

\put(13.3,2.5){\rotatebox{90}{No convection}}
\put(13.8,2.3){\rotatebox{0}{Wall modes}}
\put(15.7,3.8){\rotatebox{90}{Subcritical}}
\put(16.4,4){\rotatebox{0}{Bulk}}
\put(16.6,3.6){\rotatebox{0}{+}}
\put(16.4,3.2){\rotatebox{0}{BZF}}
\end{picture}
\caption{Midplane temperature field maximum (averaged in precessing frame at maximum with respect to $\phi$) $T(r)$ versus $r/R$ for (a) $\Ra/10^7$ = 3, 10, and 30. 
\oo
Dashed colored lines indicate the temperature eigenfunction for a planar wall \cite{Herrmann1993}  (as in Fig.\ \ref{ThetaEigenGamma}(a)) adjusted for approximate sidewall mean temperature. Mean width $\delta_{u_z}$ and its variation (lateral error bars) in our computations are shown.
(b) $\Ra/10^7 = $ 40, 60, and 80. The corresponding temperature eigenfunction for $\Ra = 4 \times 10^8$ and  
the mean width $\delta_{u_z}$ are shown. 
(c) maximum radial value (at $r/R=1$) 
$T_\text{max}$ (black circles) and $T_\text{rms}$
(red squares) versus $\Ra/10^7$ with different regions labeled by vertical dashed lines. 
The long-dashed lines denote transitions from weakly nonlinear growth of $T$ (solid line is 
$6.0\times10^{-5}(\Ra-\Ra_\text{w})^{1/2} - 7.6\times10^{-10}(\Ra-\Ra_\text{w})$ 
to decreasing amplitude at $\Ra \approx 8.5 \times 10^7$ as the vertical temperature profile in the wall mode steepens and from periodic to aperiodic (chaotic) time dependence of $\Nu$ at $\Ra \approx 6.5 \times 10^8$.
\bb
} 
\label{TempRadial}
\end{figure}
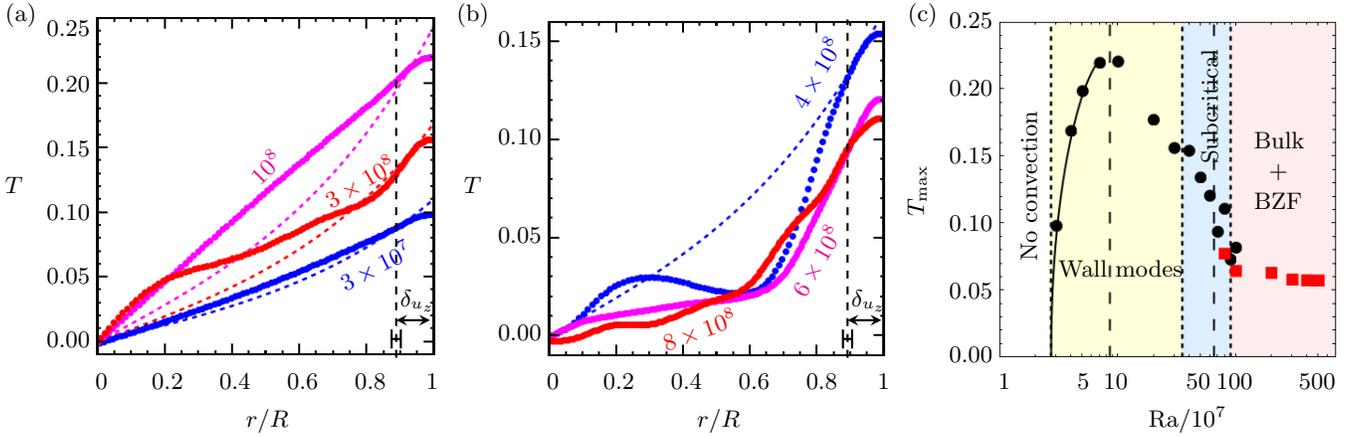

The other dominant field contributing to the heat transport is the temperature field which, as noted above, has a much broader radial distribution and is strongly affected by small $\Gamma$ as was suggested by considerations of the temperature eigenfunction \cite{Herrmann1993} shown in Fig.\ \ref{ThetaEigenGamma}(a). To get an overall flavor of the wall mode states for $\Ra \leq 3 \times 10^8$, we show in Fig.\ \ref{Contours} 
some representative azimuthal mid-plane contours and corresponding fields; in all of them $T$ extends significantly over most of the radial domain. Fig.~\ref{Contours}(a)  shows the nearly sinusoidal mean contour for $\Ra = 3 \times 10^7$ ($\epsilon = \Ra/\Ra_\text{w} -1 = 0.07$) 
with counter clockwise rotation and clockwise precession. For $\Ra = 2 \times 10^8$ ($\epsilon$ = 6.1) in Fig.\ \ref{Contours}(b), the contour is more nonlinear but the precession remains steady whereas for $\Ra=5\times 10^8$ ($\epsilon = 17)$ there is increased distortion of the mean contour and the state is time dependent.  By going to the precessing frame one can obtain well-converged statistical averages of the radial profiles of different fields.  We show in Fig.\ \ref{TempRadial}, the radial distribution of the max value of $T$ (averaged in precessing frame) in a mid-plane ($z = 1/2$) cross section.  For $\Ra \leq 10^8$, the profiles are remarkably linear, more so than the eigenfunction, indicating that the finite geometry is having a significant effect on $T(r)$.  For larger $\Ra > 10^8$, $T(r)$ becomes increasingly nonlinear with radial undulations owing to the transition to time-dependent states and bulk/BZF convection described below. In Fig.\ \ref{TempRadial}(c), we show the $\Ra$ variation of the max value of $T$ which indicates the 
weakly nonlinear growth of the wall mode amplitude but for higher $\Ra \approx 9 \times 10^7$ this reverses and $T$ decreases slowly owing to the decreasing vertical gradient of $T$ in the cell center as discussed below. This feature is correlated with the small drop in $u_{\phi_{max}}$ in Fig.~\ref{uzFBessel}(e). 
Also shown is $T_\text{rms}$ for $\Ra \gtrsim 10^9$ owing to the unsteady and intermittent precession once the bulk mode is present which makes averaging in the precessing frame much slower to converge.

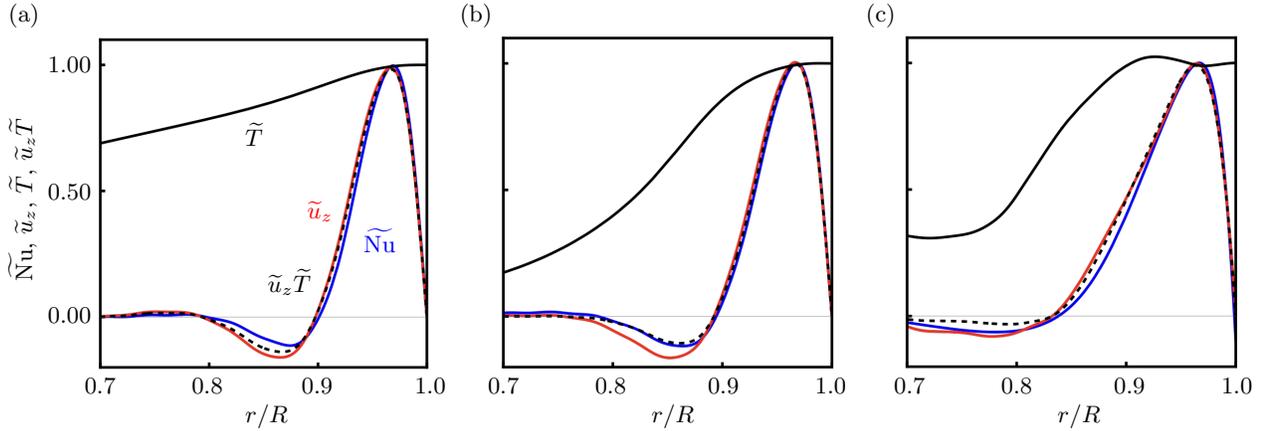
\begin{figure}[th]
\unitlength0.75truecm
\begin{picture}(18,8)
\put(-3.5,0){
\put(2.35,1.1){\includegraphics[height=4.49cm]{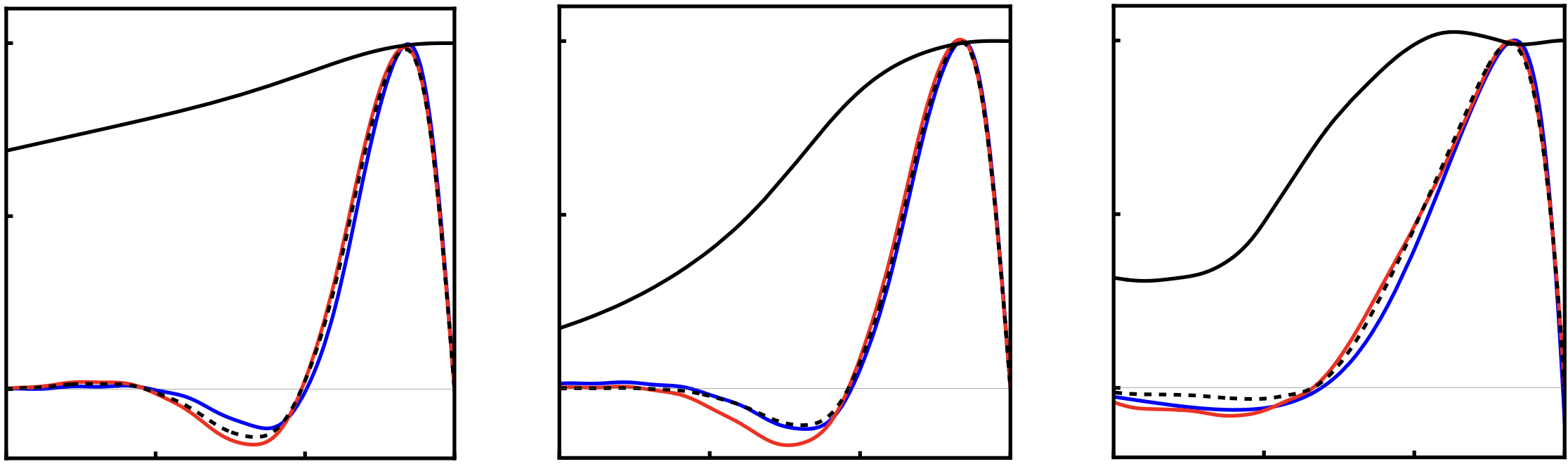}}
\put(1,-7.5){
\put(4,12.6){{\color{black}$\widetilde T$}}
\put(5.1,11.3){{\color{red}$\widetilde u_z$}}
\put(6.1,10.7){{\color{blue}$\widetilde \Nu$}}
\put(4.4,10.){{\color{black}$\widetilde u_z \widetilde T$}}

\put(15.0,14.8){(c)}
\put(15.5,8.2){0.7}
\put(17.4,8.2){0.8}
\put(19.4,8.2){0.9}
\put(18.4,7.7){$r/R$}
\put(21.3,8.2){1.0}
\put(-0.2,14.8){(a)}
\put(8.3,8.2){0.7}
\put(10.2,8.2){0.8}
\put(12.2,8.2){0.9}
\put(11.2,7.7){$r/R$}
\put(14.1,8.2){1.0}
\put(7.8,14.8){(b)}
\put(0.5,13.9){1.00}
\put(0.5,11.65){0.50}
\put(-0.2,10.2){\rotatebox{90}{$\widetilde\Nu$, $\widetilde u_z$, $\widetilde T$, $\widetilde u_z \widetilde T$}}
\put(0.5,9.45){0.00}
\put(1.15,8.2){0.7}
\put(3.05,8.2){0.8}
\put(5.0,8.2){0.9}
\put(4,7.7){$r/R$}
\put(7,8.2){1.0}
}}
\end{picture}
\caption{
$\widetilde{\Nu} (r)$ (blue), $\widetilde{u_z}(r)$ (red), $\widetilde{T}(r)$ (black), and $\widetilde{u_z}(r) \widetilde{T}(r)$ for $\Ek = 10^{-6}$, $\Pran = 0.8$. Each profile for a given $\Ra$ is the radial profile evaluated at the max value in $\phi$ in the mean precessing frame and then normalized by the maximum value. Then different ranges of $\Ra$ are averaged. (a) Averaged over $3 \leq \Ra/10^ 7\leq 30$, (b) Averaged over $40 \leq \Ra/10^7 \leq 90$. 
(c) Averaged over $100 \leq \Ra/10^7 \leq 500$ where the bulk contribution is subtracted out.
} 
\label{Nu_uz_T} 
\end{figure}

With the insights provided by the structure of $u_z(r)$ and $T(r)$, one can understand the radial structure of the heat transport $\Nu(r)$.  In Fig.\ \ref{Nu_uz_T}(a), the normalized  (peak values set to 1) fields $\widetilde{u_z}(r)$, $\widetilde{T}(r)$, $\widetilde{u_z}(r)\widetilde{T}(r)$, and $\widetilde{\Nu}(r)$ are shown for an average over time-independent wall modes $3 \leq \Ra/10^7 < 30$ .  $T(r)$ varies slowly with $r$ and the fields are time independent in the precessing frame.  
\oo 
If the azimuthal phase difference between $u_z$ and $T$ is small, i.e., the place where $u_z$ is maximally upward (downward) is the same as where $T$ is maximally positive (negative) with respect to the mean, then one
\bb 
has $\Nu(r) \sim u_z(r)$ to within about 10\%.  Taking $\Nu \sim \langle T \rangle \langle u_z \rangle$ improves the agreement to about $\pm 5\%$.  Averaging over the time-dependent wall mode states in the range $40 \leq \Ra/10^7 \leq 90$ yields similar results shown in Fig.\ \ref{Nu_uz_T}(b). More surprising is that averaging over the bulk/BZF states $100 \leq \Ra/10^7 \leq 500$, Fig.\ \ref{Nu_uz_T}(c), also produces a fairly good correspondence where here we subtract out the bulk contribution in the cell interior.  There is a widening zone of boundary zone influence in the BZF  but the form of $\Nu$ near the wall is driven predominately by the vertical velocity $u_z$ of the wall mode which remains highly localized to within a radial width of order $\Ek^{1/3}$ characteristic of the linear wall-mode eigenfunctions. A more quantitiative test of this comparison requires detailed amplitude averaging of the $u_z(\phi)$ and $T(\phi)$ for which we would need better converged statistical quantities in the bulk phase with $\Ra > \Ra_c$; we will consider this in future work.

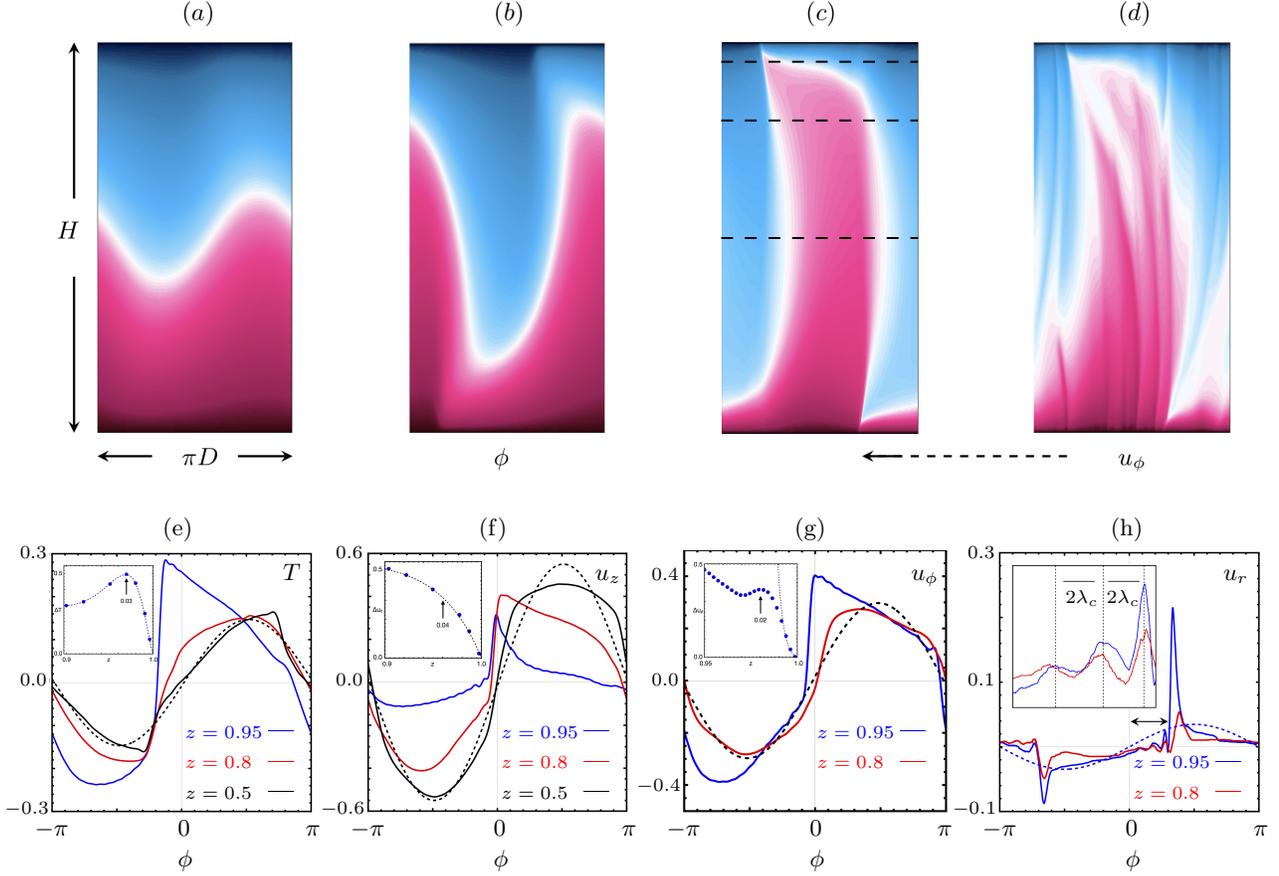
\begin{figure}[th]
\unitlength1truecm
\begin{picture}(18,12.5)
\put(0.2,5){\includegraphics[height=6.3cm]{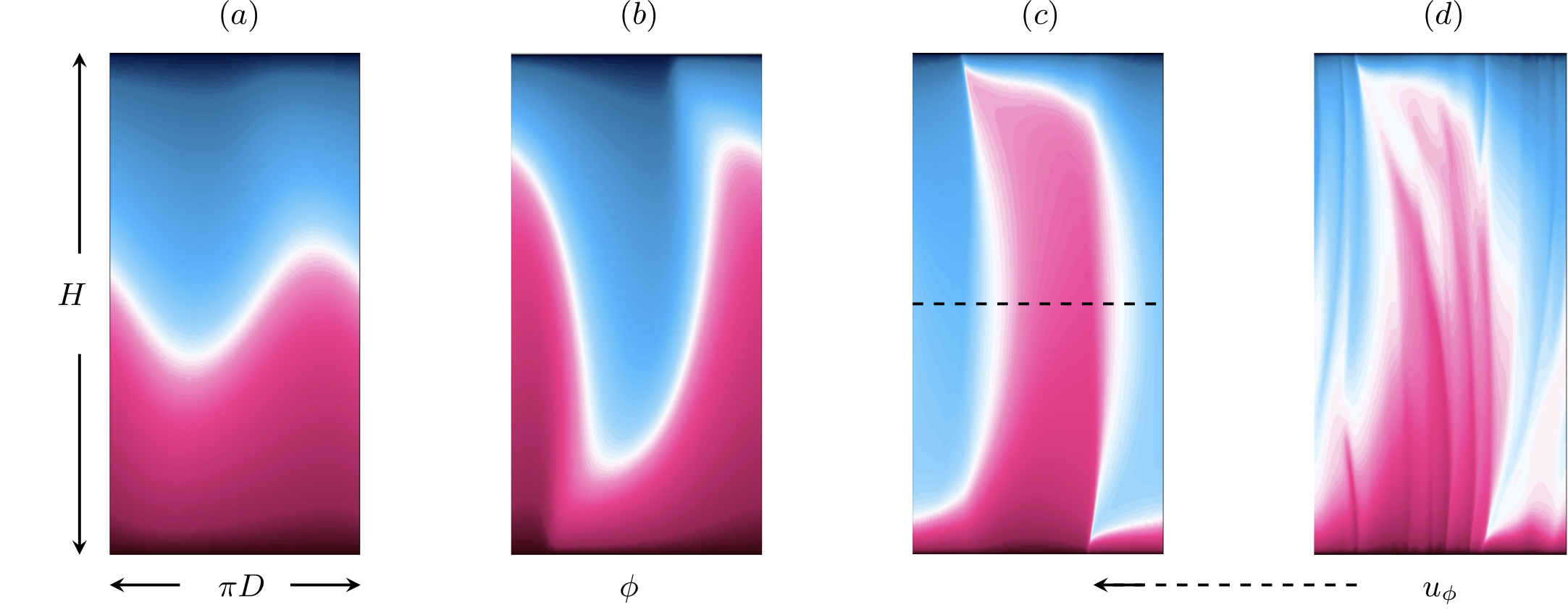}}
\put(9.625,5.55){\includegraphics[height=5.2cm]{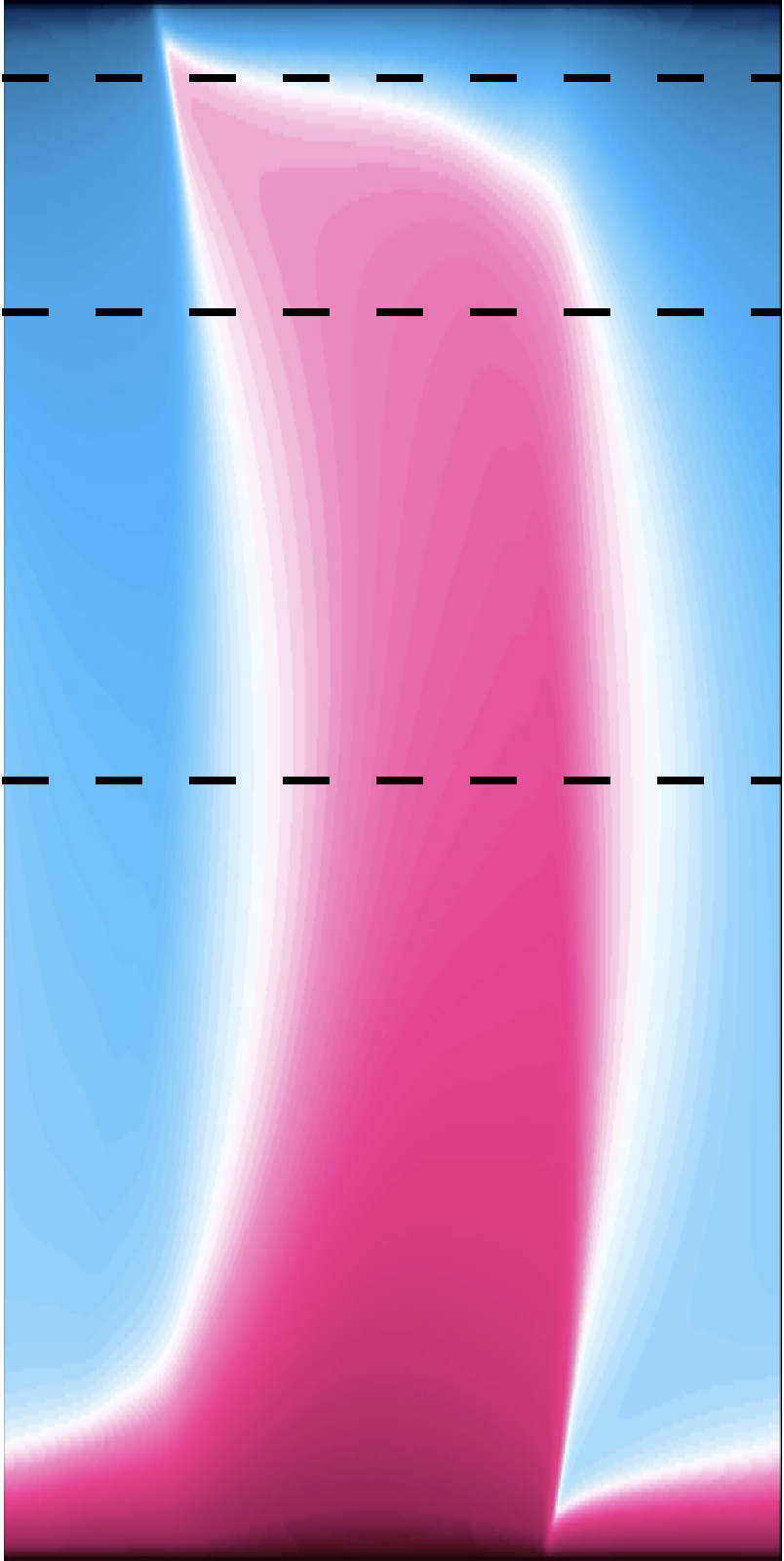}}
\put(0.2,0){
\put(0.5,0.5){\includegraphics[height=3.5cm]{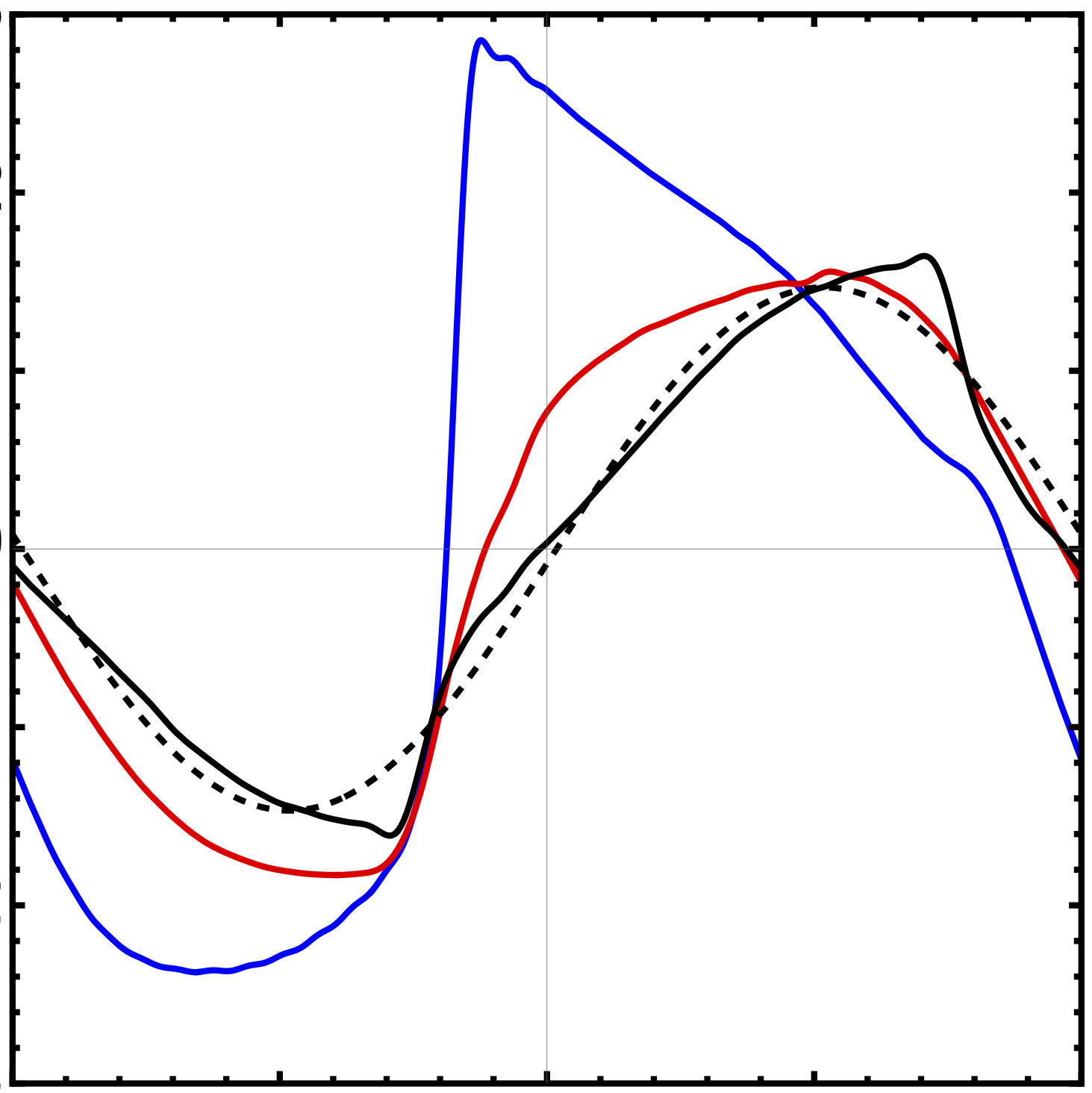}}
\put(0.55,2.5){\includegraphics[height=1.3cm]{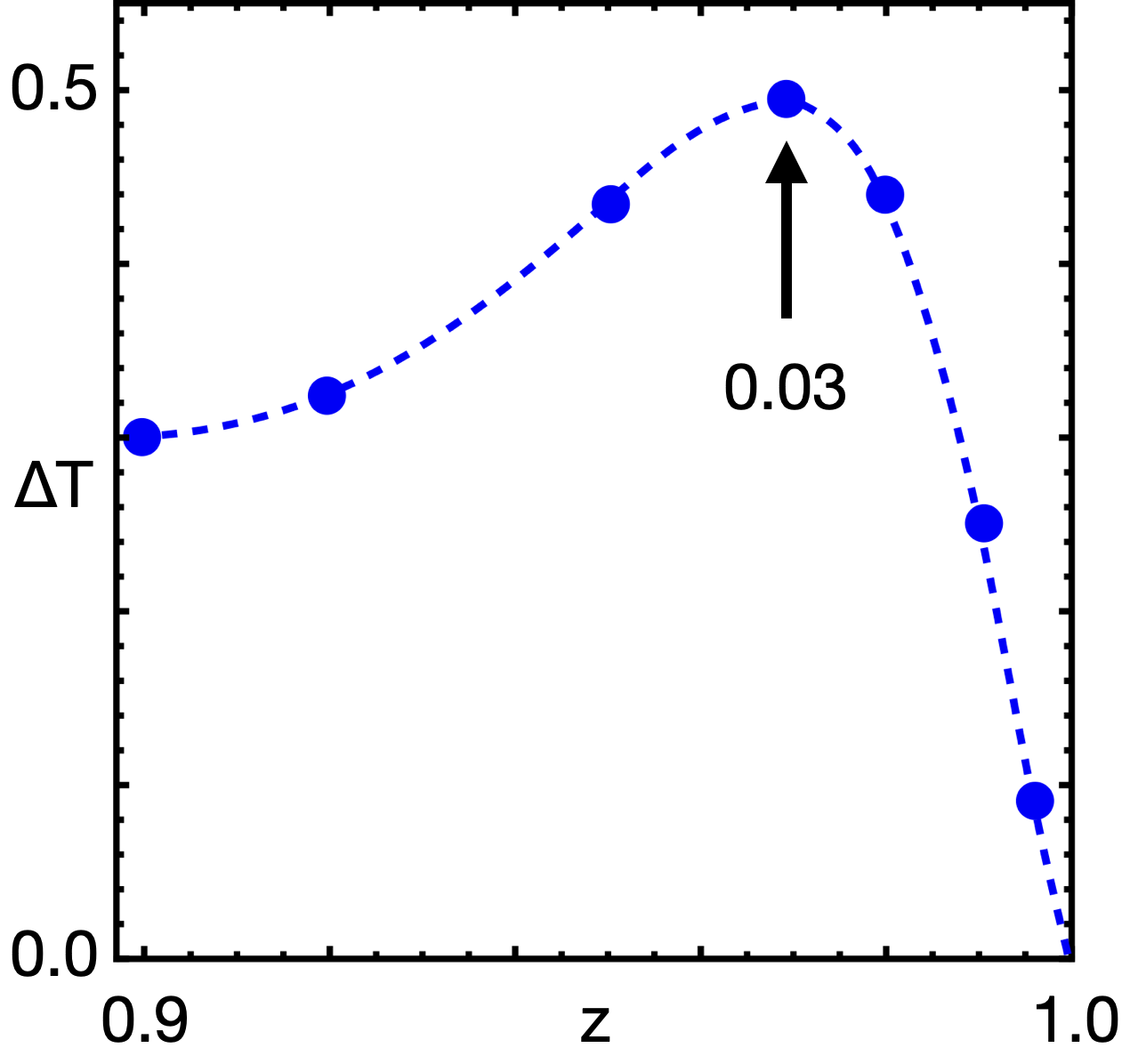}}
\put(2,4.2){(e)}
\put(-0.1,0.5){\scriptsize $-0.3$}
\put(0.1,2.2){\scriptsize 0.0}
\put(0.1,3.9){\scriptsize 0.3}
\put(3.6,3.6){$T$}
\put(0.3,0.2){$-\pi$}
\put(2.2,0.2){0}
\put(3.9,0.2){$\pi$}
\put(2.2,-0.2){$\phi$}
\put(2.3,1.5){\color{blue}{\scriptsize $z=0.95$}}
\put(3.4,1.4){\color{blue}{\huge --}}
\put(2.3,1.1){\color{red}{\scriptsize $z=0.8$}}
\put(3.4,1){\color{red}{\huge --}}
\put(2.3,0.7){\color{black}{\scriptsize $z=0.5$}}
\put(3.4,0.6){\color{black}{\huge --}}
\put(4.7,0.5){\includegraphics[height=3.5cm]{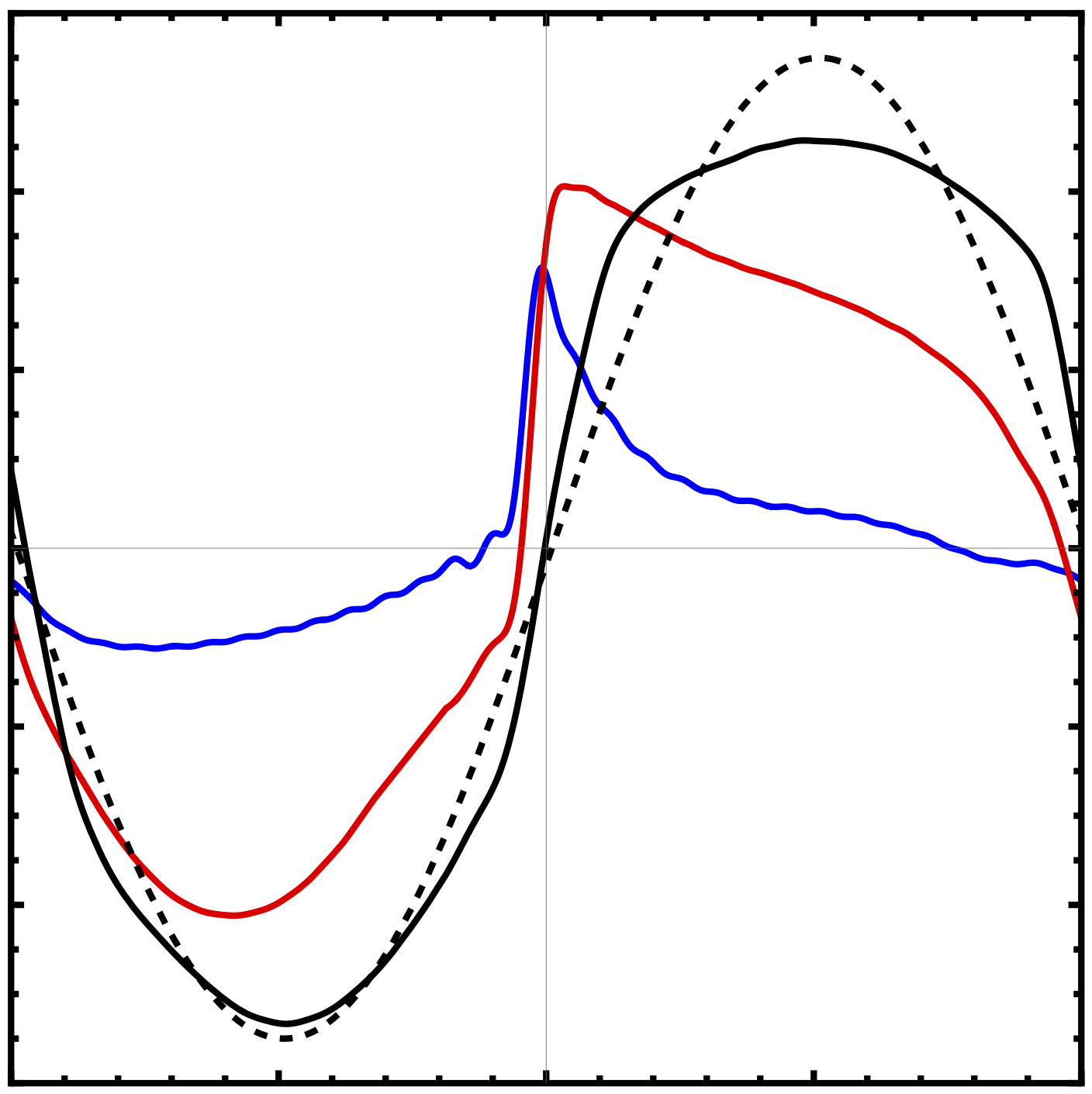}}
\put(4.8,2.45){\includegraphics[height=1.4cm]{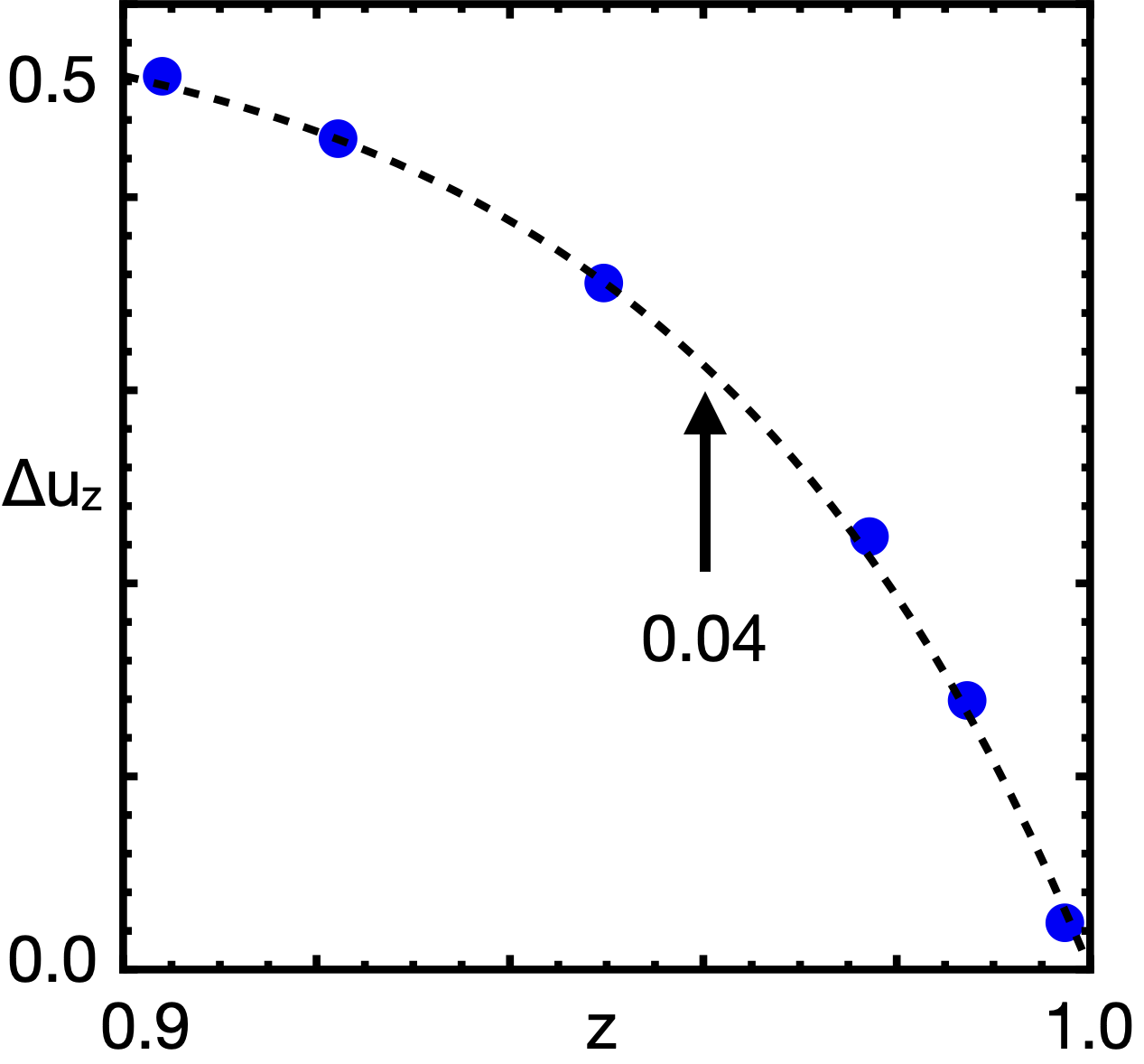}}
\put(6.2,4.2){(f)}
\put(4.1,0.5){\scriptsize $-0.6$}
\put(4.3,2.2){\scriptsize 0.0}
\put(4.3,3.9){\scriptsize 0.6}
\put(7.75,3.6){$u_z$}
\put(4.5,0.2){$-\pi$}
\put(6.4,0.2){0}
\put(8.1,0.2){$\pi$}
\put(6.4,-0.2){$\phi$}
\put(6.5,1.5){\color{blue}{\scriptsize $z=0.95$}}
\put(7.6,1.4){\color{blue}{\huge --}}
\put(6.5,1.1){\color{red}{\scriptsize $z=0.8$}}
\put(7.6,1){\color{red}{\huge --}}
\put(6.5,0.7){\color{black}{\scriptsize $z=0.5$}}
\put(7.6,0.6){\color{black}{\huge --}}
\put(8.9,0.5){\includegraphics[height=3.5cm]{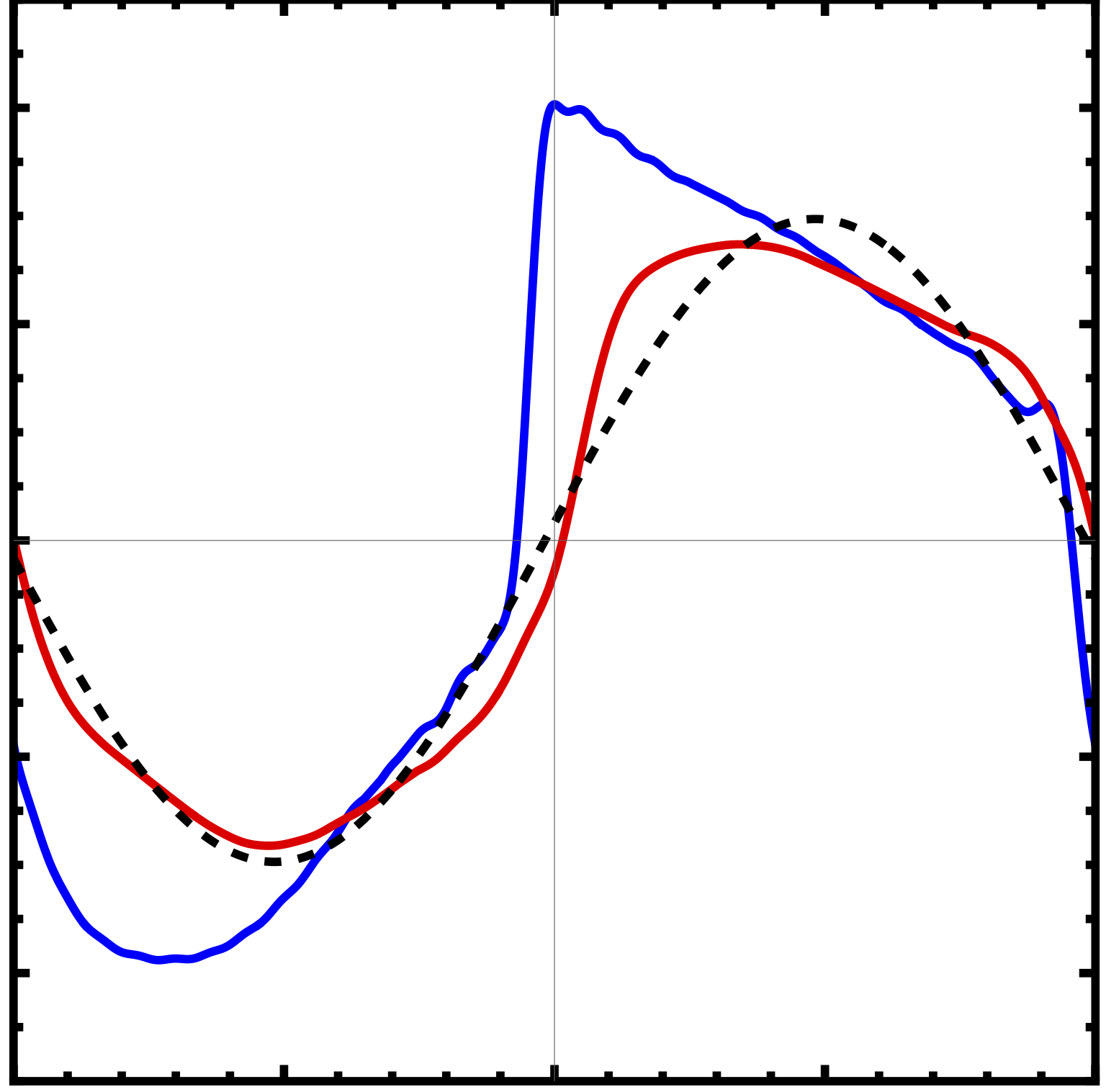}}
\put(9.05,2.45){\includegraphics[height=1.4cm]{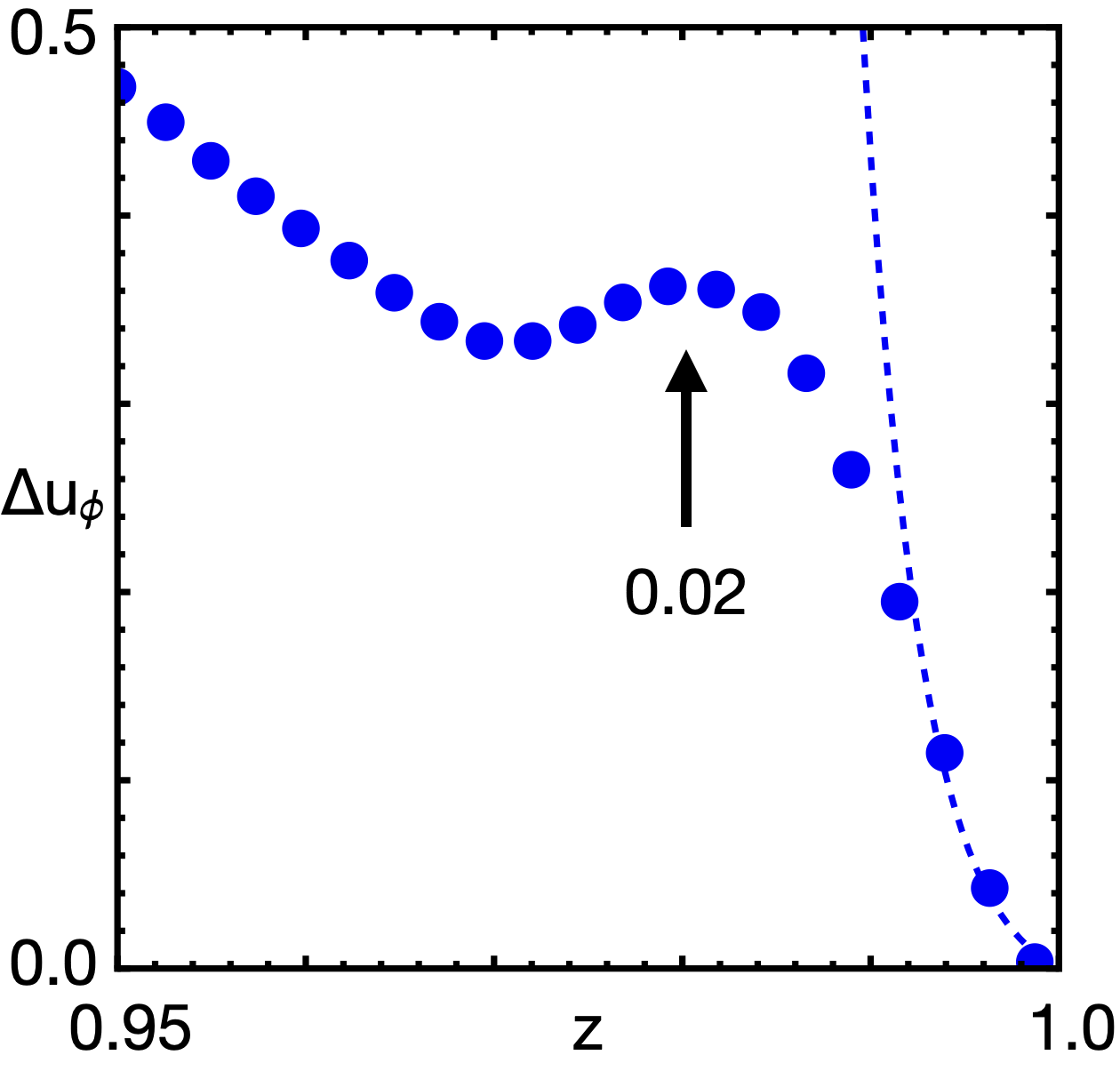}}
\put(10.4,4.2){(g)}
\put(8.3,0.8){\scriptsize $-0.4$}
\put(8.5,2.2){\scriptsize 0.0}
\put(8.5,3.6){\scriptsize 0.4}
\put(11.95,3.6){$u_\phi$}
\put(8.7,0.2){$-\pi$}
\put(10.6,0.2){0}
\put(12.3,0.2){$\pi$}
\put(10.6,-0.2){$\phi$}
\put(10.7,1.5){\color{blue}{\scriptsize $z=0.95$}}
\put(11.8,1.4){\color{blue}{\huge --}}
\put(10.7,1.1){\color{red}{\scriptsize $z=0.8$}}
\put(11.8,1){\color{red}{\huge --}}
\put(13.1,0.5){\includegraphics[height=3.5cm]{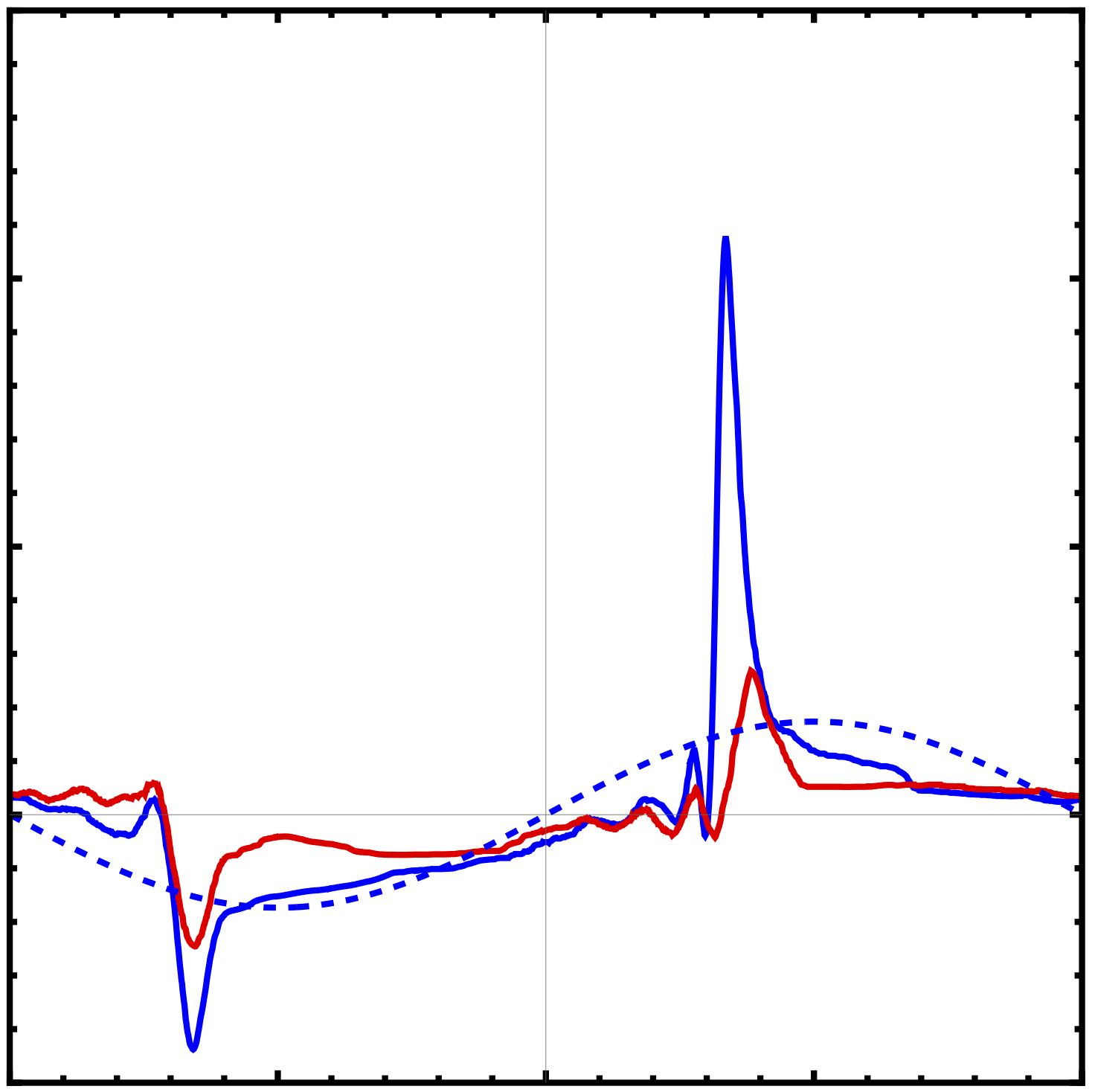}}
\put(14.6,4.2){(h)}
\put(12.5,0.5){\scriptsize $-0.1$}
\put(12.7,2.2){\scriptsize 0.1}
\put(12.7,3.9){\scriptsize 0.3}
\put(16.1,3.6){$u_r$}
\put(12.9,0.2){$-\pi$}
\put(14.8,0.2){0}
\put(16.5,0.2){$\pi$}
\put(14.8,-0.2){$\phi$}
\put(14.9,1.1){\color{blue}{\scriptsize $z=0.95$}}
\put(16,1.0){\color{blue}{\huge --}}
\put(14.9,0.7){\color{red}{\scriptsize $z=0.8$}}
\put(16,0.6){\color{red}{\huge --}}
\put(13.3,1.9){\includegraphics[height=1.9cm]{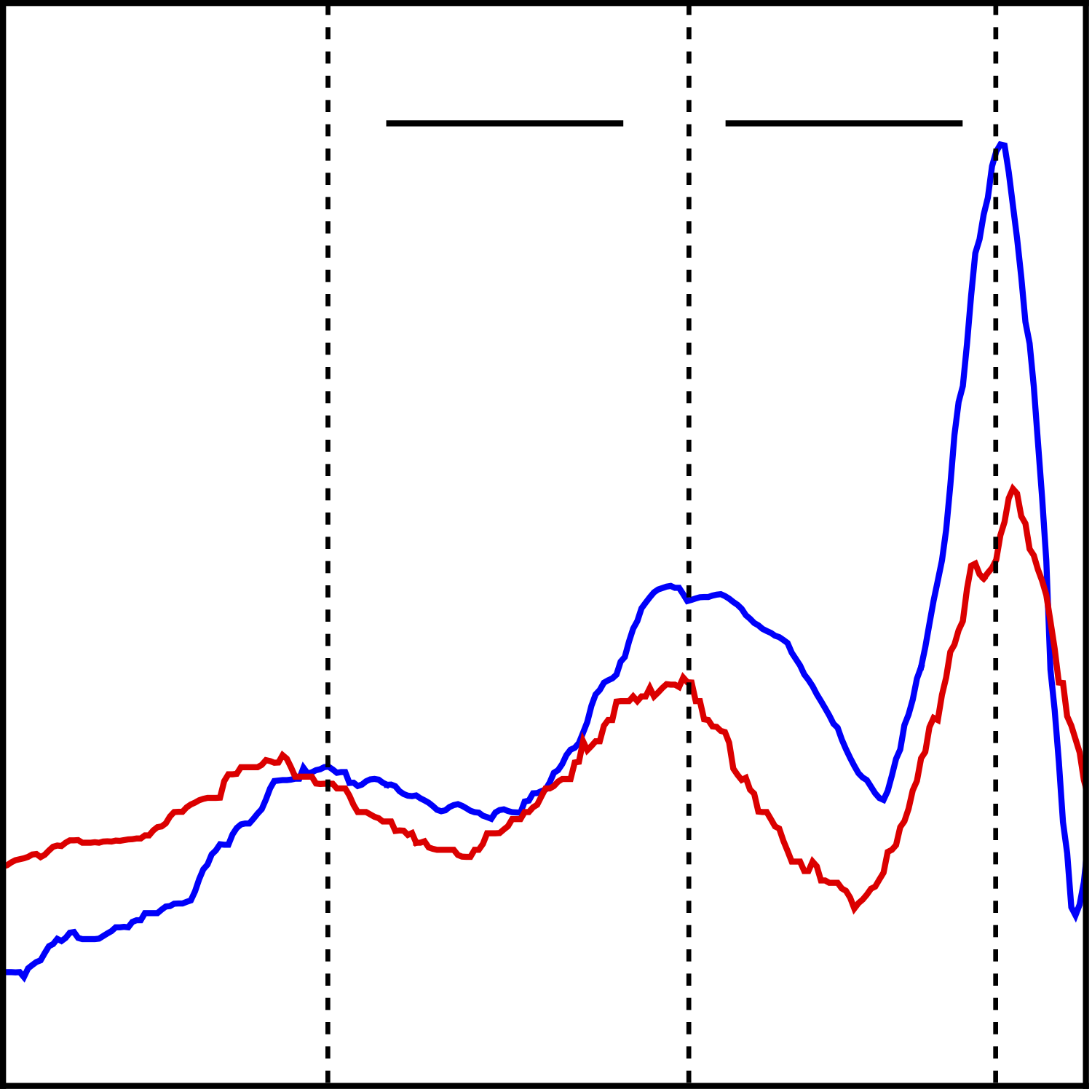}}
\put(14,3.3){\scriptsize $2\lambda_c$}
\put(14.55,3.3){\scriptsize $2\lambda_c$}
}
\end{picture}
\unitlength1truecm
\begin{picture}(18,1)
\unitlength1truecm
\put(14.95,2.75){
\begin{tikzpicture}
\unitlength1truecm
    \draw [stealth-stealth](14.2,1.8) -- (14.7,1.8);
\end{tikzpicture}
}
\end{picture}
\caption{
\oo
(a-d)
\bb
Vertical temperature field $T(r = 0.98R, \phi, z)$ for $\Ek = 10^{-6}$, $\Ra$: (a) $3 \times 10^7$, (b) $5 \times 10^7$, (c) $5 \times 10^8$, (d) $10^9$. 
Precession direction is right to left as indicated by the arrow. 
\oo
(e-h)
\bb
Averaged angular profiles (temporally averaged in the precessing frame, i.e., $\phi_p = \phi_0+\omega t$) for $\Ra = 5 \times 10^8$ \oo (arbitrary amplitude scaled approximately to $\pm 1/2$)\bb. 
The first Fourier mode corresponding to the linear eigenfunction in $\phi$ is black dashed line whereas different $z = 0.5$, 0.8, and 0.95 are indicated by black, red, and blue solid lines, respectively. 
\oo
The insets of (e, f, g) are the approximate amplitudes $\Delta T, \Delta u_z, \Delta u_\phi$ of the shock-like feature, respectively. The dashed blue lines are (e) guide to the eye with local maximum at $1-z_{\max} = 0.03$, (f) exponential saturation fit to $\sim 1-e^{-(1-z)/z_0}$ with $z_0 = 0.04$, and (g) exponential growth fit to $\sim e^{(1-z)/z_1}$ with $z_1 = 0.003$.
\bb
(e) $\langle T(r = 0.98R, \phi, z) \rangle_{\phi_p}$,  (f) $\langle u_z(r = 0.98R, \phi, z=H/2) \rangle_{\phi_p}$, (g) $\langle u_\phi(r = 0.98R, \phi, z=H/2) \rangle_{\phi_p}$, and (h) $\langle u_r(r = 0.98R, \phi, z=H/2) \rangle_{\phi_p}$. 
For $T (z=1/2)$ and $u_z(z=1/2)$, the profiles are close to the linear eigenfunction as is $u_\phi (z=0.8)$ ($u_\phi \approx 0$ at $z=1/2$) and display very sharp nonlinear fronts at their advancing edge for $z=0.95$, i.e., close to the upper boundary. 
On the other hand, $u_r$ has a pulse-like shape with very little of the profile being described by the first Fourier mode.  The inset shows a expanded view at the advancing edge (region indicated by black arrow) of the narrow peak at $\phi \approx 0.3 \pi$ with a spatial oscillation with length about $2 \lambda_c$. 
} 
\label{VertTemp} 
\end{figure}

In addition to the radial structure of the fields, we consider their azimuthal structure.  
In Figs.~\ref{VertTemp}~(a-d), we show instantaneous vertical temperature fields $T(r = 0.98R, \phi, z)$ for values of $\Ra$ in the interval $3 \times 10^7 \leq \Ra \leq 5 \times 10^9$. 
A very nonlinear profile develops with increasing $\Ra$ with only the smallest $\Ra=3 \times 10^7$ being well approximated by a single-mode sinusoidal function. 
To characterize this nonlinearity for $\Ra = 5 \times 10^8$, we show in Figs.~\ref{VertTemp}~(e-h) azimuthal profiles of $T$, $u_z$, $u_{\phi}$, and $u_r$, respectively, averaged in the precessing frame ($\phi_p = \phi_0 + \omega t$) at $r=0.98R$ and at three different $z$ = 0.5, 0.8, and 0.95 (horizontal dashed lines in Fig.~\ref{VertTemp}~(c)). 
For $T$ and $u_z$, the mid-plane profiles are rather close to the linear eigenfunction, i.e., to the lowest Fourier mode. 
The eigenfunction solutions for the $\phi$ dependence of $T$ and $u$ are discrete Fourier series 
\oo
$\sum_{n=1}^N a_n \sin (n m \phi) e^{i \omega t}$ with the first term being $\sin (m \phi + \phi_p)$. 
\bb
For $u_\phi$ and $u_r$, the mid-plane is problematic because it corresponds to a zero-crossing. 
Considering $z=0.8$, one also sees reasonable correspondence with the lowest Fourier mode for $u_\phi$ but $u_r$ is very different with almost no weight in the lowest Fourier mode.   
Close to the top (or bottom) boundary with $z = 0.95$, the nonlinearity of the wall mode is revealed where there is a sharp front-like feature appearing in $T$, $u_z$, and $u_\phi$ and a pulse-like structure for $u_r$. 
The emerging contribution of bulk-like excitations are visible in the azimuthal oscillations of order $2 \lambda_c$ at the leading edge of $u_r$ shown in the inset of Fig.~\ref{VertTemp}(h). \oo The shock-like features for $T$, $u_z$, and $u_\phi$ vanish at the upper boundary consistent with rigid and isothermal boundary conditions as shown in the insets of (f-g):  $T$ approaches zero consistent with global heat transport constraint, $u_z$ because the zero normal velocity BC, and $u_\phi$ because of the zero tangential BC.  The latter should be controlled by an Ekman boundary layer condition of order $\delta_E = \Ek^{1/2} = 0.001$ but more quantitatively there is a coefficient relating $\delta_E$ to the viscous boundary layer of order $3$ \cite{Guzman2022}. The blue dashed line in (g) is proportional to $e^{(1-z)/0.003}$ consistent with an Ekman boundary layer. \bb 
More detail on the $z$-dependence is provided in the Appendix (Fig.~\ref{RMSzprofile}) where we compute for $\Ra = 5 \times 10^8$: $\langle X \rangle_\text{rms}$ and $\langle X-X^{(1)} \rangle_\text{rms}$ where $X$ is the convective field and $X^{(1)}$ is its lowest Fourier mode. 

\oo
\subsection{Evolution of wall modes: nonlinearity and instability}
\label{subsec-Evolution}
Wall modes for $\Ek > 10^{-4}$ \cite{Zhong1993,Ning1993,Liu1999} have a narrow $\Ra$-window between their onset and the onset of bulk convection $\Ra_c/\Ra_\text{w} < 4$ whereas for $\Ek = 10^{-6}$ one has $\Ra_c/\Ra_\text{w} \approx 30$. 
Thus, the wall mode state becomes increasingly nonlinear and eventually time-dependent over that interval as noted previously \cite{Favier2020,Madonia2021}.  
In Figs.\ \ref{VertTemp}(a-d), representative instantaneous temperature fields $T(r=0.98R,\phi, z)$ are shown for increasing $\Ra$ at $\Ek = 10^{-6}$.  As $\Ra$ increases, the vertical profile is initially sinusoidal near onset ((a) $\Ra = 3 \times 10^7$) but becomes highly nonlinear ---  square-wave like and non-single-valued at the leading (left) and trailing (right) edge  ((b) $\Ra = 5 \times 10^7$ and (c) $\Ra = 5 \times 10^8$). Above $\Ra = 7 \times 10^8$, there are strong lateral spatial striations, ((d) $\Ra = 10^9$, near bulk onset) superimposed on the overall single mode structure.

\begin{figure}[th]
\unitlength1truecm
\begin{picture}(18,6)
\put(1.5,1.8){
\begin{tikzpicture}[style=thick]
\put(1,0.5){\includegraphics[height=4.6cm]{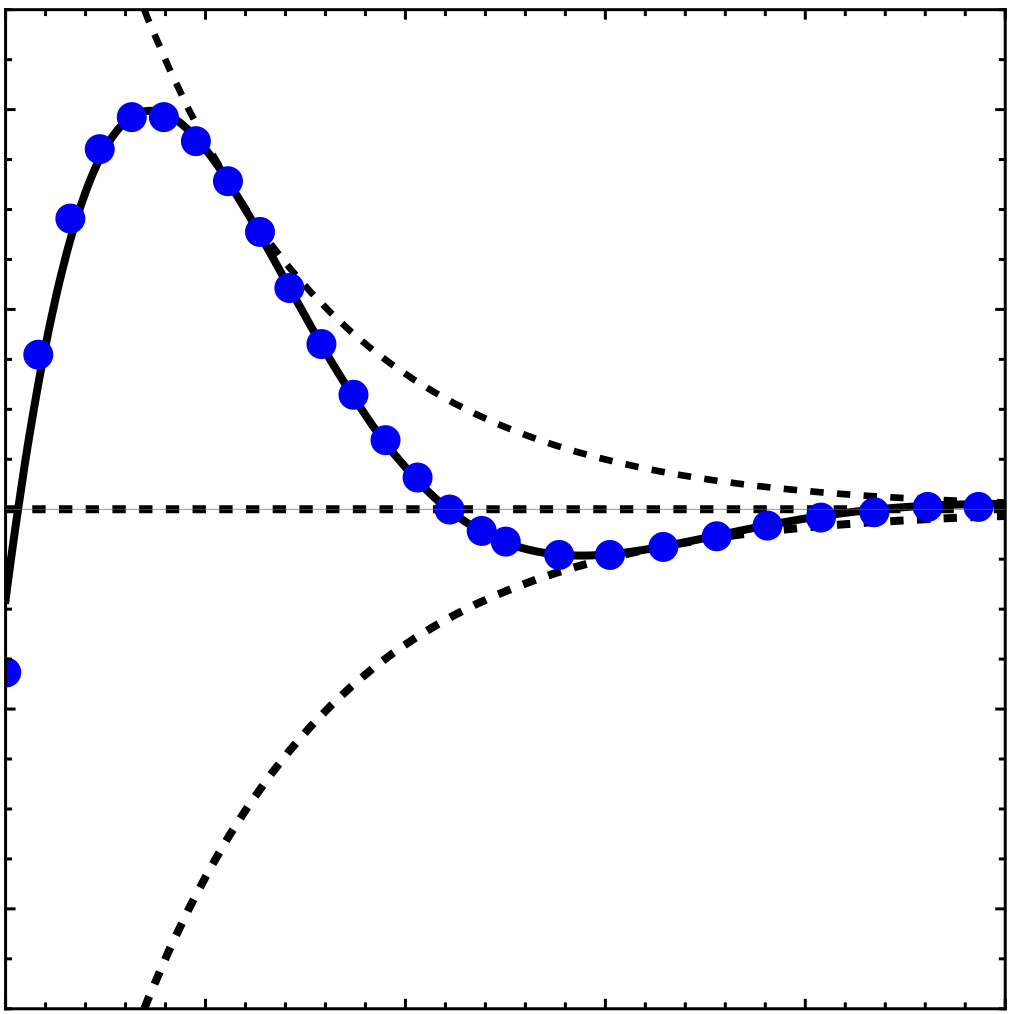}}
\put(7,0.5){\includegraphics[height=4.6cm]{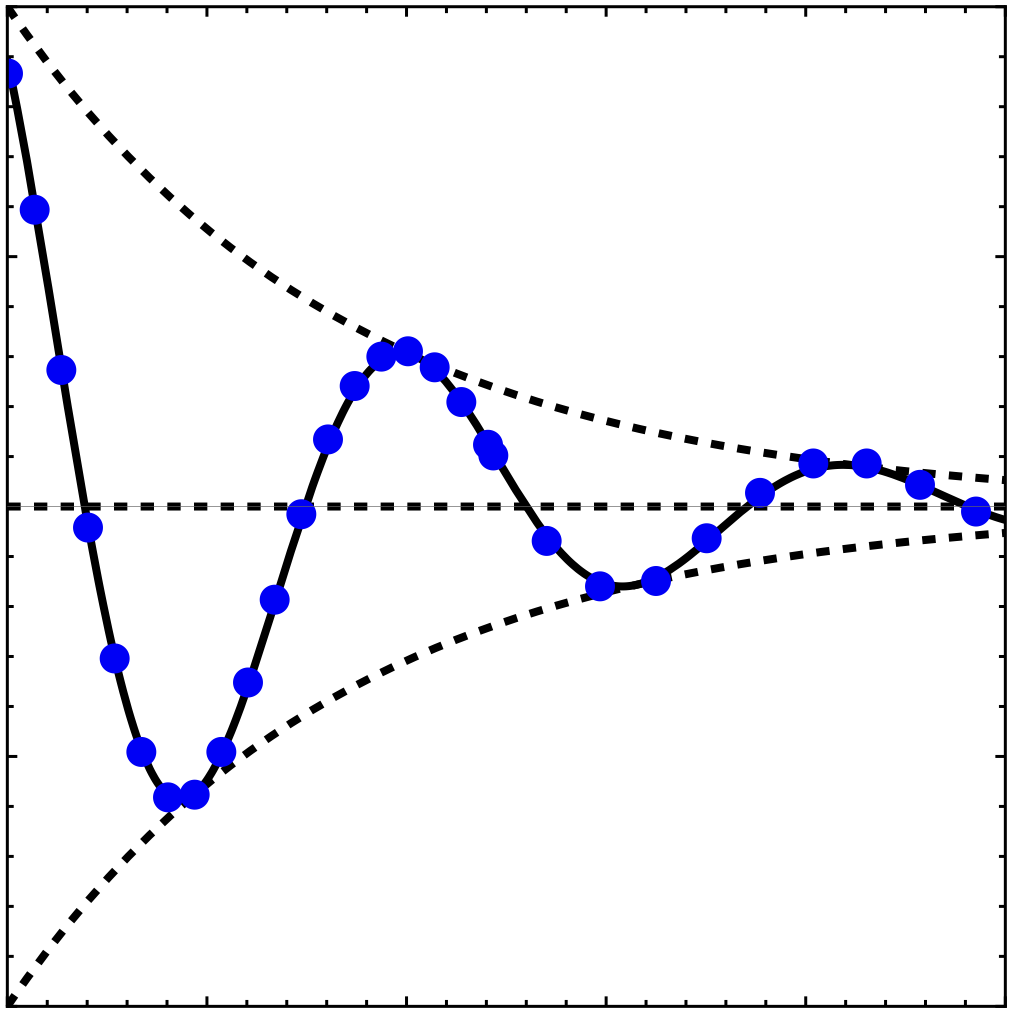}}
\put(13,0.5){\includegraphics[height=4.6cm]{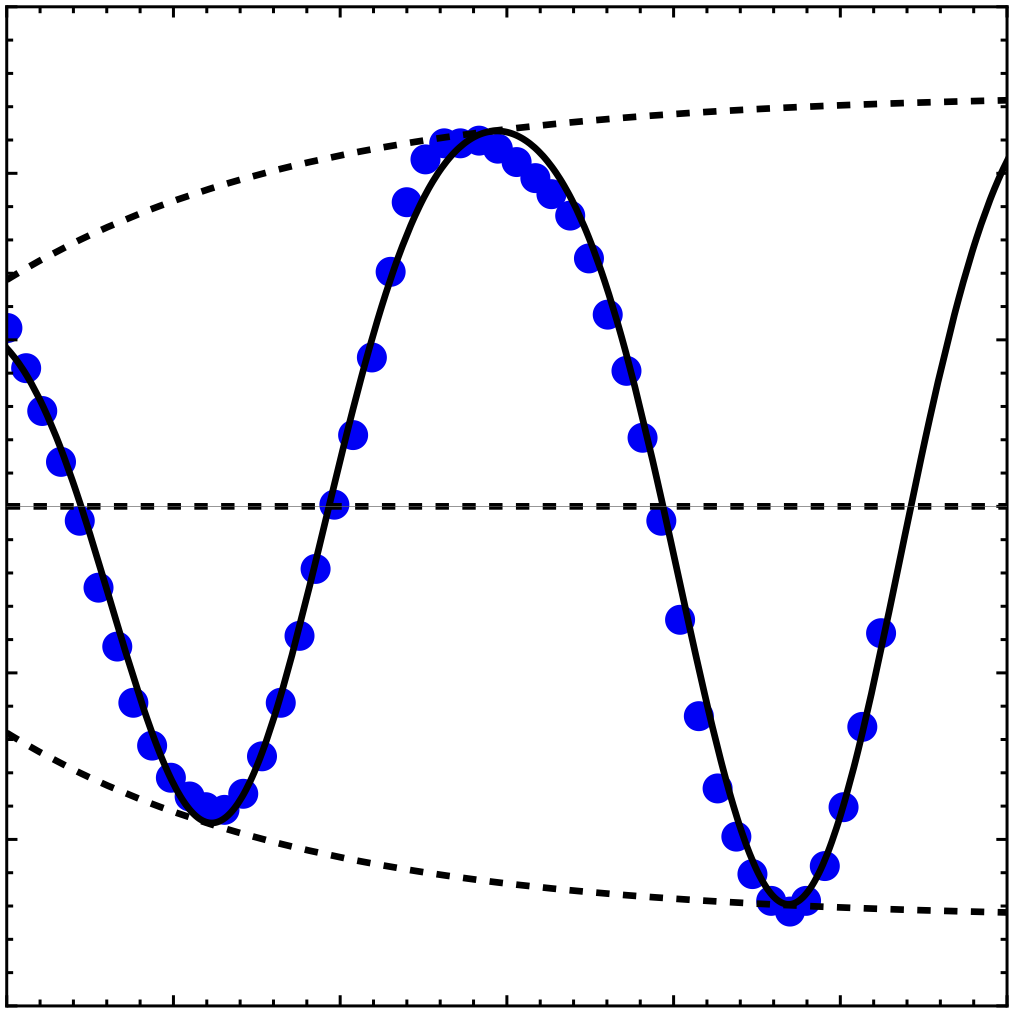}}
    \draw [|-|, color=blue](1.5,0.75) -- (5.05,0.75) ;
    \draw [|-|, color=blue](7.9,0.75) -- (9.9,0.75) ;
    \draw [|-|, color=blue](13.95,0.75) -- (16.62,0.75) ;
    \draw [|-|, color=black](2.8,4.85) -- (3.79,4.85) ;
    \draw [|-|, color=black](8.35,4.85) -- (9.8,4.85) ;
    \draw [|-|, color=black](13.25,4.85) -- (14.59,4.85) ;
    \end{tikzpicture}
}
\put(-0.2,6){(a)}
\put(-0.2,2.5){\rotatebox{90}{$\Nu(t)/\Nu(\infty)-1$}}
\put(0.5,5.3){0.4}
\put(0.5,4.4){0.2}
\put(0.5,3.5){0.0}
\put(0.3,2.6){$-0.2$}
\put(0.3,1.7){$-0.4$}
\put(1.1,1.2){0}
\put(1.8,1.2){100}
\put(2.7,1.2){200}
\put(3.62,1.2){300}
\put(4.53,1.2){400}
\put(5.45,1.2){500}
\put(3.3,0.7){$t$}

\put(6,6){(b)}
\put(6.5,6.0){1.0}
\put(6.5,4.9){0.5}
\put(6.5,3.78){0.0}
\put(6.3,2.65){$-0.5$}
\put(6.3,1.55){$-1.0$}
\put(7.1,1.2){0}
\put(7.8,1.2){100}
\put(8.73,1.2){200}
\put(9.64,1.2){300}
\put(10.55,1.2){400}
\put(11.46,1.2){500}
\put(9.3,0.7){$t$}

\put(12,6){(c)}
\put(12.5,6.03){1.5}
\put(12.5,5.28){1.0}
\put(12.5,4.55){0.5}
\put(12.5,3.8){0.0}
\put(12.3,3.0){$-0.5$}
\put(12.3,2.25){$-1.0$}
\put(12.3,1.55){$-1.5$}
\put(13.1,1.2){0}
\put(14.45,1.2){100}
\put(15.95,1.2){200}
\put(17.45,1.2){300}
\put(15.3,0.7){$t$}

\put(2.8,2){\color{blue}{$\tau_{p}=430$}}
\put(8.8,2){\color{blue}{$220$}}
\put(14.2,2){\color{blue}{$174$}}

\put(2.85,5.45){$\tau_0=100$}
\put(9,5.45){$170$}
\put(13.9,5.45){$91$}
\end{picture}
\caption{
Time dependence of the heat transport $\Nu(t)$ for $\Ra$: (a) $10^8$, (b) $2 \times 10^8$, and (c) $5 \times 10^8$. 
The dashed lines in (a) and (b) have the form $\Nu(\infty) + \Nu_0 \cos{(\omega_2 (t -t_0)}) e^{-t/\tau_0}$ where $\omega_2 =2\pi/\tau_2$ is the oscillatory frequency and $\tau_0$ is the decay time.  For (c) one has a more complicated relationship that accounts for large/small asymmetry $\sim\cos{^4(\omega_2(t-t_0))}$ and a transient growth $\sim 1-e^{-t/\tau_0}$.
The dashed lines are fits to the data using these forms, fitted values are indicated, and horizontal (blue/black) bars indicate the $\tau_2$ and $\tau_0$, respectively. 
}
\label{scoscill}
\end{figure}
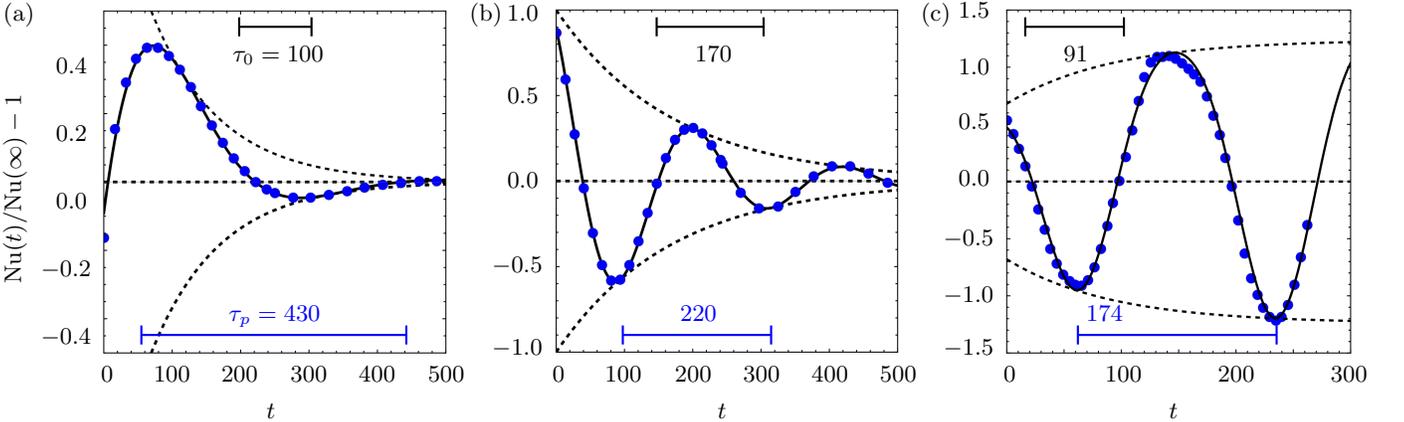

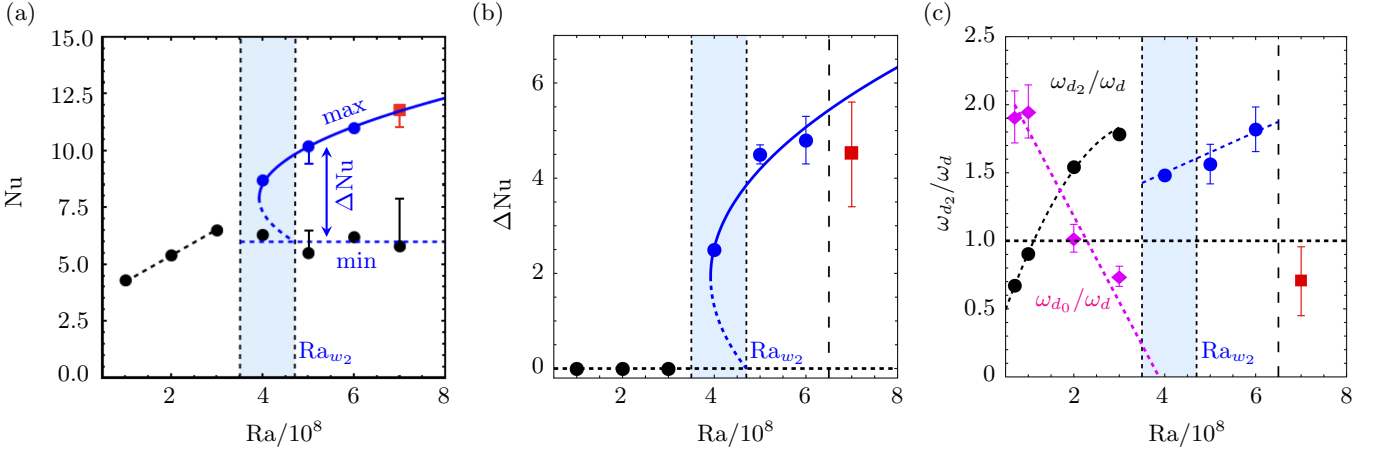
\begin{figure*}[th]
\unitlength1truecm
\begin{picture}(18,6)
\put(3.95,3.1){
\begin{tikzpicture}[style=thick]
\put(1,0.5){\includegraphics[height=4.6cm]{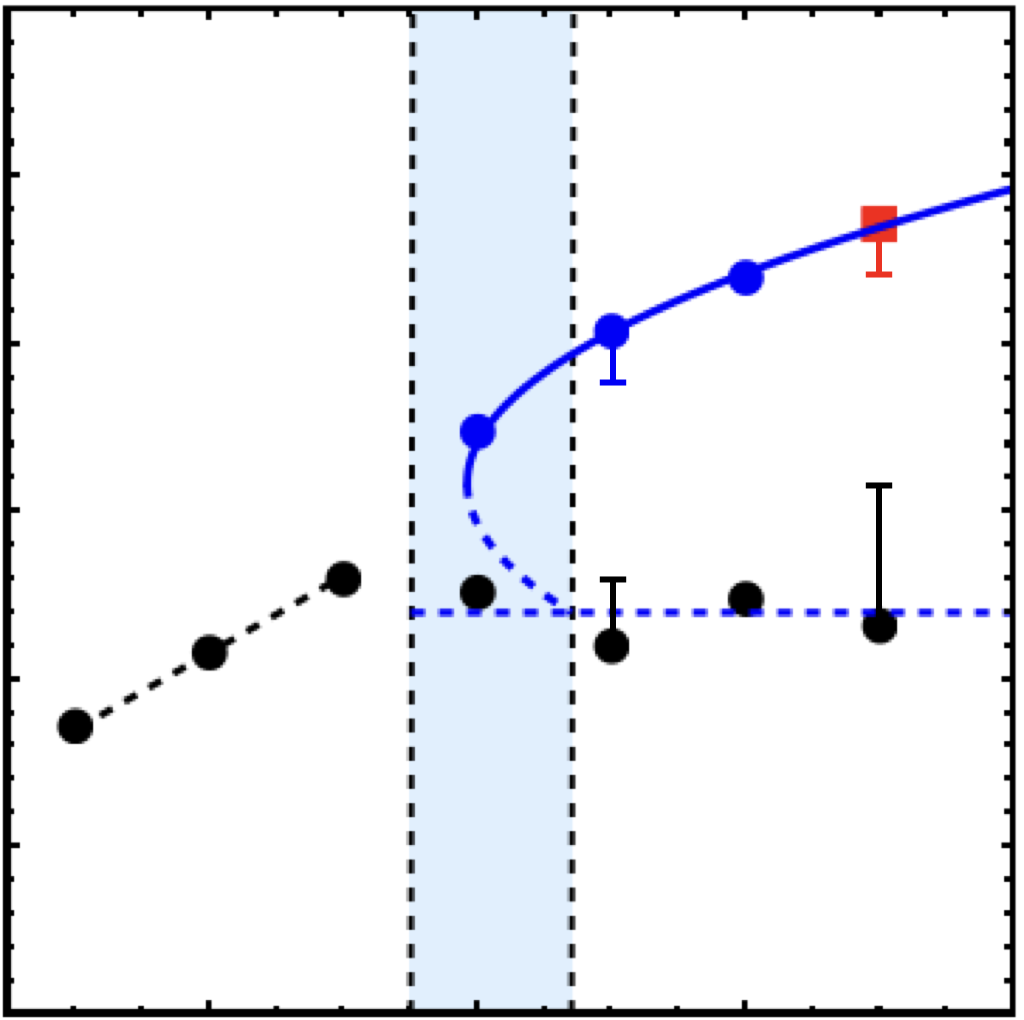}}
\put(7,0.5){\includegraphics[height=4.6cm]{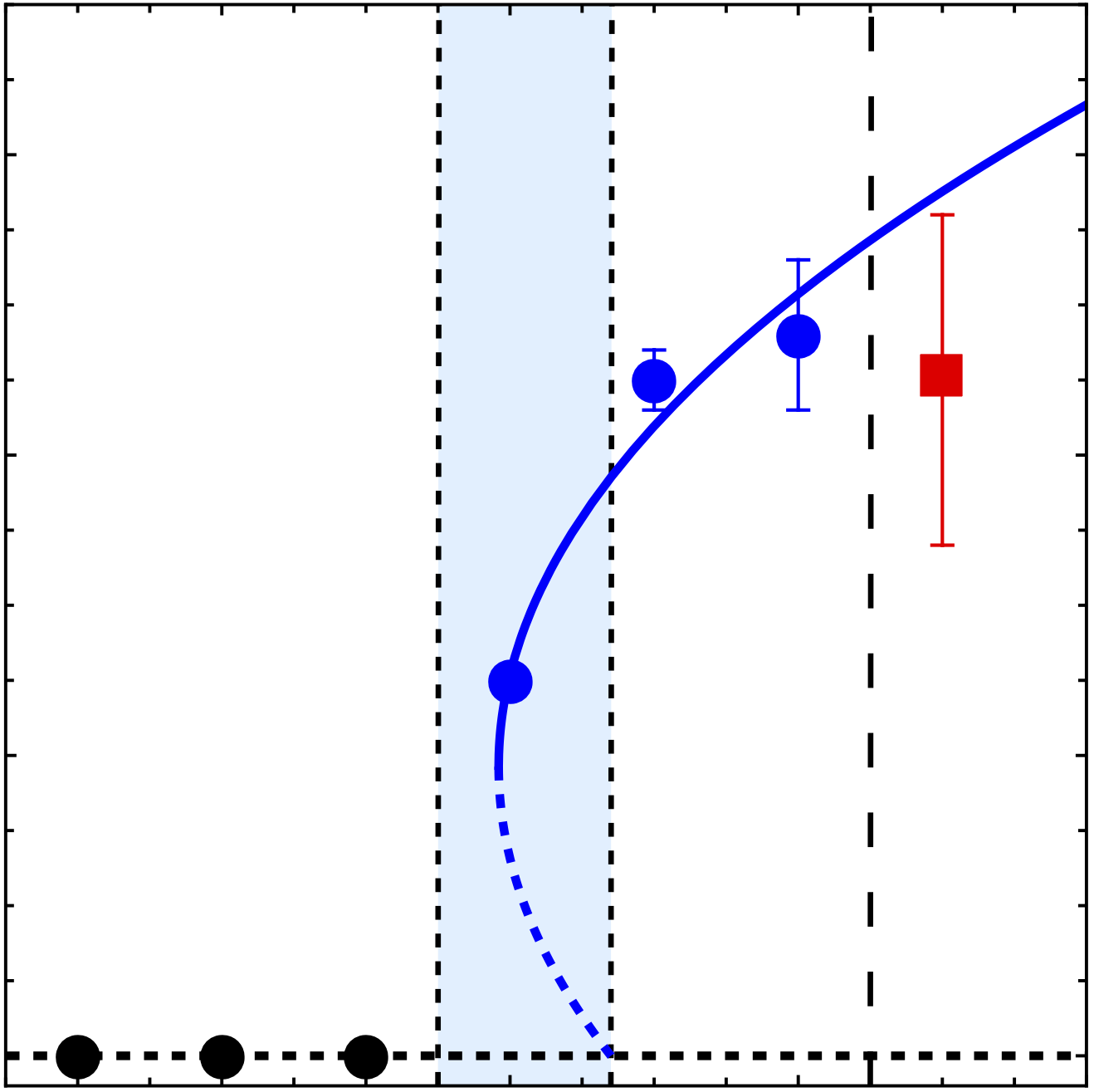}}
\put(13,0.5){\includegraphics[height=4.6cm]{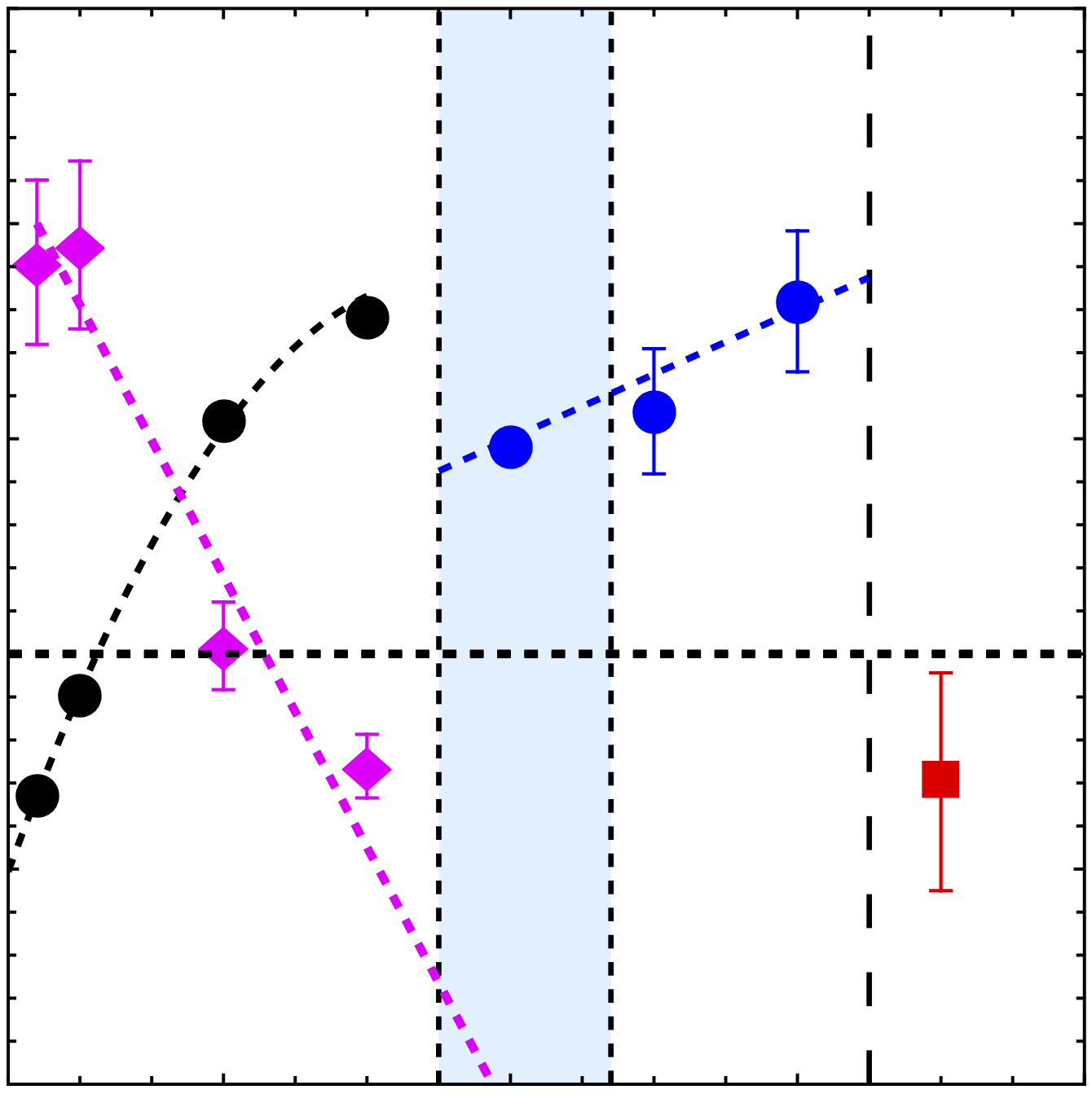}}
    \draw [stealth-stealth, color=blue](4,2.4) -- (4,3.6) ;
    \end{tikzpicture}
}
\put(-0.2,6){(a)}
\put(-0.2,3.5){\rotatebox{90}{$\Nu$}}
\put(0.4,5.65){15.0}
\put(0.4,4.9){12.5}
\put(0.4,4.15){10.0}
\put(0.5,3.4){7.5}
\put(0.5,2.65){5.0}
\put(0.5,1.9){2.5}
\put(0.5,1.2){0.0}
\put(1.9,0.9){2}
\put(3.15,0.9){4}
\put(4.35,0.9){6}
\put(5.55,0.9){8}
\put(3,0.4){$\Ra/10^8$}
\put(3.7,1.5){\color{blue}{$\Ra_{w_2}$}}
\put(4,4.6){\rotatebox{20}{\color{blue}{max}}}
\put(4.2,2.7){\rotatebox{0}{\color{blue}{min}}}
\put(4.2,3.45){\rotatebox{90}{\color{blue}{$\Delta\Nu$}}}

\put(6,6){(b)}
\put(6.3,3.2){\rotatebox{90}{$\Delta\Nu$}}
\put(6.8,5.075){6}
\put(6.8,3.8){4}
\put(6.8,2.55){2}
\put(6.8,1.3){0}
\put(7.9,0.9){2}
\put(9.1,0.9){4}
\put(10.35,0.9){6}
\put(11.58,0.9){8}
\put(9,0.4){$\Ra/10^8$}
\put(9.7,1.5){\color{blue}{$\Ra_{w_2}$}}

\put(12,6){(c)}
\put(12.1,3.2){\rotatebox{90}{$\omega_{d_2}/\omega_{d}$}}
\put(12.55,5.71){2.5}
\put(12.55,4.8){2.0}
\put(12.55,3.9){1.5}
\put(12.55,2.99){1.0}
\put(12.55,2.08){0.5}
\put(12.8,1.2){0}
\put(13.93,0.9){2}
\put(15.1,0.9){4}
\put(16.35,0.9){6}
\put(17.55,0.9){8}
\put(15,0.4){$\Ra/10^8$}
\put(15.7,1.5){\color{blue}{$\Ra_{w_2}$}}
\put(13.7,5.1){\color{black}{$\omega_{d_2}/\omega_{d}$}}
\put(13.5,2.2){\color{magenta}{$\omega_{d_0}/\omega_{d}$}}
\end{picture}
\caption{
(a) Extreme maximum and minimum values of $\Nu (t)$ (minima: black solid circles, maxima: blue solid circles and solid red square for chaotic state),
\oo
where the error bars depict the range of less extreme max/min values,
\bb
(b) $\Delta \Nu$ (mean peak-to-peak), and (c) normalized oscillation frequency $\omega_{d_2}/\omega_d$ (solid circles: black - stable state and blue - unstable state) and normalized decay frequency (scaled by 1/2 for comparison)  (1/2)$(2 \pi/\tau_{d_0})/\omega_d$ (magenta diamonds) vs $\Ra/10^8$ where $\omega_d$ is the wall mode precession frequency. Error bars denote variability of $\Delta \Nu$, $\omega_{d_2}/\omega_{d}$, and $(2 \pi/\tau_{d_0})/\omega_d$ owing to shorter or longer lengths of time series and/or unsteady oscillations. The blue shaded region denotes an approximate zone of subcritical instability.  The solid (dashed) lines are schematic suggestions for stable (unstable) behavior. The square (red) indicates a state where the oscillations have become chaotic with increased fluctuations and variable frequency. The dashed magenta line represents the stable state transient inverse time constant which is expected to vary as $\tau_0^{-1} \sim (\Ra_{w_2} -\Ra)$. }
\label{subcrit}
\end{figure*}

We characterize the transition to an oscillatory state through the time dependence of the heat transport which is time-independent for precessing wall modes.  For $10^8 \leq \Ra\leq 3 \times 10^8$, $\Nu(t)$ is a decaying oscillation of the form $\Nu \approx \Nu_\infty  + \Nu_0 \cos{\left(\omega_2 \left(t - t_0\right)\right)} e^{-t/\tau_0}$ starting from some random initial condition 
as shown in Fig.\ \ref{scoscill}(a-b). For $4 \times 10^8 \leq \Ra\leq 6 \times 10^8$, the waveform is approximately stationary (or slightly increasing saturation, see Appendix~\ref{subsec-Transients})  with amplitude $\Delta \Nu$ (mean peak-to-peak) and an oscillation frequency $\omega_2$ as in Fig.\ \ref{scoscill}~(c). \back The nature of the onset of time dependence is a subcritical bifurcation as shown in Fig.\ \ref{subcrit} where we plot  the max-min envelope of $\Nu(t)$, the difference $\Nu_{\max} - \Nu_{\min} = \Delta \Nu$, the oscillation frequency $\omega_{d_2}/\omega_d$, and the decay frequency $\omega_{d_0}/\omega_d$ versus $\Ra$ where $\omega_d$ is the precession frequency in units of $\Omega^{-1}$. 

We show in Fig.\ \ref{subcrit}~(a), the $\Nu(t)$ envelope defined by its maximum and minimum values as a function of $\Ra$.  For $\Ra \leq 3 \times 10^8$, $\Nu$ is time-independent and $\Nu_{\max} = \Nu_{\min}$. There is a rapid increase in $\Nu_{\max}$ for $\Ra \geq 4 \times 10^8$ whereas the minima stay roughly constant at $\Nu_{\min} \approx 6.3$, very close to its value at $\Ra = 3 \times 10^8$. The solid (dashed) blue line is an 3 parameter adjustable fit to the stable (unstable) portions of a subcritical bifurcation. This  schematic illustration of the nonlinear time-dependent amplitude (solid line) and the unstable subcritical region (dashed line) is presented in a conventional bifurcation diagram in Fig.\ \ref{subcrit}~(b). 
In Fig.\ \ref{subcrit}~(c), we show $\omega_{d_2}/\omega_d$ where the black circles and corresponding black dashed line indicate the oscillatory part of the decaying solutions associated with the time-independent state  whereas the blue circles and dashed line  show $\omega_{d_2}/\omega_d \sim  \Ra \approx 1.7$. The decay frequency $\omega_{d_0}/\omega_d \sim (\Ra_{w_2}-\Ra)$ (magenta diamonds and dashed line), where its zero intercept indicates the Rayleigh number $\Ra_{w_2}$ at which the wall mode state becomes unstable to infinitesimal perturbations. For $\Ra \geq  7 \times 10^8$, the waveform becomes chaotic with increasing fluctuations and irregular oscillation frequency. The inference from Fig.\ \ref{subcrit} (a) is that the minimum value  $\Nu_{\min}$ is when the system is close to the pure wall mode state whereas $\Nu_{\max}$ occurs when the lateral jet makes the maximal contribution to the heat transport.  We discuss this in more detail below when we consider the full range of $\Nu$. 

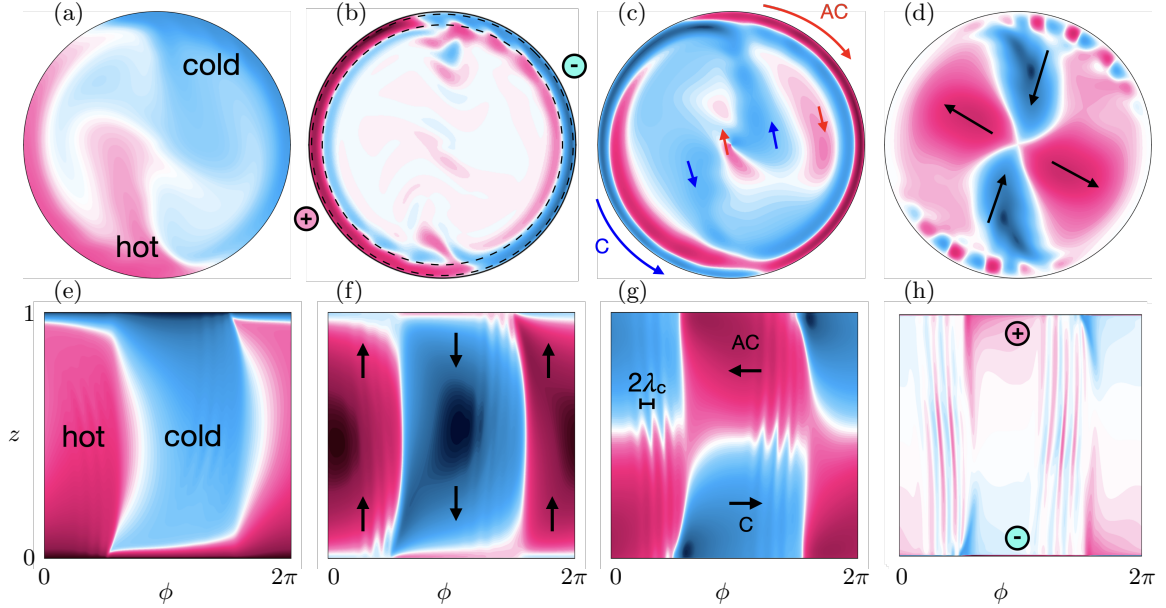
\begin{figure*}
\begin{center} 
\unitlength1truecm
\begin{picture}(18, 8)
\put(1.5,0.2){\includegraphics[width=15cm]{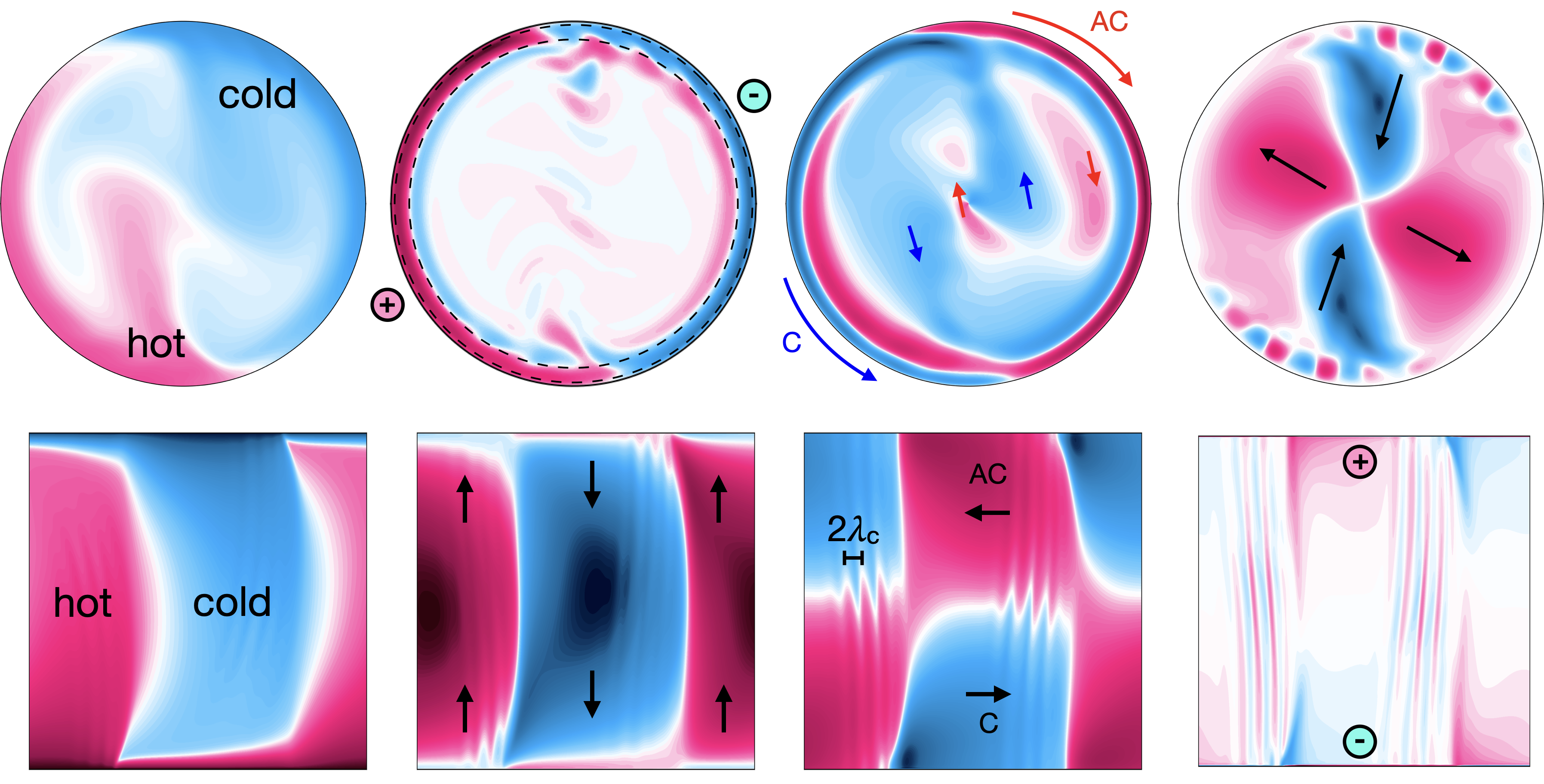}}
\put(1.9,7.5){(a)}
\put(5.65,7.5){(b)}
\put(9.4,7.5){(c)}
\put(13.2,7.5){(d)}
\put(1.9,3.8){(e)}
\put(5.65,3.8){(f)}
\put(9.4,3.8){(g)}
\put(13.2,3.8){(h)}
\put(1.5,3.5){1}
\put(1.3,1.9){$z$}
\put(1.5,0.25){0}
\put(1.7,-0.){0}
\put(4.8,-0.){$2\pi$}
\put(3.3,-0.2){$\phi$}
\put(5.55,-0.){0}
\put(8.65,-0.){$2\pi$}
\put(7,-0.2){$\phi$}
\put(9.3,-0.){0}
\put(12.4,-0.){$2\pi$}
\put(10.7,-0.2){$\phi$}
\put(13.1,-0.){0}
\put(16.2,-0.){$2\pi$}
\put(14.4,-0.2){$\phi$}
\end{picture}
\end{center}
\caption{For $\Ra=5 \times 10^8$, $\Ek = 10^{-6}$, and $\Pran=0.8$: horizontal cross sections of (a) $T(r,\phi)$ 
and (b-d) $u_z$ 
\oo
(dashed lines are at radii $r/R = 0.98$ and $0.90$, corresponding to the peak and the first zero crossing, respectively),
\bb
$u_\phi$, and $u_r$, respectively, for $z=0.8$. Red (blue) corresponds to hot (cold) and up (down),
\oo
AC -- anticyclonic (C -- cyclonic),
\bb
and radially out (radially in), respectively.  Vertical profiles (e-h) for corresponding fields ($\phi$, $z$) evaluated at $r/R=0.98$ 
\oo
(see (b) outer dashed line).
\bb
Arrows and $\pm$ symbols indicate qualitative motion directions. Spatial oscillations of order $2 \lambda_c$ are visible in horizontal cross-sections of $u_z$ and $u_r$ and in vertical azimuthal profiles of $u_z$, $u_\phi$, and $u_r$. Note that the major ejections of jets (and return flow) represented in $u_r$ happen near the top and bottom boundaries. }
\label{Ra5e8hs_V}
\end{figure*}

The time dependence of the wall mode was noted earlier \cite{Favier2020} for $\Pran =1$, $\Gamma = 3/2$, and $\Ek = 10^{-6}$ where  the emission of horizontally-propagating thermal plumes originating within the wall-mode region and moving into the interior were observed for $\Ra \gtrsim 5 \times 10^8$, consistent with our results
\oo
(see also \cite{Madonia2021, Wit2023}).
The mechanism for these fluctuations was hypothesized to result from a shear instability of a Stewartson layer of width $\sim\Ek^{1/4}$ \cite{Favier2020,Kunnen2011} that develops a net mean flow for $\Ra \gg \Ra_\text{w}$.  The criterion for instability in a differentially sheared rotating layer (not a wall-bounded flow) is $\Rey \approx 10$ to $20$ \cite{Fruh1999} which is qualitatively consistent with the computational results \cite{Favier2020}.  For comparison we get $\Rey$ values of 6 and 12 for $\Ra=2 \times 10^8$ and $5 \times 10^8$, respectively, based on the wall-bounded layer and 15 and 30 based on the outer free shear layer. Although the qualitative $\Rey$ are about right for shear instability of a zonal barotropic mode \cite{Busse1968}, a more detailed analysis was beyond their scope \cite{Favier2020}.  Our results show that small scale fluctuation structures appear in the wall-bounded zone of order $\Ek^{1/3}$, see Figs.\ \ref{Ra5e8hs_V} (d,g,h), rather than in the outer shear layer of order $\Ek^{1/4}$ (see also below for length scale of striations).  Further analysis and characterization of this instability is also beyond the scope of our work.
\bb
Here we elucidate the onset of this time-dependent degree of freedom for $\Pran = 0.8$, $\Gamma = 1/2$, and $\Ek = 10^{-6}$. 
In Fig.~\ref{Ra5e8hs_V}, we show instantaneous fields of horizontal and vertical profiles, respectively: (a,e) $T$, (b,f) $u_z$, (c, g) $u_\phi$, and (d, h) $u_r$ for $\Ra = 5 \times 10^8$, above the subcritical onset of time dependence. 
Figs.~\ref{Ra5e8hs_V}(a-d) show horizontal cross sections at (a) $z=1/2$ and (b-d) $z=0.8$ whereas (e-f) illustrate the wall mode region $r/R=0.98$ with vertical $0 \leq z \leq 1$ and horizontal $0 \leq \phi \leq 2\pi$. 
The horizontal fields show the lateral plume emission into the interior near the crossing point of positive/negative $u_\phi$ and $u_z$ whereas the vertical fields show a spatially oscillatory signature of order $2 \lambda_c$ that generates in the build-up to the emission. 
The oscillations are strongest for $u_\phi$ and $u_r$ compared to less obvious features in $T$ and $u_z$.  In the bulk region represented in (a-d), $u_z$ is quite small compared to its value near the wall whereas the other fields have significant contributions in the interior.  
Thus, although the contribution to $\Nu \propto u_z$ seems small; the perturbations in the interior region probably act as the foundation for the nucleation of bulk convection.  
At the same time, the vertical striations of order $2 \lambda_c$ seem to be related to bulk instability implying that the underlying wall mode acts as the foundation for the nucleation of bulk instability. 
Thus, the jet instability and the $2 \lambda_c$ bulk structures seem inextricably 
\oo
linked. 
The striation length scale (proportional to $\Ek^{1/3}$) suggests that the origins of the instability arise within the wall shear zone rather than in the outer free shear zone of order $\Ek^{1/4}$ \cite{Favier2020}.  
\bb
Other characteristics of the sub-critical state include unsteady precession and the qualitative change in the shape of the contours of $T$: compare Figs.~\ref{Contours}(b,~c).  

\subsection{Transition to bulk modes}\label{subsec-Transition}

\begin{figure*}
\begin{center} 
\unitlength1truecm
\begin{picture}(18, 8)
\put(1.5,0){\includegraphics[width=15cm]{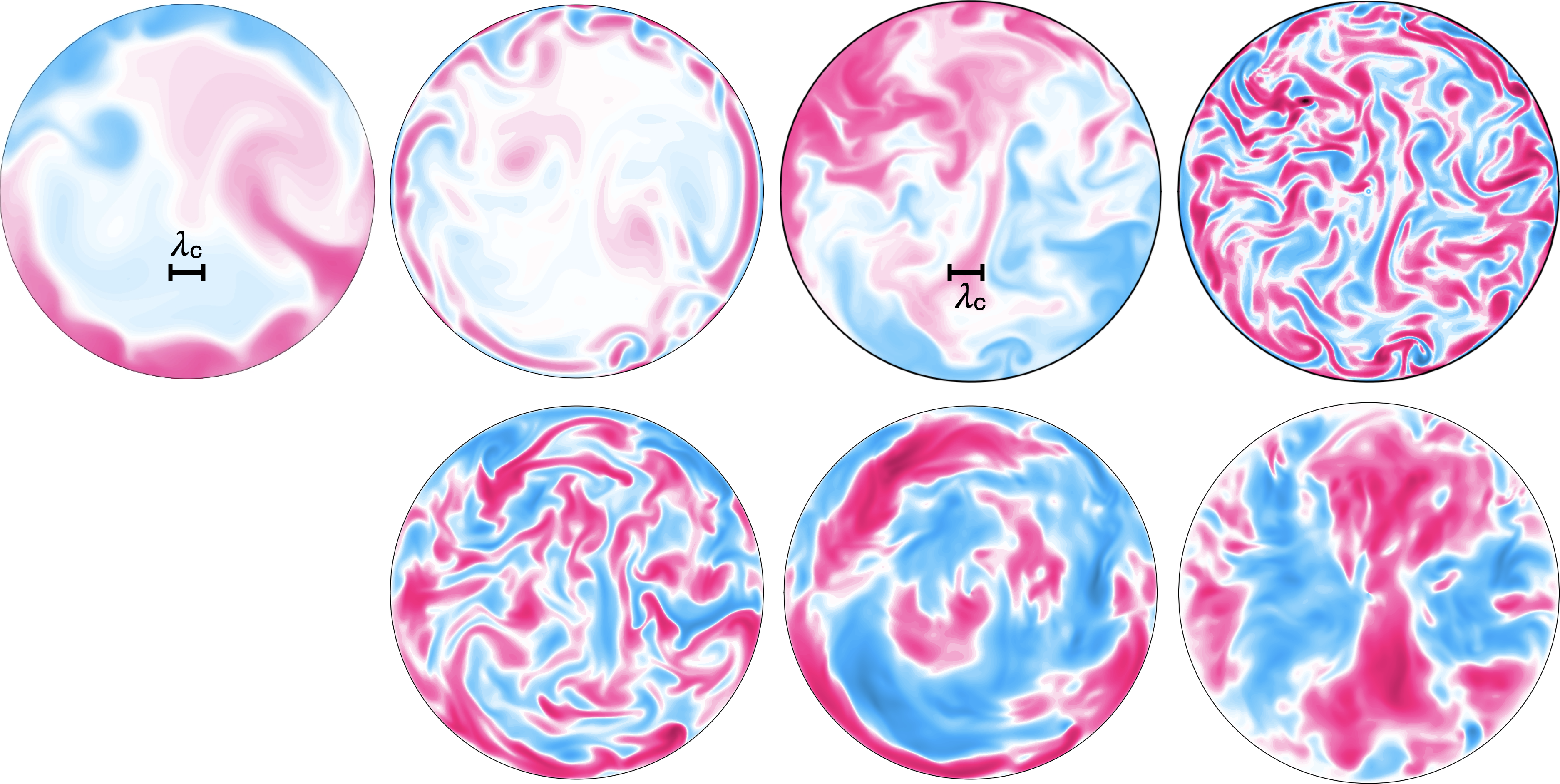}}
\put(1.9,7.5){(a)}
\put(4.5,7.5){$T$}
\put(5.6,7.5){(b)}
\put(8.2,7.5){$\omega_z$}
\put(9.3,7.5){(c)}
\put(11.9,7.5){$T$}
\put(13,7.5){(d)}
\put(15.6,7.5){$\omega_z$}
\put(16.7,5.6){$z=0.5$}
\put(5.6,3.6){(e)}
\put(8.2,3.6){$u_z$}
\put(9.3,3.6){(f)}
\put(11.9,3.6){$u_\phi$}
\put(13,3.6){(g)}
\put(15.6,3.6){$u_r$}
\put(16.7,1.8){$z=0.8$}
\put(1,1.8){
\begin{tabular}{|c|c|c|}
 \hline
$T$& {\color{magenta}warmer}&{\color{cyan}colder}\\
 \hline
$\omega_z$& {\color{magenta}cyclonic}&{\color{cyan}anti-cyclonic}\\
 \hline
$u_z$& {\color{magenta}up}&{\color{cyan}down}\\
 \hline
$u_\phi$& {\color{magenta}cyclonic}&{\color{cyan}anti-cyclonic}\\
 \hline
$u_r$& {\color{magenta}outwards}&{\color{cyan}inwards}\\
 \hline
\end{tabular}
}
\end{picture}
\end{center}
\caption{Horizontal cross-sections 
\oo with fields indicated and $z=0.5$ (top) and $z=0.8$ (bottom) as labeled on the far right. 
\bb
(a,c) $T$  (red -- warmer, blue -- colder 
than the arithmetic mean of the top and bottom temperatures),
(b,d) $\omega_z$  (red -- cyclonic, blue -- anti-cyclonic),
and (e-g) $u_z$ (red -- up, blue -- down), $u_\phi$ (red -- cyclonic, blue -- anticyclonic), $u_r$ (red -- outwards, blue -- inwards), respectively. 
$\Ra$: (a,b) $10^9$; (c-g) \oo $4 \times 10^9$ at the same time.\bb (a, c) Length scale depicted is bulk $\lambda_c$.}
\label{HorBulk}
\end{figure*}
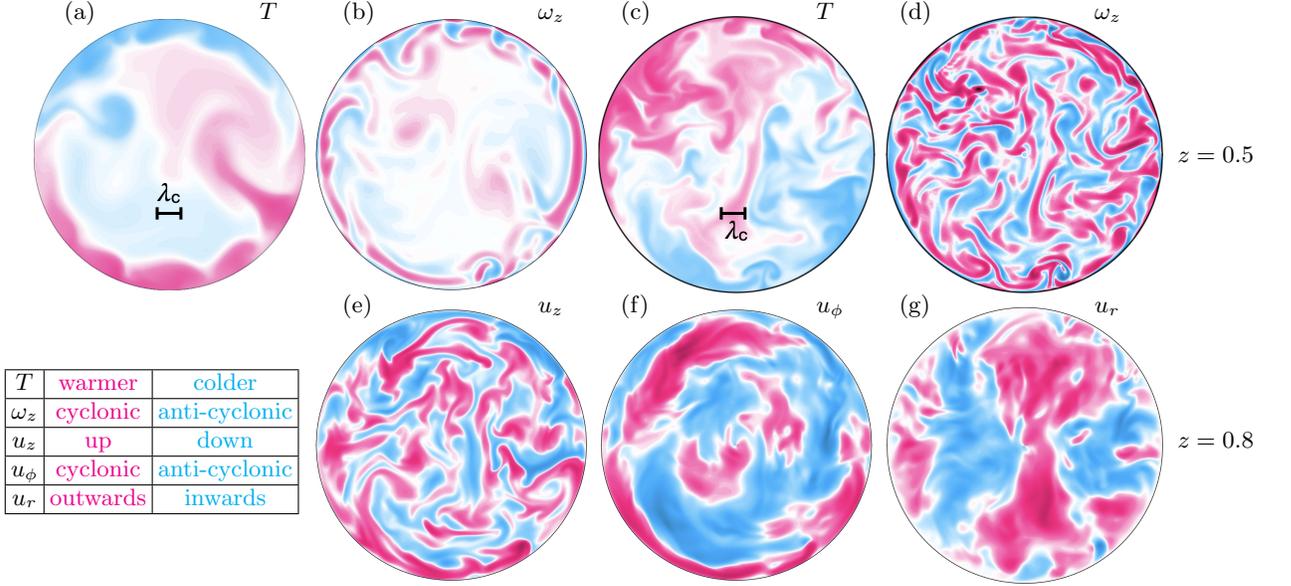

The transition to finite amplitude bulk convection occurs for $\Ra \approx 10^9$ as compared to the theoretical prediction of $\Ra_c \approx 7.8 \times 10^8$ for an infinite layer without sidewall effects. 
\oo
For comparison with the theoretical predictions of the NHQGS equations (see also below), the range  $8 \times 10^8 < \Ra \leq 5 \times 10^9$ corresponds to $1 \lesssim \Ra/\Ra_c \lesssim 6$ ($8 \lesssim \Ra \Ek^{4/3} \lesssim 60$) where the plume state is predicted to be prevalent for $\Pran =1$ \cite{Julien2012a,Julien2012b}.
\bb
In Fig.~\ref{HorBulk}, representative horizontal cross sections for instantaneous fields of $T$, $\omega_z$, $u_z$, $u_\phi$, and $u_r$ are shown ($z=1/2$ for $T$ and $\omega_z$ and $z=0.8$ for $u$). 
At $\Ra = 10^9$, there are strong lateral jets of contrasting thermal signature for $T$ in (a) which are generated near the intersection of cold and hot zones, i.e., where $T \approx \langle T \rangle$ and $u_z \approx 0$ (see Fig.\ \ref{VertTemp}). There are corresponding regions of cyclonic vorticity generation as shown for $\omega_z$ in (b).  Around the sidewall boundary, there are azimuthal variations of $T$ and $\omega_z$ with some spatially-intermittent structure.  When bulk convection is well established at $\Ra=5 \times 10^9$ as in Figs.\ \ref{HorBulk}(c,d) for $T$ and $\omega_z$, respectively, the BZF remains visible in the $T$ field but there are significant small scale temperature and vorticity fluctuations with spatial scales of the order of the linear length scale $\lambda_c = 4.8 \Ek^{1/3}$ (0.051 for $\Ek=10^{-6}$, see Eq. 7). In Figs.\ \ref{HorBulk}(e-g), $u_z$, $u_\phi$, and $u_r$, evaluated for $\Ra=4 \times 10^9$ and at the same time, show that $u_z$ displays little remnant of the horizontal jet flow but $u_\phi$ and $u_r$ show that even at $\Ra/\Ra_c \approx 4$, there are large scale motions suggesting residual emanations from the BZF region --- note in particular the in/out radial velocity structure in Fig.\ \ref{HorBulk}(g).

\begin{figure*}
\begin{center} 
\unitlength1truecm
\begin{picture}(18, 8)
\put(1.5,0){\includegraphics[width=15cm]{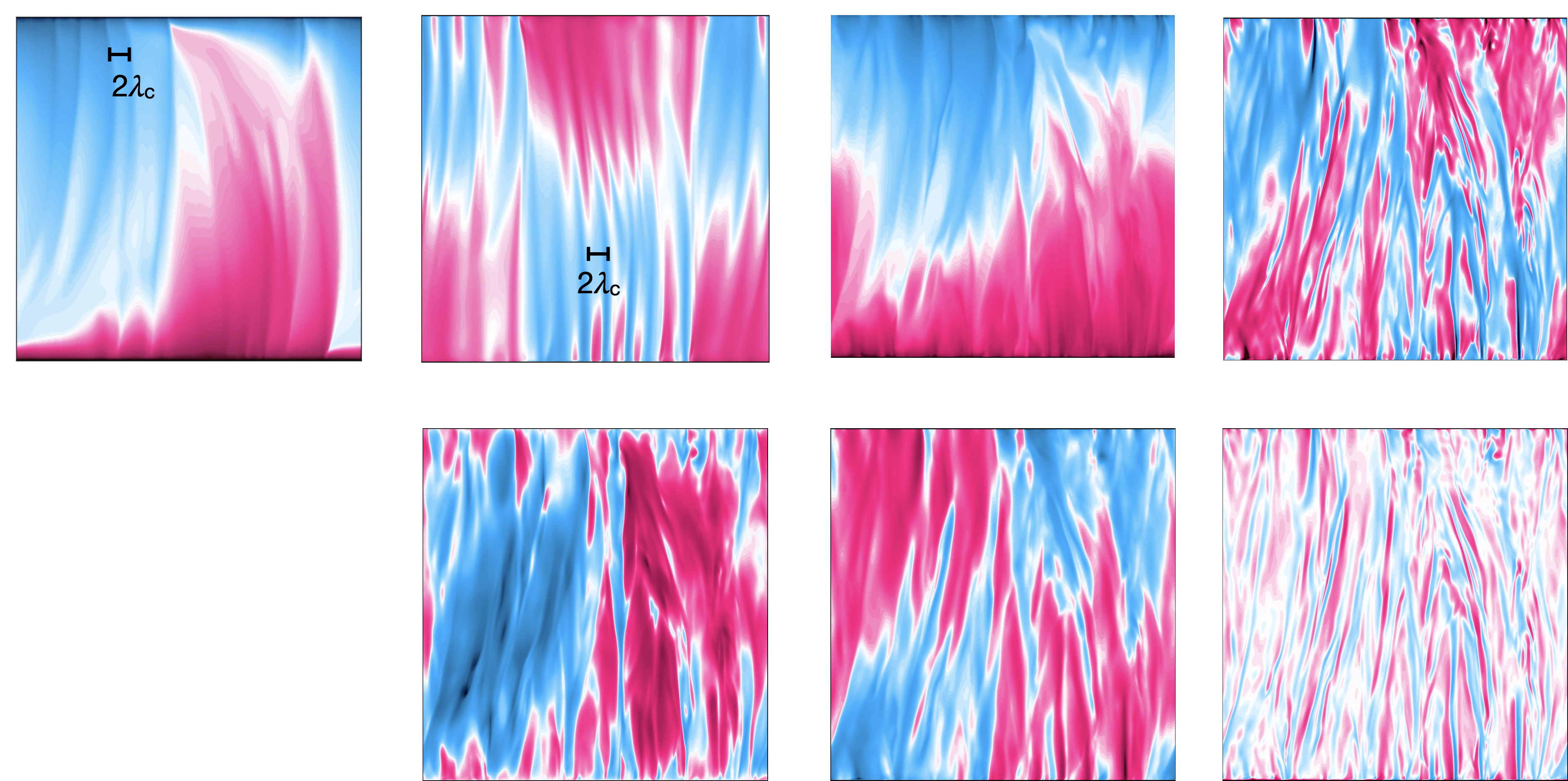}}
\put(1.8,7.5){(a)}
\put(4.4,7.5){$T$}
\put(5.7,7.5){(b)}
\put(8.3,7.5){$\omega_z$}
\put(9.7,7.5){(c)}
\put(12.3,7.5){$T$}
\put(13.4,7.5){(d)}
\put(16,7.5){$\omega_z$}
\put(16.8,5.6){$z=0.5$}
\put(5.7,3.5){(e)}
\put(8.3,3.5){$u_z$}
\put(9.7,3.5){(f)}
\put(12.3,3.5){$u_\phi$}
\put(13.4,3.5){(g)}
\put(16.,3.5){$u_r$}
\put(16.8,1.8){$z=0.8$}
\put(1,1.8){
\begin{tabular}{|c|c|c|}
 \hline
$T$& {\color{magenta}warmer}&{\color{cyan}colder}\\
 \hline
$\omega_z$& {\color{magenta}cyclonic}&{\color{cyan}anti-cyclonic}\\
 \hline
$u_z$& {\color{magenta}up}&{\color{cyan}down}\\
 \hline
$u_\phi$& {\color{magenta}cyclonic}&{\color{cyan}anti-cyclonic}\\
 \hline
$u_r$& {\color{magenta}outwards}&{\color{cyan}inwards}\\
 \hline
\end{tabular}
}
\end{picture}
\end{center}
\caption{
Vertical azimuthal slice at $r/R=0.98$ 
\oo with fields indicated and $z=0.5$ (top) and $z=0.8$ (bottom) as labeled on the far right.
\bb 
(a, c) $T$  (red -- warmer, blue -- colder), (b, d) $\omega_z$ , and (e) $u_z$ ,  (f) $u_\phi$, (g) $u_r$. Color: red (blue): warmer (colder), cyclonic (anti-cyclonic), up (down), cyclonic (anti-cyclonic), out (in), respectively.  $\Ra$: (a,b) $1 \times 10^9$ at the same time; (c-g) \oo $4 \times 10^9$ at the same time \bb; (a, c) length scale depicted is bulk $2 \lambda_c$.} 
\label{VertHorBulk}
\end{figure*}

The vertical profiles of the same fields evaluated for $r/R = 0.98$, Figs.\ \ref{VertHorBulk}(a-g) further characterize interactions of the boundary flow and the bulk interior flow. The vertically-coherent structures of $T$ and $\omega_z$ in (a,b), respectively, at $\Ra=10^9$ have a characteristic length of order $2 \lambda_c$ shown in both plots and the $m=1$ BZF remains quite coherent.  At $\Ra = 5 \times 10^9$, the BZF is instantaneously less coherent in $T$ (c) but with vertical structures with remaining vertical correlation. In the non-hydrostatic, quasi-geostrophic description \cite{Julien2012a,Julien2012b}, the state at $\Ra = 10^9$ corresponds to the cellular regime of bulk convection whereas $2 \times 10^9 \leq \Ra \leq 5 \times 10^9$ corresponds to the plume regime (ending at about $\Ra/\Ra_c \approx 6$, i.e., for $\Ra = 6 \times 10^9$ for our data at $\Ek = 10^{-6}$).  
The vertical profiles of the velocity fields (e-g) for $\Ra = 4 \times 10^9$ show similar spatial structure with some vertical coherence and a range of horizontal spatial scales. The BZF structure is again most visible in $u_z$ (e)  compared to $u_\phi$ (f) and $u_r$ (g). The spatial structure of $u_\phi$ is similar to $u_z$ whereas $u_r$ has  much finer spatial structures.  The $u_r$ field in particular shows the strong radially inward and outward exchange with the bulk interior flow.  To make these statements involving spatial length scales more quantitative, we compute the different measures of spatial length scale covering the range from the wall modes up to full bulk convection.

\subsection{Length scales of wall modes and BZF}\label{subsec-Length}

\begin{figure}[th]
\unitlength1truecm
\begin{picture}(18,7)
\put(5.7,1.1){\includegraphics[height=6cm]{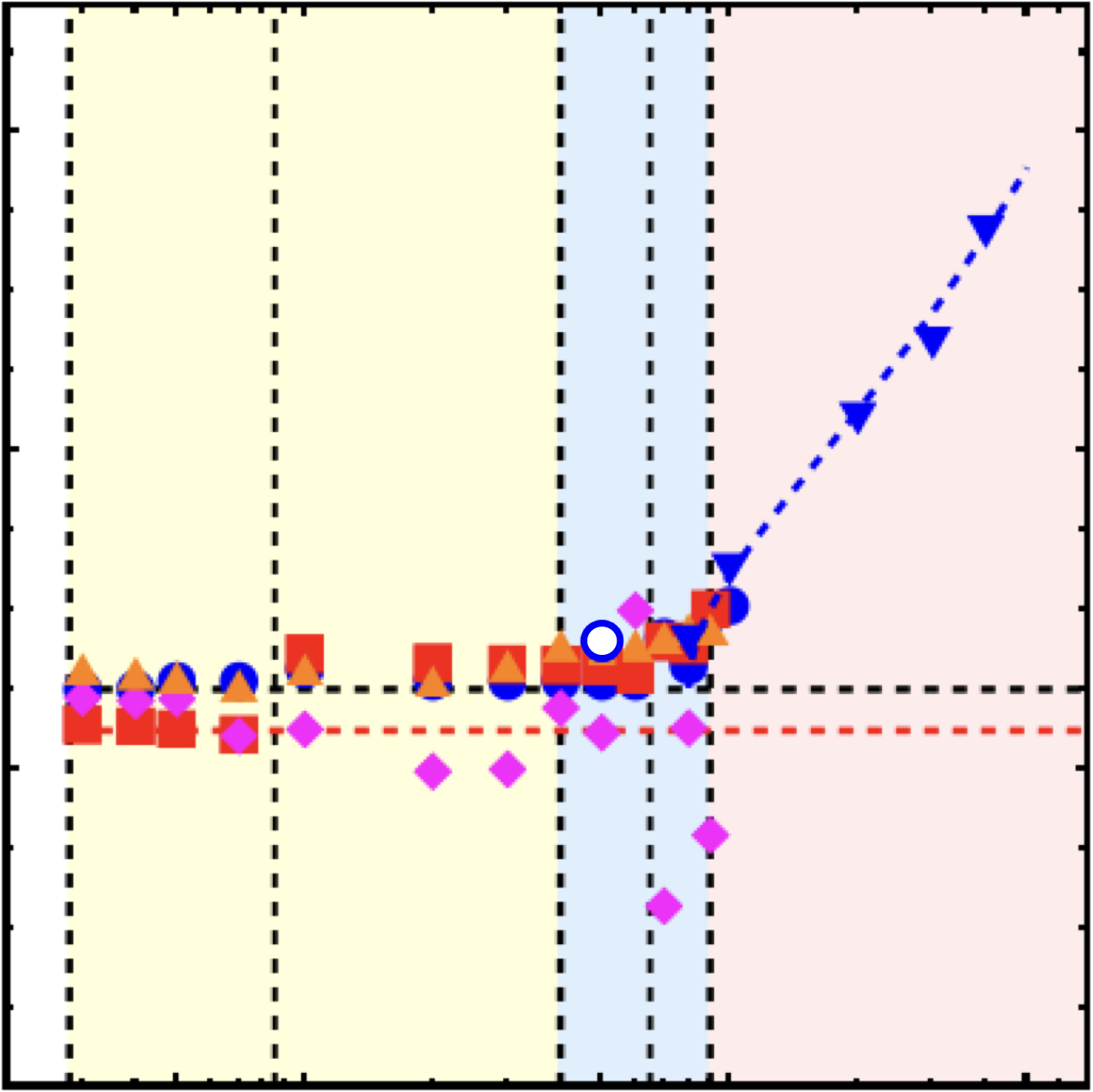}}
\put(4.5,-7.5){
\put(0.8,13.8){6}
\put(0.8,12.05){4}
\put(0.8,10.3){2}
\put(0.8,8.6){0}
\put(-0.1,10.7){\rotatebox{90}{$\delta \,\Ek^{-1/3}$}}
\put(2.1,8.2){5}
\put(2.7,8.2){10}
\put(4.35,8.2){50}
\put(4.9,8.2){100}
\put(3.7,7.6){$\Ra/10^7$}
\put(6.6,8.2){500}
}
\end{picture}
\caption{Radial widths $\delta\, \Ek^{-1/3}$ versus $\Ra/10^7$ for $\Ek=10^{-6}$: $\Gamma=1/2$ (closed symbols) and $\Gamma=2$ 
\oo
(open symbol); 
\bb
$\delta_{u_z}$ (blue circles), $\delta_{u_{\phi}}$ (red squares), $\delta_{u_r}$ (magenta diamonds), and $\delta_{\omega_z}$ (orange triangles), and $\delta_{u_{z_\text{rms}}}$ (blue triangles). 
The three regions are steady wall modes (yellow), sub-critical unsteady wall modes (blue), and BZF/bulk modes (pink).  The lateral (blue) dashed line is the length scale of the eigenfunction of $u_z$ of 2.5, the lateral red dashed line is the length scale of the eigenfunction of $u_\phi$ of 2.24, and the orange lateral line is for the radial width of $\omega_z$ of 3.3.  The values for $\delta_{u_z}$ for $\Ra \geq 9 \times 10^8$ are the rms values and 
\oo
the blue dashed curve is the function $\delta_{u_z} = 2.5 +0.16 (\Ra/10^7-80)^{1/2}$.
\bb
}
\label{PIC18}
\end{figure}
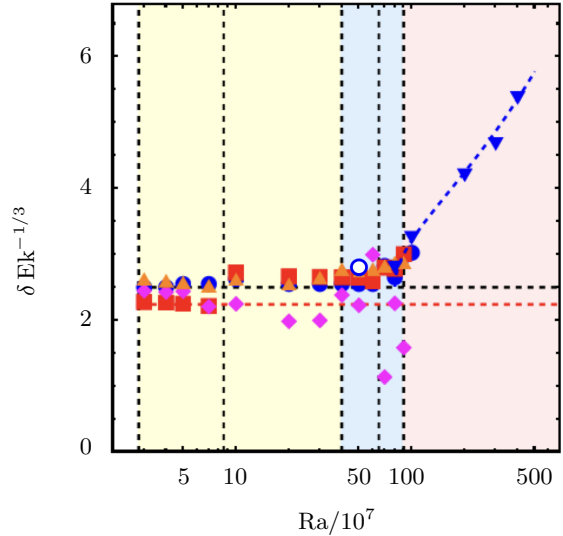

There are multiple different length scales to consider when spanning the regimes of steady wall modes, $3 \times 10^7 \leq \Ra \leq 3 \times 10^8$, unsteady sub-critical wall modes $4 \times 10^8 \leq \Ra \leq 8 \times 10^8$, and bulk modes $9 \times 10^8 \leq \Ra \leq 5 \times 10^9$. For our $\Gamma = 1/2$, the largest length scale is the $m=1$ wall mode state with $\lambda_1 = \pi/2$ which persists over the full range considered here.  The radial length scale of the wall-mode/BZF states can be found in different ways \cite{Zhang2020,Zhang2021}. Here we take a radial length defined either by the first zero-crossing $r_0$ of each field in the radial direction starting from the sidewall boundary ($r/R=1$),  for $u_r$, the radial position of the peak maximum (see Fig.\ \ref{uzFBessel}(d)), and for $\omega_z$ we take the second zero crossing.  Note that $\delta_u/H = (1-r_0/R)\Gamma/2 = (1-r_0/R)/4$.  We compute this in the precessing frame at the maximum  of the field in the azimuthal direction; the approach we used above for comparing with linear eigenfunctions \cite{Herrmann1993}.  
Previously, we measured the radial length scale of the wall-mode/BZF states by the first zero-crossing of $\langle u_\phi \rangle_{\phi,t}$ at the midplane \cite{Zhang2020,Zhang2021}. This does not work well for the wall mode state because $u_\phi (z)$ has an approximate 0 at the midplane, a radial length scale $\delta_{u_\phi}$ is about 1/2 of its value at, for example, $z=0.8$.  This  explains the apparent scaling $\sim \Ek^{2/3}$ of $\delta_0$ 
\oo
found earlier for a range of $\Ek$ \cite{Zhang2020,Zhang2021} instead of the expected $\Ek^{1/3}$ scaling. 
This approach works well even for unsteady precession in the sub-critical regime but there are insufficient data to get statistical convergence for the highly-intermittent BZF state in this study where strong bulk flow is present.  
In that case, we use the first minimum of the root-mean-square (rms) field $\langle u_z \rangle_{\phi, t}$ which should be at the approximate location of the first zero crossing used for the wall-mode regime.  
The results are shown in Fig.~\ref{PIC18} for $u_z$, $u_\phi$, $u_r$, and $\omega_z$ where the widths $\delta$ are normalized by $\Ek^{1/3}$ for comparison with the wall-mode linear eigenfunctions.
In the weakly nonlinear regime with $\Ra \leq 7 \times 10^7$, $\delta_{u_\phi} \approx 2.25\Ek^{1/3}$ consistent with the linear eigenfunction whereas for $\Ra \geq 10^8$, we have $\delta_{u_\phi} \approx \delta_{u_z} \approx 2.5\Ek^{1/3}$.
\bb
In contrast, $\delta_{\omega_z} \approx 3.3 \Ek^{1/3} \approx 1.3 \delta_{u_z}$ for $\Ra \leq 10^9$.  
Both $\delta_{u_\phi}$ and $\delta_{u_z}$ start to increase slightly in the region of subcritical instability.  
For $\Gamma = 2$ (blue, open circle), $\delta_{u_z} \approx \delta_{u_\phi}$ at $\Ra = 5 \times 10^8$. For $\Ra \geq 10^9$, $\delta_{{u_z}_\text{rms}}$ increases rapidly with $\Ra$. 
$\delta_{u_r}$ shows more variability, perhaps owing to the smaller amplitude of $u_r$ which might require better statistical averaging. 

\begin{figure}[th]
\unitlength1truecm
\begin{picture}(18,7)
\put(2.0,1.1){\includegraphics[height=4.9cm]{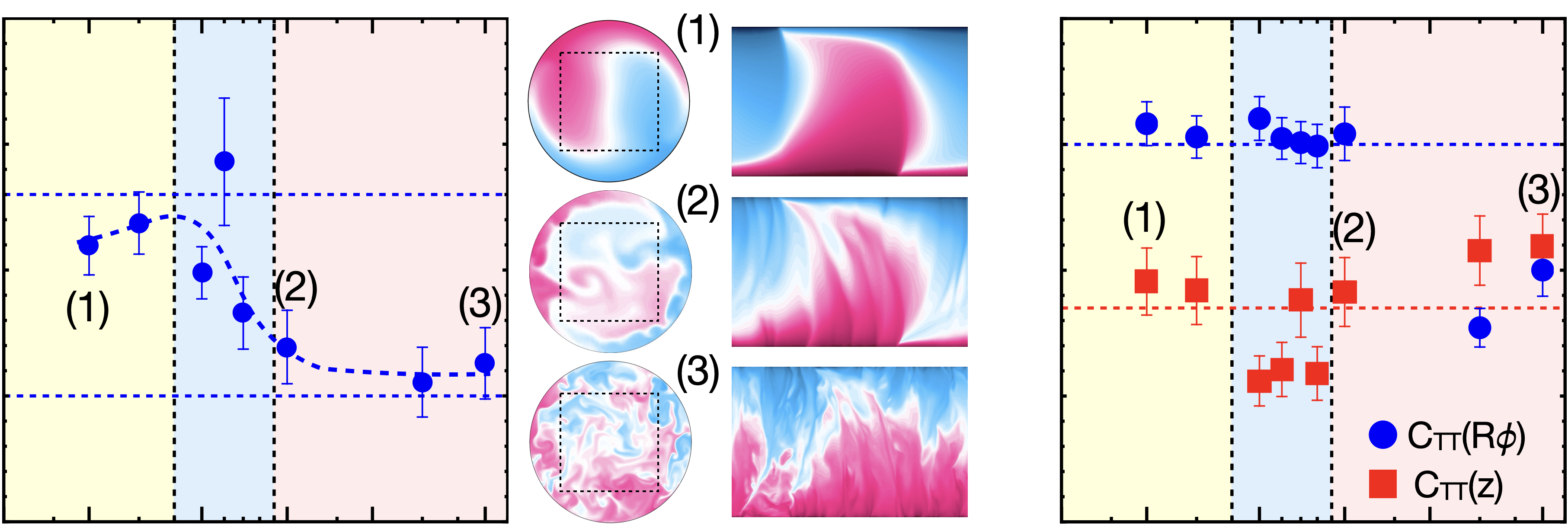}}
\put(1,3.2){\rotatebox{90}{$\delta_T \,\Ek^{-1/3}$}}
\put(4.1,0.4){$\Ra/10^7$}
\put(14,0.4){$\Ra/10^7$}
\put(4.2,6.15){(a)}
\put(9.3,6.15){(b)}
\put(14.0,6.15){(c)}
\put(1.8,1.1){0}
\put(1.8,2.2){5}
\put(1.6,3.35){10}
\put(1.6,4.5){15}
\put(1.6,5.7){20}
\put(11.65,1.1){0}
\put(11.45,2.2){10}
\put(11.45,3.35){20}
\put(11.45,4.5){30}
\put(11.45,5.7){40}
\put(1.9,0.85){10}
\put(2.7,0.85){20}
\put(3.7,0.85){50}
\put(4.4,0.85){100}
\put(5.2,0.85){200}
\put(6.25,0.85){500}
\put(11.75,0.85){10}
\put(12.55,0.85){20}
\put(13.55,0.85){50}
\put(14.25,0.85){100}
\put(15.05,0.85){200}
\put(16.1,0.85){500}
\end{picture}
\caption{(a) 
\oo 
Bulk length scales 
\bb 
$\delta_T \Ek^{-1/3}$ versus $\Ra/10^7$ for horizontal cross sections of $T$ at $z=1/2$ in a square domain with 46\% of the circular  area (shown on images in (b)). The labels (1), (2), (3) correspond to $\Ra$ values of images in (b). (b) False color images of $T$: (left) $T(x, y, z=1/2)$ and (right) $T(r/R=0.98, \phi, z)$. (1) $\Ra = 2 \times 10^8$, (2) $\Ra = 1 \times 10^9$, and (3) $\Ra = 5 \times 10^9$. (c) 
\oo
Wall-mode/BZF scales 
\bb
$\delta_T \Ek^{-1/3}$ vs. $\Ra/10^7$ for vertical sections $T(r/R=0.98, \phi, z)$: $C_{TT} (0.98 R \phi)$ (horizontal) and $C_{TT}(z)$ (vertical).
\oo
Unless labeled in the figure dashed lines (with corresponding colors) are guides to the eye. 
\bb
}
\label{deltaCorrTHV}
\end{figure}
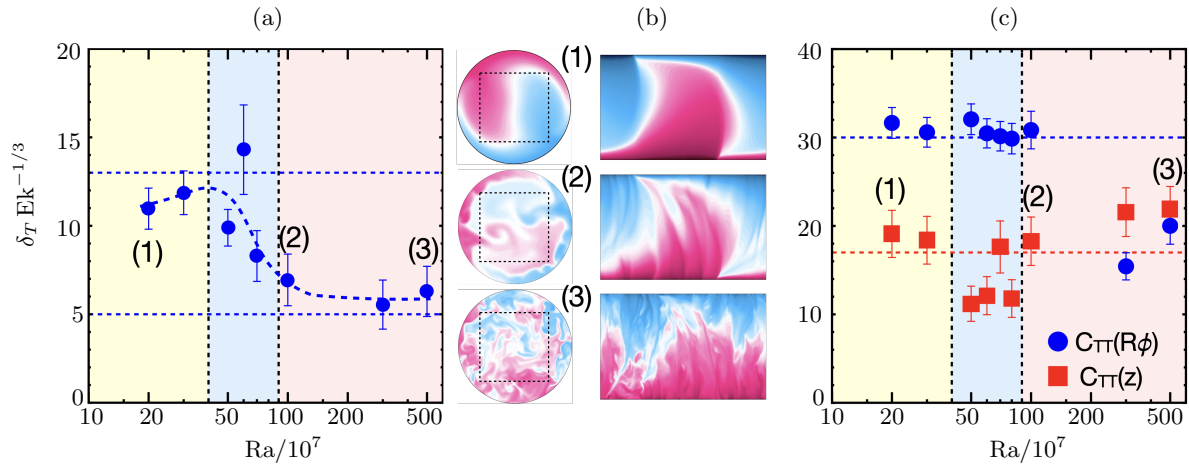

Other length scales are the horizontal length scales in the interior of the cell as measured in horizontal cross-sections and the horizontal and vertical length scales in the azimuthal and vertical direction around the circumference in the sidewall boundary region. The first is characteristic of bulk convection whereas the second is reflective of BZF--bulk interactions. For horizontal cross sections, we compute the 2D auto-correlation function of the field $F$ as $C_{FF}(r) \equiv \langle F({\bf x}) F({\bf x+r})\rangle_{A,t}/\langle F^2({\bf x})\rangle_{A,t}$ in a centered rectangular domain with dimensions $0.6 D \times 0.6 D$, where $\langle\cdot\rangle_{A,t}$ means averaging over a horizontal cross-section and in time.
We define the length scale $\delta$ such that $C_{FF}(\delta) = 0.25$ with uncertainty defined by $d\delta = \pm (d C(r)/dr)^{-1} \Delta C = \pm 0.05 (d C(r)/dr)^{-1}$ (as compared to \cite{Madonia2021} who use $\delta = \int_0^\infty C(r) dr$ so our lengths are a bit larger).  
In Fig.~\ref{deltaCorrTHV}(a), we show $\delta_T \Ek^{-1/3}$ versus $\Ra$ with corresponding representative images (left) in (b) for (1) $\Ra = 2 \times 10^8$, (2) $\Ra = 1 \times 10^9$, and (3) $\Ra = 5 \times 10^9$.  The size of the interior domain of area $0.36 D^2$ is shown.  In the steady wall-mode region, $\delta_T \approx 13$ which corresponds roughly to the exponential decay shown in Fig.\ \ref{ThetaEigenGamma}(a). In the unsteady subcritical regime, $\delta_T$ decreases owing to an increasing fraction of lateral spatial structures with smaller length scale.  In the bulk regime for $\Ra \geq 1 \times 10^9$, $\delta_T \Ek^{-1/3}  \approx \lambda_c \Ek^{-1/3} = 5.1$ characteristic of bulk rotating convection.

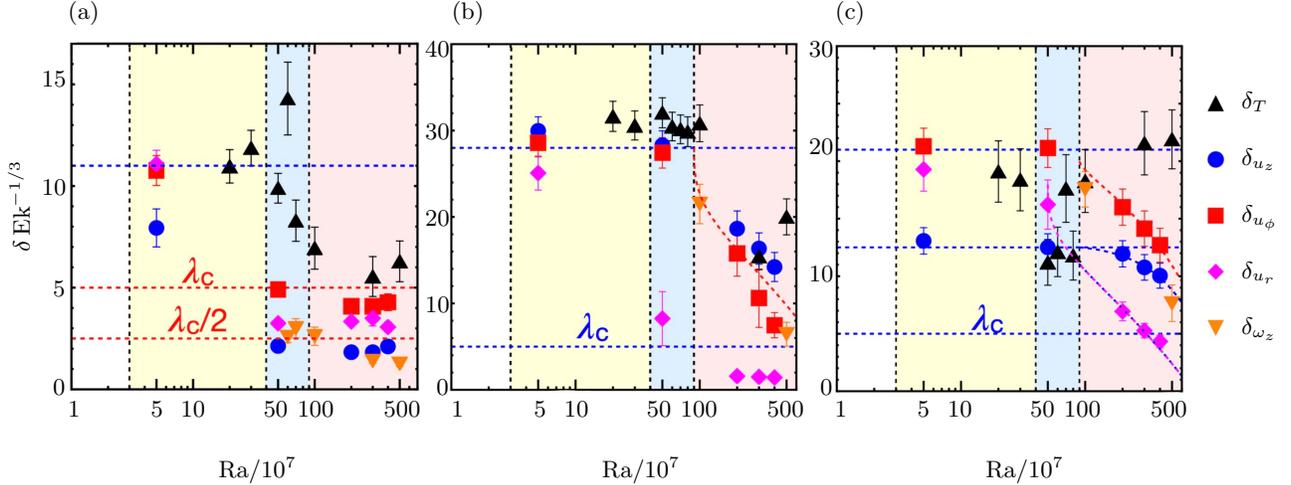
\begin{figure}[th]
\unitlength1truecm
\begin{picture}(18,7)
\put(1.3,1.3){\includegraphics[height=4.7cm]{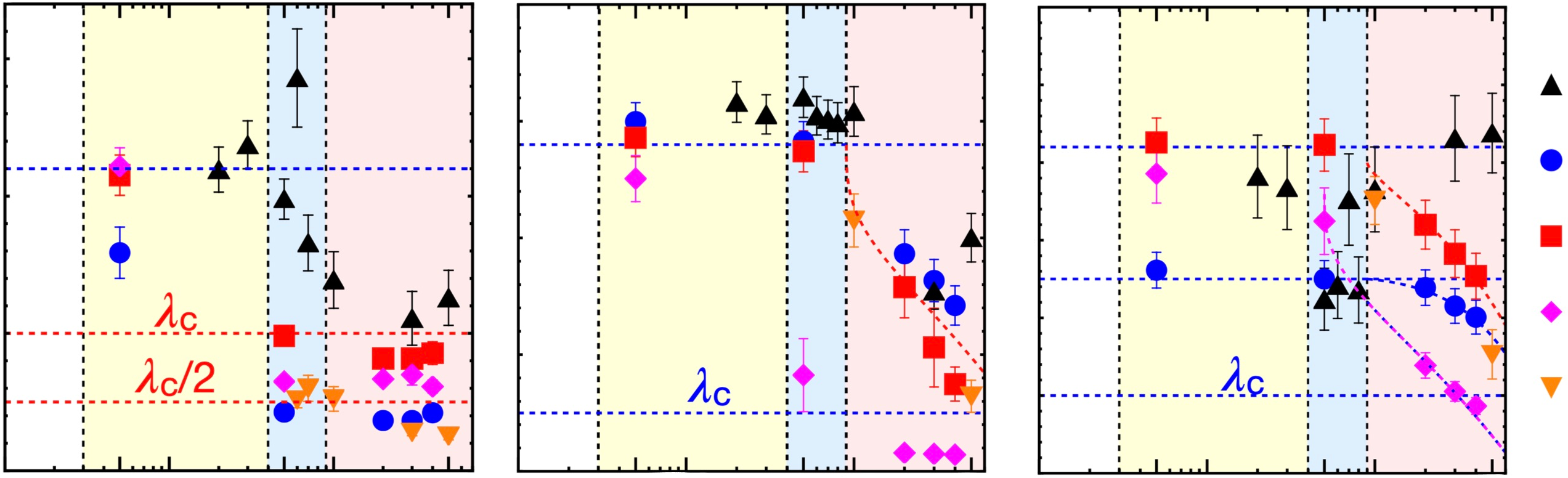}}
\put(0.5,3.2){\rotatebox{90}{$\delta \,\Ek^{-1/3}$}}
\put(3.3,0.2){$\Ra/10^7$}
\put(8.4,0.2){$\Ra/10^7$}
\put(13.5,0.2){$\Ra/10^7$}
\put(1.3,6.3){(a)}
\put(6.4,6.3){(b)}
\put(11.5,6.3){(c)}
\put(16.9,5.1){$\delta_T$}
\put(16.9,4.35){$\delta_{u_z}$}
\put(16.9,3.6){$\delta_{u_\phi}$}
\put(16.9,2.85){$\delta_{u_r}$}
\put(16.9,2.1){$\delta_{\omega_z}$}
\put(1.1,1.3){0}
\put(1.1,2.6){5}
\put(0.95,3.95){10}
\put(0.95,5.4){15}
\put(6.2,1.3){0}
\put(6.05,2.4){10}
\put(6.05,3.55){20}
\put(6.05,4.7){30}
\put(6.05,5.8){40}
\put(11.3,1.3){0}
\put(11.15,2.8){10}
\put(11.15,4.5){20}
\put(11.15,5.8){30}

\put(1.3,1){1}
\put(2.4,1){5}
\put(2.8,1){10}
\put(3.9,1){50}
\put(4.35,1){100}
\put(5.5,1){500}

\put(6.4,1){1}
\put(7.5,1){5}
\put(7.9,1){10}
\put(9,1){50}
\put(9.45,1){100}
\put(10.6,1){500}

\put(11.5,1){1}
\put(12.6,1){5}
\put(13,1){10}
\put(14.1,1){50}
\put(14.55,1){100}
\put(15.7,1){500}

\end{picture}
\caption{
Auto-correlation lengths $\delta\,\Ek^{-1/3}$ vs $\Ra/10^7$: $\delta_T$ (black triangles), $\delta_{u_z}$ (blue circles),  $\delta_{u_\phi}$ (red squares), $\delta_{u_r}$ (magenta  diamonds), $\delta_{\omega_z}$ (orange inverted triangles). (a) 
\oo
Bulk length scales from 
\bb
horizontal cross sections of $T$ and $\omega_z$ at $z=1/2$ and $u_z$, $u_\phi$, and $u_r$ at $z=0.8$ in a square domain with 46\% of the circular  area,  (b) 
\oo 
wall-mode/BZF horizontal auto-correlation lengths 
\bb 
for vertical sections at $r/R=0.98$ and (c) wall-mode/BZF vertical correlation lengths for vertical sections $r/R=0.98$. Averaging as in Fig.~\ref{deltaCorrTHV}. 
\oo
Unless labeled in the figure the dashed lines (with corresponding colors) are guides to the eye. 
\bb
}
\label{deltaCorrTuHV}
\end{figure}

We also compute the 1D correlation functions for the anisotropic directions in a vertical azimuthal slice $T(r/R=0.98, \phi, z)$:  $C_{TT}(R \phi) =\langle \langle T(r/R=0.98, \theta, z) T(r/R=0.98,(\theta+\phi,z))\rangle_z/\langle T^2\rangle_z \rangle_t$ and $C_{TT}(z) =\langle \langle T(r/R=0.98, \phi, y) T(r/R=0.98, \phi, y+z)\rangle_\phi/\langle T^2\rangle_\phi \rangle_t$.  The results are shown in Fig.~\ref{deltaCorrTuHV}(c) with corresponding representative images in (b) (right). The azimuthal length scale is dominated by the $m=1$ sinusoidal character of the wall mode and is about 30 up to the onset of bulk convection --- a $m=1$ cosine function has a correlation length of about 20 in these units. In the bulk regime, it drops to about 17.  The vertical length scale is roughly independent of $\Ra$ which is consistent with the main vertical variation of $T$ being a linear gradient with a corresponding slope independent correlation length of about 25 (the correlation length of a linear gradient is independent of its slope).  Instead, the value is somewhat lower, about 18-20.

To obtain a more complete description of the length scales defined above, we apply the auto-correlation analysis to the velocity and vertical vorticity fields for horizontal cross sections and for the horizontal and vertical directions in a azimuthal slice near the sidewall.  In particular, we use for the horizontal sections, 
$z=0.8$
for $u_z$, $u_\phi$, $u_r$, and $\omega_z$ to avoid the approximate vertical zero crossing of $u_\phi$ at 
$z=1/2$. We show in Figs.\ \ref{deltaCorrTuHV}(a-c) the correlation lengths $\delta \Ek^{-1/3}$ versus $\Ra/10^7$ for (a) horizontal cross sections, (b) horizontal variations in a vertical azimuthal profile, and (c) vertical variations in a vertical azimuthal profile. The length scales for horizontal cross sections all show the same overall trend with a decrease from about 10 starting at $\Ra \approx  5 \times 10^8$ to between 2 and 5  corresponding to $\lambda_c/2$ and $\lambda_c$, respectively, for higher $\Ra$. Whereas, $\delta_T$ decreases gradually over the subcritical regime, the velocities and $\omega_z$ abruptly decrease and maintain roughly constant values through the subcritical and bulk regimes $5 \times 10^8 \leq \Ra \leq 5 \times 10^9$. This suggests that the subcritical wall-mode instability is related to the bulk instability with structures of the order of the linear stability length scale.  The slower decline of $\delta_T$ is related to the longer (exponential) width of the temperature field relative to the radially-confined velocity and $\omega_z$ fields.

The horizontal and vertical lengths in the sidewall boundary region, Figs.\ \ref{deltaCorrTuHV}~(b,c), are more complicated to interpret.  The length scales of $T$, $u$ and $\omega_z$ are consistently about 28 for $\Ra \leq \Ra_c \approx 1 \times 10^9$ with the exception of $\delta_{u_r} \approx 8$ in the subcritical regime. 
For $1 \times 10^9 \leq \Ra \leq 5 \times 10^9$, $\delta_T \approx \delta_{u_z} \approx \delta_{u_\phi} \approx \delta_{\omega_z} \sim F(\Ra/\Ra_c -1)$ (assuming a dependence on $\Ra/\Ra_c -1$).  An empirical approximation to the data is $F(x) \sim x^{1/3}$. The exception is $u_r \approx 1.5$ which reflects the small scale nature of the boundary-interior interactions, see Fig.\ \ref{VertHorBulk}~(g).  The vertical correlations are somewhat different although $\delta_{u_z}$ and $\delta_{u_\phi}$ are reasonably described by $F(x) \sim x^\gamma$ with $\gamma = 1.2, 0.7$, respectively. Comparatively, $\delta_{u_r}$ starts to decrease in the subcritical regime, similar to the horizontal component.  The quantitative trends shown here are consistent with the qualitative impressions provided by representative images in Figs.\ \ref{VertHorBulk}.

\subsection{Heat transport $\Nu$}\label{subsec-Heat}

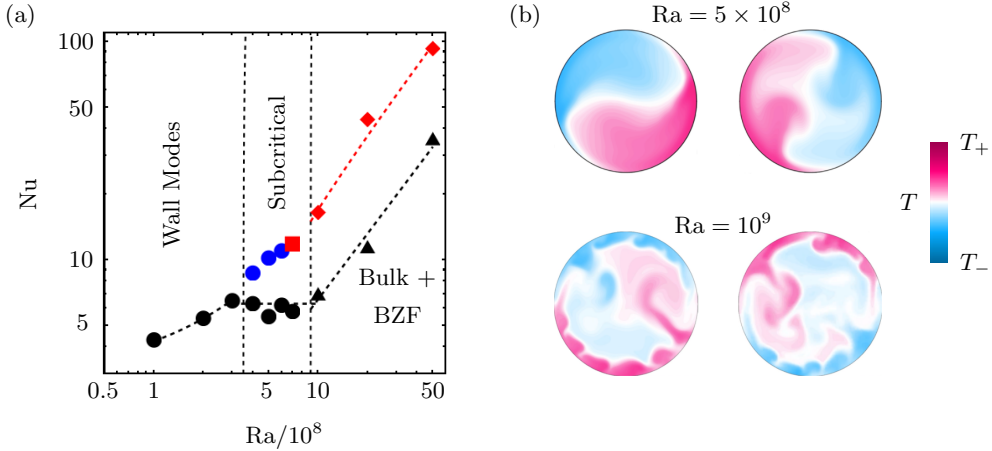
\begin{figure*}[th]
\unitlength1truecm
\begin{picture}(18,6)
\put(3,0){
\put(0.1,0.8){
\put(1,0.5){\includegraphics[height=4.6cm]{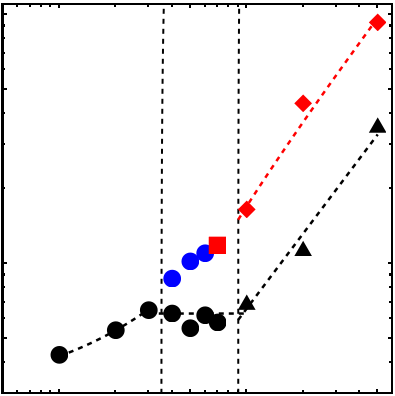}}
\put(7,0.5){\includegraphics[height=4.6cm]{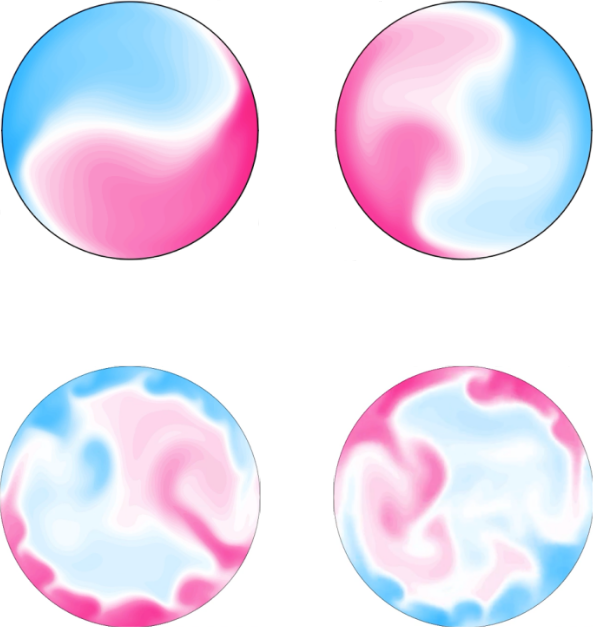}}
\put(12,2){\rotatebox{90}{\includegraphics[width=1.6cm]{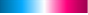}}}
}
\put(-0.2,6){(a)}
\put(0,3.5){\rotatebox{90}{$\Nu$}}
\put(0.5,5.65){100}
\put(0.65,4.75){50}
\put(0.65,2.75){10}
\put(0.75,1.9){5}
\put(0.9,1){0.5}
\put(1.7,1){1}
\put(3.2,1){5}
\put(3.8,1){10}
\put(5.33,1){50}
\put(3,0.4){$\Ra/10^8$}
\put(1.9,3){\rotatebox{90}{\color{black}{Wall Modes}}}
\put(3.3,3.5){\rotatebox{90}{\color{black}{Subcritical}}}
\put(4.5,2.5){\rotatebox{0}{\color{black}{Bulk +}}}
\put(4.7,2){\rotatebox{0}{\color{black}{BZF}}}
\put(6.5,6){(b)}
\put(8.4,6){$\Ra=5\times10^8$}
\put(8.7,3.2){$\Ra=10^9$}
\put(12.5,4.3){$T_+$}
\put(12.5,2.7){$T_-$}
\put(11.7,3.5){$T$}
}
\end{picture}
\caption{
(a) $\Nu_{\max}$ (blue solid circles, red solid square and diamonds) and $\Nu_{\min}$ (black, solid circles and triangles) vs. $\Ra/10^8$.  Regions are labeled with vertical dashed lines separating them.
Red and black dashed lines are linear fits 
\oo
(of the form $a + b \Ra$)
\bb
to $\Nu$ vs. $\Ra$ for $\Ra \geq 10^9$. 
(b) $T$ fields for (top) $\Ra = 5 \times 10^8$ and (bottom) $\Ra = 10^9$ showing example at the minimum (left image) and maximum (right image) of $\Nu$. 
\oo
Color bar of $T$ is shown.
\bb
}
\label{NuminmaxAll}
\end{figure*}

There are heat transport contributions from the wall modes and their remnants in the BZF and from the bulk flow. As we have seen above, the two states are not independent. The wall-mode jet instability for $\Ra \geq 4 \times 10^8$ injects temperature and velocity perturbations into the interior prior to and continuing into the region of bulk instability, see Figs.\ \ref{HorBulk}. In the other direction, the finite convective amplitude near the sidewall boundary appears to be the nucleation site for bulk-like perturbations with spatial separation of order $\lambda_c$ and with strong vertical coherence, see Figs.\ \ref{VertHorBulk}.  Thus, cleanly separating the individual contributions, especially for convections cells with $\Gamma <1$, is a complicated proposition. In Fig.\ \ref{subcrit}a, we noted that the time series of $\Nu$ in the subcritical regime $4 \times 10^8 \leq \Ra \lesssim 10^9$ had minima corresponding to a more quiescent state and maxima corresponding to the lateral jet ejection process.  Extending the range of this analysis to include bulk convection up to $\Ra =  5 \times 10^9$ yields Fig.\ \ref{NuminmaxAll}(a) where both the maxima and minima of $\Nu$ vary linearly with $\Ra$ (the red and black dashed lines are 
\oo
of the form \back $a + b \Ra$)
\bb
so the difference $\Delta \Nu$ also varies linearly with $\Ra$.  What then is the base state for the onset of bulk interior convection?  In Fig.\ \ref{NuminmaxAll} (b), $T$ fields corresponding to minimum and maximum $\Nu$ for $\Ra = 5 \times 10^8$ and $1 \times 10^9$ are shown.  For $\Ra = 5 \times 10^8$, the minimum $\Nu$ state corresponds quite closely to a nonlinear time-independent state whereas the maximum $\Nu$ state involves the emission of lateral jet structures.  The situation near the onset of bulk instability at $\Ra = 1 \times 10^9$ is more complex with lateral jets visible in both images; the timing and strength of the jets is probably important and not characterized here.

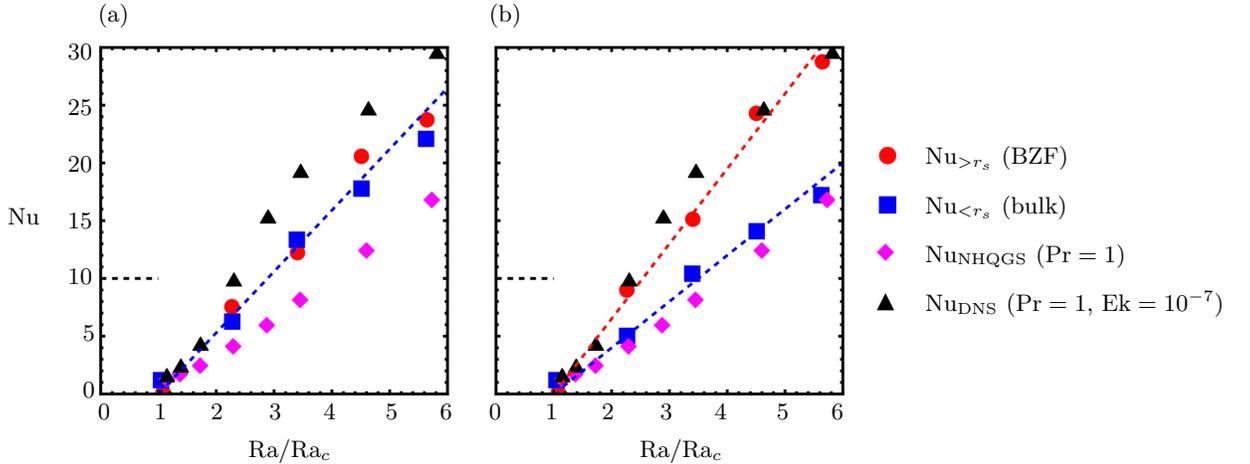
\begin{figure}[th]
\unitlength1truecm
\begin{picture}(18,6)
\put(3,0.8){\includegraphics[height=4.7cm]{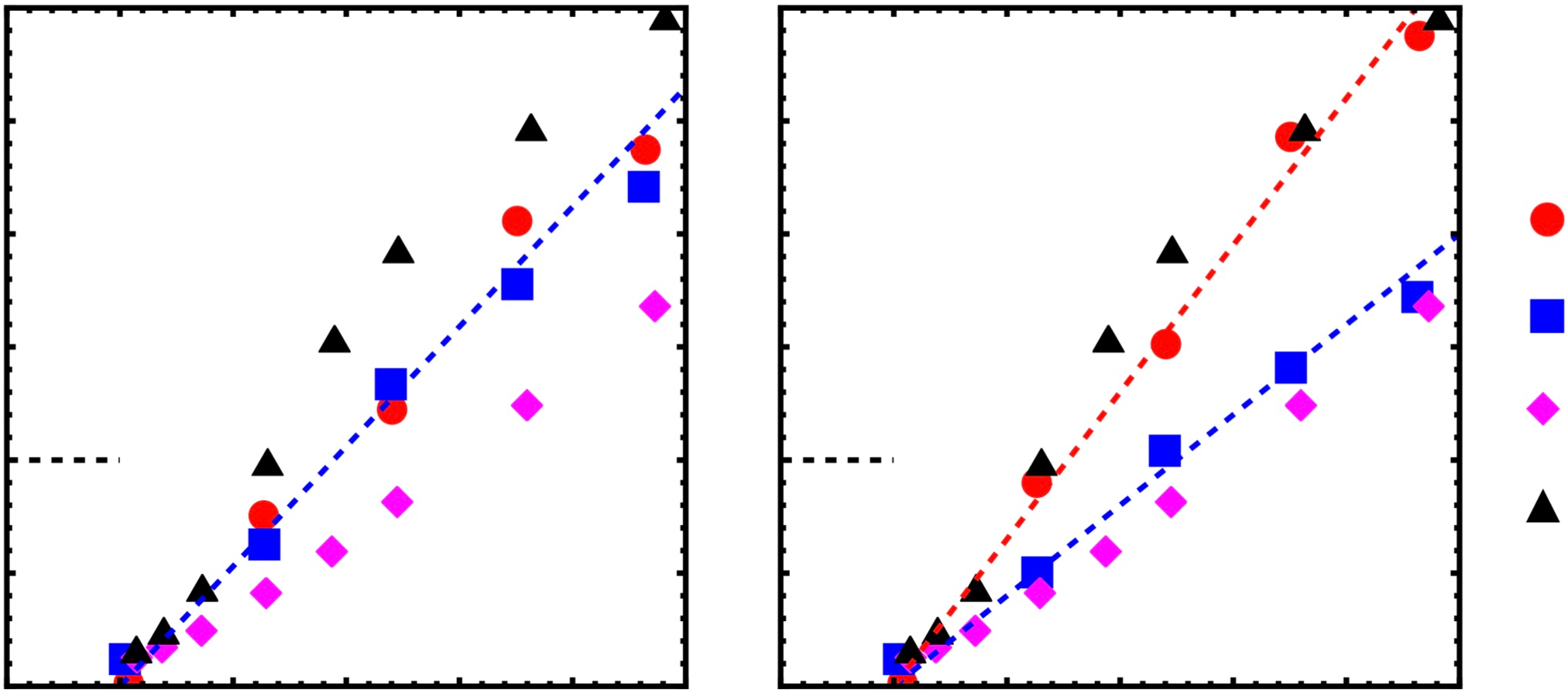}}
\put(3,0.5){0}
\put(3.75,0.5){1}
\put(4.5,0.5){2}
\put(5.275,0.5){3}
\put(5,0){$\Ra/\Ra_c$}
\put(6,0.5){4}
\put(6.825,0.5){5}
\put(7.55,0.5){6}
\put(8.2,0.5){0}
\put(8.95,0.5){1}
\put(9.7,0.5){2}
\put(10.475,0.5){3}
\put(10.2,0){$\Ra/\Ra_c$}
\put(11.2,0.5){4}
\put(12.025,0.5){5}
\put(12.75,0.5){6}
\put(2.8,0.8){0}
\put(2.8,1.55){5}
\put(2.6,2.3){10}
\put(2.6,3.05){15}
\put(1.8,3.1){$\Nu$}
\put(2.6,3.8){20}
\put(2.6,4.55){25}
\put(2.6,5.3){30}
\put(3,5.8){(a)}
\put(8.2,5.8){(b)}
\put(14,3.9){$\Nu_{>r_s}$ (BZF)}
\put(14,3.25){$\Nu_{<r_s}$ (bulk)}
\put(14,2.6){$\Nu_\text{NHQGS}$ ($\Pran=1$)}
\put(14,1.95){$\Nu_\text{DNS}$ ($\Pran=1$, $\Ek=10^{-7}$)}
\end{picture}
\caption{
$\Nu$ versus $\Ra/\Ra_c$: (a) $r_s/R=0.8$, $\Nu_{>r_s} - \Nu_\text{wm}$ BZF (red circles),  $\Nu_{<r_s}$ Bulk (blue squares),  $\Nu_\text{NHQGS}$ $\Pran$ = 1 \cite{Julien2012b} (magenta diamonds), and $\Nu_\text{DNS}$ $\Pran=1$, $\Ek=10^{-7}$ (black triangles) \cite{Stellmach2014}. Horizontal dashed line for $\Ra/\Ra_c \leq 1$ is the wall mode contribution $\Nu_\text{wm}$. 
The blue dashed line is a linear approximation to $\Nu_{>r_s} - \Nu_\text{wm} \approx \Nu_{<r_s}$.
(b) same data with $r_s/R=0.7$. 
\oo
The blue and red dashed lines are linear fits to the bulk (blue squares) and BZF (red circles), respectively.
\bb
}
\label{NuvsRaovRac}
\end{figure}

It appears from the analysis presented in Figs.~\ref{NuminmaxAll}(a,~b) that the lateral jets from the wall region contribute substantially to the heat transport at the onset of bulk convection at $\Ra \approx 1 \times 10^9$.  
With that in mind, we look at the average $\Nu$ and consider how best to separate the wall and bulk contributions.
Previously, we have done so in several ways.  
First, owing to the rather unique radial structure of the wall modes and their remnants, we defined the BZF contribution as $2 \pi  \int_{r_0}^1 \Nu(r) r dr$ where $r_0$ is the first zero crossing of $\Nu(r)$ \cite{Zhang2020,Zhang2021}.  
But this misses the negative contribution in $\Nu(r)$ as discussed above, see Figs.~\ref{Nu_uz_T}(a,~b).  
Recently, we generalized this approach by dividing the total $\Nu$ into inner and outer portions $(2/r_s^2) \int_0^{r_s}  \Nu(r) r dr$ and $2/(1 - r_s^2) \int_{r_s}^1 \Nu(r) r dr$, respectively, where $r_s$ was varied \cite{Ecke2022}. 
Here we choose $r_s = 0.8$ (in units of $R$) which is a good approximation to where the radial signature of the wall-mode/BZF seems to have become small, see Fig.~\ref{Nu_uz_T}(b). 
So from this perspective there are 3 contributions to the heat transport: pure wall mode (with jet instability), BZF, and bulk.  
In Fig.~\ref{NuvsRaovRac}(a), these different pieces are plotted versus $\Ra/\Ra_c$. 
For $r_s$ = 0.8, $\Nu$ contributions of BZF and bulk are about the same and both vary quite linearly with $\Ra/\Ra_c$.
For comparison, $\Nu$ from the non-hydrostatic quasi-geostrophic model (NHQGS) \cite{Julien2012a,Julien2012b} for $\Ek \rightarrow 0$ and from $\Nu$ from a laterally-periodic DNS for $\Pran=1$ and $\Ek = 10^{-7}$ \cite{Stellmach2014} 
\oo 
which includes Ekman pumping contributions (see also \cite{Plumley2016, Plumley2017}) 
\bb 
are plotted.  
The NHQGS is smaller owing to ignoring the contributions of Ekman pumping whereas the DNS is a bit higher, perhaps owing to larger $\Ek$, i.e., $\Ek > 0$. 

Another possible separation strategy is to define $r_s (\Ra)$ based on some feature of, for example, $u_z(r)$ or $Nu(r)$ rather than having it be a constant value. Varying $r_s$ slightly for different $\Ra$ introduces a net upward (downward) curvature to $\Nu$ versus $\Ra$ for BZF (bulk) pieces. We choose instead to evaluate the separation for several constant $r_s$, recognizing that there is no unambiguous, testable separation strategy using the data we have here. In Fig.\ \ref{Nu_uz_T}(c), the features of $\widetilde{\Nu}$, $\widetilde{u_z}$, and $\widetilde{T}$ suggest $0.7 < r_s/R < 0.8$.  In Fig.\ \ref{NuvsRaovRac}(b), we show results using $r_s/R=0.7$ to contrast with $r_s/R=0.8$ in Fig.\ \ref{NuvsRaovRac}(a).  The trends are both consistent with a linear increase of $\Nu$ with $\Ra/\Ra_c$ but with a greater (lesser) slope for BZF (bulk) contributions. Here the bulk contribution is quite comparable to the NHQGS data with little Ekman pumping correction whereas the seeming correspondence of the BZF contribution with the DNS results is purely coincidental.

\begin{figure*}[th]
\unitlength1truecm
\begin{picture}(18,8)
\put(0,0){\includegraphics[width=\textwidth]{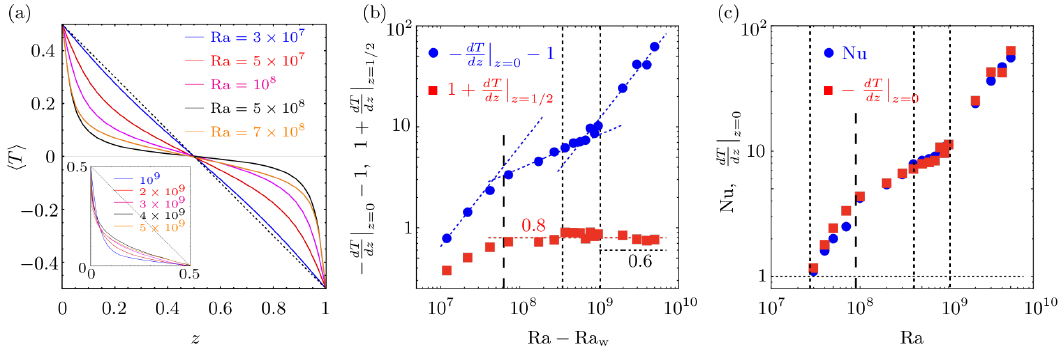}}
\put(12.228,2.95){\rotatebox{90}{-}}
\end{picture}
\caption{
(a) Temperature profiles $\langle T (r=0.98R) \rangle_{\phi,t}$ for $3\times10^7  \leq \Ra \leq 7\times10^8$ as indicated.  The diagonal black dashed line is the linear conductivity profile of the thermally conducting state. 
Inset shows $\langle T (r=r_{\max}) \rangle_{\phi,t}$ for $10^9  \leq \Ra \leq 5\times10^9$  where $r_{\max}$ is the radial position of the maxima of $\langle u_z(r) \rangle_{\phi, t}$. (b) $-dT/dz |_{z=0}-1$ (blue, solid circles) and $1+ dT/dz |_{z=1/2}$ (red, solid squares) at the same $r$ as in (a). (c) Comparison of $-dT/dz |_{z=0}$ at the same $r$ as in (a) (blue, solid circles) and $\Nu$ (red, solid squares) showing close correspondence.
}
\label{VertTempProfile}
\end{figure*}

Although the procedure above is a tempting one that offers a more or less clean separation, there are concerns about this approach that can be elucidated by considering the vertical temperature profile $\langle T(r=r_{\max}, \phi, z) \rangle_{\phi, t}$, where $r_{\max}$ is the value of, and its variation with, $\Ra$. 
In Fig.\ \ref{VertTempProfile}(a), profiles of $\langle T \rangle$ are shown as functions of $z$ for $3\times10^7 \leq \Ra \leq 7\times10^8$ and in the inset for $10^9 \leq \Ra \leq 5\times10^9$. 
There is a continual increasing slope at the top and bottom boundaries but the interior gradient is not monotonic with a smaller slope at $\Ra = 5\times10^8$ compared to that at $\Ra = 7\times10^8$. 
This reversal is also observed in the inset where higher $\Ra$ have steeper interior slope. These variations are features associated with the different states of wall modes for $\Ra \leq \Ra_c \approx 9 \times 10^8$ and the development of the bulk instability. 
An important point here is that the slope is continuously varying (see Appendix Fig.\ \ref{VertTempProfile}) so that the definition of a thermal BL thickness is quite problematic. Perhaps there is localized Ekman pumping associated with the BZF -- a topic for further investigation.  

In Fig.\ \ref{VertTempProfile}(b), we present the slope at the lower boundary $-dT/dz$ as $-dT/dz(z=0, r=r_{\max}) -1$ for which there are 3 distinct regions: weakly nonlinear growth $\sim (\Ra - \Ra_\text{w})$, the nonlinear state including the periodic time-dependent state up to $\Ra = 6 \times 10^8$, and the combined BZF/bulk state with increasing slope $\sim \Ra$.  If one averaged over $r$ as well as over $\phi$ and $t$, one would expect an exact correspondence between global heat transport and local temperature slope $\Nu-1=-dT/dz(z=0, r=r_{\max}) -1$. In Fig.\ \ref{VertTempProfile}(c), there is very close agreement between these two quantities except in the weakly nonlinear regime.  Returning to the interior slope, we plot in  Fig.\ \ref{VertTempProfile}(b), $dT/dz(z=1/2)+1$ which is 0 at $\Ra_\text{w}$ and would approach 1 with increasing $\Ra$ for non-rotating convection but has a finite slope for bulk rotating convection \cite{Ecke2022}. Here we have between 0.8 and 0.9 compared to a value of 0.6 for pure bulk geostrophic convection \cite{Julien2012a}. The difference is not surprising given the coexistence of bulk fluctuations on top of the wall-localized state as previously documented above.  

Perhaps the most interesting feature of these profiles of $\langle T \rangle$ is the close agreement of the localized (in $r$) measure of $\Nu_\text{local} = -dT/dz(z=0, r_{\max})$ and the global heat transport $\Nu$ that averages over the whole domain. If there was a clean separation between wall and bulk modes, $\Nu_\text{local}$ would not increase rapidly and in close correspondence to $\Nu$.  One can then infer that the bulk-like fluctuations in the wall zone continue to increase in strength and dominate the heat transport with the average wall mode foundation playing a minor role. One cannot escape the realization that there 
\oo
are
\bb
strong interactions among the bulk and coexisting BZF that plays a major role in the heat transport in some transition region from wall modes to bulk rotating convection.

\section{Conclusion}\label{sec-Conclusion}

We employed DNS using the Goldfish code of RRBC in a cylindrical geometry with aspect ratio $\Gamma=1/2$, insulating sidewall boundary conditions, $\Ek=10^{-6}$, and $\Pran=0.8$ to characterize the progression of states over a range $3 \times 10^7 \leq \Ra \leq 5 \times 10^9$. We also computed a smaller set of properties for the same parameters but with $\Gamma = $1, 2 and for  $\Ek = 10^{-4}$ with $\Gamma$ = 0.3, 1/2, 1, 2, 5, primarily to determine the mode number $m$. The set of values  for $\Ek = 10^{-6}$ are tabulated in the Appendix Table \ref{TABEk10m6}). We elucidated how the mode number $m$ observed for different $\Gamma$ was consistent between wall modes and the BZF state, how the observed wall mode critical $\Ra_w$ varies with $\Ek$, $\Gamma$, and $m$, and how small $\Gamma \lesssim 0.7$ always has $m=1$ owing to the 
\oo
azimuthal periodicity.
\bb
We further demonstrate the decoupling of radial length scales for $T \sim (k_\text{w})^{-1}$ and $u \sim \Ek^{-1/3}$ in the steady wall mode regime $3 \times 10^7 \leq \Ra \leq 3 \times 10^8$ and show that the eigenfunctions of the planar wall linear solutions \cite{Herrmann1993} provide an excellent representation of the data, particularly for the radial dependence of $u(r)$ where, for $\Ek=10^{-6}$, the radial localization near the sidewall of about 0.1 makes the sidewall curvature effects very small, i.e., $0.1/(2 \pi) \approx 0.02$. On the other hand, for $T$ the linear eigenfunction extends over a significant fraction of $R$ and deviations owing to  finite curvature and growing nonlinearity are more keenly felt. Similarly, the azimuthal and vertical eigenfunctions are also more effected by nonlinearity and curvature but give insight into the development of the steady wall mode state.

In the steady wall-mode regime, $\Nu$ is constant despite the traveling wave nature of the wall mode. For $4 \times 10^8 \leq \Ra \leq 8 \times 10^8$, the steady wall mode undergoes a subcritical bifurcation to a state of time-dependent $\Nu$ through a nonlinear mechanism of lateral jet ejection from the wall mode into the bulk interior (see \cite{Favier2020}, \cite{Madonia2021}).  Throughout the linear (steady) and nonlinear (time dependent) regimes for $\Ek=10^{-6}$,  $\Nu(r) \sim \langle u_z(r) \rangle \langle T(r) \rangle$ owing to the rapid radial variation of $u_z$ compared to the very slow spatial variation of $T$ and the fixed small phase difference between them. The jet instability strengthens as $\Ra \rightarrow \Ra_c \approx 1 \times 10^9$ and  fine scale spatial (and temporal oscillations, see Appendix) structures form around the azimuth in the wall localized region. We show that the radial length scale  $\delta_{u_z}$ and $\delta_{u_\phi}$ defined by first zero-crossing is about 2.5 $\Ek^{1/3}$ whereas   $\delta_{\omega_z} \approx 3.3 \Ek^{1/3}$ using its second zero crossing.  
We also use auto-correlation analysis to extract correlation lengths for different cross sections of fields, namely horizontal cross sections and vertical surfaces of $\{\phi, z\}$ at constant $r/R=0.98$ (the maximum value of $u_z(r)$).  The correlation lengths are fairly constant in the wall mode regime and decrease rapidly once the bulk mode sets in for $\Ra \gtrsim 10^9$.  In horizontal regions in the cell center and $\Ra \gtrsim 10^9$,  $\delta_T, \delta_{u_\phi} \approx \lambda_c$ whereas $\delta_{u_z}, \delta_{u_r}, \delta_{\omega_z} \approx \lambda_c/2$. 
In the wall localized regime, the azimuthal correlation lengths are are roughly constant at about a half wavelength of the $m=1$ wall mode but decrease to 2-4 $\lambda_c$ in the bulk mode region.  The vertical correlation lengths are more varied among the different fields but systematically decrease in the bulk mode region.

When considered in the context of heat transport scaling, the analysis of the structure of wall modes and their remnant robust features in the presence of bulk convection - the BZF -  play a key role in attempting to separate bulk influence from that of the wall-localized portion. The negative feature in $\Nu$ is understood as arising from the oscillatory radial structure of $u_z(r)$ and the much slower radial dependence of $T$.  Once bulk convection begins in the interior, the BZF spreads out radially and different features of $\Nu$, $u_z$ and $T$ including the lateral jet instability makes separation of the 2 components complicated. The resulting separation using $r_s/R$ = 0.8 and 0.7 and subtracting the wall mode contribution \cite{Zhang2021,Ecke2022}, yields $\Nu \sim \Ra/\Ra_c$ but with different slopes: one that gives agreement with the NHQGS theory and the other with Ekman pumping corrections.  There are also subtle changes in the $\Ra$ dependence for the two $r_s$ values. The average vertical temperature profiles reinforce the understanding that the BZF carries strong bulk fluctuations that contribute heavily to the total heat transport.  There seems to be no unambiguous way to disentangle the two components of $\Nu$ in small aspect ratios such as studied here. A closer inspection of the interplay of thermal boundary layers in the transitioning wall mode and for bulk convection with coexisting BZF states is certainly warranted. 

\section{Acknowledgement}\label{sec-Acknowledgement}
The authors acknowledge the support from the German Research Foundation (DFG), grants Sh405/20 and Sh405/22, and the Leibniz Supercomputing Centre (LRZ) for providing computing time. REE acknowledges support from the Los Alamos National Laboratory LDRD Program.

\section{Appendix}\label{sec-Appendix}

In this section, we provide tabulated data from our DNS, empirically expand the set of critical parameters to higher order in $\Ek$, and show some additional features of wall modes in the BZF regime.  In addition, we tabulate movies that we present in supplemental material.

\subsection{Data}\label{subsec-Data}

 In Table \ref{TABEk10m6}, the computed values of control parameters $\Ek = 10^{-6}$, $\Pran=0.8$, $\Gamma$, $\Ra$, and $\epsilon = \Ra/\Ra_\text{w}-1$ are listed with the resultant values of $\Nu$, $\omega_d$, $\omega_{d_2}$, and $\omega_{d_0}/\omega_d$. 
 
\begin{table}[h]
\begin{center}
\begin{tabular}[t]{lcccccccc}
\toprule
 $\Ek$ & $\Gamma$ & $\Ra$ 		& $\epsilon$ 	& $\Ro$ 	& $\Nu$ 	&	$\omega_d$	& $\omega_{d_2}$ & $\tau_0 \omega_d/(2 \pi)$\\
\hline
$10^{-6}$	&1/2& $3.0 \times 10^7$ 	& 0.07 	& 0.0061	& 1.1 	& $1.9 \times 10^{-4}$ & &						\\
&& 			$4.0 \times 10^7$ 	& 0.43 	& 0.0071	& 1.6		& $2.2 \times 10^{-4}$ & &						\\
&& 			$5.0 \times 10^7$ 	& 0.79	& 0.0079	& 2.0		& $2.4 \times 10^{-4}$ & &					  	\\
&& 			$7.0 \times 10^7$ 	& 1.5		& 0.0094	& 2.5 	& $2.8 \times 10^{-4}$ &	$1.7 \times 10^{-4}$ &	-0.25	\\
&& 			$1.0 \times 10^8$ 	& 2.6		& 0.011	& 4.2 	& $3.7 \times 10^{-4}$ & 	$3.3 \times 10^{-4}$ &	-0.25	\\
&& 			$2.0 \times 10^8$ 	& 6.1		& 0.016	& 5.4 	& $6.0 \times 10^{-4}$   & 	$8.9 \times 10^{-4}$ &	-0.42	\\
&& 			$3.0 \times 10^8$ 	& 9.7		& 0.019	& 6.5 	& $8.1 \times 10^{-4}$   & 	$1.4 \times 10^{-3}$ &	-0.56	\\
&& 			$4.0 \times 10^8$ 	& 13.3	& 0.022	& 7.9 	& $9.7 \times 10^{-4}$   & 	$1.5 \times 10^{-3}$ &		\\
&& 			$5.0 \times 10^8$ 	& 16.9	& 0.025	& 8.4		& $1.2 \times 10^{-3}$   & 	$1.8 \times 10^{-3}$ &	0.21	\\
&& 			$6.0 \times 10^8$ 	& 20.4	& 0.027	& 8.7		& $1.4 \times 10^{-3}$   & 	$2.4 \times 10^{-3}$ &	0.34	\\
&& 			$7.0 \times 10^8$ 	& 24.		& 0.030	& 9.1		& $2.9 \times 10^{-3}$ & 	$1.7(5)  \times 10^{-3}$ &		\\
&& 			$8.0 \times 10^8$ 	& 27.6	& 0.032	& 10.3	& $2.7 \times 10^{-3}$   &					&		\\
&& 			$9.0 \times 10^8$ 	& 27.6	& 0.034	& 10.8	& $3.4 \times 10^{-3}$   &					&		\\
&& 			$1.0 \times 10^9$ 	& 34.7	& 0.035	& 11.2	& $4.4 \times 10^{-3}$ & 					& 		\\
&& 			$2.0 \times 10^9$ 	& 70		& 0.05	& 24.1	& $7.4 \times 10^{-3}$ & 					&		\\
&& 			$3.0 \times 10^9$ 	& 106	& 0.061	& 36.4	& $9.6 \times 10^{-3}$    & 				&		\\
&& 			$4.0 \times 10^9$ 	& 142	& 0.071	& 47.1	& $1.2 \times 10^{-3}$    & 				&		\\
&& 			$5.0 \times 10^9$ 	& 178	& 0.079	& 55.8 	& $1.5 \times 10^{-2}$ & 					&		\\
\hline
 \end{tabular}
\caption{Data from DNS with $\Pr = 0.8$ indicating $\Ek$, $\Gamma$,  $\Ra$, $\epsilon$, $\Ro$, $\Nu$, $\omega_d$, $\omega_{d_2}$, and  $\tau_0 \omega_d/(2 \pi)$. }
\label{TABEk10m6}
\end{center}
\end{table}

\subsection{Wall Mode Critical Parameters and Aspect Ratio Effects}\label{subsec-WallMode}

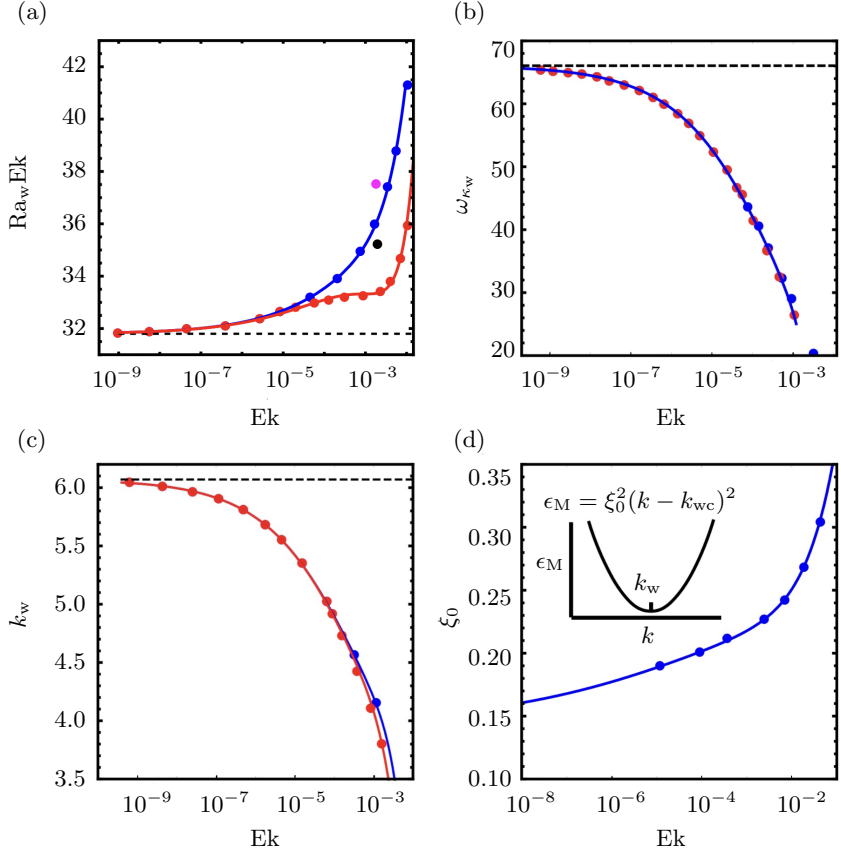
\begin{figure}[th]
\unitlength1truecm
\begin{picture}(18,11)
\put(4,0.8){\includegraphics[height=10cm]{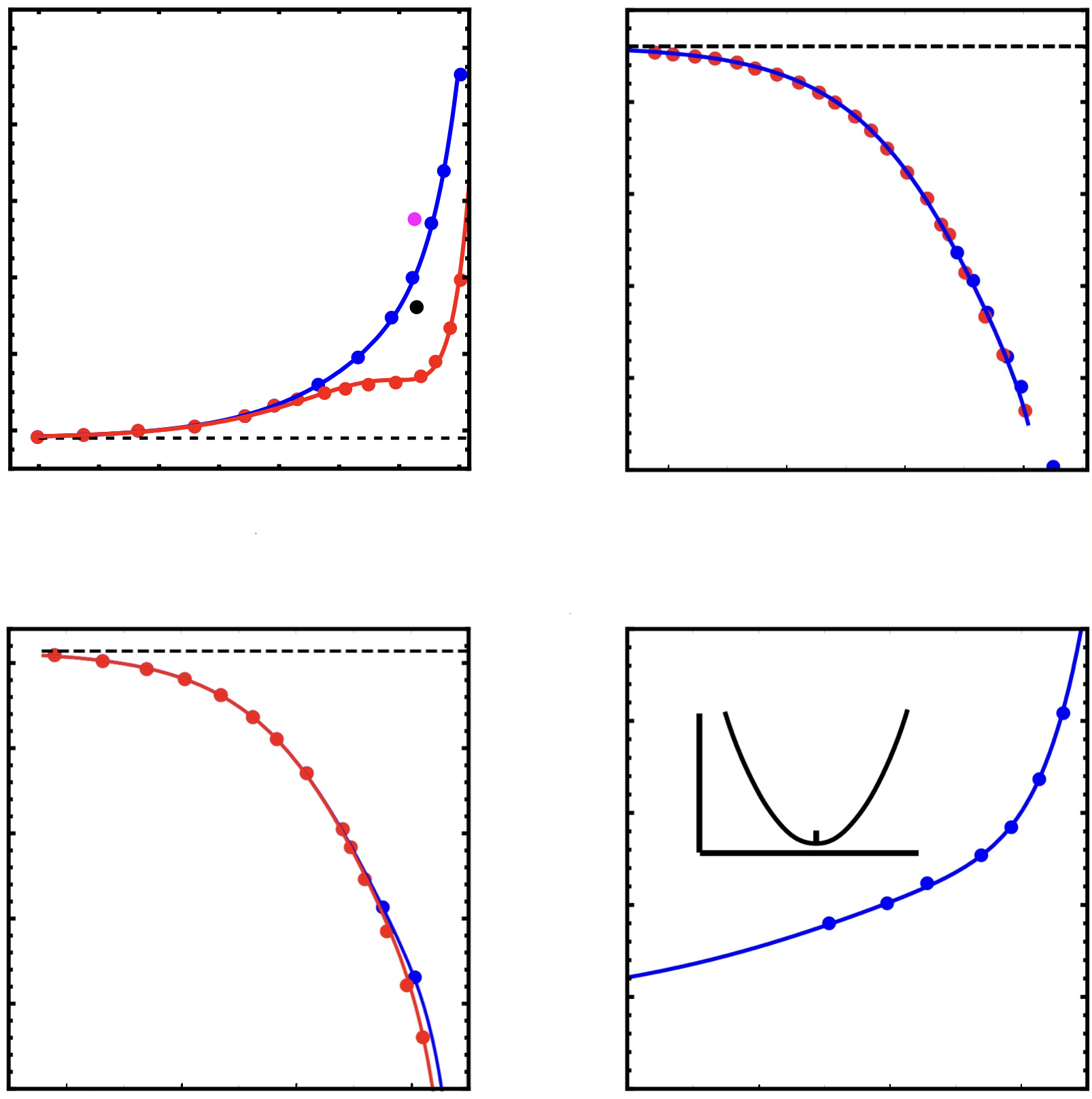}}
\put(3.5,0.8){3.5}
\put(3.5,1.575){4.0}
\put(3.5,2.35){4.5}
\put(3.5,3.125){5.0}
\put(2.9,2.9){\rotatebox{90}{$k_\text{w}$}}
\put(3.5,3.9){5.5}
\put(3.5,4.675){6.0}
\put(9.0,0.8){0.10}
\put(9.0,1.6){0.15}
\put(9.0,2.4){0.20}
\put(8.6,2.9){\rotatebox{90}{$\xi_0$}}
\put(9.0,3.2){0.25}
\put(9.0,4.0){0.30}
\put(9.0,4.8){0.35}
\put(3.6,6.8){32}
\put(3.6,7.5){34}
\put(3.6,8.2){36}
\put(3.6,8.9){38}
\put(3.6,9.6){40}
\put(3.6,10.3){42}
\put(2.9,8.2){\rotatebox{90}{$\Ra_\text{w}\Ek$}}
\put(9.3,6.5){20}
\put(9.3,7.3){30}
\put(9.3,8.1){40}
\put(9.3,8.9){50}
\put(9.3,9.7){60}
\put(9.3,10.5){70}
\put(8.85,8.4){\rotatebox{90}{$\omega_{\kappa_{\text{w}}}$}}
\put(4,6.1){$10^{-9}$}
\put(5.2,6.1){$10^{-7}$}
\put(6.3,6.1){$10^{-5}$}
\put(7.4,6.1){$10^{-3}$}
\put(6.1,5.6){$\Ek$}
\put(9.7,6.1){$10^{-9}$}
\put(10.9,6.1){$10^{-7}$}
\put(12,6.1){$10^{-5}$}
\put(13.1,6.1){$10^{-3}$}
\put(11.5,5.6){$\Ek$}
\put(4.4,0.45){$10^{-9}$}
\put(5.425,0.45){$10^{-7}$}
\put(6.45,0.45){$10^{-5}$}
\put(7.5,0.45){$10^{-3}$}
\put(6.1,0){$\Ek$}
\put(9.5,0.45){$10^{-8}$}
\put(10.7,0.45){$10^{-6}$}
\put(11.9,0.45){$10^{-4}$}
\put(13.1,0.45){$10^{-2}$}
\put(11.5,0){$\Ek$}
\put(11.2,3.4){$k_\text{w}$}
\put(11.3,2.7){$k$}
\put(9.9,3.7){$\epsilon_\text{M}$}
\put(10,4.5){$\epsilon_\text{M}=\xi_0^2(k-k_\text{wc})^2$}
\put(3,11){(a)}
\put(8.8,11){(b)}
\put(3,5.3){(c)}
\put(8.8,5.3){(d)}
\end{picture}
\caption{\oo
Critical parameters of a planar wall mode with perfectly insulating sidewall boundary conditions versus $\Ek$ using digitized data from Fig.~3 of \cite{Herrmann1993}. Blue and red data points and corresponding fits in powers of $\Ek^{1/3}$ are for $\Pr = 7$ and $\Pr = 0.7$, respectively. 
(a) $\Ra_\text{w}$ where black \cite{Goldstein1993} and magenta \cite{Liu1999} data points for $\Pran \approx 7$ (below and above the $\Pran = 7$ (blue) curve) show the effects, respectively, of finite curvature ($\Gamma = 1$) and finite wall conductivity (plexiglass walls). (b) $\omega_{\kappa_{\text{w}}}$ (recall that $\omega_d = 2\Ek/\Pr \omega_\kappa$), and 
(c) 
\bb
$k_\text{w}$. 
\oo
(d)
\bb
The curvature parameter $\xi_0$ of the marginal stability boundary $\epsilon_\text{M} = \xi_0^2 (k-k_\text{w})^2$ (schematically shown in the inset) from data at $\Pr = 7$ and weakly conducting sidewall boundary conditions \cite{Kuo1993}.
}
\label{CritFit}
\end{figure}

The asymptotic approximations for the critical values of $\Ra_\text{w}$, $\omega_\text{w}$, and $k_\text{w}$ are valid over varying ranges of $\Ek$ but the approximations are not adequate to accurately estimate these critical values for larger $\Ek$.  Using the numerical data presented in \cite{Herrmann1993}, one can empirically extend the asymptotic results to larger $\Ek$ using fits to the data and expansions in higher powers of $\Ek^{1/3}$.  In Fig.\ \ref{CritFit}, we fit the data for the critical values assuming insulating sidewall boundary conditions at $\Pr = 0.7$, 7 (see Fig.\ 3 \cite{Herrmann1993}). \oo Only  $\Ra_\text{w}$ has significant variation with $\Pran$. Finite wall curvature ($\Gamma = 1$ \cite{Goldstein1993}) is shown to slightly decrease $\Ra_\text{w}$ whereas finite wall conductivity slight increases  $\Ra_\text{w}$ \cite{Liu1999} as shown in Fig.\ \ref{CritFit}(a).\bb We also fit the curvature parameter $\xi_0$ of the marginal stability boundary computed for $\Pr =7$ and for partially insulating sidewalls \cite{Kuo1993}. The results are shown in Fig.\ \ref{CritFit}.

with the resulting fit parameters:
\begin{eqnarray*}
\Ra_\text{w} &=&31.82  \Ek^{-1} + 46.5 \Ek^{-2/3} - 465  \Ek^{-1/3} + 1564 \ \  (\Pran=0.7)\\
&=&31.82  \Ek^{-1} + 46.5 \Ek^{-2/3} - 194  \Ek^{-1/3} + 845  \ \  (\Pran=7),\\
k_\text{w} &=&6.07  - 35 \Ek^{1/3} + 250  \Ek^{2/3} - 990 \Ek \ \  (\Pran=0.7)\\
&=&6.07  - 35 \Ek^{1/3} + 250  \Ek^{2/3} - 870 \Ek  \ \  (\Pran=7),\\
\omega_{\kappa_\text{w}} &=&66.05  - 732 \Ek^{1/3} + 5700  \Ek^{2/3} - 2300 \Ek\\
\xi_0 &=& 0.142 + 0.447 \Ek^{1/6} - 1.08 \Ek^{1/3} + 1.34 \Ek^{1/2} \ \ (\Pran=7).
\end{eqnarray*}

\subsection{Transients}\label{subsec-Transients}

\begin{figure*}[th]
\unitlength1truecm
\begin{picture}(18,6)
\put(0,-0.5){
\put(1.5,1.5){
\begin{tikzpicture}[style=thick]
\put(1,0.5){\includegraphics[height=4.6cm]{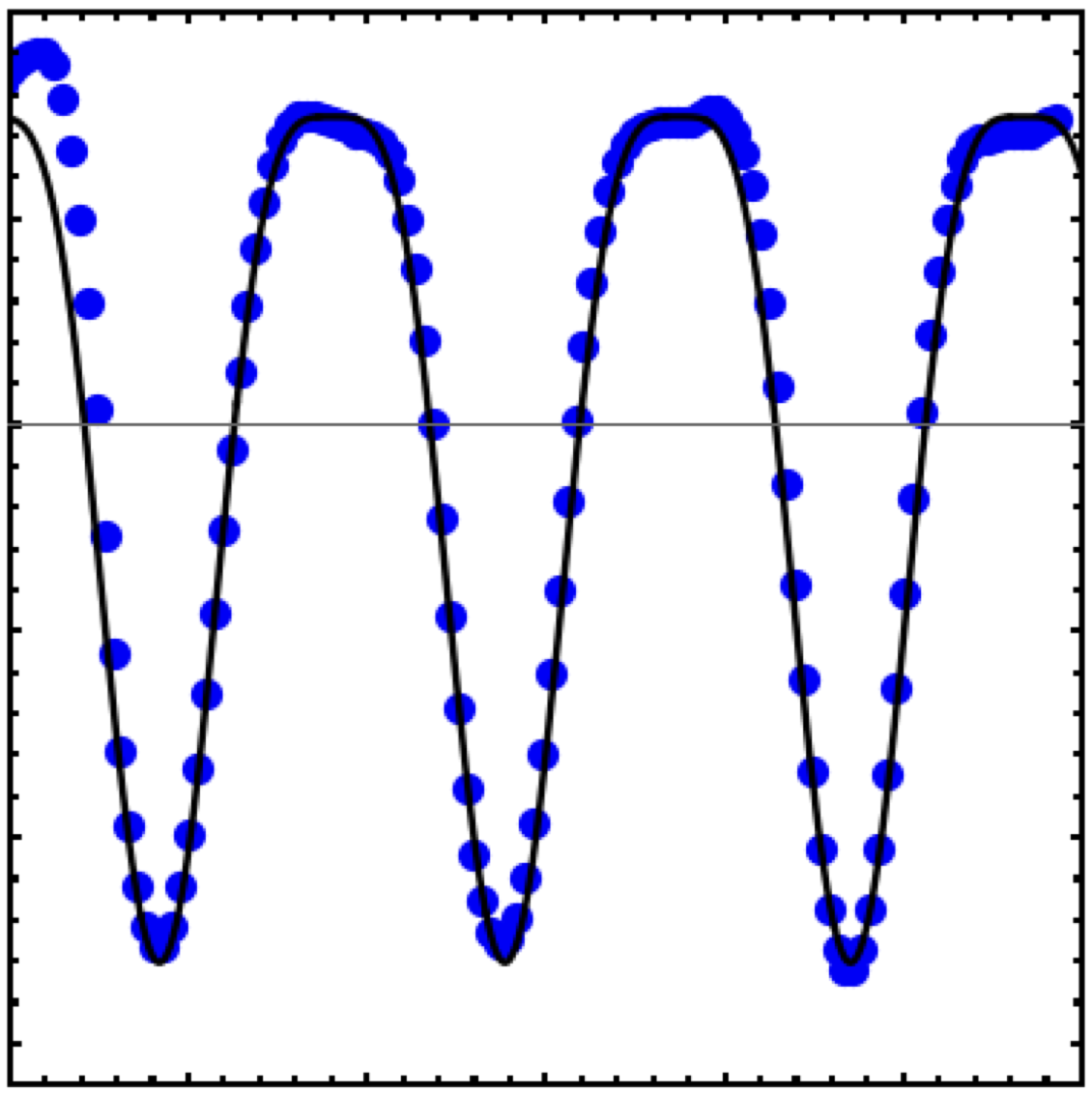}}
\put(7,0.5){\includegraphics[height=4.6cm]{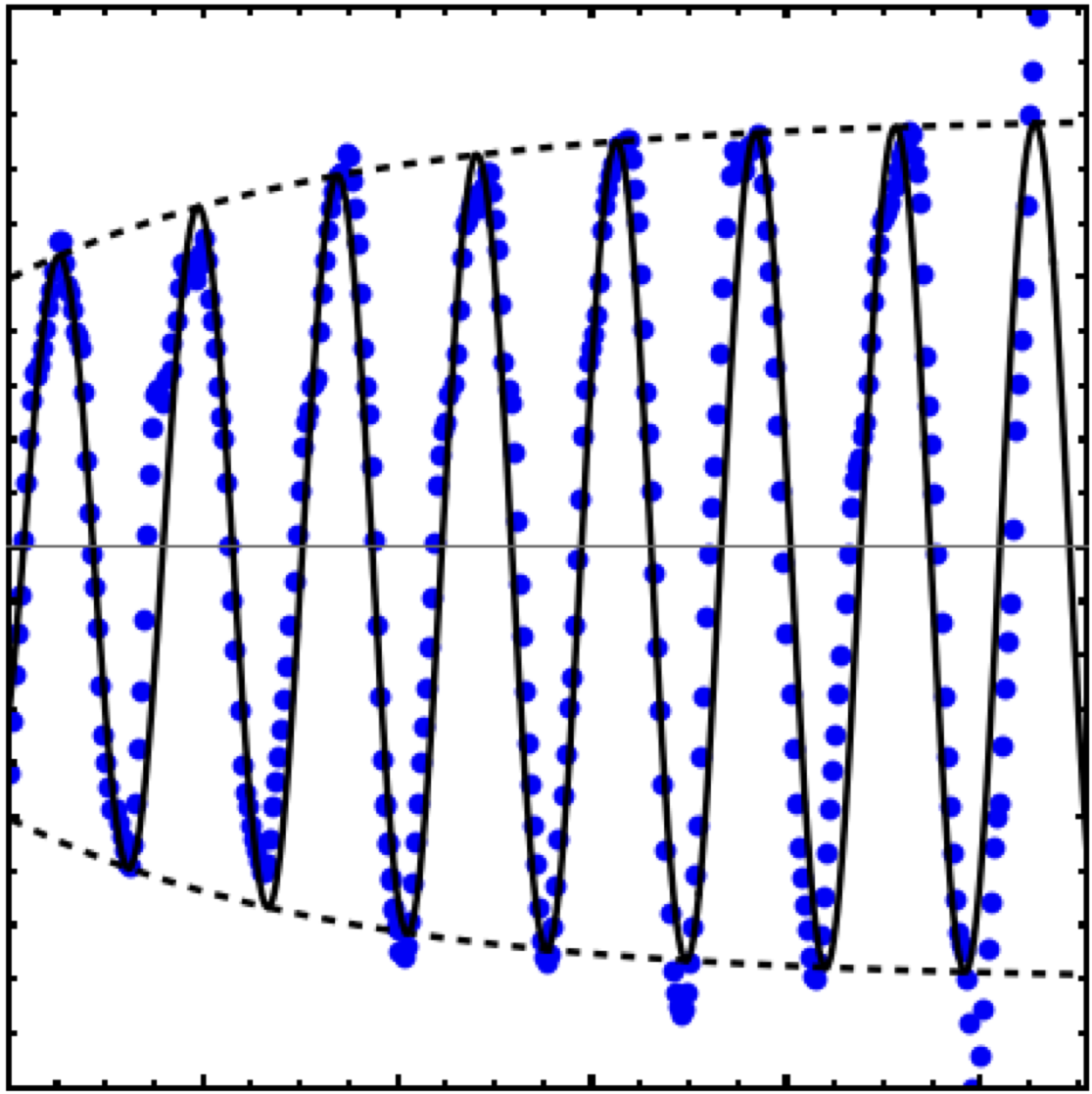}}
\put(13,0.5){\includegraphics[height=4.6cm]{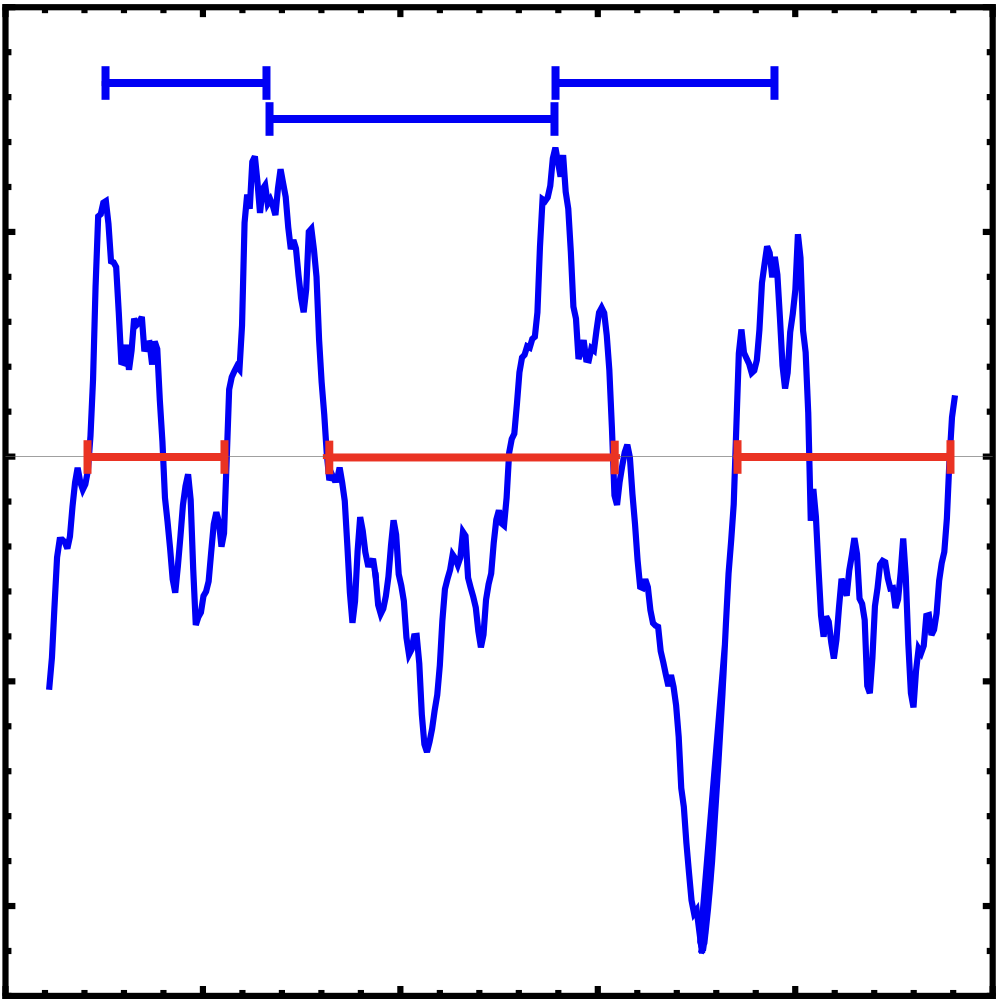}}
    \draw [|-|, color=blue](1.65,0.8) -- (3.15,0.8) ;
    \end{tikzpicture}
}
\put(-0.2,6){(a)}
\put(6,6){(b)}
\put(12,6){(c)}

\put(-0.2,2.5){\rotatebox{90}{$\Nu(t)/\Nu_n-1$}}
\put(0.5,5.65){1.0}
\put(0.5,4.8){0.5}
\put(0.5,3.9){0.0}
\put(0.25,3.05){$-0.5$}
\put(0.25,2.2){$-1.0$}
\put(0.25,1.3){$-1.5$}
\put(0.9,0.9){0}
\put(1.5,0.9){100}
\put(2.25,0.9){200}
\put(3,0.9){300}
\put(3.75,0.9){400}
\put(4.5,0.9){500}
\put(5.25,0.9){600}
\put(3.2,0.4){$t$}
\put(2.1,1.65){\rotatebox{0}{\color{blue}{194}}}

\put(6.5,5.6){1.0}
\put(6.5,4.5){0.5}
\put(6.5,3.4){0.0}
\put(6.25,2.3){$-0.5$}
\put(6.25,1.2){$-1.0$}
\put(6.9,0.9){0}
\put(7.55,0.9){200}
\put(8.4,0.9){400}
\put(9.2,0.9){600}
\put(10.05,0.9){800}
\put(10.8,0.9){1000}
\put(9.2,0.4){$t$}
\put(8.1,1.65){\rotatebox{0}{\color{blue}{194}}}

\put(12.5,5.7){1.0}
\put(12.5,4.65){0.5}
\put(12.5,3.6){0.0}
\put(12.25,2.55){$-0.5$}
\put(12.25,1.5){$-1.0$}
\put(12.9,0.9){0}
\put(13.7,0.9){100}
\put(14.6,0.9){200}
\put(15.5,0.9){300}
\put(16.4,0.9){400}
\put(17.25,0.9){500}
\put(15.2,0.4){$t$}
}
\end{picture}
\caption{
Time dependence of the heat transport $\Nu(t)$ for (a) $\Ra=4 \times 10^8$ showing large/small $\Nu$ asymmetry 
$\sim\cos^4(\omega_2/2 t)$ 
and (b) $\Ra=6 \times 10^8$  where there is a long transient relaxation of the form $(1-e^{-t/\tau_0})$. Dashed lines in (a) and (b) are fits with these functional forms with fitted values for oscillation period $\tau_2$ and $\tau_0$ indicated as horizontal (blue/black) bars, respectively. (c) $\Ra = 7 \times 10^8$ chaotic time series with some time intervals from peak (blue) and zero-crossing (red): mean is $110 \pm 30$.}
\label{A2}
\end{figure*}

Whereas the form of the time dependence of $\Nu$ is very precisely sinusoidal (e.g., Fig.\ \ref{A2}(b), the subcritical nonlinear state has some interesting additional features.  
At $\Ra = 4 \times 10^8$, there is a very clear large/small $\Nu$ asymmetry in $\Nu(t)$ as shown in Fig.\ \ref{A2}(a), i.e., the system prefers to be in a larger $\Nu$ state relative to a smaller one. 
The solid line is a fit with $a-b [\cos^4{((\omega_2/2) t - \phi_0)}]$ to the normalized $\Nu(t)$ where $\Nu_n$ is its approximate mean value.  In Fig.\ \ref{scoscill}(c) ($\Ra = 5 \times 10^8$), there is some remaining asymmetry with both $\cos{(\omega t)}$ and $\cos{((\omega_2/2) t)}^4$ contributions.  Whereas the amplitude $\Delta \Nu$ at $\Ra = 4 \times 10^8$ is very constant over several oscillation periods, the amplitude increases slowly for $\Ra = 6 \times 10^8$ as a saturation of the form $1-e^{-t/\tau_0})$; this is also observed for $\Ra = 5 \times 10^8$, see Fig.\ \ref{scoscill}(c), but the single oscillation period makes the saturation difficult to measure accurately. The time series for $\Ra = 7 \times 10^8$ in Fig.\ \ref{A2}(c) is quite chaotic with an aperiodic oscillation and an intermittent amplitude.

\subsection{$\Gamma = 1/3$ prograde dynamics}\label{subsec-Third}

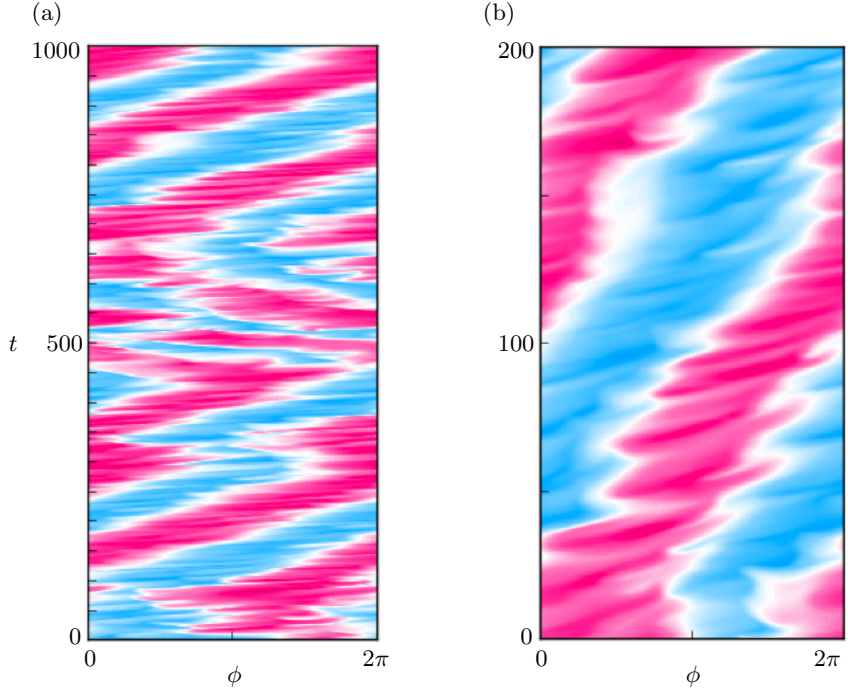
\begin{figure*}[th]
\unitlength1truecm
\begin{picture}(18,9)
\put(4,0.5){\includegraphics[height=8cm]{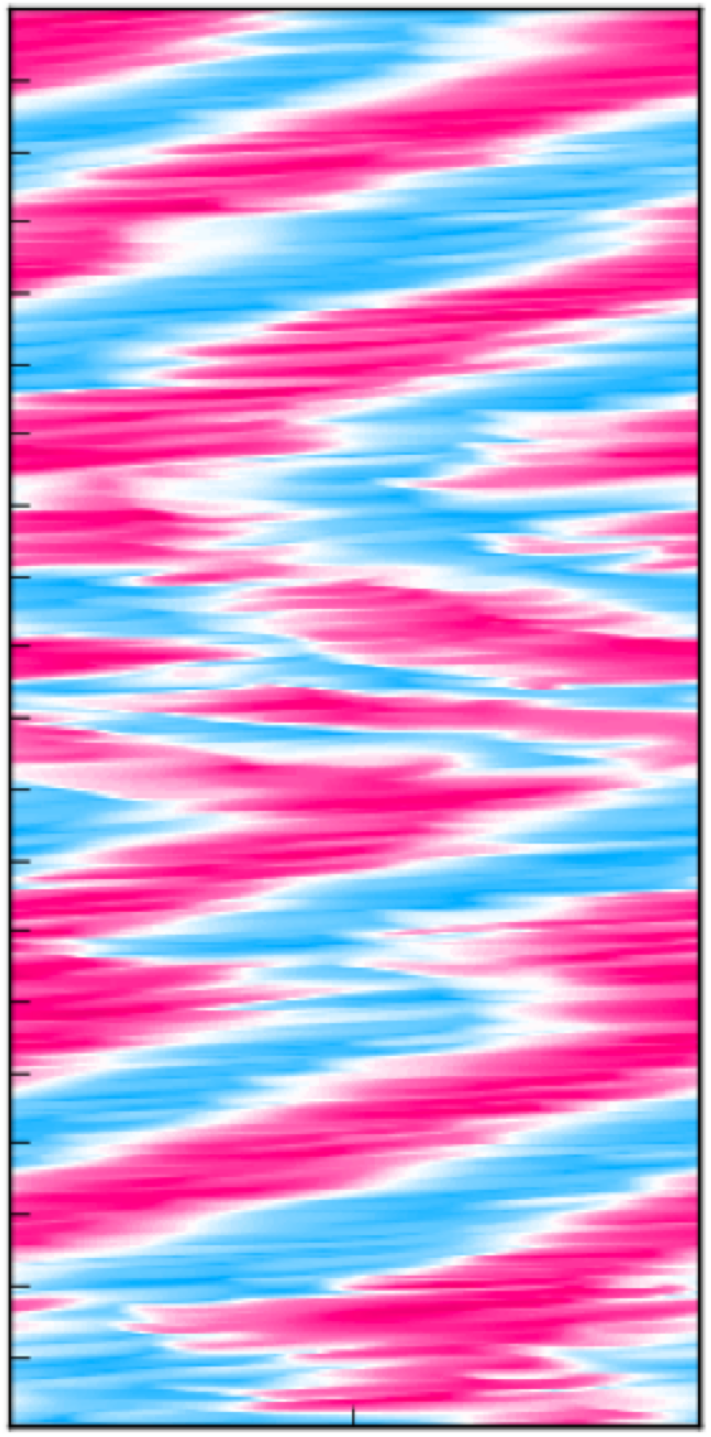}}
\put(10,0.5){\includegraphics[height=8cm]{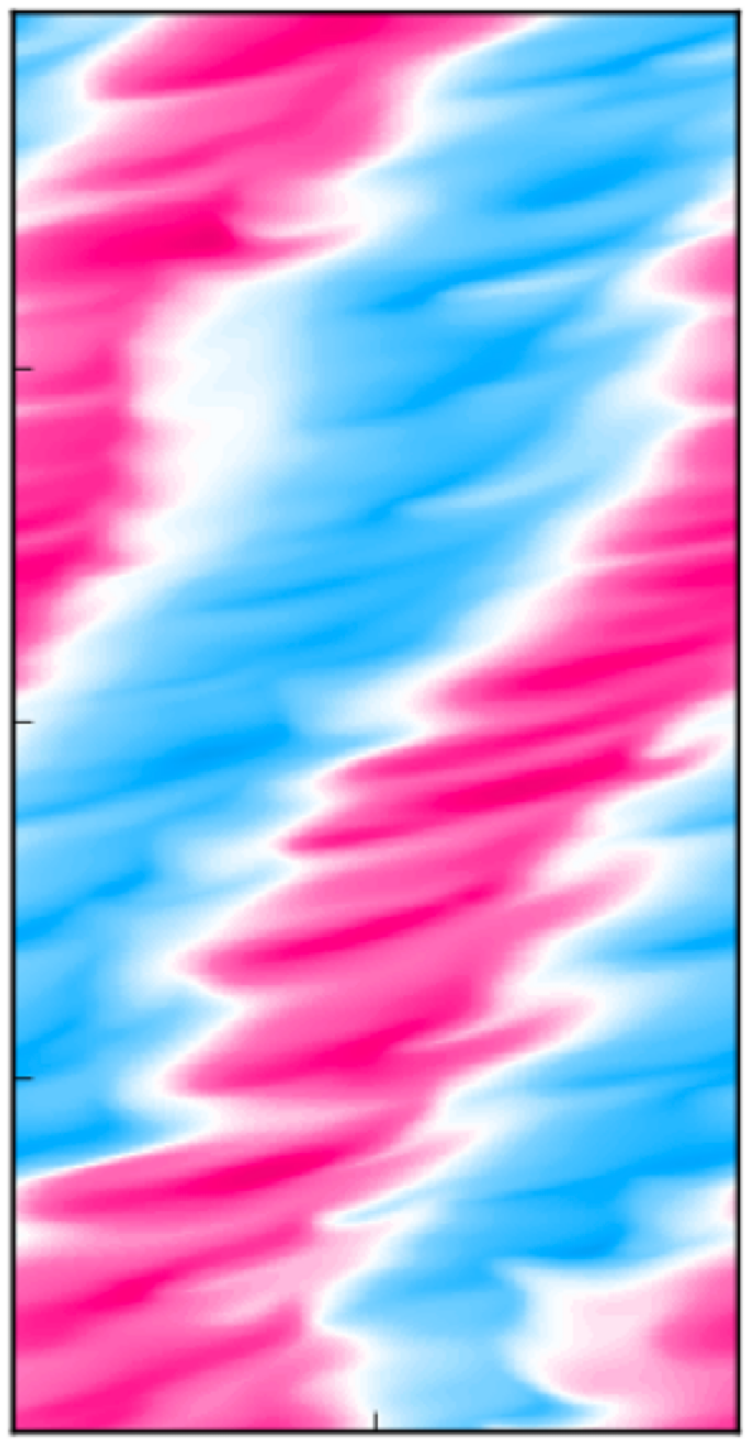}}
\put(3,4.4){$t$}
\put(3.3,8.8){(a)}
\put(3.3,8.3){1000}
\put(3.5,4.4){500}
\put(3.8,0.5){0}
\put(4,0.2){0}
\put(5.9,0){$\phi$}
\put(7.7,0.2){$2\pi$}
\put(9.3,8.8){(b)}
\put(9.5,8.3){200}
\put(9.5,4.4){100}
\put(9.8,0.5){0}
\put(10,0.2){0}
\put(12,0){$\phi$}
\put(13.7,0.2){$2\pi$}
\end{picture}
\caption{
Angle-time plot for $\Ra = 10^8$, $\Gamma=1/3$, and $\Ek = 1.1 \times 10^{-5}$ of $T(r=0.98R, \phi(t), z=1/2)$
\oo
and (a) the whole time series of 1000 time units, (b) the last 200 time units.
\bb
Hotter (red), cooler (blue).
}
\label{AppRetro}
\end{figure*}

Most of the BZF states that we explored had retrograde (anti-cyclonic) precession. The exception is for $\Gamma = 1/3$ with $\Ra=10^8$, and $\Ek = 1.1 \times 10^{-5}$, where we find prograde (cyclonic) precession and alternations between prograde and retrograde directions,
\oo
see Figs.~\ref{AppRetro} (a,b).  
At early $t \lesssim 400$ (Fig.~\ref{AppRetro}(b) shows an expanded interval) and late times $t \gtrsim  700$, the precession is prograde but for intermediate times, an interval of retrograde precession is observed.  It is not determined from our data whether the prograde state is stable at long times. 
\bb 
Uniform cyclonic precession dynamics for wall mode states was predicted theoretically for small $\Gamma$ and $\Pr < 1$ \cite{Goldstein1993} and observed in previous simulations \cite{Horn2017} for $\Gamma = 1/2$, $\Pr = 0.8$, $\Ra = 10^5$ and $\Ek = 1.4 \times 10^{-3}$.  The prograde direction of precession depended sensitively on a combination of $\Gamma$ and $\Ek$ as we also find here whereas retrograde precession is observed in almost all cases;  we see prograde precession for $\Gamma = 1/3$ but not for $\Gamma = 1/5, 1/2, 3/4, 1$ over similar ranges of $\Ra$ and $\Ek$.

\subsection{Eigenfunction RMS}\label{subsec-Eigenfunction}

\begin{figure*}[th]
\unitlength1truecm
\begin{picture}(18,4.5)
\put(0.2,0){
\put(0.5,0.5){\includegraphics[height=3.5cm]{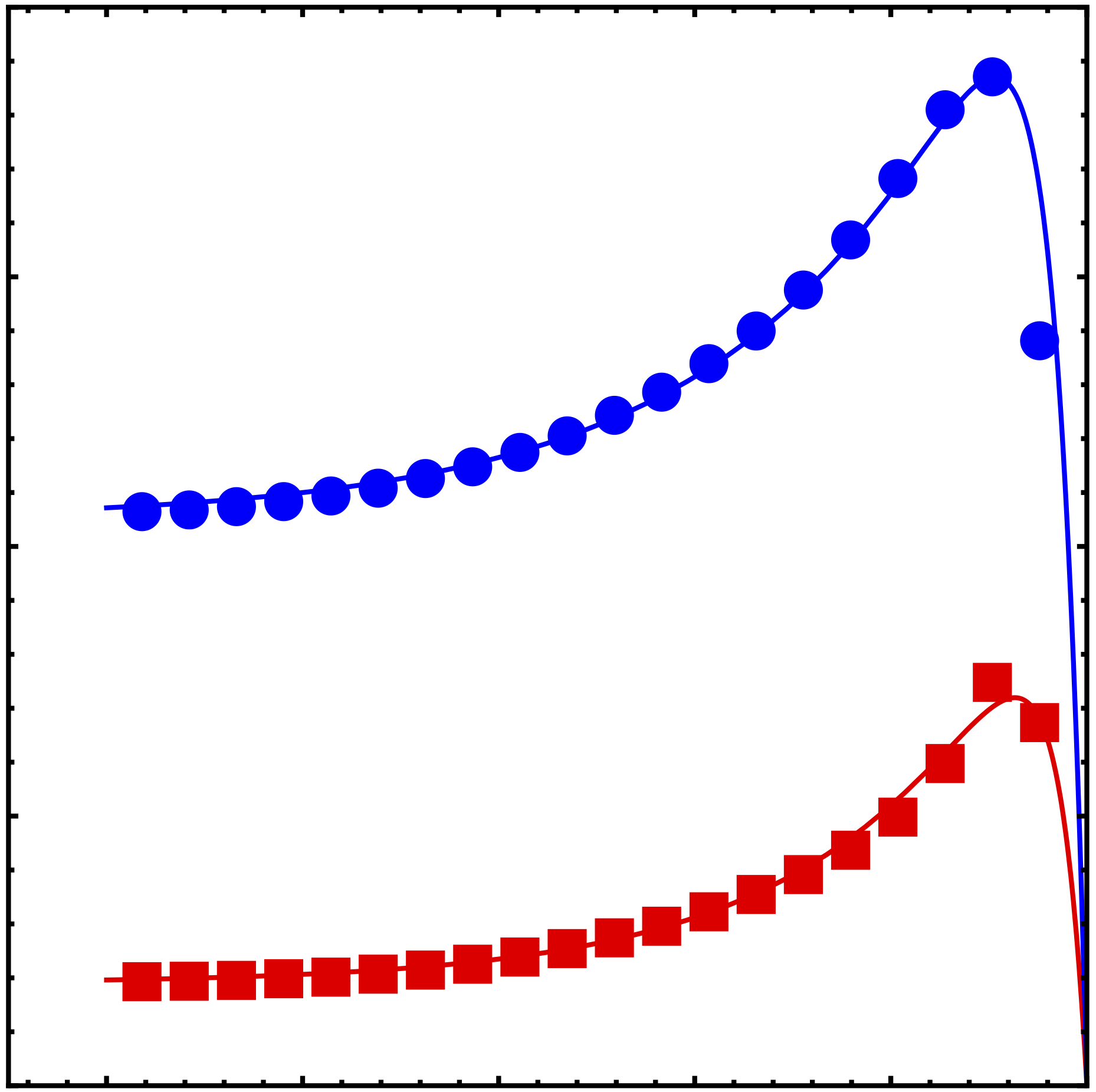}}
\put(2,3.6){(a)}
\put(0.1,0.5){\scriptsize 0.0}
\put(0.1,2.2){\scriptsize 0.1}
\put(0.1,3.9){\scriptsize 0.2}
\put(0.8,2.7){\color{blue}{$\langle T\rangle_{\text{rms}}$}}
\put(0.8,1.2){\color{red}{$\langle T-T^{(1)}\rangle_{\text{rms}}$}}
\put(2.9,2){$\Delta_T^{(1)}$}
\put(4.7,0.5){\includegraphics[height=3.5cm]{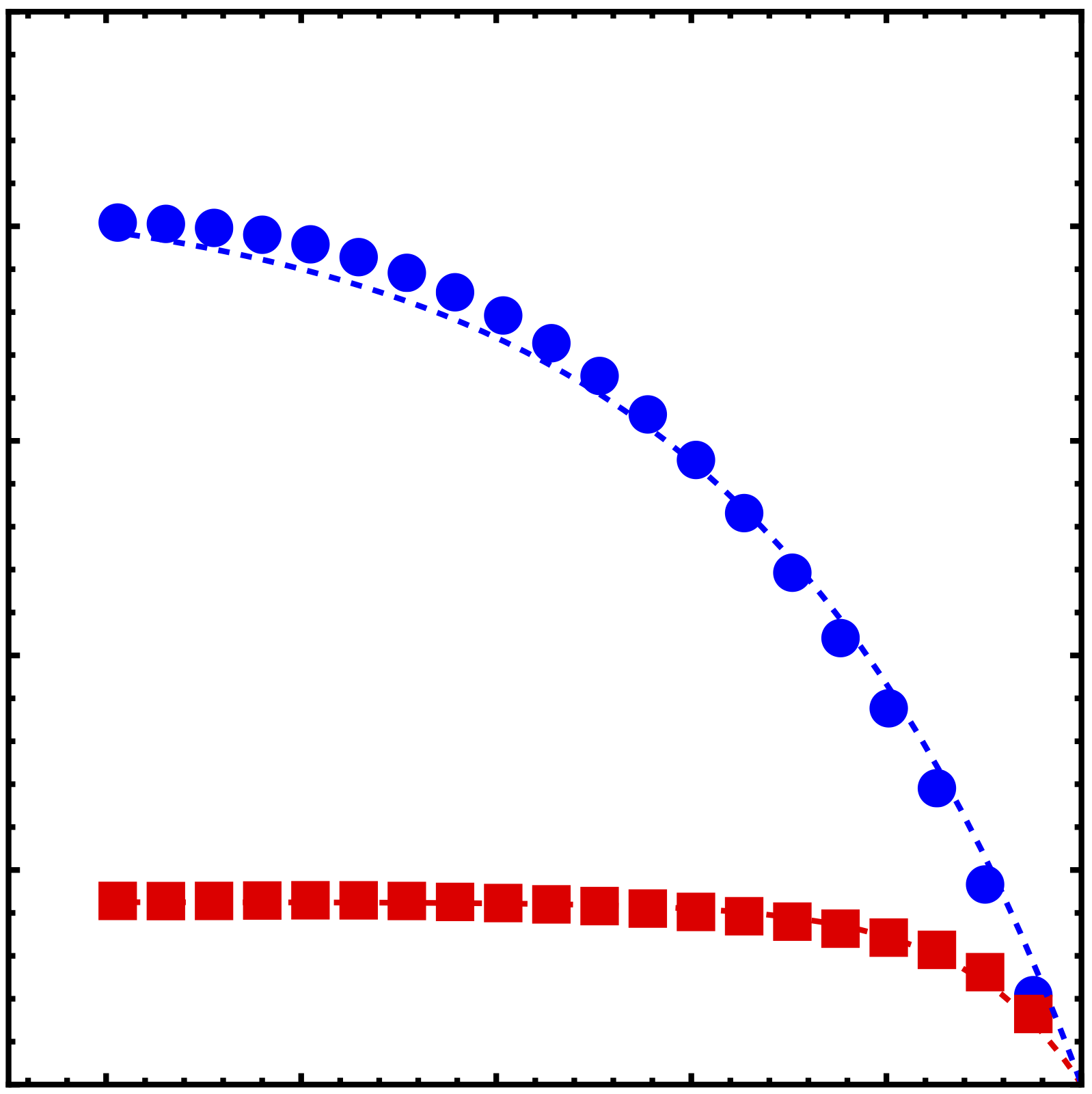}}
\put(6.2,3.6){(b)}
\put(4.3,0.5){\scriptsize 0.0}
\put(4.3,1.17){\scriptsize 0.1}
\put(4.3,1.84){\scriptsize 0.2}
\put(4.3,2.51){\scriptsize 0.3}
\put(4.3,3.18){\scriptsize 0.4}
\put(4.3,3.9){\scriptsize 0.5}
\put(5,2.8){\color{blue}{$\langle u_z\rangle_{\text{rms}}$}}
\put(5,1.4){\color{red}{$\langle u_z-u_z^{(1)}\rangle_{\text{rms}}$}}
\put(8.9,0.5){\includegraphics[height=3.5cm]{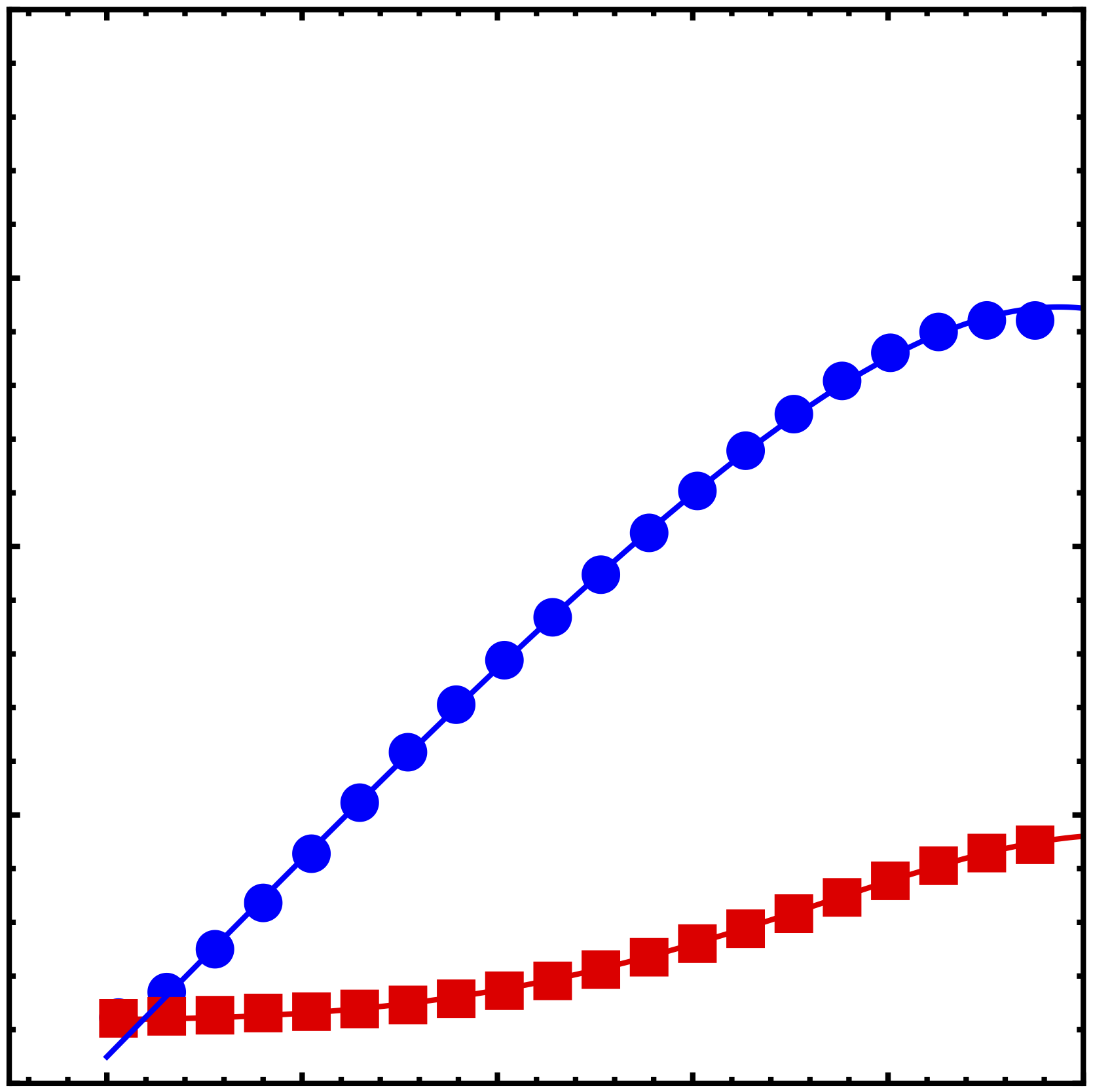}}
\put(10.4,3.6){(c)}
\put(8.5,0.5){\scriptsize 0.0}
\put(8.5,1.35){\scriptsize 0.1}
\put(8.5,2.2){\scriptsize 0.2}
\put(8.5,3.05){\scriptsize 0.3}
\put(8.5,3.9){\scriptsize 0.4}
\put(10,2){\rotatebox{45}{\color{blue}{$\langle u_\phi\rangle_{\text{rms}}$}}}
\put(10,1){\rotatebox{15}{\color{red}{$\langle u_\phi-u_\phi^{(1)}\rangle_{\text{rms}}$}}}
\put(13.1,0.5){\includegraphics[height=3.5cm]{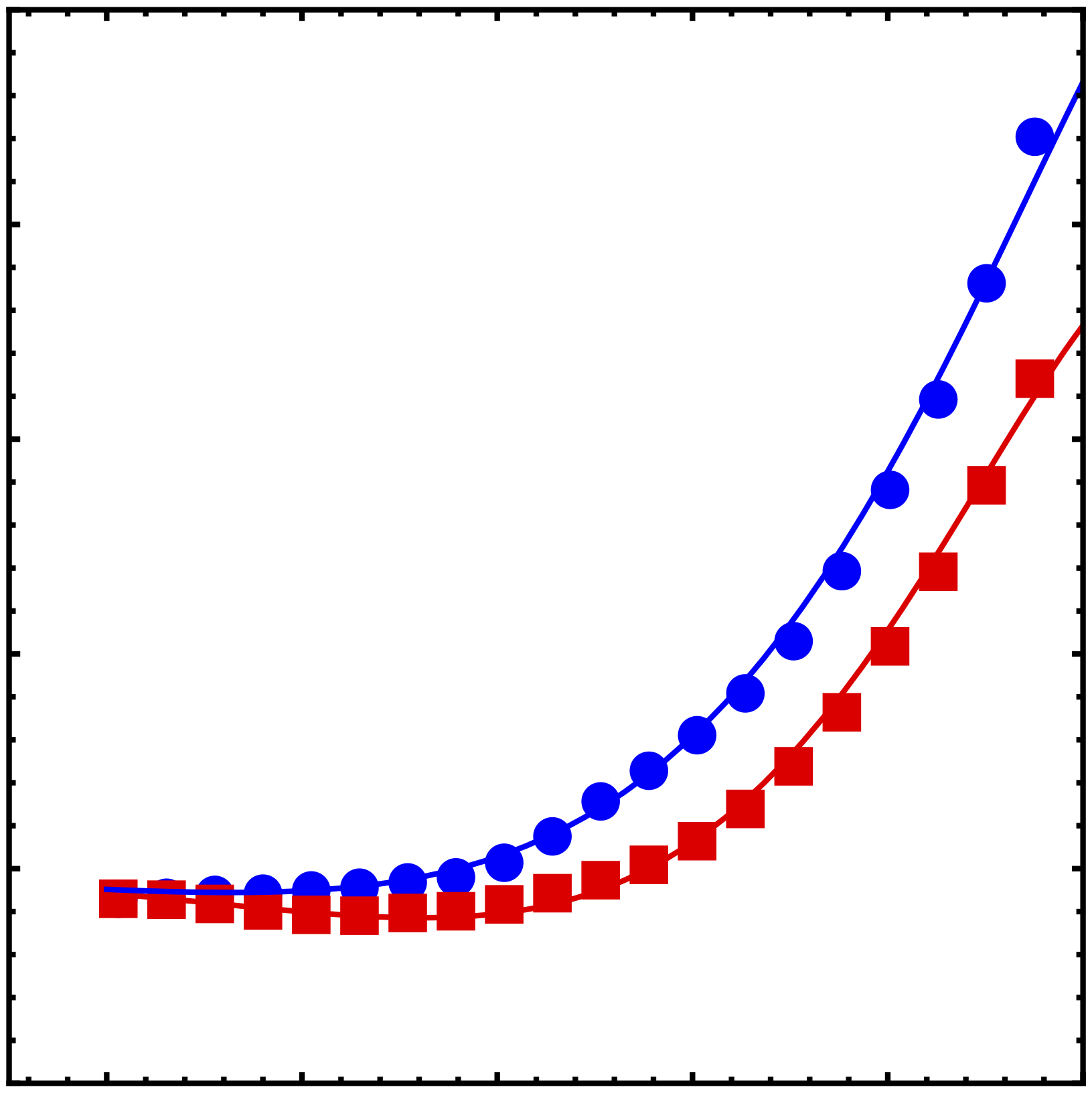}}
\put(14.6,3.6){(d)}
\put(12.6,0.5){\scriptsize 0.00}
\put(12.6,1.17){\scriptsize 0.01}
\put(12.6,1.84){\scriptsize 0.02}
\put(12.6,2.51){\scriptsize 0.03}
\put(12.6,3.18){\scriptsize 0.04}
\put(12.6,3.9){\scriptsize 0.05}
\put(15,2){\rotatebox{53}{\color{blue}{$\langle u_r\rangle_{\text{rms}}$}}}
\put(14.3,0.7){\rotatebox{0}{\color{red}{$\langle u_r-u_r^{(1)}\rangle_{\text{rms}}$}}}
}
\end{picture}
\begin{picture}(18,4.5)
\put(0.2,0){
\put(0.5,1.3){\includegraphics[height=3.5cm]{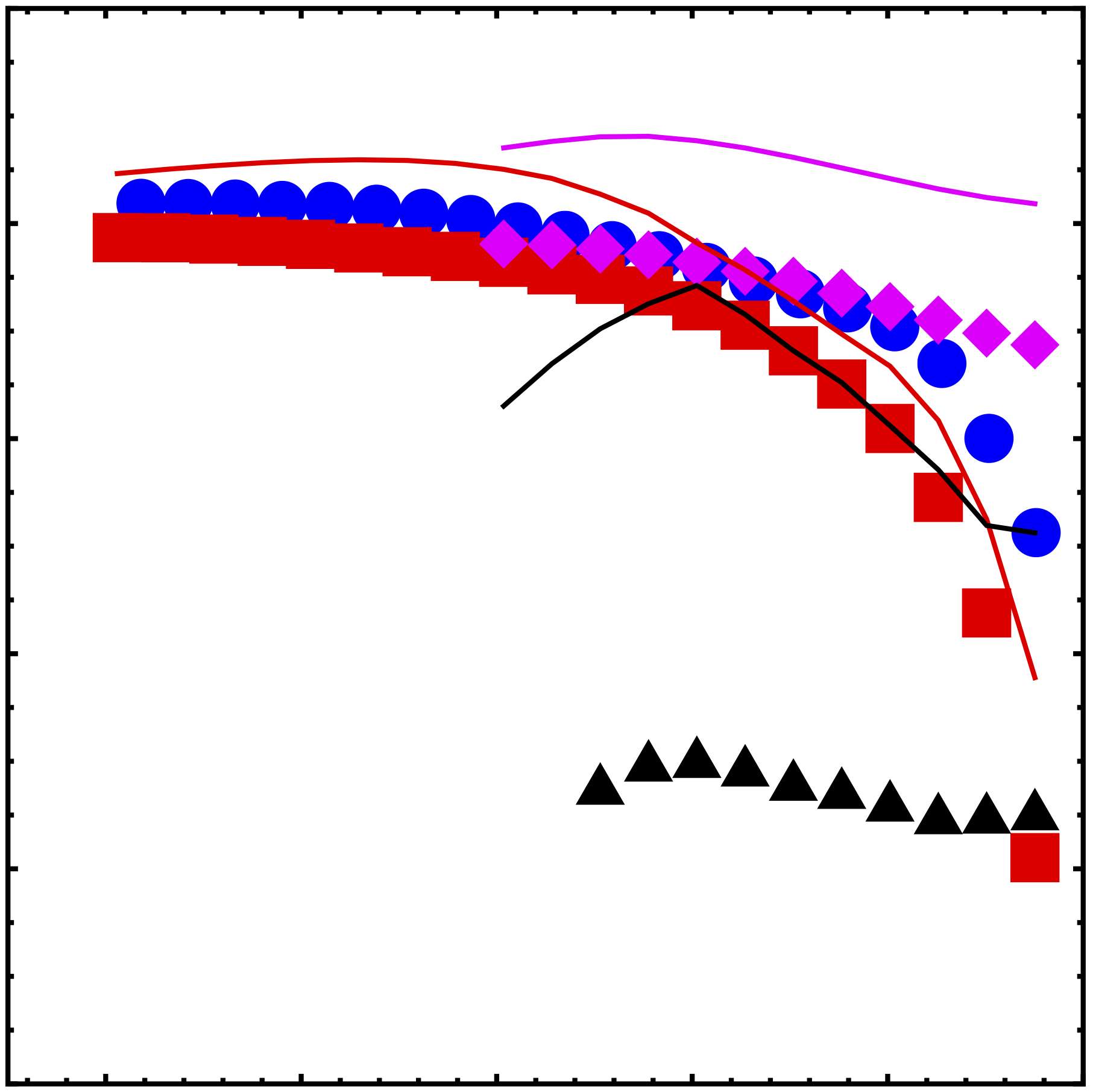}}
\put(0.7,1.8){\includegraphics[height=1.3cm]{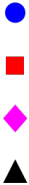}}
\put(2,4.4){(e)}
\put(0.1,1.3){\scriptsize 0.0}
\put(0.1,1.97){\scriptsize 0.2}
\put(0.1,2.64){\scriptsize 0.4}
\put(0.1,3.31){\scriptsize 0.6}
\put(0.1,3.98){\scriptsize 0.8}
\put(0.1,4.7){\scriptsize 1.0}
\put(2.2,2.9){$\dfrac{\Delta_x^{(1)}}{\langle x\rangle_{\text{rms}}}$}
\put(1,2.925){\color{blue}{$T$}}
\put(1,2.575){\color{red}{$u_z$}}
\put(1,2.175){\color{magenta}{$u_\phi$}}
\put(1,1.8){\color{black}{$u_r$}}
\put(0.65,1.0){\scriptsize 0.5}
\put(1.275,1.0){\scriptsize 0.6}
\put(1.9,1.0){\scriptsize 0.7}
\put(2.525,1.0){\scriptsize 0.8}
\put(3.15,1.0){\scriptsize 0.9}
\put(3.75,1.0){\scriptsize 1.0}
\put(2.25,0.6){$z$}
\put(4.7,1.3){\includegraphics[height=3.5cm]{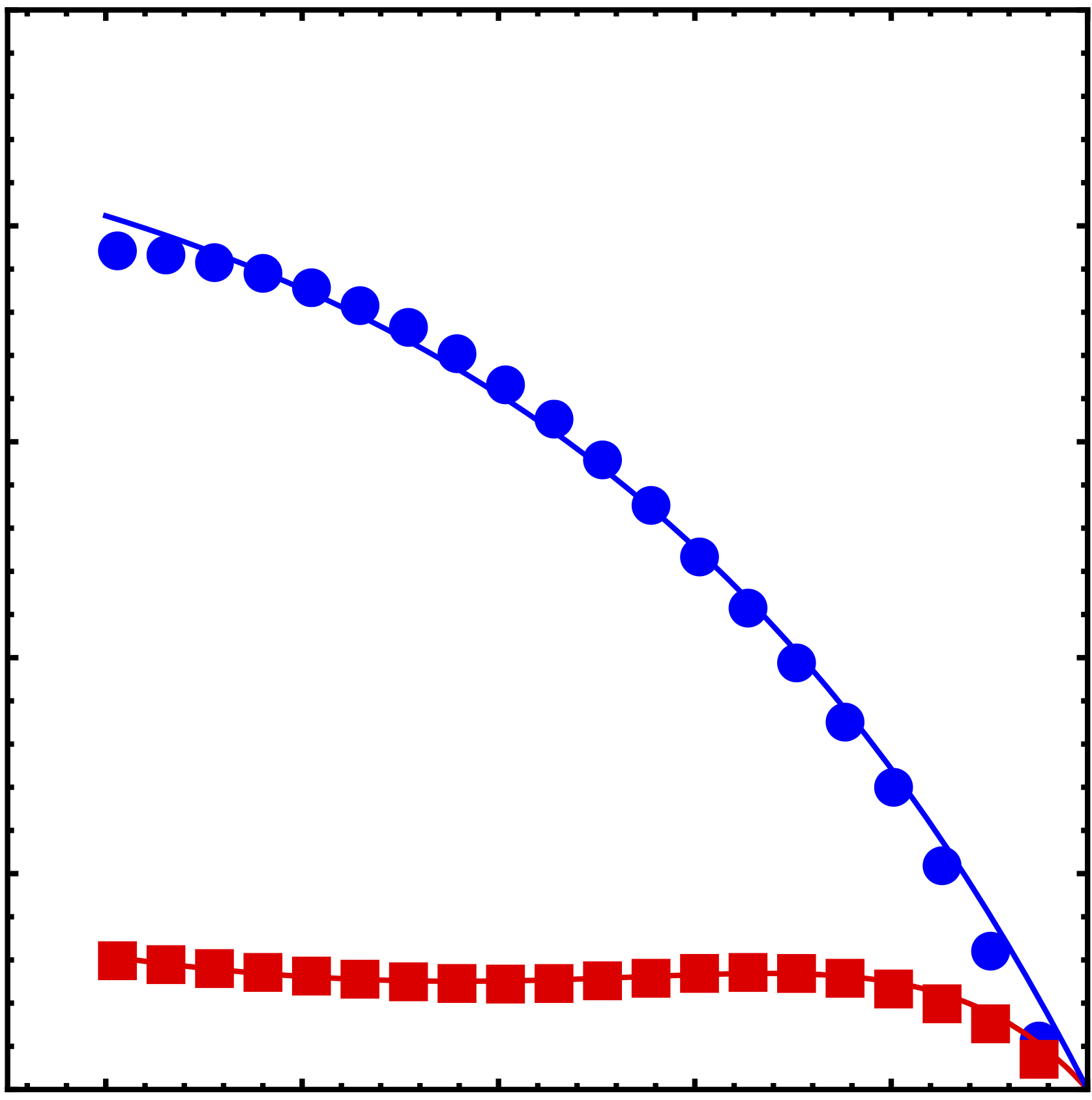}}
\put(6.2,4.4){(f)}
\put(4.3,1.5){\scriptsize 0.0}
\put(4.3,1.97){\scriptsize 0.1}
\put(4.3,2.64){\scriptsize 0.2}
\put(4.3,3.31){\scriptsize 0.3}
\put(4.3,3.98){\scriptsize 0.4}
\put(4.3,4.7){\scriptsize 0.5}
\put(5,3.4){\color{blue}{$\langle u_z\rangle_{\text{rms}}$}}
\put(5,2.0){\color{red}{$\langle u_z-u_z^{(1)}\rangle_{\text{rms}}$}}
\put(4.85,1.0){\scriptsize 0.5}
\put(5.475,1.0){\scriptsize 0.6}
\put(6.1,1.0){\scriptsize 0.7}
\put(6.725,1.0){\scriptsize 0.8}
\put(7.35,1.0){\scriptsize 0.9}
\put(7.95,1.0){\scriptsize 1.0}
\put(6.45,0.6){$z$}
\put(8.9,1.3){\includegraphics[height=3.5cm]{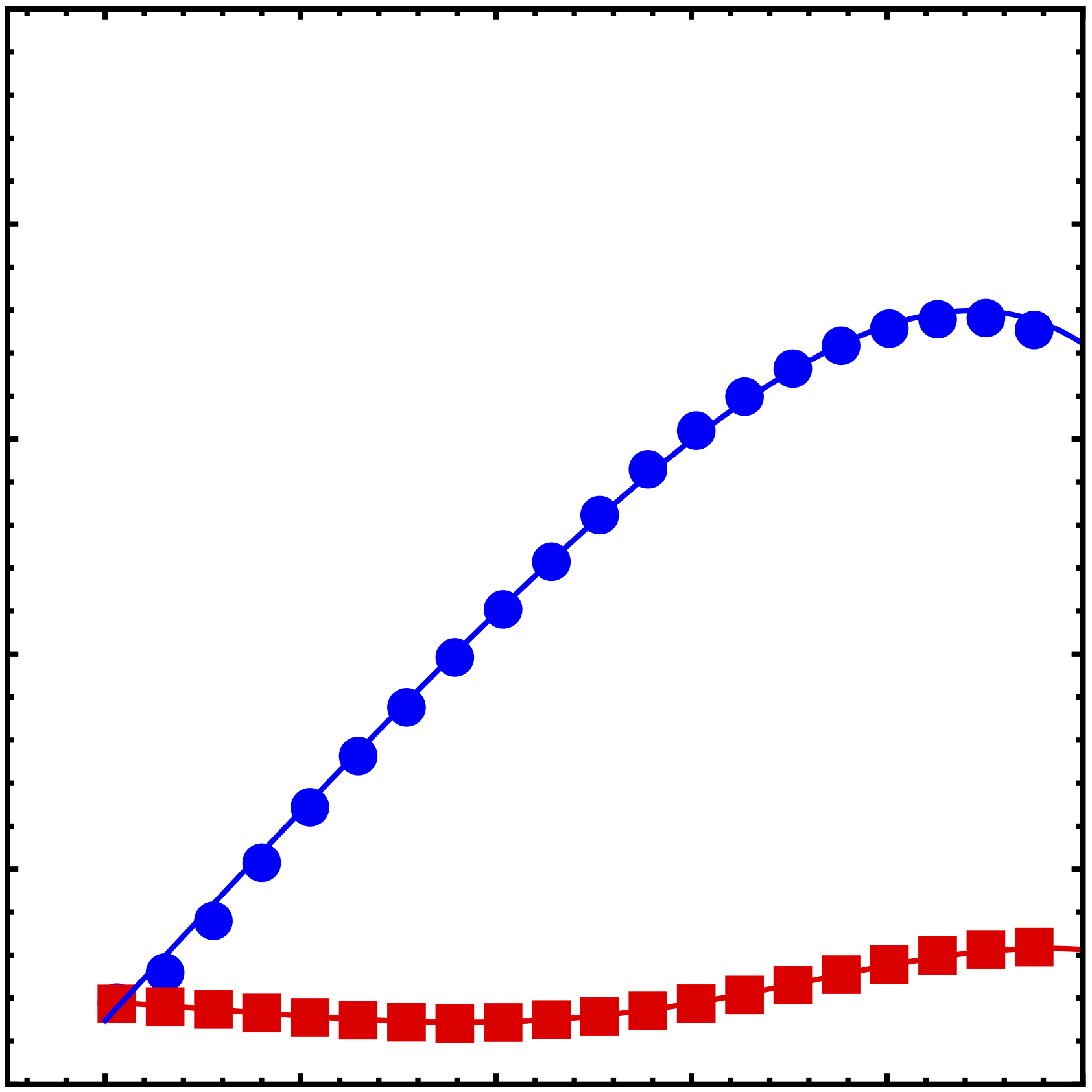}}
\put(10.4,4.4){(g)}
\put(8.5,1.3){\scriptsize 0.0}
\put(8.5,1.97){\scriptsize 0.1}
\put(8.5,2.64){\scriptsize 0.2}
\put(8.5,3.31){\scriptsize 0.3}
\put(8.5,3.98){\scriptsize 0.4}
\put(8.5,4.7){\scriptsize 0.5}
\put(10,2.9){\rotatebox{45}{\color{blue}{$\langle u_\phi\rangle_{\text{rms}}$}}}
\put(10,1.75){\rotatebox{8}{\color{red}{$\langle u_\phi-u_\phi^{(1)}\rangle_{\text{rms}}$}}}
\put(9.05,1.0){\scriptsize 0.5}
\put(9.675,1.0){\scriptsize 0.6}
\put(10.3,1.0){\scriptsize 0.7}
\put(10.925,1.0){\scriptsize 0.8}
\put(11.55,1.0){\scriptsize 0.9}
\put(12.15,1.0){\scriptsize 1.0}
\put(10.55,0.6){$z$}
\put(13.1,1.3){\includegraphics[height=3.5cm]{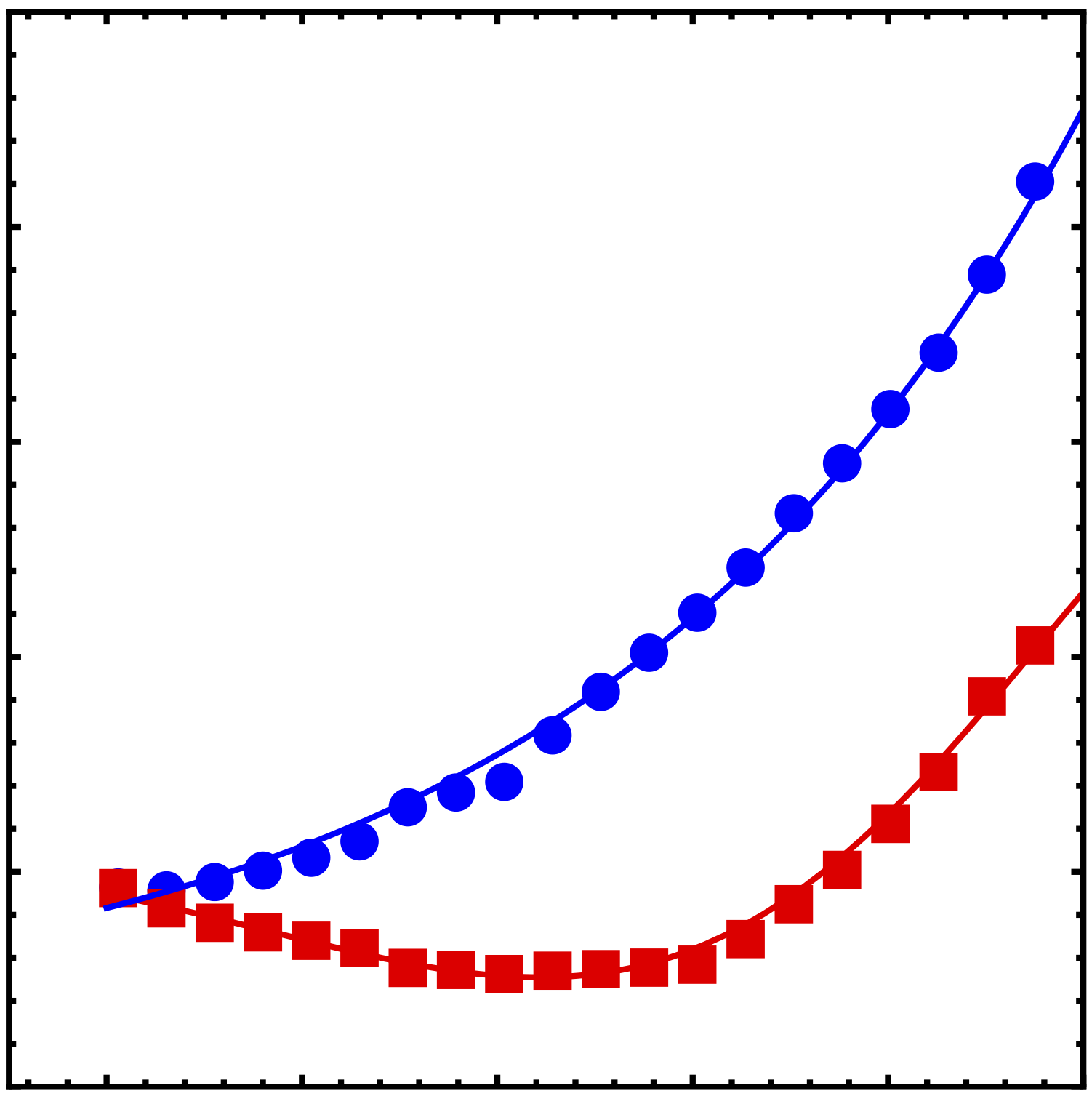}}
\put(14.6,4.4){(h)}
\put(12.6,1.3){\scriptsize 0.00}
\put(12.6,1.97){\scriptsize 0.01}
\put(12.6,2.64){\scriptsize 0.02}
\put(12.6,3.31){\scriptsize 0.03}
\put(12.6,3.98){\scriptsize 0.04}
\put(12.6,4.7){\scriptsize 0.05}
\put(15,3.0){\rotatebox{50}{\color{blue}{$\langle u_r\rangle_{\text{rms}}$}}}
\put(14.8,1.8){\rotatebox{43}{\color{red}{$\langle u_r-u_r^{(1)}\rangle_{\text{rms}}$}}}
\put(13.25,1.0){\scriptsize 0.5}
\put(13.875,1.0){\scriptsize 0.6}
\put(14.5,1.0){\scriptsize 0.7}
\put(15.125,1.0){\scriptsize 0.8}
\put(15.75,1.0){\scriptsize 0.9}
\put(16.35,1.0){\scriptsize 1.0}
\put(14.75,0.6){$z$}
}
\end{picture}
\begin{picture}(18,0.1)
\unitlength1truecm
\put(2.9,6.1){
\begin{tikzpicture}[style=thick]
\unitlength1truecm
    \draw [stealth-stealth](5,8) -- (5,9.5);
\end{tikzpicture}
}
\end{picture}
\caption{
 $\langle X(\phi, z) \rangle_ {{\phi_{rms}}_t}$ (blue solid circles) and $\langle X(\phi,z)-X^{(1)} (z) \rangle_{{\phi_{rms}}_t }$ (red solid squares vs. $z$ where $X^{(1)}$ is first Fourier mode of field $X$ \oo where solid and dashed lines are guides to the eye unless otherwise specified.\bb  (a) $T$, (b) $u_z$, (c) $u_\phi$, and (d) $u_r$ for $\Ra = 5 \times 10^8$ and (f) $u_z$, (g) $u_\phi$, and (h) $u_r$ for $\Ra = 5 \times 10^7$.  \oo ($u_{{\phi}_{rms}}$ and $u_{r_{rms}}$ go to 0 at $r/R = 1$ for $0.99 \lesssim z < 1.0$ consistent with zero tangential BC - not shown). \bb
(e) The normalized difference $\langle X - X^{(1)} \rangle_\text{rms}/\langle X \rangle_\text{rms} - 1$. 
For most of the height (excluding the center $z=1/2$ for $u_\phi$ and $u_r$), $X^{(1)}$ captures about 80\% of the $\phi$ variation of $T$ (blue solid circles), $u_z$ (red solid squares), and $u_\phi$ (magenta solid diamonds) with little variation for $\Ra \approx \Ra_\text{w}$ (solid blue, red, and magenta lines). $u_r$ (black solid triangles and black line) is quite different with only about 20\% for the higher $\Ra$ and $50-70$\% for lower $\Ra$.}
\label{TempRMSz}
\end{figure*}
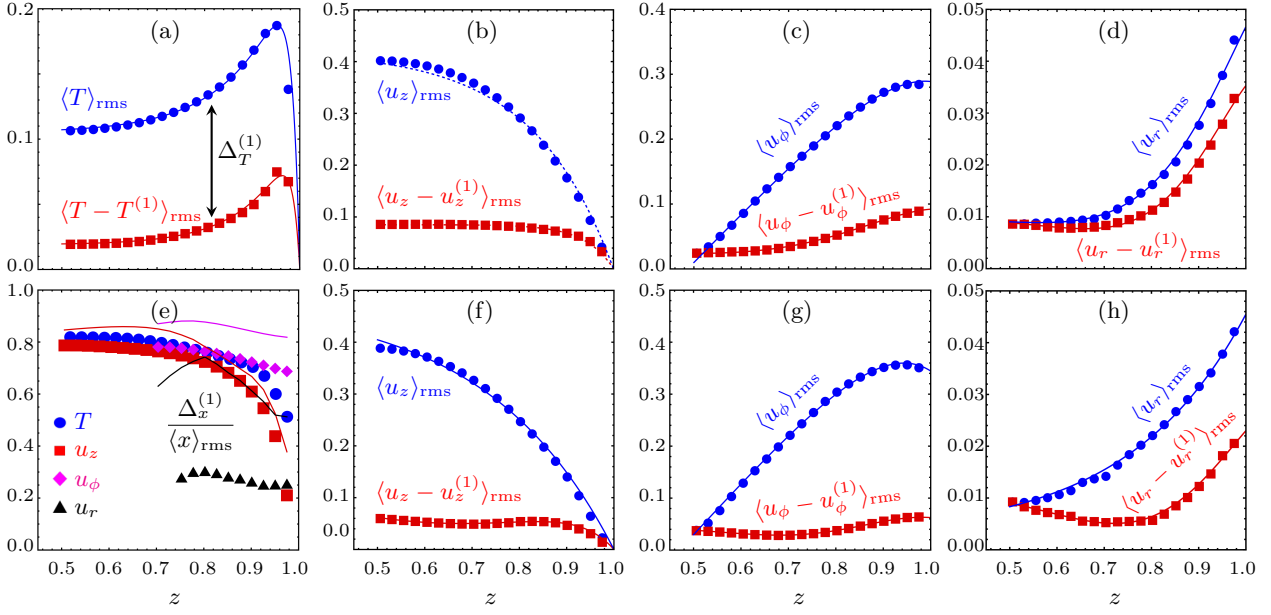

The qualitative features for the field shapes at $\Ra = 5 \times 10^8$ seen in Fig.\ \ref{VertTemp}(e-h) show that the development of the nonlinearity of the wall mode is near regions close to the horizontal isothermal top/bottom boundaries. To make this more quantitative, we compute $X(\phi, z, r=0.98)$  and its first Fourier mode $X^{(1)}(\phi, z, r=0.98)$ at fixed $z$. We then find the rms value with respect to $\phi$ of $X$ and of $X-X^{(1)}$ and average in time.  Figure \ref{TempRMSz} (a-d) shows the $z$ variation of the rms values for $T$, $u_z$, $u_\phi$, and $u_r$, respectively, for $\Ra = 5 \times 10^8$ whereas (f-h) show $u$ fields for $\Ra = 3 \times 10^7$; 
\oo 
the differences with $\Ra$ are not large.
\bb  
The difference $\Delta_X(z)$ (as shown in (a) for $T$) indicates the degree to which the $X^{(1)}(z)$ captures the main features of its $\phi$ dependence.  Figure \ref{TempRMSz} (e) shows the normalized ratio $\Delta_X/X-1$. Except for $u_r$,  over much of the vertical extent almost 80\% of the full RMS value is captured by $X^{(1)}(z)$ with only a small enhancement for $\Ra \approx \Ra_w$.  On the other hand, $u_r$ has very little weight - about 25\% - in the first Fourier mode.

\begin{figure*}[th]
\unitlength1truecm
\begin{picture}(18,7.5)
\put(0.5,0){
\put(2,0.8){\includegraphics[height=6cm]{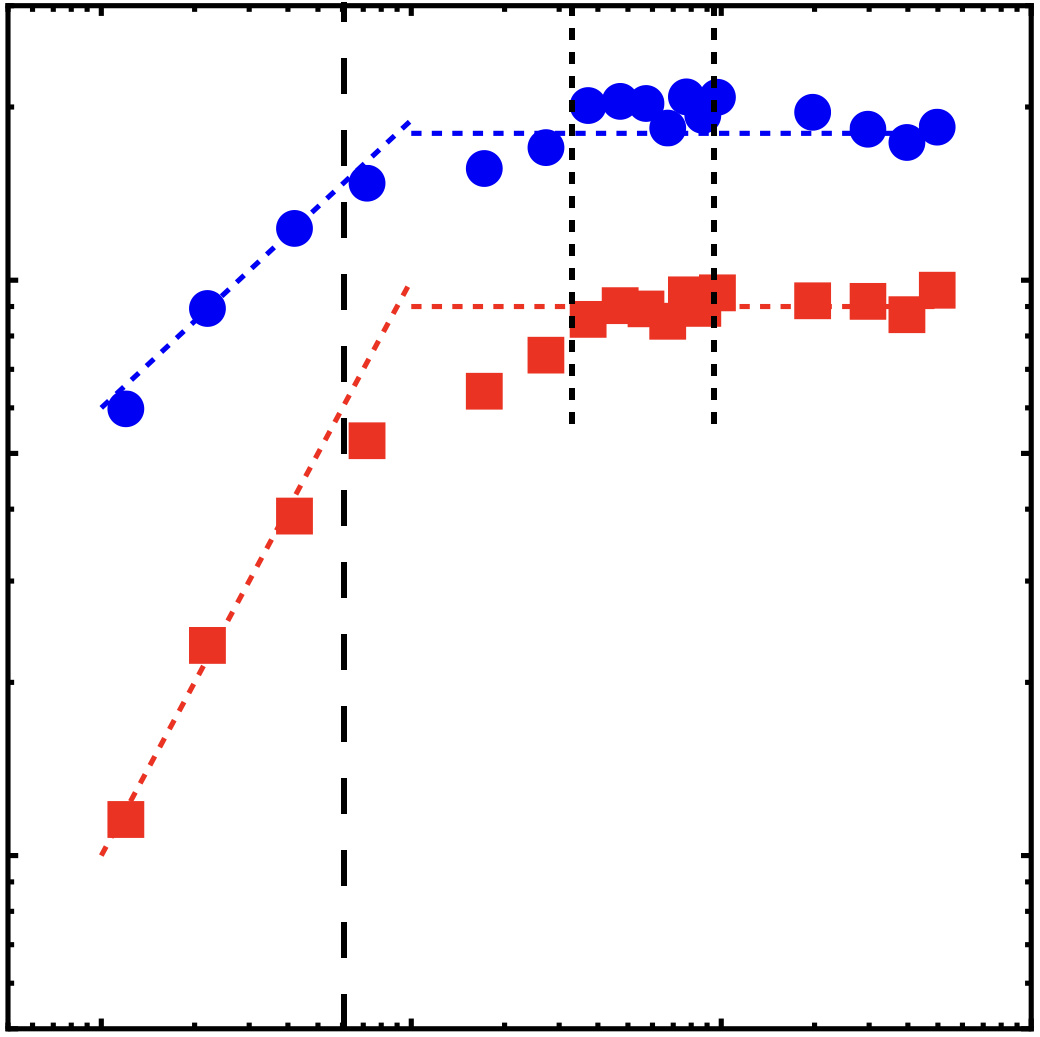}}
\put(4.8,1.3){\includegraphics[height=3cm]{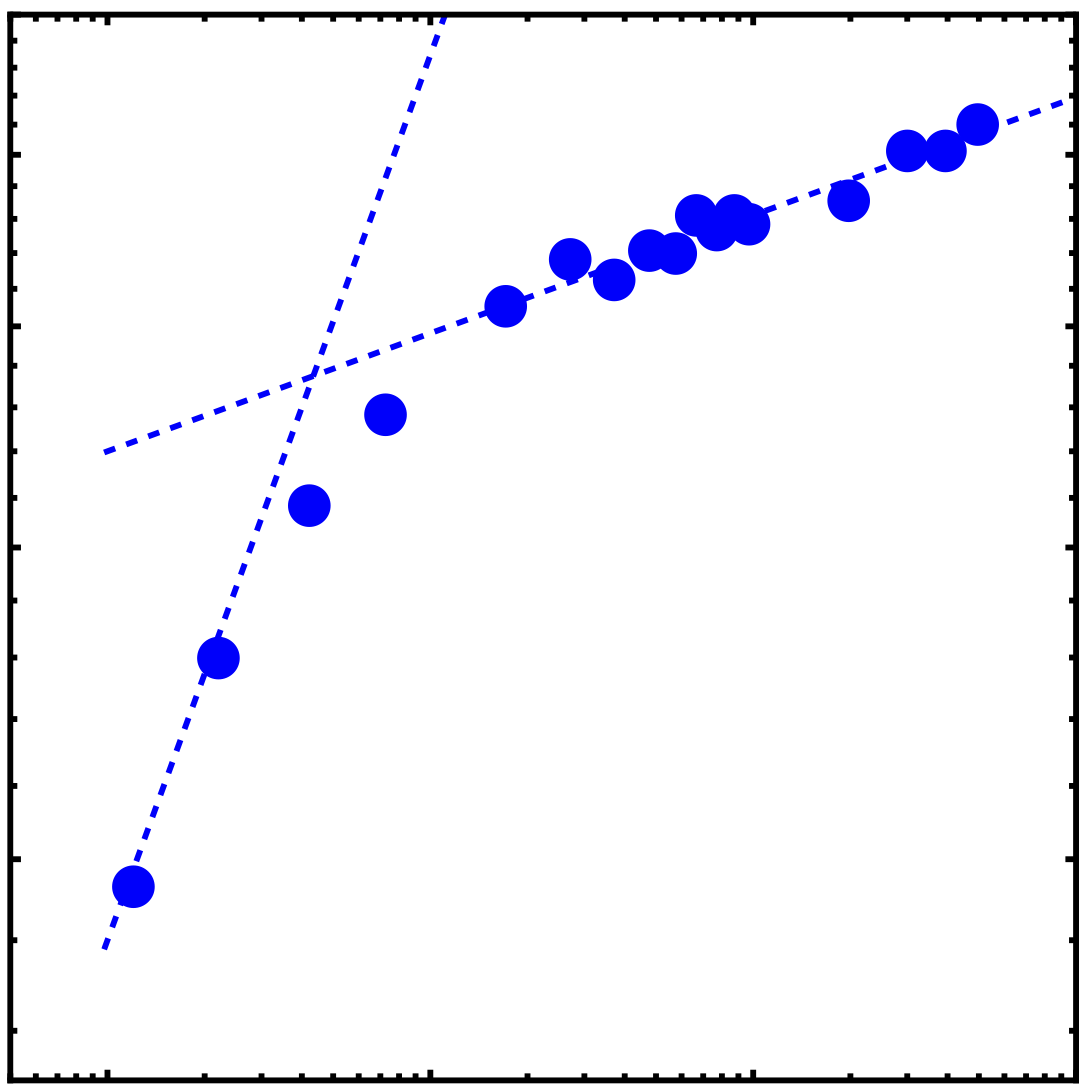}}
\put(10,0.8){\includegraphics[height=6cm]{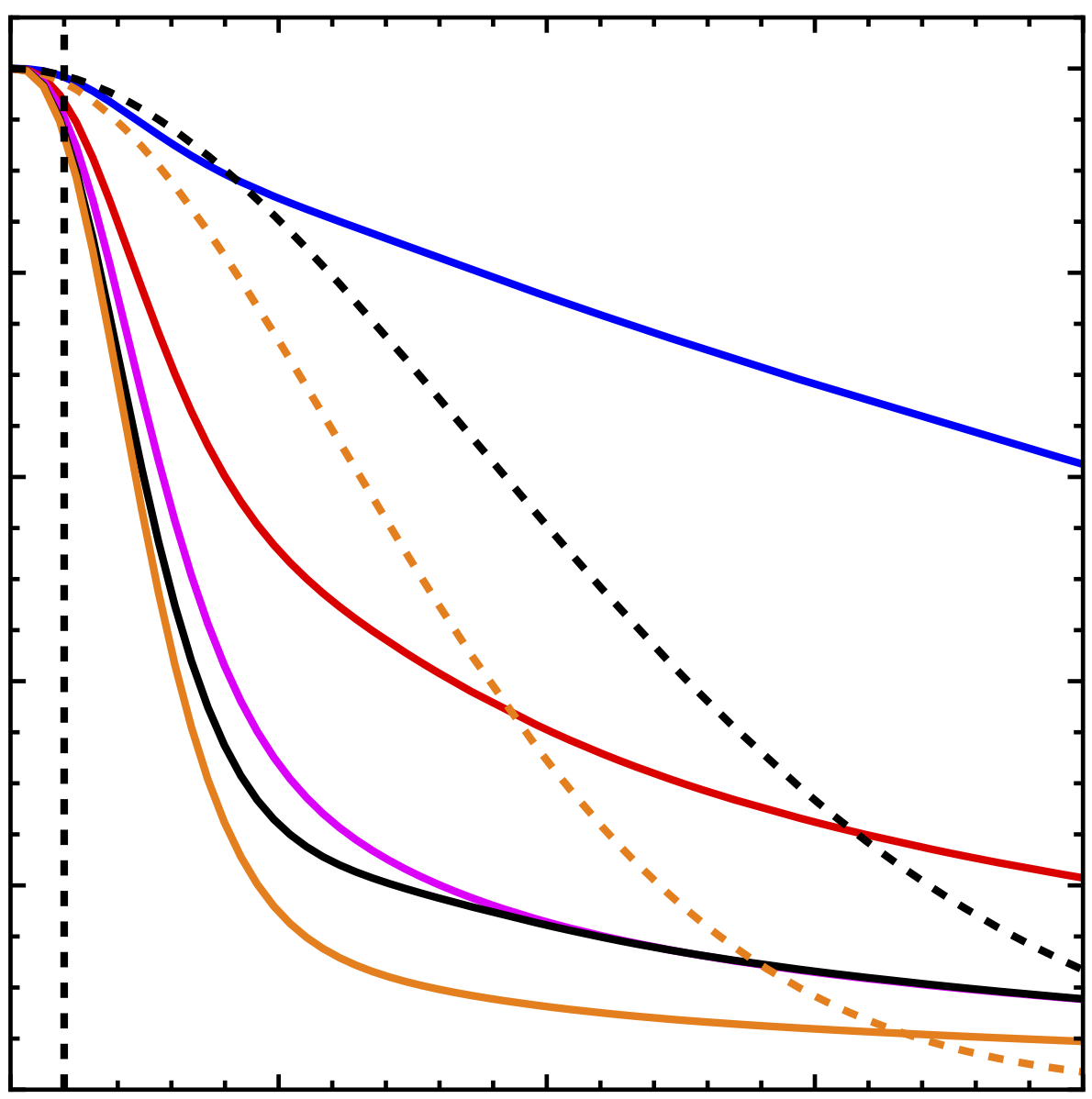}}
\put(0.6,2.3){\rotatebox{90}{$\langle T \rangle_\text{rms}$,\; $\langle T-T^{(1)}\rangle_\text{rms}$}}
\put(1,7){(a)}
\put(1.4,5.1){0.10}
\put(1.4,4.05){0.05}
\put(1.4,1.75){0.01}
\put(1.2,0.75){0.005}
\put(2.4,0.4){$10^7$}
\put(4.2,0.4){$10^8$}
\put(6,0.4){$10^9$}
\put(7.7,0.4){$10^{10}$}
\put(4.7,0){$\Ra-\Ra_\text{w}$}
\put(4.9,1){\scriptsize $10^{7}$}
\put(5.8,1){\scriptsize $10^{8}$}
\put(6.7,1){\scriptsize $10^{9}$}
\put(7.4,1){\scriptsize $10^{10}$}
\put(4.4,3.775){\scriptsize $0.5$}
\put(4.4,3.3){\scriptsize $0.4$}
\put(4.4,2.7){\scriptsize $0.3$}
\put(4.4,1.85){\scriptsize $0.2$}
\put(4.1,2.3){\rotatebox{90}{Ratio}}
\put(2.8,5.6){WM}
\put(5.5,5.6){SC}
\put(6.5,5.6){Bulk}
\put(8.8,2.3){\rotatebox{90}{$(dT/dz)/(\left.dT/dz\right|_{z=0})$}}
\put(9,7){(b)}
\put(9.5,6.4){1.0}
\put(9.5,5.25){0.8}
\put(9.5,4.15){0.6}
\put(9.5,3){0.4}
\put(9.5,1.85){0.2}
\put(9.5,0.75){0.0}
\put(9.95,0.4){$0$}
\put(11.2,0.4){$0.005$}
\put(12.7,0.4){$0.01$}
\put(14.2,0.4){$0.015$}
\put(15.7,0.4){$0.02$}
\put(12.7,5.5){\rotatebox{-18}{\color{blue}{$\Ra/10^7=100$}}}
\put(13.2,2.85){\color{red}{200}}
\put(11.5,2.7){\color{magenta}{300}}
\put(12.3,1.5){\color{black}{400}}
\put(11.5,1.2){\color{orange}{500}}
\put(12.9,0){$z$}
}
\end{picture}
\caption{
\oo
(a) $\langle \langle T(z)\rangle_{\phi,t} \rangle_\text{rms}$ (blue, solid circles) and $\langle \langle T(z) \rangle_{\phi,t} - T^{(1)}(z) \rangle_\text{rms}$ (red, solid squares) vs. $\Ra-\Ra_\text{w}$; $T^{(1)(z)}$ is the first Fourier component of $\langle T(z)\rangle_{\phi,t}$.  
Inset shows the ratio of $\langle T-T^{(1)(z)} \rangle_\text{rms}/\langle T \rangle_\text{rms}$. 
Vertical dashed lines show boundaries of different regions. 
(b) $(dT/dz)/(dT/dz(z=0))$
\bb
vs. $z$ for values of $\Ra/10^7$ in the range of bulk rotating convection: 100, 200, 300, 400, 500. There is no clear demarcation of a thermal boundary layer with an approximately constant slope only for $z$ values within an Ekman boundary layer $\delta_{\Ek} = \Ek^{1/2} = 0.001$ indicated by dashed black vertical line. The dashed orange and black curves are the corresponding profiles for non-rotating convection with the same $\Nu$ as for the corresponding solid orange and black curves.\
}
\label{RMSzprofile}
\end{figure*}

For the vertical profiles $T(z)$, we computed the RMS analysis of the deviation of $T(z)$ from a linear profile, i.e., $\theta(z) = T(z) - (1-z)$, with respect to its first Fourier mode for different $\Ra$ as shown in 
\oo
Fig.~\ref{RMSzprofile}(a)
\bb
where one sees the weakly nonlinear growth $\sim (\Ra-\Ra_w)^{1/2}$ (blue dashed) for the $\theta_{rms}$. Subtracting out the dominant linear eigenfunction one gets second order scaling of $\sim \Ra-\Ra_w$ (red dashed) for $\Ra \leq 1 \times 10^8$. The saturation at higher $\Ra$ results from increasing gradients near the top/bottom plates but saturating amplitude of $\theta$. 
Finally, we show in Fig.~\ref{AppRetro} (b) the boundary layer profiles $dT/dz (z)$ near the bottom boundary in the bulk/BZF region $1 \leq \Ra/10^9 \leq 5$. The curve (blue) for $\Ra = 1 \times 10^9$ has a form still dominated by the wall-mode whereas the higher $\Ra$ curves fall off much faster. 
For comparison the the boundary layer profiles of $dT/dz$ for non-rotating convection for the same total $\Nu$ are shown for $\Ra/10^9$ = 4 and 5. 
The RRBC profiles are quite different than those of non-rotating RBC.  

\subsection{Movies}\label{subsec-Movies}

\begin{figure*}[th]
\unitlength1truecm
\begin{picture}(18,3.8)
\put(1,0){
\put(0,0){\includegraphics[height=3cm]{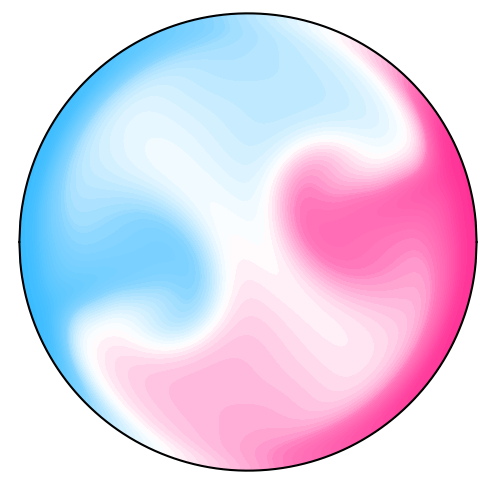}}
\put(3.2,0){\includegraphics[height=3cm]{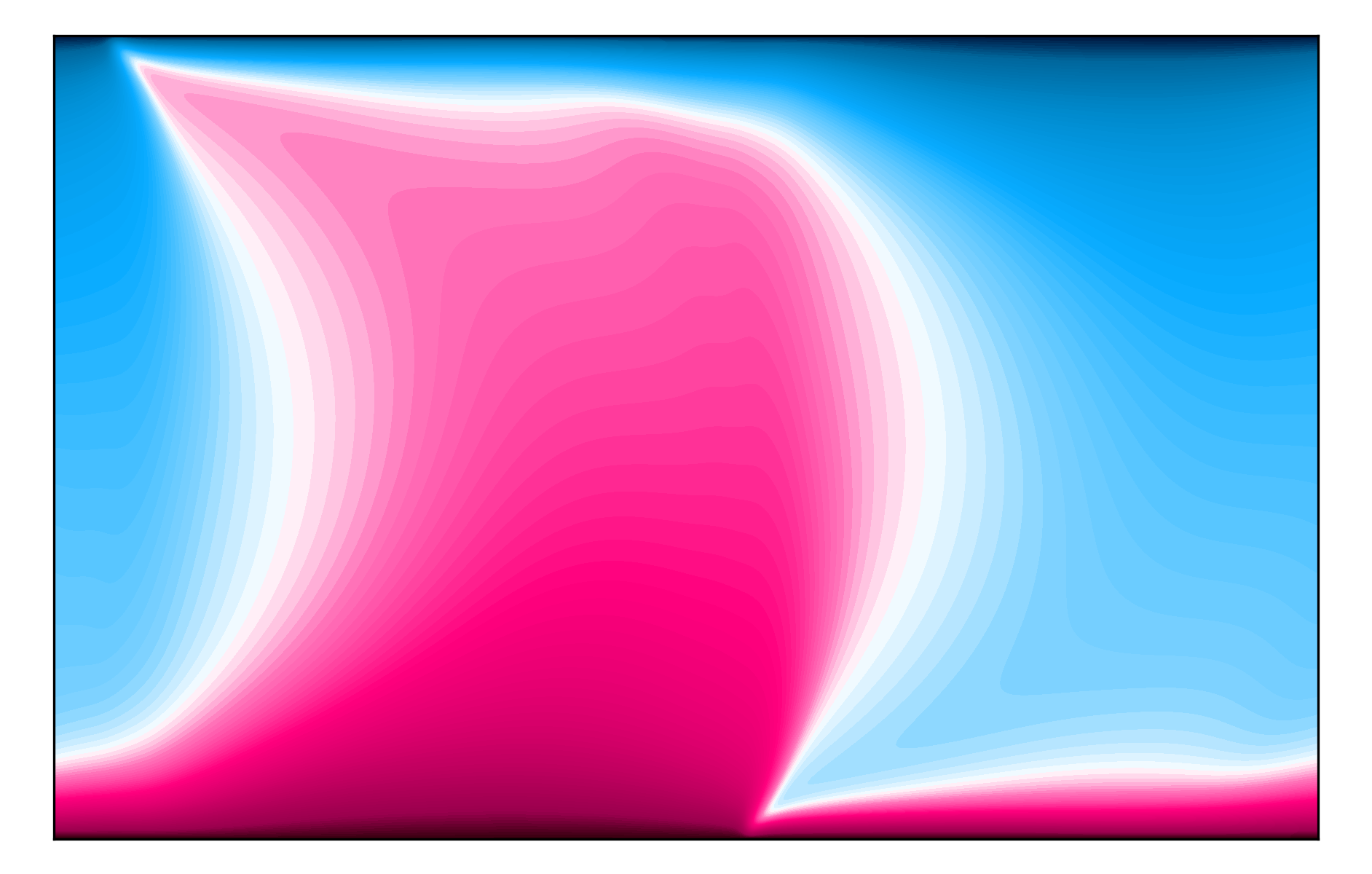}}
\put(8,0){\includegraphics[height=3cm]{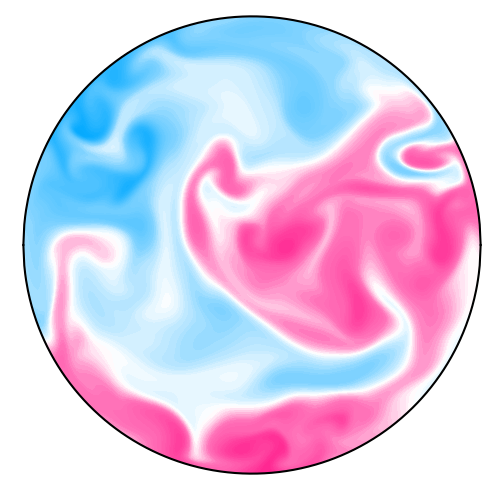}}
\put(11.2,0){\includegraphics[height=3cm]{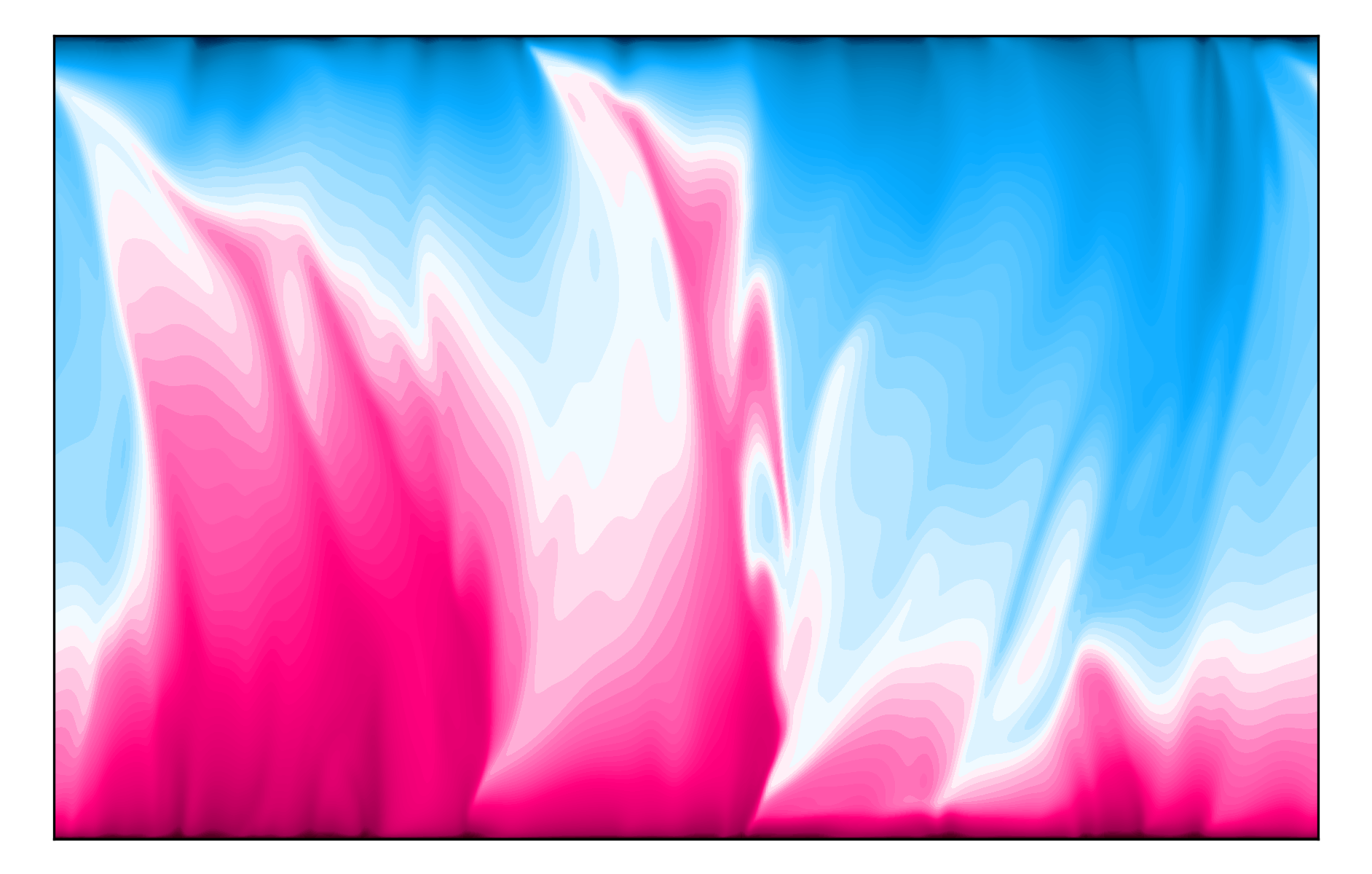}}
\put(1.25,3.2){(a)}
\put(5.3,3.2){(b)}
\put(9.25,3.2){(c)}
\put(13.3,3.2){(d)}
}
\end{picture}
\begin{picture}(18,3.8)
\put(1,0){
\put(0,0){\includegraphics[height=3cm]{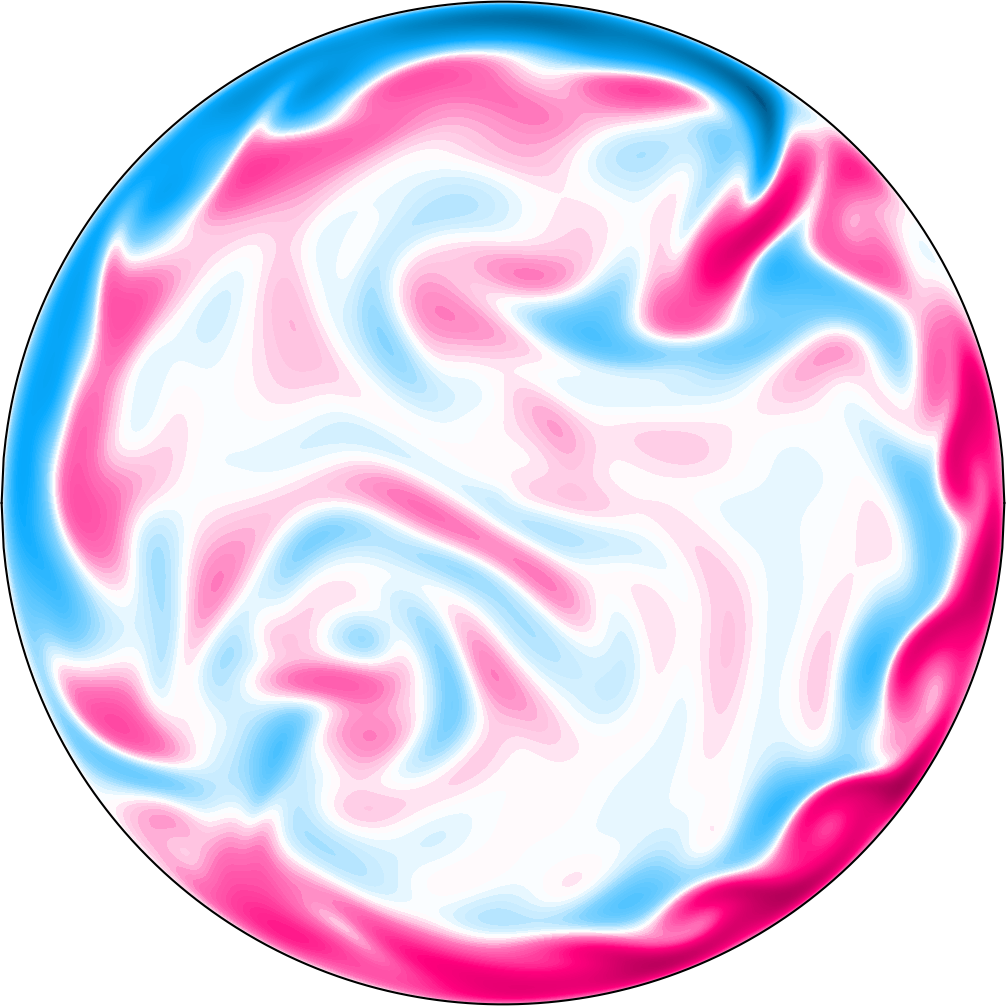}}
\put(3.2,0){\includegraphics[height=3cm]{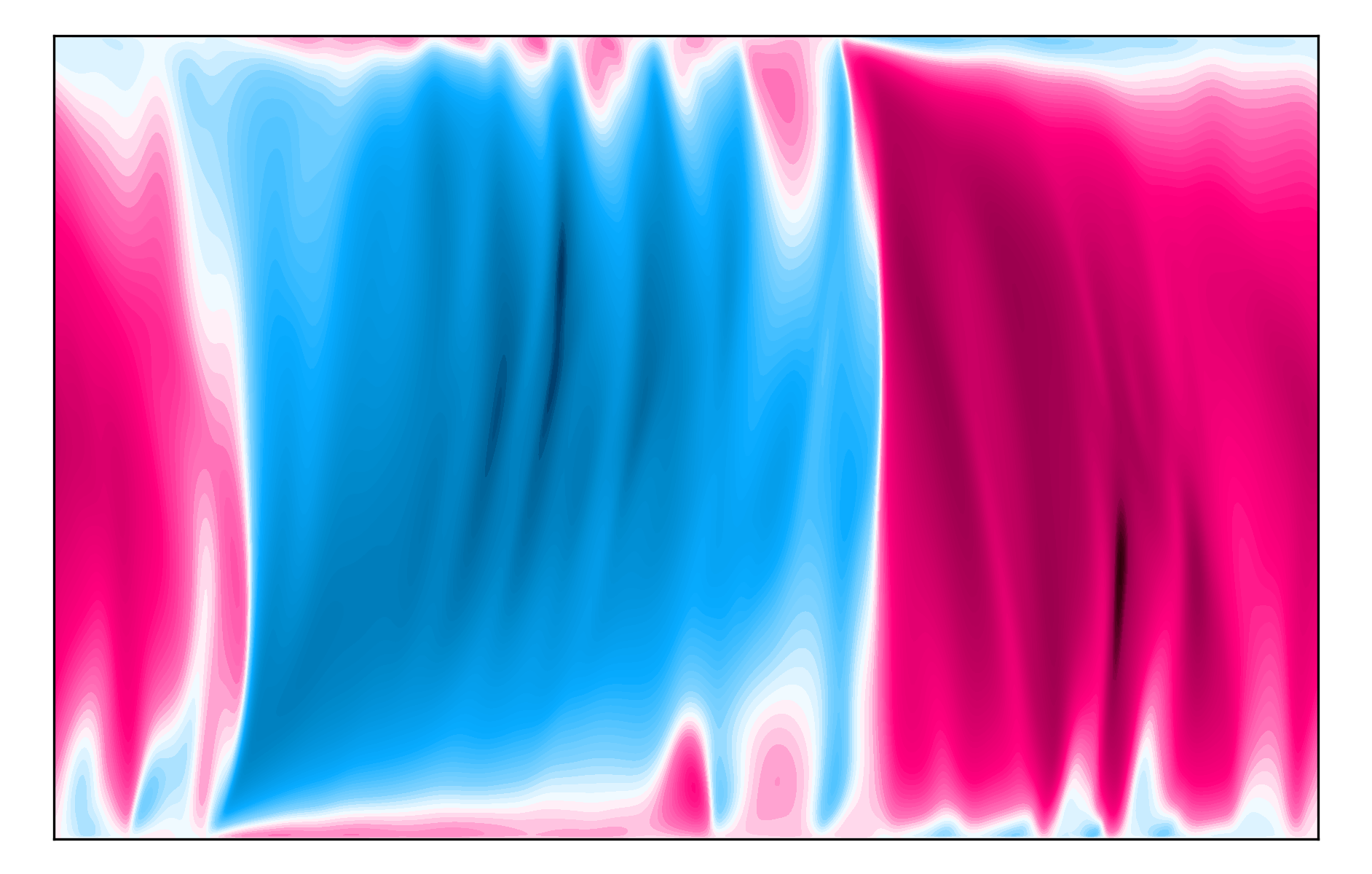}}
\put(8,0){\includegraphics[height=3cm]{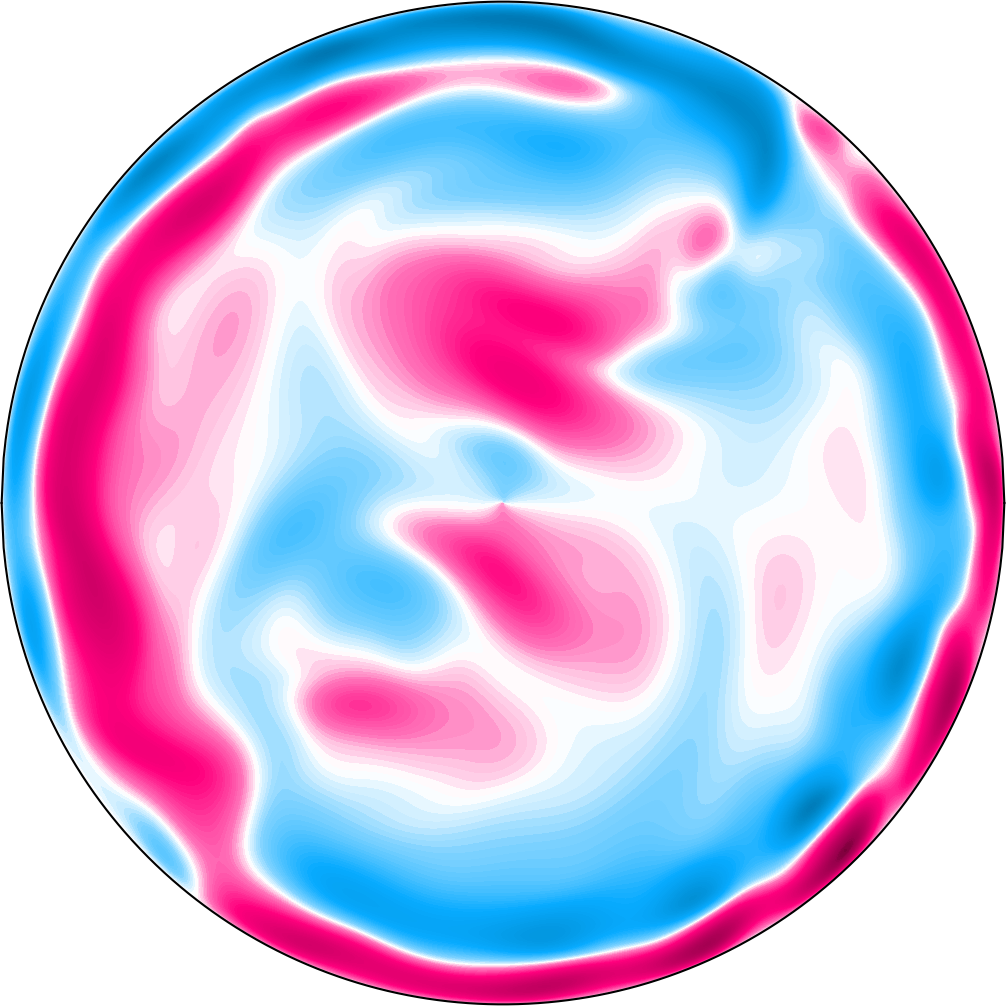}}
\put(11.2,0){\includegraphics[height=3cm]{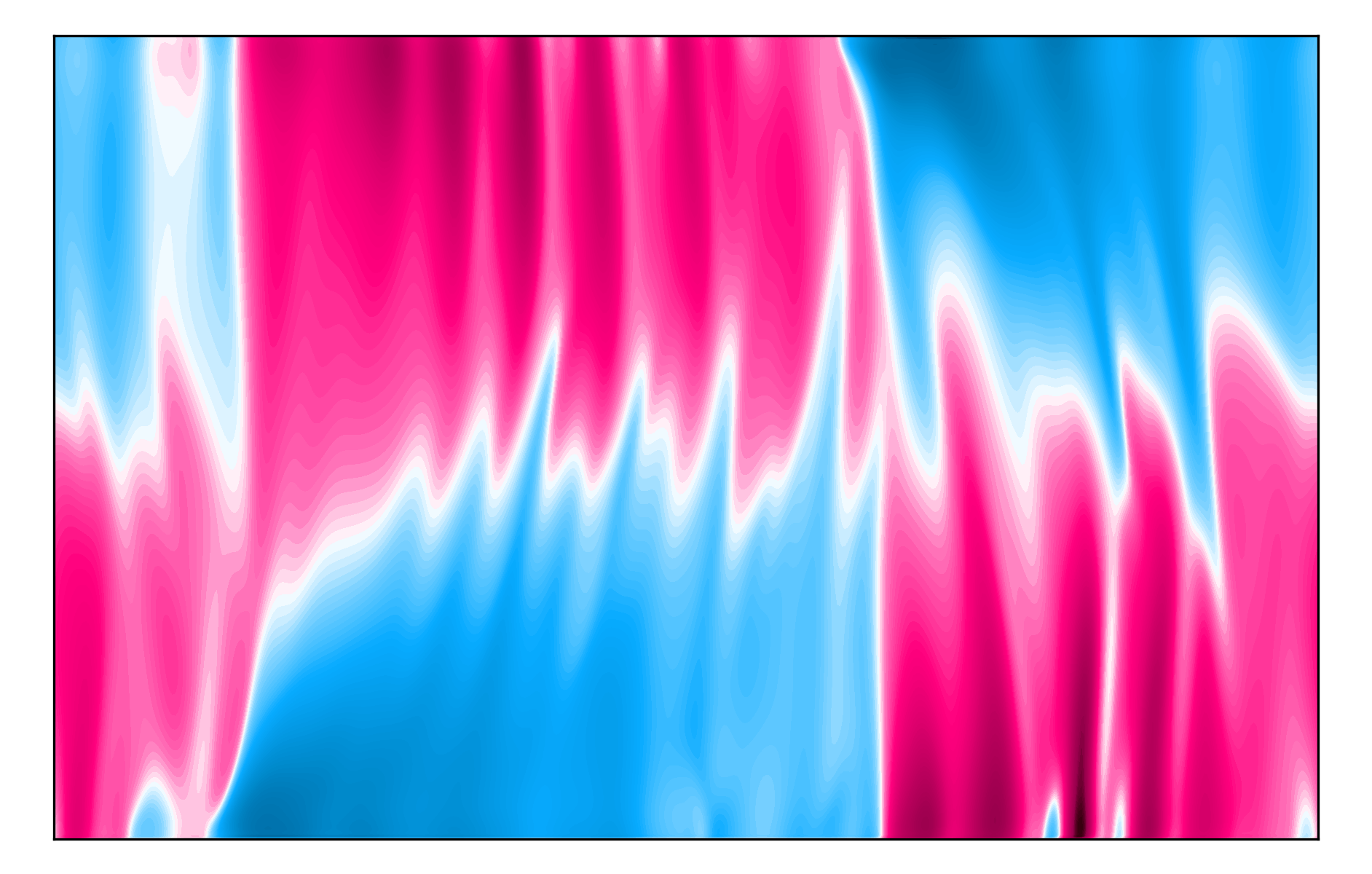}}
\put(1.25,3.2){(e)}
\put(5.3,3.2){(f)}
\put(9.25,3.2){(g)}
\put(13.3,3.2){(h)}
}
\end{picture}
\begin{picture}(18,3.8)
\put(1,0){
\put(0,0){\includegraphics[height=3cm]{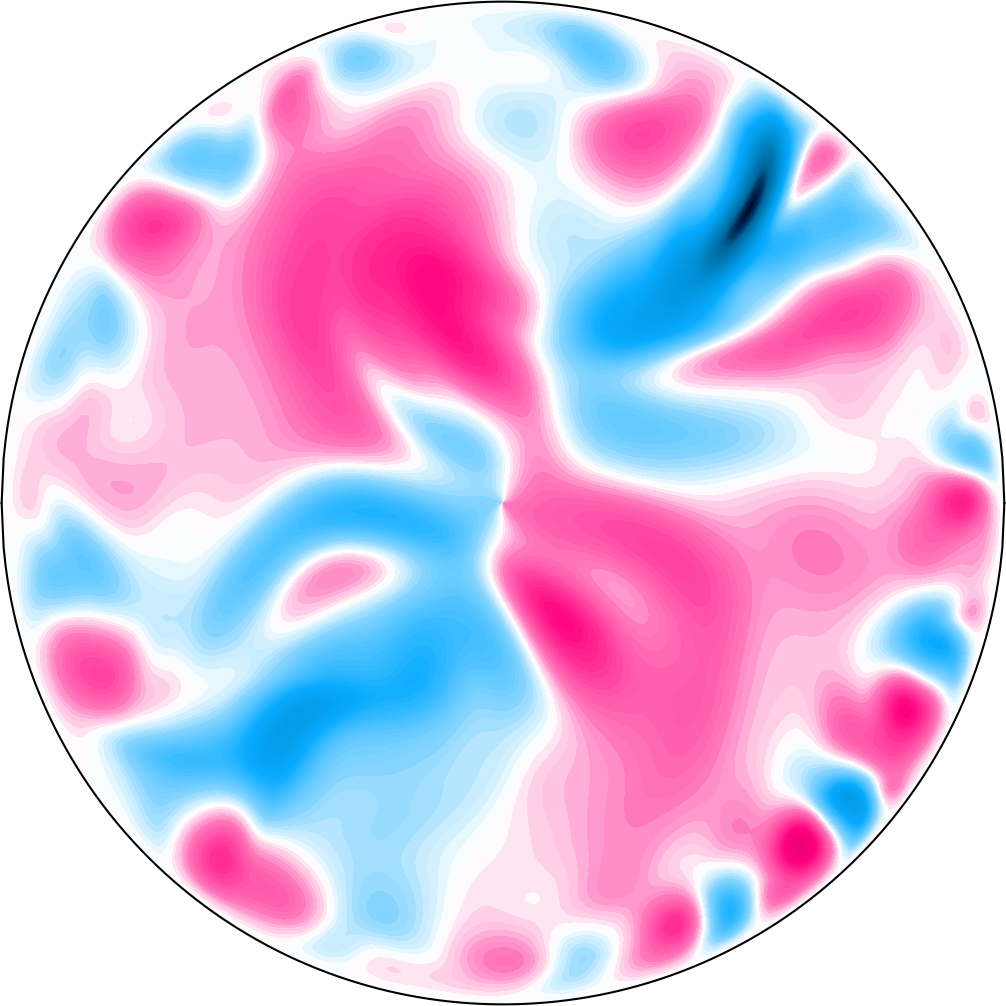}}
\put(3.2,0){\includegraphics[height=3cm]{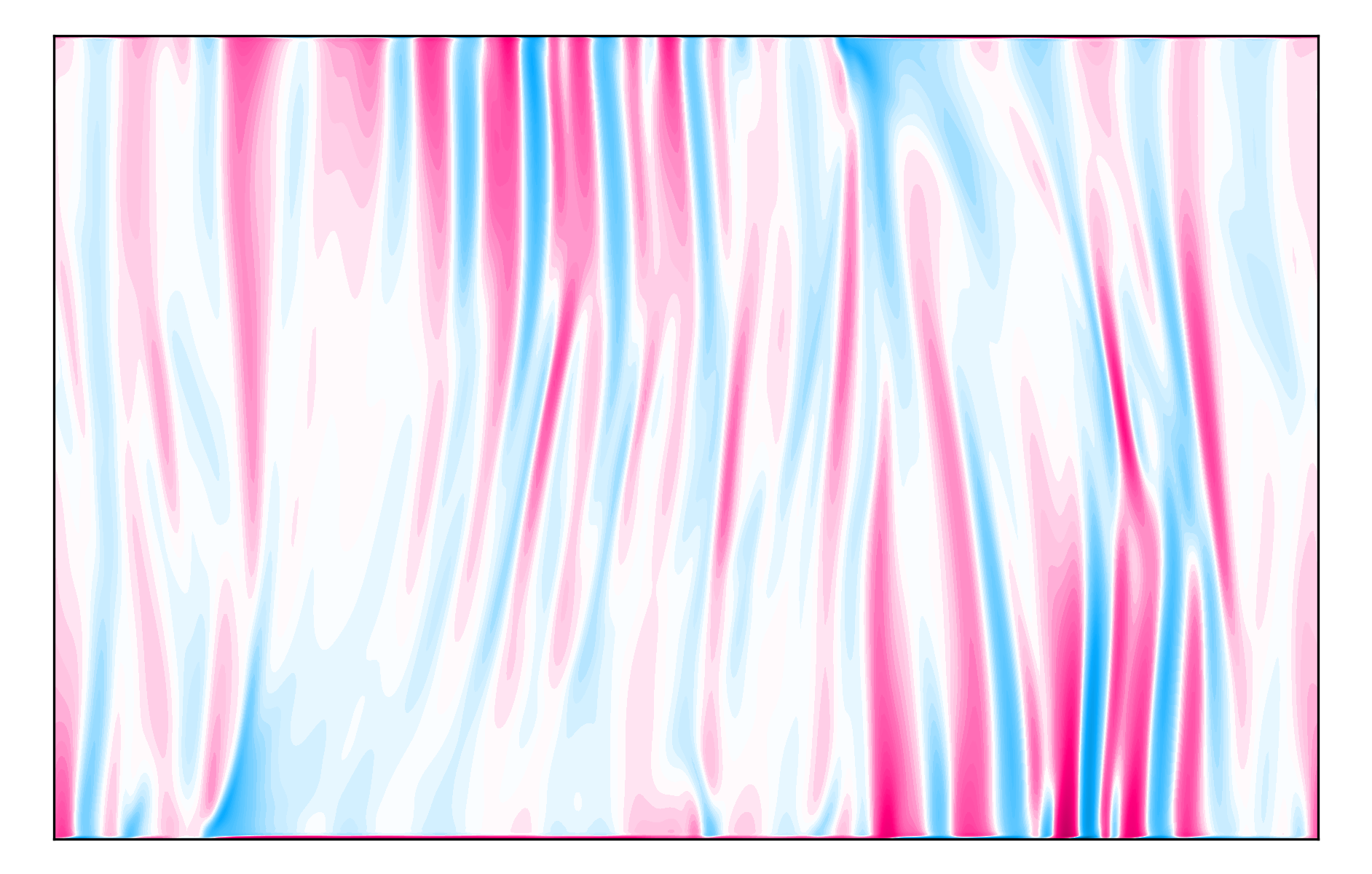}}
\put(8,0){\includegraphics[height=3cm]{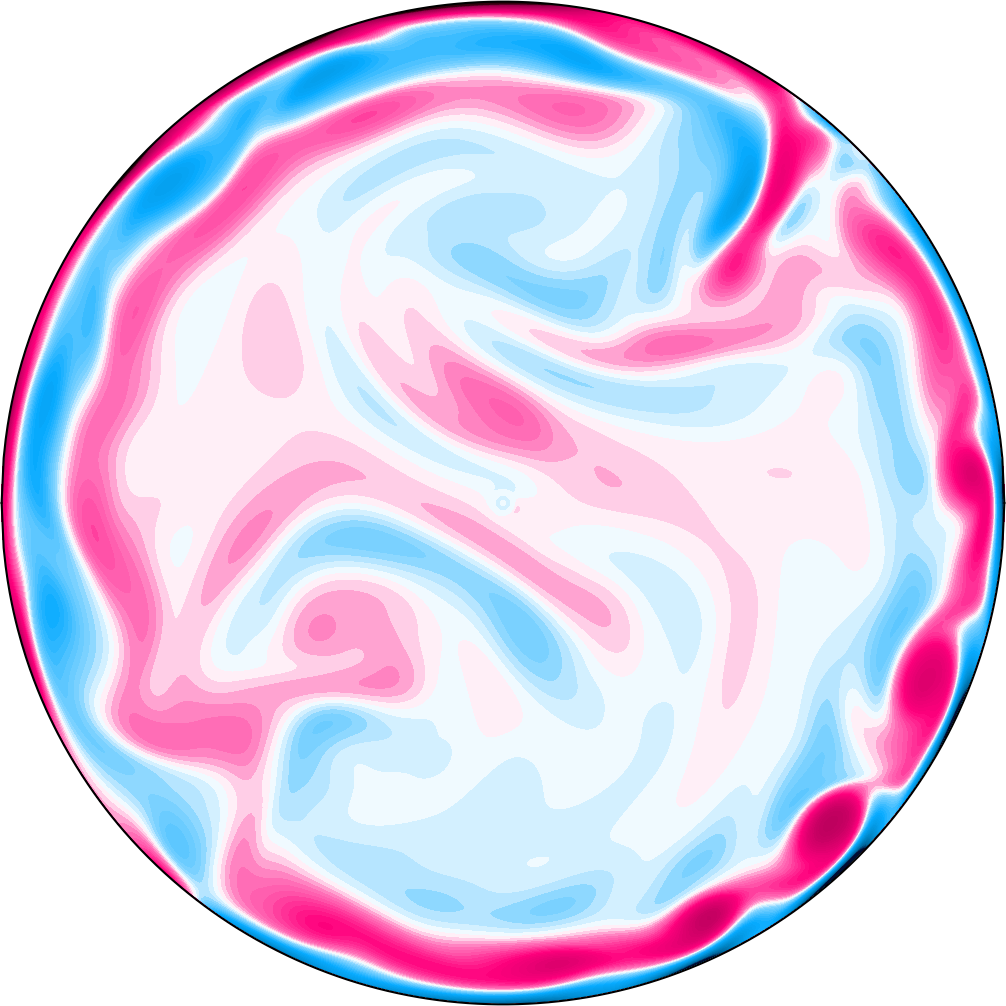}}
\put(11.2,0){\includegraphics[height=3cm]{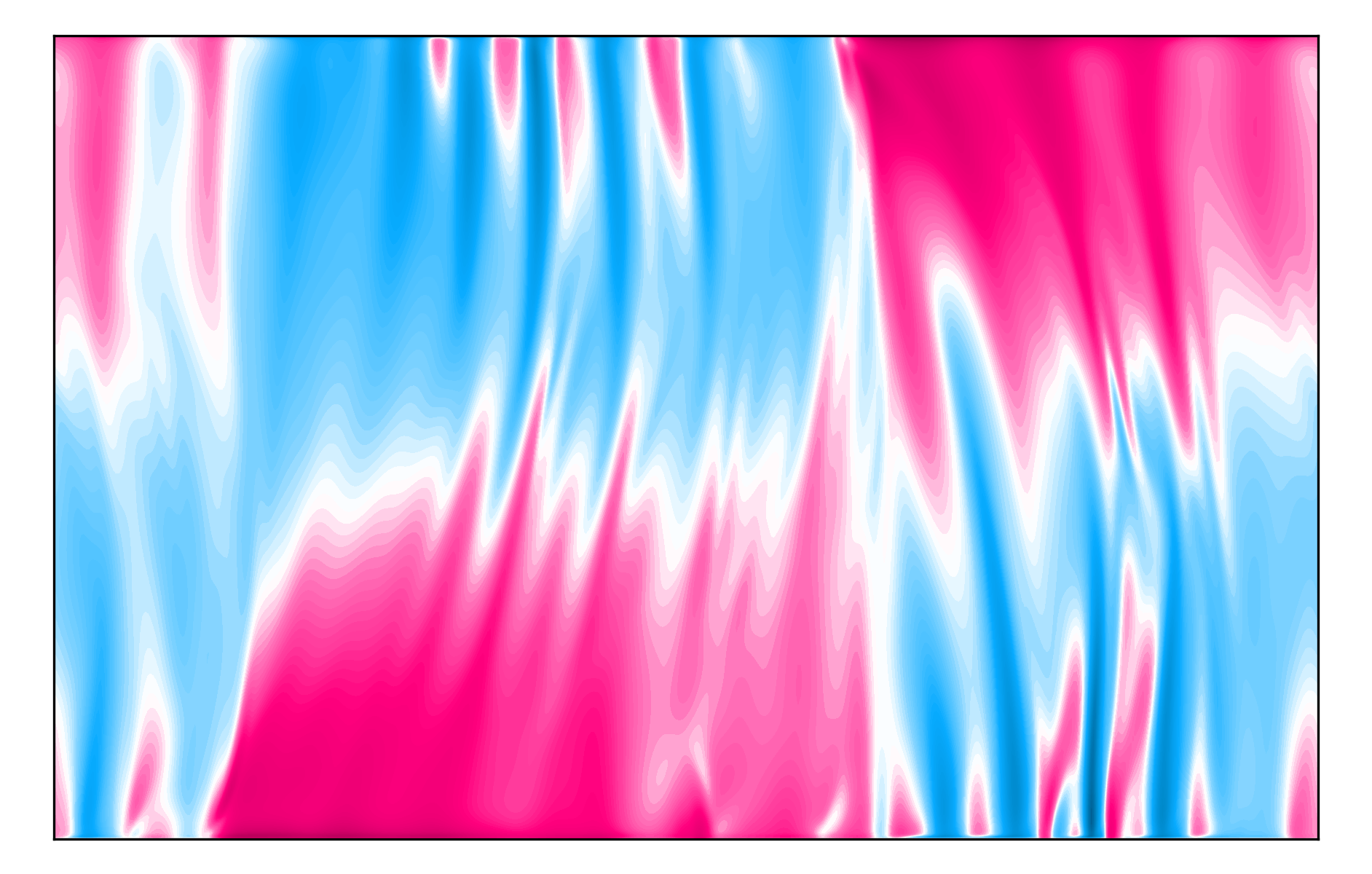}}
\put(1.25,3.2){(i)}
\put(5.3,3.2){(j)}
\put(9.25,3.2){(k)}
\put(13.3,3.2){(l)}
}
\end{picture}
\caption{
Representative images of movies presented in Supplemental Material. $\Ek=10^{-6}$, $\Pran=0.8$, and $\Gamma = 1/2$. $\Ra = 5 \times 10^8$: (a) $T(r, \phi, z=1/2)$ (b)  $T(r=0.98 R, \phi, z)$; $\Ra=2 \times 10^9$: (c)  $T(r, \phi, z=1/2)$, (d) $T(r=0.98 R, \phi, z)$; $\Ra = 10^9$:  (e) $u_z(r, \phi, z=0.8)$, (f) $u_z(r=0.98R, \phi, z)$, (g) $u_\phi(r, \phi, z=0.8)$, (h) $u_\phi(r=0.98R, \phi, z)$, (i) $u_r(r, \phi, z=0.8)$, (j) $u_r(r=0.98R, \phi, z)$, (k) $\omega_z(r, \phi, z=0.8)$, (l) $\omega_z(r=0.98R, \phi, z)$.
}
\label{AppMovies}
\end{figure*}

The dynamics and structure of the wall modes and the bulk/BZF states are hard to fully characterize through static single frame images.  In the Supplementary Material, we provide representative movies that illustrate a variety of states.  Figure \ref{AppMovies} shows a representative frame for each movie with details for each in the caption.

 \clearpage


%

\end{document}